\documentclass[a4paper,12pt]{book}
\usepackage[english]{babel}
\usepackage{epsfig}
\newcommand{\be}{\begin{equation}}
\newcommand{\ee}{\end{equation}}
\newcommand{\beq}{\begin{eqnarray}}
\newcommand{\eeq}{\end{eqnarray}}

\newcommand{\ba}{\begin{array}}
\newcommand{\ea}{\end{array}}
\newcommand{\mc}{\mathcal}
\newcommand{\ra}{\rightarrow}
\newcommand{\lra}{\leftrightarrow}
\def\a{\alpha}
\def\b{\beta}
\def\d{\delta}
\def\D{\Delta}
\def\r{\rho}
\def\s{\sigma}
\def\S{\Sigma}
\def\t{\tau}
\def\m{\mu}
\def\n{\nu}
\def\c{\chi}
\def\p{\psi}
\def\g{\gamma}
\def\th{\theta}
\def\z{\zeta}
\def\vth{\vartheta}
\def\k{\kappa}
\def\w{\wedge}
\def\G{\Gamma}
\def\S{\Sigma}

\def\O{\Omega}
\def\f{\phi}
\def\l{\lambda}
\def\L{\Lambda}
\def\e{\epsilon}
\def\btd{\bigtriangledown}
\def\part{\partial}
\def\tld{\tilde}
\def\bra{\rangle}
\def\ket{\langle}

\begin{document}

\pagestyle{empty}
\begin{flushright}
ROM2F-01/35
\end{flushright}
\vskip 10pt
\begin{center}
{\large  \bf Universit\`a degli studi di Roma ``Tor Vergata''} \\
{\large  \bf Facolt\`a di Scienze Matematiche Fisiche e Naturali}
\vskip 0.8in
{\large Tesi di Laurea in Fisica}
\vskip 0.8in
{\Large \bf Stringhe, Brane e Campi Magnetici Interni}
\end{center}

\vskip 1in

\begin{flushleft}
Candidata    \hskip 7.2cm            Relatore\\
Marianna Larosa  \hskip 6cm     Prof. Augusto Sagnotti\\
Matricola n. 000688
\end{flushleft}
\vskip 0.5in
{\it [Unabridged version of the Thesis presented to the University of
Rome ``Tor Vergata'', in partial fulfillment of the requirements for the
``Laurea'' degree in Physics, May 2001.]}
\vskip 1in
\begin{center}
Anno Accademico 1999-2000
\end{center}
\vfill\eject
\begin{flushright}
{\it Ai miei genitori, Elio e Paola,\\ 
e a Marcello,\\ 
senza il sostegno e l'affetto dei quali\\ 
questo lavoro non sarebbe mai\\ 
stato portato a termine.}
\end{flushright}
\vfill\eject
\pagestyle{plain}
\pagenumbering{roman} 
\tableofcontents\vfill\eject
\pagenumbering{arabic}    
{\LARGE \bf Prefazione}\\

Questa Tesi \`e stata svolta presso il Dipartimento di Fisica 
dell'Universit\`a di Roma ``Tor Vergata'', che ringrazio per avermi offerto
un am\-biente di lavoro  amabile e molto stimolante.  
Vorrei rivolgere in questa sede 
un  ringraziamento particolare al Prof. Augusto Sagnotti e
al Dr. Gianfranco Pradisi,
per avermi introdotto al vasto mondo della Teoria delle Stringhe,
e per avermi offerto la possibilit\`a di un'istruttiva collaborazione 
che mi ha consentito, tra l'altro, di giungere a dei risultati originali. 

\vskip .7 cm
\begin{flushleft}
Roma, Maggio 2001 \hskip 6.7cm   Marianna Larosa
\end{flushleft}
\vfill\eject
\chapter{Introduzione}
La non rinormalizzabilit\`a della Relativit\`a Generale di Einstein
\cite{Feynman:1963ax, Dewitt:1967yk, 'tHooft:1974bx, Grisaru:1976ua,
Goroff:1986th, 
Marcus:1985ei, vandeVen:1992gw, Bern:2000fm}
ha fornito la motivazione pi\`u forte per riconsiderare la teoria delle 
stringhe \cite{Green:1987sp}, nata inizialmente per descrivere le risonanze 
adroniche, come un candidato per far rientrare l'interazione 
gravitazionale nello schema di una teoria quantistica unificata che comprenda 
tutte le particelle e le forze attualmente note \cite{Scherk:1974ca,
Yoneya:1974jg}. 
L'idea di base che sottende alla teoria delle stringhe \`e che le entit\`a 
elementari non sono particelle che occupano un singolo punto dello spazio, ma
oggetti unidimensionali dotati di una lunghezza caratteristica $l_s$.
La loro storia in uno spazio-tempo a D-dimensioni \`e quindi descritta da 
superfici bidimensio\-nali chiamate superfici d'universo 
(o \emph{world-sheets}).
Nella teoria delle stringhe dunque, quelle che venivano concepite come 
particelle, vengono raffigurate come onde che si propagano lungo la stringa.
All'emissione o assorbimento di una particella da parte di un'altra 
corrisponde la congiunzione o la divisione di corde. In questo contesto
quindi, le eccitazioni puntiformi della teoria dei campi vengono sostituite
da eccitazioni unidimensionali che intera\-giscono, in ``maniera geometrica'',
mediante la divisione e la fusione dei loro {\it world-sheets}.

Ai modi di vibrazione di una stringa restano associati infiniti stati di spin 
e massa via via crescenti \cite{Veneziano:1968yb, Virasoro:1970zu,
Shapiro, Virasoro}. 
Nella formulazione convenzionale,  la loro massa 
tipica \`e dell'ordine della massa di Planck ($\sim 10^{19} GeV$), ma a
distanze confrontabili con la dimensione della stringa ($r \sim l_s$), 
l'interazione diviene ``soffice'' \cite{Scherk:1975fm}.
La fisica di bassa energia pu\`o essere cos\`\i \  descritta  da una teoria di
campo effettiva, ed \`e dominata da modi a massa nulla che comprendono, 
genericamente, un tensore antisimmetrico di rango due, 
$B_{\m\n}$, un campo scalare $\f$, detto \emph{dilatone}, 
il cui valore di vuoto definisce la costante di accoppiamento
della stringa, ed infine un campo di spin due definito da un tensore simmetrico
 a traccia nulla $h_{\m\n}$, ed identificato con le fluttuazioni del tensore 
metrico \cite{Gliozzi:1977qd}.
La risoluzione del problema delle divergenze a corta distanza della gravit\`a  
quantistica e la presenza inevitabile nello spettro di particelle di spin due 
prive di massa, hanno condotto ad intravedere nella teoria delle stringhe  una
 promettente teoria quantistica finita della gravit\`a.  

La quantizzazione della teoria pu\`o essere effettuata, ad esempio, me\-diante
l'uso del formalismo funzionale \cite{Polyakov:1981re}. 
Questo consente di descrivere la propagazione di una
stringa in maniera del tutto analoga alla propagazione di una particella,
a patto di sostituire alla somma su tutti i possibili cammini (linee 
d'universo) la somma sulle superfici di Riemann che collegano gli stati
iniziali e finali, e connette effettivamente la teoria delle stringhe
 alle  teorie di campo conformemente invarianti in due dimensioni
\cite{Belavin, Friedan:1986ge}. 
Nel caso della stringa bosonica, ad esempio, che realizza 
la generalizzazione pi\`u semplice di una particella ad un oggetto con 
un'estensione spaziale, e dove i campi bidimensionali sono le coordinate 
spazio-temporali 
della stringa, la simmetria conforme pone delle restrizioni  sulla dimensione 
dello spazio-tempo, fissandola a $D=26$ \cite{Goddard:1973qh}. 
La stringa bosonica  risulta apparentemente
inadeguata alla descrizione di una teoria  delle interazioni fondamentali, a 
causa dell'assenza, nel suo spettro perturbativo, di eccitazioni 
fermioniche. Tale problema ha trovato una soluzione con l'introduzione della
 superstringa, una teoria che generalizza la stringa bosonica mediante 
l'inclusione di campi fermionici bidimensionali anticommutanti. 
In questo caso si ha 
un'invarianza superconforme sulla superficie d'universo, che fissa la 
dimensione {\it critica} dello spazio-tempo a $D=10$ \cite{Neveu:1971rx}. 
Questa simmetria superconforme
\`e un ingre\-diente essenziale dei mo\-delli fermionici, i pi\`u semplici dei
quali esibiscono una o pi\`u supersimmetrie spazio-temporali nei loro
spettri.

Le teorie di stringa supersimmetriche consistenti a dieci dimensioni sono 
cinque: la superstringa di Tipo IIA, il cui limite di bassa energia \`e la
supergravit\`a non chirale di Tipo IIA; la superstringa di Tipo IIB 
\cite{Nahm:1976dj, Green:1984sg},  il cui
limite di bassa energia \`e descritto dalla supergravit\`a chirale di
Tipo IIB; la superstringa di Tipo I \cite{Green:1983tk, Sagnotti:1987tw},
che ha come limite di bassa energia la
supergravit\`a minimale accoppiata ad una teoria di 
super-Yang-Mills con gruppo di gauge $SO(32)$; ed infine i due modelli 
eterotici, con gruppi di gauge $SO(32)$ ed $E_8 \times E_8$ 
\cite{Gross:1985fr}, che hanno come limiti di bassa energia 
la supergravit\`a mini\-male a dieci 
dimensioni accoppiata a teorie di super-Yang-Mills con gruppi di gauge 
$SO(32)$ ed $E_8 \times E_8$.

La Tipo IIA e la Tipo IIB sono teorie di (super)stringhe chiuse ed orientate.
Per le stringhe \emph{chiuse}, i campi hanno origine da modi destri e sinistri
indipendenti, e di conseguenza le superfici che entrano nella loro serie 
perturbativa sono tutte e sole le superfici di Riemann compatte ed 
orientabili. 
Ad un determinato ordine $g$ della serie perturbativa corrisponde quindi una 
sola classe di diagrammi legati al cosiddetto {\it multitoro} con $g$ manici. 
Come verr\`a evidenziato nel quarto capitolo, un ruolo di 
particolare rilievo nella descrizione di questi modelli \`e rivestito 
dal calcolo delle {\it Funzioni di Partizione},
ovvero delle ampiezze di vuoto a 
genere uno, quando la superficie spazzata dalle stringhe \`e equivalente ad un 
toro. L'espressione corrispondente  deve risultare invariante sotto l'azione 
del cosiddetto \emph{gruppo modulare} \cite{Nahm:1986gf, Witten:1985mj, 
Cardy:1986ie,Seiberg:1986by}, il gruppo discontinuo $SL(2,Z)/Z_2$.
 
Per le stringhe \emph{aperte}, la formulazione perturbativa coinvolge, in
gene\-rale, la somma su superfici di Riemann con bordi e non orientate
\cite{Alessandrini:1971cz}, 
in quanto modi destri e modi sinistri vengono mescolati. 
Inoltre la corrispondente Funzione di Partizione non gode dell'invarianza 
modulare, ma risulta da contributi di diverse superfici. 
Nel quinto capitolo verr\`a discussa detta\-gliatamente la 
soluzione di questo problema. Essa trae origine dall'osserva\-zione che 
 la stringa di Tipo I possa essere ottenuta dalla stringa 
di Tipo IIB. 
La caratteristica fondamentale della teoria di Tipo IIB \`e, non
a caso, quella di essere simmetrica sotto lo scambio dei modi destri e 
sinistri.
Questo consente di proiettare lo spettro chiuso con $\O$, l'operazione di 
parit\`a sulla superficie d'universo, ottenendo cos\`\i \ uno spettro chiuso
non orientato. Ma la Funzione di Partizione  richiede, per consistenza, 
l'introduzione di un nuovo settore, quello \emph{twistato} con $\O$, 
che corrisponde appunto alle stringhe aperte. 
Questa procedura, nota come \emph{metodo dei discendenti aperti} 
\cite{Bianchi:1990yu}  
(detti anche \emph{orientifolds} di corrispondenti stringhe chiuse),
ha consentito di costruire ampie classi di modelli di stringhe aperte non
orientate in varie dimensioni e di definire i corrispondenti sviluppi 
perturbativi.

Come emerger\`a dalla discussione nel quarto capitolo, le (super)stringhe a cui
abbiamo accennato sono immerse in uno spazio-tempo la cui dimensione critica
(10 o 26) non corrisponde direttamente al nostro  mondo quadridimensionale. 
Questo problema pu\`o essere affrontato estendendo, in 
questo contesto, l'usuale procedura di riduzione \`a la Kaluza-Klein della 
teoria dei campi \cite{Appelquist:1988fh}. 
Si immagina quindi che un adeguato numero di dimensioni 
spazio-temporali vadano a formare una variet\`a compatta di volume molto 
piccolo, risultando cos\`\i \ invisibili a bassa energia \cite{Scherk:1979ta,
Cremmer:1974mg, Candelas:1985en, Narain:1986jj, Narain:1987am}.
A seguito della compattificazione emergono una gran variet\`a di modelli adatti
alla descrizione di  vuoti perturbativi. Questo rappresenta il problema 
pi\`u serio per l'approccio perturbativo della teoria delle stringhe, in quanto
 allo stato attuale non esiste un principio primo che consenta di individuare 
un unico vuoto fisico. \`E noto che il vuoto risulta stabile se la teoria 
possiede  una o pi\`u supersimmetrie spazio-temporali, condizione soddisfatta 
se per la procedura di compattificazione vengono utilizzate particolari 
variet\`a complesse, note come spazi di Calabi-Yau \cite{Calabi, Yau:1977ms},
  che comportano anche l'assenza di una costante cosmologica in quattro 
dimensioni. 
Ogni compattificazione di questo tipo \`e per\`o caratterizzata da un numero 
variabile di parametri liberi, detti {\it moduli}, che nel limite di bassa 
energia si presentano nella lagrangiana effettiva come campi scalari con 
accoppiamenti eslusivamente derivativi, e quindi con potenziali piatti e 
valori di vuoto indeterminati. 
Non si \`e ancora in grado di comprendere come la teoria possa selezionare
 un particolare vuoto di questo tipo. D'altra parte, vuoti non supersimmetrici
tipicamente fissano alcuni moduli, ma danno luogo a spazi-tempo non piatti.

L'idea di una sostanziale unicit\`a della teoria ha fornito, negli ultimi anni,
 un impulso decisivo per una sua pi\`u  completa comprensione. 
In particolare, la scoperta e la formulazione di equivalenze tra due o pi\`u 
teorie di stringhe, note come \emph{dualit\`a} \cite{Schwarz:1996du}, 
ha portato in via definitiva 
all'idea che i cinque mo\-delli supersimmetrici siano in realt\`a vuoti diversi 
di un'unica e pi\`u fondamentale teoria \cite{Sen:1997yy}. 
Questa, usualmente detta 
\emph{ M-teoria} \cite{Witten:1995ex, Hull:1995mz},  
\`e una teoria quantistica in undici dimensioni, 
della quale  attualmente si conosce solo la descrizione effettiva di bassa 
energia, la supergravit\`a in undici dimensioni di Cremmer, Julia e Scherk
\cite{Cremmer:1978km}.
In concreto, compattificando la M-teoria su un cerchio di raggio $R$, si pu\`o
ottenere la stringa di Tipo IIA, mentre compattificandola su un intervallo 
$I$, si pu\`o ottenere la stringa eterotica $E_8 \times E_8$ 
\cite{Horava:1996ma}. 
Queste ultime sono legate alle rimanenti tre teorie supersimmetriche da 
importanti relazioni di dualit\`a, e quindi il numero di modelli indipendenti 
a dieci dimensioni viene ad essere ridotto. 
Ad esempio, si pu\`o verificare che la
teoria di Tipo IIB compattificata su un cerchio di raggio $R$, nel limite
$R \ra 0$, diviene la teoria di Tipo IIA e viceversa. In queste condizioni
si dice che le due teorie sono T-duali. Analogamente, sono T-duali 
la stringa eterotica $E_8 \times E_8$ e la stringa eterotica $SO(32)$
\cite{Giveon:1994fu}.
Quindi la T-dualit\`a collega teorie 
compattificate su spazi di piccolo volume con altre compattificate su spazi
di grande volume. Un'altra forma di dualit\`a \`e la S-dualit\`a, che 
collega una teoria ad accoppiamento forte con una ad accoppiamento debole.
Essa agisce sulla costante d'accoppiamento della stringa, invertendola. Ad
esempio, la superstringa di Tipo I e l'eterotica $SO(32)$ sono S-duali, 
mentre la Tipo IIB \`e S-autoduale \cite{Sen:1994fa, Font:1990gx}.
L'ultima relazione che chiude le dualit\`a a dieci e undici dimensioni \`e la
relazione di discendente aperto (o di orientifold) menzionata in precedenza.

Il ruolo delle dualit\`a in teorie di stringhe \`e duplice.
Da un lato esse hanno messo in luce il legame tra nozioni geometriche ed 
algebriche, ed in particolare la T-dualit\`a ha 
evidenziato il modo inusuale in cui le stringhe possano percepire la 
geometria spazio-temporale rivelando, in alcuni casi, 
l'equivalenza tra le propagazioni su diversi spazi di piccoli e grandi volumi.
D'altra parte le dualit\`a, in generale, connettono gli stati perturbativi di
un certo modello a stati non perturbativi di un altro.
Nello studio e la comprensione delle propriet\`a non perturbative delle teorie
di stringa si \`e rivelata fondamentale l'identificazione di opportuni 
solitoni, le \emph{p-brane}, stati non perturbativi presenti nello spettro e
correttamente descritti dalla supergravit\`a di bassa energia, che appaiono 
come superfici estese in p-dimensioni \cite{Duff:1995an}.
Quando le p-brane sono cariche rispetto ai campi del settore di Ramond-Ramond 
della teoria di stringhe, esse sono dette \emph{D-brane} 
\cite{Polchinski:1995mt, Dai:1989ua}. 
Pi\`u precisamente, le D-brane possono essere pensate come 
superfici infinitamente estese sulle quali le stringhe aperte possono 
terminare, e nelle varie teorie di stringhe esse si presentano con diverse 
dimensioni spaziali. La T-dualit\`a scambia condizioni al bordo di Neumann
con condizioni al bordo di Dirichlet, e quindi connette D-brane con diverse
dimensioni. Pertanto, \`e possibile, in generale, dare varie formulazioni
T-duali di teorie assegnate con diversi contenuti di D-brane 
\cite{Polchinski:1996fm}.

Abbiamo accennato in precedenza al ruolo  giocato dalla
supersimmetria per assicurare la stabilit\`a di un vuoto perturbativo.
Essa inoltre risulta fondamentale per la verifica quantitativa delle dualit\`a.
Ma per ottenere dalla teoria delle stringhe una descrizione fisicamente 
consistente e completa della natura, \`e necessario considerare il problema 
della rottura della supersimmetria. Questo argomento viene affrontato in parte
nel quinto ed in parte nel sesto capitolo di questa tesi.

Un meccanismo di rottura spontanea tanto della supersimmetria quanto delle
simmetrie di gauge \`e stato ottenuto estendendo alla teoria delle stringhe
la procedura introdotta da Scherk e Schwarz in teoria dei campi. Tale 
meccanismo \`e essenzialmente una generalizzazione della riduzione dimensionale
 di Kaluza-Klein, dove la rottura della supersimmetria \`e indotta dai
diversi comportamenti dei modi fermionici e bosonici nello spazio interno.
Esso quindi pu\`o essere interpretato in teoria di stringhe come una 
deformazione del reticolo di un toro sul quale la teoria viene compattificata.
 Quando questo meccanismo viene applicato a modelli di stringa chiusa, la scala
di rottura della supersimmetria \`e inversamente proporzionale al volume della
variet\`a compatta. Quindi per avere rottura spontanea nella regione del $TeV$
occorre considerare variet\`a interne molto grandi, con corrispondenti
eccitazioni di Kaluza-Klein molto vicine alle energie attualmente accessibili.
 Inoltre, un esame pi\`u accurato mostra che questi modelli sono tipicamente
in regione di accoppiamento forte, e quindi la descrizione perturbativa non
\`e affidabile \cite{Dine:1989vu, Antoniadis:1988jn, Antoniadis:1990ew}. 
Quando il meccanismo  viene 
applicato a modelli di stringhe aperte, nuovi scenari sono possibili, 
che suscitano interesse sia dal punto di vista teorico che fenomenologico.

Modelli di Tipo I, dove le interazioni di gauge e la materia sono loca\-lizzate 
su D-brane, mentre l'interazione gravitazionale invade l'intero spazio a 
dieci dimensioni, offrono nuove ed importanti possibilit\`a per rompere la 
supersimmetria, che possiamo essenzialmente riassumere  in tre classi:
\begin{enumerate}
\item rottura per compattificazione \cite{Antoniadis:1999ep, Antoniadis:1998ig,
Arkani-Hamed:1998rs, Dudas:2000bn, Shiu:1998pa}: 
in questo caso, accanto alla 
convenzionale rottura \`a la Scherk-Schwarz, in cui la direzione utilizzata
per separare bosoni e fermioni \`e parallela alla D-brana, \`e possibile
considerare rotture in direzioni ortogonali.
In questo caso nello spettro di bassa energia degli stati che vivono sulla
brana sopravvivono una o pi\`u supersimmetrie globali. 
Il fenomeno \`e noto come \emph{rottura di M-teoria} 
(o \emph{brane supersymmetry}), poich\'e pu\`o essere associato alla rottura 
della supersimmetria \`a la Scherk-Schwarz lungo l'undicesima direzione della 
M-teoria;
\item \emph{brane supersymmetry breaking} \cite{Antoniadis:1999xk}: 
in queste costruzioni le condizioni di consistenza, ovvero la conservazione
di cariche di simmetria associate alla stringa, richiedono che la 
supersimmetria sia rotta, alla scala di stringa, ma che
rimanga esatta, su un set di brane, al livello ad albero, nel settore chiuso 
e sulle restanti brane;
\item rottura per introduzione di campi magnetici interni: in questo caso
la rottura della supersimmetria \`e legata all'inserimento di campi ma\-gnetici
di background $H$, nel caso pi\`u semplice  su un toro interno 
$T$ di raggi $R_1, R_2$.
Questi, accoppiandosi agli estremi delle stringhe aperte, che sotto $H$
portano cariche $q_L$ e $q_R$, danno origine ad accoppiamenti
diversi per particelle di spin diverso, che acquistano cos\`\i \
masse diffe\-renti dando luogo alla rottura della supersimmetria. 
La separazione in massa degli stati di stringa pu\`o essere riassunta dalla 
formula:
$\d m^2 \sim (2 n+1) |H_i| + 2 \S_i H_i$, dove $n$ \`e l'ordine dei
livelli di Landau e $\S_i$ sono le elicit\`a interne degli stati 
proiettati. Lo spettro, generalmente, contiene dei tachioni
che derivano dagli scalari con  $\S_i=-1$ ($\S_i=1$), per campi magnetici
positivi (negativi), e quindi l'analisi del vuoto \`e piuttosto complessa.
\end{enumerate}
I primi modelli di tori magnetizzati \cite{Fradkin:1985qd} vennero studiati 
nel tentativo di 
ottenere, dalla teoria di campo effettiva della superstringa, spettri
chirali a quattro dimensioni \cite{Ferrara:1993sq, Witten:1984dg}. 
Pi\`u recentemente, l'analisi di deformazioni
magnetiche abeliane per le teorie di stringhe aperte ha rivelato come i
corrispondenti accoppiamenti magnetici possano condurre a modelli chirali
con rottura della supersimmetria \cite{Bachas:1995ik}. 
Queste costruzioni sono state effettuate assumendo che il numero 
istantonico per il campo magnetico 
interno fosse nullo. In realt\`a \`e  possibile compensare una densit\`a 
istantonica non nulla mediante l'introduzione di ulteriori brane
\cite{Angelantonj:2000hi, Angelantonj:2000rw}. Questa possibilit\`a segue
dai peculiari accoppiamenti delle brane ai campi di R-R 
\cite{Douglas:1995bn, Green:1997dd} .
Nel sesto capitolo vengono quindi analizzati gli effetti di deformazioni 
magnetiche, con numero istantonico diverso da zero, su alcuni modelli
di stringhe di tipo I a  sei e quattro dimensioni. 
Questi rivelano la presenza di
nuovi vuoti perturbativi con supersimmetria non rotta o con \emph{brane 
supersymmetry breaking}, e dove le D9-brane possono effettivamente emulare 
delle (anti)D5-brane. 

In questa tesi viene descritto come il meccanismo di rottura \`a la 
Scherk-Schwarz possa essere combinato con opportune configurazioni di campi
ma\-gnetici interni. Mediante tale procedura sono stati quindi costruiti 
nuovi mo\-delli in quattro dimensioni, le cui configurazioni di vuoto 
presentano rottura  della supersimmetria e nelle quali le 
D5-brane sono sostituite da D9-brane magnetizzate.
\chapter{La stringa bosonica}

La stringa bosonica \`e il modello pi\'u semplice di stringa relativistica.
Storicamente, il primo principio d'azione per una teoria di stringa venne 
formulato da Nambu \cite{nambu1}  e Goto \cite{Goto:1971ce} negli anni '70,
generalizzando le nozioni della teoria relativistica per una particella
puntiforme.
Sappiamo che una particella libera relativistica, durante la sua evoluzione 
in uno spazio-tempo D-dimensionale, descrive una traiettoria parametrizzata 
in termini di una sola variabile, l'ascissa curvilinea lungo la \emph{linea} 
d'universo: $x^{\mu}(\tau)$.
Analogamente una stringa, un oggetto esteso in una dimensione spaziale, 
nel suo moto
 in uno spazio-tempo a D dimensioni, descrive una \emph{superficie} 
d'universo parametrizzata da due variabili: $\t$  e  $\s$.   
$\tau$  pu\`o essere pensata come la coordinata temporale per un qualsiasi 
punto lungo la stringa nella posizione $ \sigma$, e convenzionalmente si 
sceglie $ \sigma  \in \, [0,\pi].$
Quindi, la superficie d'universo (o \emph {world-sheet}) della stringa \`e
descritta specificando la posizione  $X^{\mu}( \tau, \sigma )$, per alcuni
 valori di  $ \tau \,  e \, \sigma$.
In altri termini, considerando uno spazio-tempo Minkowskiano $ \mathcal{M}$ 
a D dimensioni, di metrica  $\eta_{\mu\nu}=diag(-1,1,\ldots,1)$, la coordinata
di stringa  $X^{\mu}(\xi)$  \`e la mappa
\begin{center}
$X^{\mu}:   \;  \mathcal{S} \longrightarrow  \:  \mathcal{M} \, , \qquad  
\m = 0,\dots,D-1 \, ,  $
\end{center}
dove $\mathcal{S}$  \`e lo spazio dei parametri bidimensionale e
$\xi = ( \tau,\, \sigma ) \in \mathcal{S}$ .

L'azione di Nambu-Goto ha la forma seguente:
\be  
S_{NG}  = - T \int dA =  - T \int d^2\xi \sqrt {-g(\xi)} \, ,
\label{ng}
\ee
ovvero \`e proporzionale all'elemento d'area invariante $dA$ della superficie
spazzata dalla stringa, proprio come l'azione che descrive il moto di una 
particella massiva \`e proporzionale alla lunghezza invariante della linea 
d'universo $ds$ da essa descritta.
Nella (\ref{ng})  $ g(\xi) = \det g_{\a\b}(\xi) $ , dove
 $ g_{\a \b} $  \`e la metrica che descrive la geometria del ``world-sheet'', 
ovvero la metrica indotta  sulla superficie d'universo della stringa
\be
g_{\a\b} =\eta_{\mu\nu} \part_{\a}X^\mu \part_{\b}X^\nu .   
\ee
Inoltre \,  $ T = \frac{1}{2\pi\a^\prime} $ \,  \`e la tensione di 
stringa, e $\a^\prime $ \`e detta \emph{pendenza di Regge} $([T]=L^{-2}=M^2)$, 
e $ d^2\xi = d\tau d\sigma $.

L'azione di Nambu-Goto pu\`o essere posta in forma  diversa, ma 
classicamente equivalente, considerando la metrica $g_{\a\b}$ come un 
moltiplicatore di Lagrange. L'azione risultante
\be
S= - T\int{d^d\xi( \frac{1}{2} \sqrt{-g}g^{\a\b}(\xi)\eta_{\mu\nu} \part_\a{X^\mu}\part_\b{X^\nu})} ,
\label{p}
\ee
\`e detta usualmente \emph{azione di Polyakov}, anche se inizialmente 
\`e stata formulata da Brink, Di Vecchia, Howe, Deser e Zumino 
\cite{Brink:1976sc,Deser:1976rb}.
L'azione (\ref{p}), pi\'u semplice, \`e pi\'u conveniente per la 
quantizzazione.
Essa ha le seguenti simmetrie:
\begin{enumerate}
\item \`e \emph{globalmente}  invariante sotto le trasformazioni di Poincar\`e
\beq
\d{X^\mu} = \Omega^{\mu\nu}X_\nu + a^\mu  \, , \\
\d{g_{\a\b}} = 0 \, ,
\eeq
dove \, $ \Omega^{\mu\nu} $ \`e una matrice antisimmetrica della 
rappresentazione D-dimensionale del gruppo pseudo ortogonale SO(1,D-1) ed 
$ a^{\mu}$  \`e un vettore costante.
Tale invarianza riflette la simmetria dello spazio di Minkowski in cui si 
propaga la stringa.

\item  \`e  \emph{localmente} invariante sotto:      
\begin{itemize}
\item diffeomorfismi (o riparametrizzazioni) del world-sheet:
\beq
\d{X^\mu} = \zeta^{\a} \part_{\a}X^{\mu} \, , \\
\d{g_{\a\b}} = \btd_{\a}\zeta_{\b} + \btd_{\b}\zeta_{\a} \, ,
\eeq
ovvero \`e indipendente dalla particolare scelta di coordinate $ \xi^{\a}$ .\\ 
Infatti $ d^2{\xi} \sqrt{-g(\xi)} $ \`e l'elemento di area invariante, mentre 
il termine cinetico \`e invariante grazie alla totale contrazione degli 
indici tensoriali. L'equazione (\ref{p}) descrive D 
campi scalari accoppiati alla gravit\`a in due dimensioni.

\item trasformazioni di Weyl, ovvero riscalamenti locali della metrica sul 
world-sheet:
\beq
\d{X^{\mu}} = 0 \, ,\\
\d{g_{\a\b}} = 2\Lambda(\xi) g_{\a\b} \, .
\eeq
\end{itemize}
\end{enumerate}
In realt\`a \`e possibile aggiungere all'azione (\ref{p}) ulteriori termini 
compatibili sia con l'invarianza di Poincar\`e D-dimensionale che con la 
rinormalizzabilit\`a, 
\beq
S_1 =  - \lambda   T \int{d^2{\xi} \sqrt{-g(\xi)}} \ , \\
S_2 =  - \int d^2 \xi B_{\m,\n}(X) \e^{\a\b} \part_\a X^\m \part_\b X^\n \ ,\\
S_3 =  -  \frac{1}{2\pi} \int{d^2{\xi}} \sqrt{-g(\xi)} R^{(2)}(\xi) \ ,
\eeq
dove $ S_1$ \`e un termine di ``costante cosmologica'' bidimensionale che, 
non essendo invariante sotto trasformazioni di Weyl, a $D=2$ d\`a origine ad 
un contributo ad albero alle equazioni di moto. Nel seguito poniamo 
$\lambda = 0$. $S_2$ \`e un termine di Wess-Zumino.
 $ S_3$, in cui  $R^{(2)}(g)$ rappresenta la curvatura scalare intrinseca del 
$world-sheet$, ha un ruolo importante nelle interazioni di stringa, 
ma non contribuisce alla dinamica locale (ovvero alle equazioni classiche 
del moto), essendo un termine topologico. Si pu\`o mostrare che l'integrando 
\`e una derivata totale, e che 
\beq
S_2 = - \frac{T}{2} \int d^2{\xi} \sqrt{-g(\xi)} R^{(2)} = \chi = 
2 - 2\emph{h} \, ,
\eeq
dove  $\chi$  \`e la \emph{caratteristica di Eulero} della superficie 
bidimensionale e $ h $  \`e il numero dei suoi manici \cite{Yau:1987tg}.

A questo punto possiamo ricavare le equazioni del moto dei campi di stringa 
$ X^{\mu} $ e della metrica $ g^{\a\b} $, variando l'azione (\ref{p}) 
rispettivamente rispetto ad $ X^{\m} $  e $ g_{\a\b} $. Quindi
\beq
\frac{\d S}{\d {X^\mu}} = 0   \iff   \part_{\a}{(\sqrt{-g} g_{\a\b} \part_{\b}{X^\mu})} = 0 \, ,
\label{eqm}
\eeq
a meno di termini di bordo che, come vedremo in seguito, determineranno 
una prima importante distinzione tra i possibili tipi di stringhe 
bosoniche (\emph{\bf{chiuse o aperte}}). Inoltre 
\beq
\frac{ \d S}{\d {g^{\a\b}}} = 0  \iff  \frac{T}{2} \sqrt{-g} (\frac{1}{2} g^{\r\s} g^{\a\b} - g^{\a\r} g^{\b\s})  \part_{\a}X^{\mu} \part_{\b}X^{\nu} = 0 ,
\label{eqc}
\eeq
dove abbiamo utilizzato le familiari relazioni
\beq
\d {\sqrt{-g}} = { \frac{1}{2} \sqrt{-g} g^{\r\s} \d {g_{\r\s}}} \ , \\
\d {g^{\a\b}} = -{ g^{\a\r} g ^{\b\s} \d {g_{\r\s}}}  \, . 
\eeq
D'altra parte, il tensore energia-impulso bidimensionale  $ T_{\a\b}$ \`e 
definito come 
\beq
T_{\a\b} = - \frac{2}{T} \frac{1}{\sqrt{-g}} \frac{ \d S}{\d {g^{\a\b}}} \, ,
\eeq
e  l'equazione di campo (\ref{eqc}) \`e quindi  equivalente al 
vincolo 
\beq
T_{\a\b} = \part_{\a}X^{\mu} \part_{\b}X_{\mu} - \frac{1}{2} g_{\a\b}g^{\r\s} \part_{\r}X^{\mu} \part_{\s}X_{\mu} = 0  \, .
\label{vin}
\eeq

\`E semplice dimostrare che $ T_{\a\b}$ gode delle seguenti propriet\`a :
\begin{enumerate}
\item \`e simmetrico : \quad  $ T_{\a\b} = T_{\b\a}  \, ;$
\item \`e conservato : \,   $ \part^{\a}T_{\a\b} = 0 $   \quad  (usando le 
equazioni del moto (\ref{eqm})) \, ;
\item  ha traccia nulla : $ g^{\a\b}T_{\a\b} = 0 $ \, (conseguenza 
dell'invarianza di Weyl).
\end{enumerate}

Una teoria di campo caratterizzata dall'avere un tensore energia impulso 
simmetrico, conservato ed a traccia nulla \`e in generale
 \emph{invariante conforme}, 
ovvero \`e invariante anche sotto trasformazioni conformi speciali.

La soluzione delle equazioni di moto dei campi di stringa (\ref{eqm}) con i 
vincoli (\ref{eqc}) \`e resa pi\`u agevole dalla scelta di un gauge opportuno.
L'invarianza sotto diffeomorfismi dell'azione, consente, almeno localmente, 
di introdurre un sistema di coordinate tali da 
rendere la metrica ``conformemente piatta'', ovvero di porla nella forma 
\beq
g_{\a\b}(\xi) = { e ^ {2\Lambda(\xi)} \eta_{\a\b}}  \, .
\label{gc}
\eeq
La (\ref{gc}) \`e nota come \emph{gauge conforme}.

Nel gauge conforme l'azione di Polyakov, i vincoli (\ref{vin}) e le equazioni 
del moto (\ref{eqm}) assumono, rispettivamente, le seguenti espressioni:
\beq
S = -{ \frac{T}{2} \int d{\t}d{\s} \, \eta^{\a\b} \eta_{\m\n} 
\part_{\a}X^{\m} \part_{\b}X^{\n}} = \frac{T}{2} \int d \t d \s \lbrack 
( \part_{\t}{X})^2 - 
( \part_{\s}{X})^2 \rbrack
\label{p2}
\eeq
dove, convenzionalmente, $\t$ e $\s$ vengono presi appartenere agli 
intervalli 
\beq
  \t  \in ( -\infty , \infty )  \qquad  \s  \in [  0 , \pi ]  \, . \nonumber
\eeq
Dalla (\ref{p2}) si deduce che nel gauge conforme l'azione di eq. (\ref{p}) 
si riduce all'azione di D campi liberi $ X^{\m}$, mentre i vincoli divengono 
\be
T_{00} = \dot{X}^2 + X^{ '2} = T_{11} = 0  \, ,
\label{vin1}
\ee
\be
T_{10} = \dot{X} X^{'} = T_{01} = 0  \, ,
\label{vin2}
\ee
dove
\beq
\dot{X} = \frac{\part{X}^{\m}}{\part{\t}} \quad,\quad    
X^{'} = \frac{\part{X}^{\m}}{\part{\s}}  \, .  \nonumber
\eeq
Infine per le equazioni del moto (\ref{eqm}) si ha 
\beq
\part_{\a} \part^{\a} X^{\m} =  \frac{\part^2{X^{\m}}}{\part{\t ^2}} - 
\frac{\part^2{X^{\m}}}{\part{\s ^2}} = 0  \, ,
\label{eqm2}
\eeq
ovvero le coordinate $ X^{\m}$  soddisfano le equazioni delle onde in 
(1+1)-dimensioni, ma a meno di termini di superficie, che nel gauge conforme 
sono 
\beq
\lbrack  \d S  \rbrack_{sup} = - T \int_{-\infty}^{+\infty} d \t [ X^{'} \d X ]_{0}^{\pi} = 0  .
\label{sup}
\eeq
Dalla eq. (\ref{sup}) segue che :
\beq
[ X^{'} \d X ]_{0}^{\pi} = 0  \, ,
\label{condbordo}
\eeq
che, come precedentemente accennato, determinano il modello di stringa a cui 
si vuole far riferimento.
Questo termine \`e assente per stringhe chiuse, per le quali $X$ \`e periodica di periodo $\pi$. Per le  stringhe bosoniche \emph{aperte} l'annullarsi del 
termine di superficie determina invece le condizioni al bordo. 

\section{ Quantizzazione della stringa bosonica}
Si definiscono {\bf chiuse} le stringhe le cui coordinate $ X^{\m}(\xi)$ 
soddisfano condizioni al contorno periodiche 
\beq
X^{\m}(\t, \s) = X^{\m}(\t, \s + \pi)\, ,
\eeq
ovvero le stringhe topologicamente equivalenti a dei cerchi.\\
Viceversa, le stringhe { \bf aperte} sono caratterizzate dall'avere degli 
estremi non coincidenti, ma tali da annullare indipendentemente i due
contributi alla (\ref{condbordo}), che ad esempio possono essere 

\emph{liberi}, e quindi soggetti a condizioni al bordo di Neumann 
\beq
X^{' \m}( \t, 0) = X^{' \m}( \t, \pi) \, ,
\eeq
{\it fissati}, e quindi soggetti a condizioni al bordo di Dirichlet
\beq
X^{ \m }( \t, 0) = a^{ \m } \ ,
X^{ \m }( \t, \pi) = b^{ \m } \ ,
\eeq
con $a^{\m}$ e $ b^{\m}$ due vettori costanti.
Mentre per le stringhe bosoniche chiuse le equazioni del moto (\ref{eqm2}) 
e la condizione di periodicit\`a di $X^{\m}(\xi)$ sono sufficienti ad 
assicurare la stazionariet\`a dell'azione (\ref{eqm2}), per le stringhe 
bosoniche aperte la (\ref{p2}) va quindi completata con la scelta di  
condizioni ai bordi.

Per le stringhe chiuse la generica soluzione dell'equazione di moto 
(\ref{eqm2}) compatibile con la condizione di periodicit\`a delle coordinate 
pu\`o essere espres\-sa come sovrapposizione di onde regressive e progressive 
(sinistre e destre)  
\beq
X^{\m}(\t,\s) = X_L^{\m}(\t + \s) + X_R^{\m}(\t - \s),
\eeq
con le seguenti espansioni in modi normali:
\beq
X_L^{\m}(\t + \s) = \frac{x^{\m}}{2} + {\a^{'}}p^{\m}(\t+\s) + i \sqrt{\frac
{{\a}^{'}}{2}} \sum_{n \neq 0} { \frac{{\bar{\a}}_n^{\m}}{n}} 
e^{-2in(\t + \s)} \, ,\\
X_R^{\m}(\t - \s) = \frac{x^{\m}}{2} + {\a^{'}}p^{\m}(\t-\s) + i \sqrt{\frac
{{\a}^{'}}{2}} \sum_{n \neq 0} { \frac{{\a}_n^{\m}}{n}} e^{-2in(\t -\s)} \, .
\eeq
Si noti che $ x^{\m}$ e  $ p^{\m} $ possono essere interpretate come la 
posizione e l'impulso del centro di massa della stringa, mentre 
$ {\bar{\a}_n^{\m}}$  e $ {\a}_n^{\m} $ sono le ampiezze dei modi di Fourier sinistri e destri. Inoltre la pendenza di Regge $\a' $ \`e legata, 
oltre che alla tensione di stringa T, alla lunghezza fondamentale di stringa
$ l_s $, dalla seguente espressione: 
\beq
l_s = \sqrt{2{\a}^{'}} \, . \nonumber
\eeq  
Per le stringhe aperte la generica soluzione dell'equazione delle onde 
(\ref{eqm2}) con condizioni al bordo di Neumann \`e invece data da  
\beq
X^{\m}(\t , \s) = x^{\m} + 2{\a}^{'} p^{\m}{\t} +i{\sqrt{2 {\a}^{'}}} 
\sum_{n \neq 0} \cos( {n \s})  \,  { \frac{{\a}_n^{\m}}{n}}  e^{-in{\t}} .
\eeq
Notiamo che una stringa bosonica aperta possiede un solo 
insieme di modi di Fourier, con frequenze dimezzate rispetto ai modi del caso 
chiuso.

Passiamo ora a considerare i vincoli (\ref{vin1}) e (\ref{vin2}) 
\beq
\dot{X}^2 + X^{' 2} = 0 ,    \qquad  \dot{X} X^{'} = 0 .  \nonumber
\eeq
Questi possono essere soddisfatti in maniera molto semplice ponendoli nella 
forma (equivalente) di  \emph{Fubini e Veneziano}  
\be
{(\dot{X} + X^{'})}^2 = 0 \, ,
\label{vin3}
\ee
\be
{(\dot{X} - X^{'})}^2 = 0 \, ,
\label{vin4}
\ee
ed introducendo i corrispondenti modi di Fourier, noti come \emph{operatori 
di Virasoro}, che per la stringa bosonica chiusa sono 
\beq
{\bar{L}}_n = \frac{T}{8} \int_0^{\pi} {( \dot{X} + X^{'} )}^2 e^{2in{\s}}  
d \s = \frac{1}{2} \sum_{k={-\infty}}^{+ \infty} {\bar{\a}}_{n-k}^{\m} 
{\bar{\a}}_{k , {\m}}  \, ,\\
L_n = \frac{T}{8} \int_0^{\pi} {( \dot{X} - X^{'} )}^2 e^{2in{\s}} d \s = 
\frac{1}{2} \sum_{k = -{\infty}}^{+{\infty}} {\a}_{n - k}^{\m} {\a}_{k ,{\m}} \, ,
\eeq
con :  \,  $  {\a}_0 = \sqrt{\frac{{\a}^{'}}{2}} p^{\m} = {\bar{\a}}_0.  $
La stringa bosonica aperta ha invece un solo tipo di operatori di Virasoro 
\beq
L_n = \frac{T}{4} \int_{ -\pi}^{\pi} {( \dot{X} + X^{'})}^2 e^{in{\s}} d \s = 
\frac{1}{2} \sum_{k = -{\infty}}^{+{\infty}} {\a}_{n - k}^{\m} {\a}_{k , {\m}}
 \ ,
\eeq
con \  $ {\a}_0 = \sqrt{2{ {\a}^{'}}} p^{\m} $, 
 dove in questo caso la variabile $ \s $ \`e stata formalmente estesa 
all'intervallo \quad $ [ - \pi , \pi ]. $
Gli operatori di Virasoro sono i coefficienti di Fourier del tensore 
energia-impulso, e quindi i vincoli (1.35) possono essere riscritti come una 
classe di infinite equazioni, note come \emph{vincoli di Virasoro} 
\be
{\bar{L}}_n = 0 \, ,
\label{vir1}
\ee
\be  
L_n = 0   \qquad  \forall n \in Z \, .
\label{vir2}
\ee
Per la stringa aperta la prima classe di equazioni \`e assente.

L'aspetto interessante da sottolineare delle eq. (\ref{vir1}) e (\ref{vir2}),
 \`e che, quando questi vincoli sono imposti sulla Hamiltoniana $ H $ del 
sistema, essi danno origine, in particolare,  alla condizione di
\emph{mass-shell} per la teoria di stringa in considerazione.

Mostriamo anzitutto come la Hamiltoniana della teoria bidimensionale sia 
legata ai modi-zero degli operatori di Virasoro $ L_n $.
Il momento canonico coniugato 
\beq
P^{\m} = \frac{\d S}{{\d {\dot{X}}^{\m}}} = T {\dot{X^{\m}}} \, ,
\eeq
consente di scrivere 
\beq
 H = \int_0^{\pi} d \s { ( \dot{X} P - \mathcal{L})} = \frac{T}{2} \int_0^{\pi} d \s {({\dot{X}}^2 + X^{' 2})} ,
\eeq
e quindi per la stringa bosonica chiusa si ha 
\beq
  H = 2({\bar{L}}_0 + L_0 ) \, , 
\eeq
 mentre per quella aperta 
\beq
  H = L_0 \ .
\eeq 
Dunque i vincoli di Virasoro per n = 0 
\beq
L_0 = \frac{1}{2} \sum_{k = -{\infty}}^{+\infty} {\a}_{-k}^{\m} \a_{k,{\m}} = 
0  \, ,\\
{\bar{L}}_0 = \frac{1}{2} \sum_{k = -{\infty}}^{+\infty} {\bar{\a}}_{-k}^{\m} 
{\bar{\a}}_k^{\m} = 0 \ ,
\eeq
implicano che 
\beq 
H = 0 \, ,
\eeq
sia nel caso chiuso che in quello aperto.
Ricordando allora che 
$$ p^{\m} p_{\m} = - m^2 = p^2  , $$
si ottengono le seguenti condizioni di \emph{mass-shell} 
\beq
  m^2 = \frac{2}{\a ^{'}} \sum_{k = 1}^{\infty} {({\a_k^{\m}} {\a_k^{\m}} + 
{\bar{\a}}_k^{\m} {\bar{\a}}_{k,{\m}})} \, ,
\eeq
per la stringa chiusa, e 
\beq
  m^2 = \frac{1}{\a^{'}} \sum_{k = 1}^{\infty} {\a}_k^{\m} {\a}_{k,{\m}} \ ,
\eeq
per la stringa aperta. Queste sono essenzialmente corrette anche dopo la 
quantizzazione, ma con una importante sottigliezza: un contributo additivo,
di segno negativo, ad $m^2$.

La quantizzazione della teoria pu\`o essere effettuata seguendo metodologie 
diverse, ma tra loro equivalenti. Due primi esempi sono: 
\begin{enumerate}
\item  \emph{\bf la quantizzazione canonica covariante (OCQ)}, che ha il 
vantaggio di rendere manifesta l'invarianza di Lorentz, ma presenta delle 
sottigliezze legate alla presenza di stati a norma negativa;
\item \emph{\bf la quantizzazione nel gauge del cono di luce}, in cui tutti 
gli stati sono fisici ma l'invarianza di Lorentz non \`e manifesta.
\end{enumerate}
Nell'ambito della OCQ, le variabili classiche quali le coordinate di stringa 
$ X^{\m}(\xi) $, vanno interpretate come operatori hermitiani agenti sullo 
spazio di Hilbert della teoria in considerazione. Tali operatori obbediscono 
quindi alle relazioni di commutazione canonica 
\beq
[X^{\m}(\t ,\s) , P^{\n}(\t ,\s  ')] = i {\eta}^{\m\n} \d {(\s - \s  ')} \, ,
\eeq
e di conseguenza i loro modi di Fourier soddisfano 
\be
[ x^{\m} , p^{\n} ] = i {\eta}^{\m\n} \, ,
\label{com1}
\ee
\beq
[ \a_{n}^{\m} , \a_{m}^{\n} ] = [ {\bar{\a}}_n^{\m} , {\bar{\a}}_m^{\n} ] = 
n {\eta}^{\m\n} \d (\s - \s') \, ,
\label{com2}
\eeq
\be
[ \a_n^{\m} , {\bar{\a}}_m^{\n} ] = 0 \, ,
\label{com3}
\ee
per la stringa chiusa. Per la stringa aperta valgono le stesse relazioni, 
salvo l'assenza degli oscillatori $ \tilde{\a} $.
Dalla hermiticit\`a di $ X^{\m} $  segue che $ {({\a}_n^{\m})}^{\dag}  = 
{\a}_{-n}^{\m} $, e gli $ {\a}_n^{\m} $ hanno quindi un'interpretazione 
naturale in termini di operatori di creazione ed annichilazione di 
oscillatori armonici, per $ n < 0 $ \, ed \,  $  n > 0 $ rispettivamente.

Lo spazio di Fock degli stati della teoria sar\`a quindi ottenuto a partire 
dallo stato fondamentale $ |0 ; p^{\m} \bra $, definito da
\beq
{\a}_n^{\m} | 0; p^{\m} \bra = {\bar{\a}}_n^{\m} | 0 ; p^{\m} \bra = 0 \quad \forall  n > 0 ,
\eeq
agendo con gli operatori di innalzamento.

La patologia di questa costruzione sta nel fatto che lo spazio di Fock
cos\`\i \, ottenuto non \`e definito positivo, in quanto le componenti 
temporali dei commutatori (\ref{com1}) e (\ref{com2}) contengono un segno 
negativo introdotto da $ {\eta}^{\m\n} $; di conseguenza uno stato della 
forma 
$$ {({\a}_n^0)}^{\dag}| 0 ; p^{\m}\bra  \, , $$
risulta avere norma negativa:  $ \ket p^{\m} ; 0 | {\a}_n^0 \  
{{\a}_m^0}^{\dag} | 0 ;  p^{\m} \bra = -1, $  se $ \ket p^{\m} ; 0 | 0 ;
  p^{\m} \bra > 0 $.

Da tutto ci\`o segue che lo spazio degli stati \emph{fisici} sar\`a un 
sottospazio dello spazio di Fock completo della teoria, ottenuto imponendo 
delle condizioni tali da eliminare gli stati di norma negativa.

Nella teoria classica si \`e visto che tali condizioni sono espresse 
dall'annullar\-si delle componenti del tensore energia-impulso $ T_{\a\b} $, 
i cui modi di Fourier forniscono i generatori di Virasoro 
\be
L_n = \frac{1}{2} \sum _{k =-{\infty}}^{\infty} {\a}_{n -k}^{\m} {\a}_{{\m},k}
\ ,
\ee
dove per ora concentriamo l'attenzione sulla teoria di stringa aperta, 
ovvero sugli operatori $ L_n $, in quanto la trattazione degli operatori 
$ {\tilde{L}}_n $ segue in maniera del tutto analoga.
\vskip .2in
Nella teoria quantistica gli $ {\a}_n^{\m} $ divengono degli operatori e 
quindi occorre risolvere le ambiguit\`a derivanti dall'ordinamento normale; 
in particolare, poich\`e $ [{\a}_{n - k}^{\m} , {\a}_{k , {\m}} ] = 0 $ per 
$ k \neq 0$, le ambiguit\`a hanno origine da $ L_0 $, e dunque 
\beq
: L_n : \,  = L_n   \quad   n \neq 0  \, ,\\
: L_0 : \,  = L_0 + a \, ,
\eeq
dove $a$ \`e una costante da determinare.
Non \`e difficile dimostrare che gli operatori $L_n$ soddisfano 
l'algebra di Virasoro 
\beq
[ L_n ,L_m ] = ( n- m ) L_{n+m} + \frac{c}{12} n ( n^2 - 1 ) {\d}_{n+m ,0} \, ,
\label{aV}
\eeq
dove la costante $ c $, detta \emph{carica centrale},  per una teoria 
di D campi bosonici liberi coincide proprio con D (gli $  \bar{L}_n $ 
soddisfano la stessa algebra (\ref{aV}) con carica centrale $ \bar{c} $) e
dove $ a = - \frac{D}{24} $ .

Si potrebbe richiedere che gli stati fisici siano tali da soddisfare 
\beq
L_n | \phi \bra = 0   \quad \forall n \, ,
\eeq
ma questa \`e una condizione troppo forte, a causa dell'estensione centrale 
dell'algebra (\ref{aV}) che impedisce che $L_n$ ed $L_{-n}$ annullino
separatamente lo stato $| \f \bra$. In realt\`a possiamo imporre che gli stati 
fisici soddisfino
\be
L_n | \phi \bra = 0 \quad n >0 ,
\label{sf0}
\ee
\be
( L_0 - a ) |\phi \bra = 0,
\label{statofis}
\ee
ovvero che solo gli operatori di abbassamento  annichilino gli stati 
fisici, i quali possono essere autostati di $L_0$. D'altra parte dallo 
studio della teoria classica \`e emerso che l'eq. (\ref{statofis}) determina 
la massa degli stati di stringa.
Infatti, riscrivendo $ L_0$ in maniera esplicita, per la teoria di stringa 
aperta si ha
\beq
L_0 = \frac{1}{2} {{\a}_0}^2 + N \, ,
\eeq
dove: $ {\a}_0 = \sqrt{2 {\a ' }} p^{\m}, $   e \ $ N = 
\sum_{k =1}^{\infty} {\a}_{-k}^{\m} {\a}_{k , {\m}} $ \ \`e l'operatore 
numero,
da cui si ricava
\be
m^2 = - p^{\m} p_{\m} = \frac{1}{{\a '}} ( N - a ) \, .
\ee

Per la stringa bosonica chiusa, la condizione 
\beq
( L_0 - a) | \phi \bra = ({\bar{ L}}_0 - a) | \phi \bra = 0
\eeq
con 
\beq
L_0 = \frac{{\a}_0^2}{2} + N ,  \quad  {\bar{L}}_0 = \frac{{\a}_0^2}{2} +
 \bar{N} ,\quad    {\bar{\a}}_0 = {\a}_0 = \sqrt{2}{{\a'}} p^{\m},   \quad 
 {\bar{N}} = \sum_{k = 1}^{\infty} {\bar{\a}}_k^{\m} {\bar{\a}}_{k , {\m}} \ ,
\nonumber
\eeq
conduce a 
\beq
m^2 = - p^{\m} p_{\m} = \frac{2}{\a'} ( N + \bar{N} - 2a ).
\eeq
Inoltre, dal fatto che $  {\bar{\a}}_0  = \a_0 $ segue che  
$$ N = \bar{N} \, , $$ nota come condizione  di \emph{raccordo dei livelli}, 
 un vincolo che lega le eccitazioni dei modi destri e sinistri della stringa 
chiusa.

Gli stati della forma
$$ |\chi \bra = \sum_{n = 1}^{\infty} L_n |{\chi}_n \bra  \quad  
( L_n^{\dag} ) = L_{-n} \, , $$
risultano ortogonali a tutti gli stati fisici, e vengono detti stati 
\emph{spuri}, mentre uno stato fisico che sia anche spurio \`e detto 
\emph{nullo}. 
Se  $|\phi \bra $  \`e fisico e $ |\phi_0 \bra $ nullo, allora $ |\phi \bra + 
|\phi_0 \bra $ \`e fisico e quindi i due stati $ |\phi \bra  $ e 
$ |\phi \bra + |\phi_0 \bra $ sono indistinguibili. Ne segue che lo spazio 
di Hilbert degli stati fisici \`e l'insieme delle classi di equivalenza: 
\ $ |\phi \bra \sim  |\phi \bra + |\phi_0 \bra $ .

\`E istruttivo a questo punto esaminare alcuni livelli per la teoria di 
stringa aperta. I primi due operatori di Virasoro sono 
\beq
L_0 & = & \a' p^2 + \a_{-1}^{\m} \a_{1 , {\m}} + \dots \\
L_1 & = & \sqrt{2 \a'} p^{\m} \a_{1 ,{\m}} + \dots
\eeq

Per  $ N = 0 $, dalla (\ref{statofis}) si ottiene: 
\ $ m^2 = - \frac{a}{\a'} $,  
ovvero esiste una sola classe di equivalenza di stati corrispondente ad una 
particella scalare, che come vedremo \`e un  tachione.

Per  $ N = 1 $, esistono D stati, $ | \e ; p\bra = 
\e_{\m} \a_{-1}^{\m} | 0 ; p\bra $, \, dove \,$\e $ \, \`e il vettore di 
polarizzazione, e dalla (\ref{statofis}) segue che per tali stati 
 $ m^2 = \frac{1 - a}{\a'} $. Dalla condizione di stato fisico (\ref{sf0}) si 
ricava quindi che 
$$ L_1 | \epsilon ; p \bra = \sqrt{2 \a'} p^{\n} \a_{1,{\n}} \epsilon_{\m} 
\a_{-1}^{m} | 0 ; p \bra = 0 \, , $$
ovvero \, $ p \cdot \epsilon $ = 0 .
D'altra parte \`e facile verificare l'esistenza di uno stato spurio 
$$ L_{-1} | 0 ; p \bra = \sqrt{2 \a'} p^{\m} \a_{-1, \m} | 0; p \bra \, , $$
che implica :\, $ \epsilon^{\m} = \sqrt{2 \a'} p^{\m} $.

Osserviamo che per  $ a = 1 , \ m^2 = 0;$ \ $ p \cdot p = 0 $ 
per lo stato spurio, che quindi risulta anche fisico.

Allora la classe di equivalenza $ \epsilon^{\m} \sim \epsilon^{\m} + p^{\m} $ 
con \, $ \epsilon \cdot p = 0 $ descrive (D-2) stati trasversi di  
norma positiva di una particella vettoriale di massa nulla.

La stessa procedura pu\`o essere applicata alla stringa bosonica chiusa.
Anche in questo caso lo stato fondamentale $ N = 0$ risulta essere tachionico,
 mentre il primo livello eccitato corrisponde a $ N = \bar{N} = 1 $, ed \`e 
descritto dagli stati 
$$ \epsilon_{\m\n} \a_{-1}^{\m} \bar{\a}_{-1}^{\n} | 0 ; p \bra $$
di massa \  $ m^2 = - \frac{4 (1 - a )}{\a'} $ \, con \ $ p^{\m} 
\epsilon_{\m\n} = 0 $.

Da un'analisi dettagliata degli stati fisici si ricava che gli stati di norma 
positiva si ottengono esclusivamente per  \ $ D \leq 26 $ ed \, $ a \leq 1 $
\cite{Thorn:1980st}. Comunque  il caso notevole \`e $D=26$ , $ a=1$, 
come visto in precedenza,
ed \`e anche l'unico per il quale la quantizzazione \`e stata effettuata in
modo soddisfacente.

\emph{La quantizzazione nel gauge del cono di luce} \cite{Goddard:1973qh} 
consente di giungere ai 
medesimi risultati ottenuti con la OCQ.
Il gauge conforme (\ref{gc}) fissa tre parametri della teoria, quelli della 
metrica sul \emph{world-sheet}, ma l'invarianza sotto riparametrizzazioni 
lascia una simmetria residua che consente di fissare  un'ulteriore 
condizione di gauge.
Pur non essendo covariante, questa risulta essere molto conveniente (si 
vedr\`a infatti come i vincoli (\ref{vin3}) e (\ref{vin4}) potranno essere 
linearizzati).

Dato un vettore in $D$-dimensioni, $ X^{\m} = ( X^0 , X^1 , X^i )$  
{\small $i= 2,\dots,D-1 $}, le coordinate di cono di luce in uno spazio-tempo
$D$-dimensionale sono definite come 
\beq
X^{\pm} = \frac{1}{\sqrt{2}} ( X^0 \pm X^1 ) \, ,
\eeq
da cui segue che 
$$ X^{\m} X_{\m} = X^i X_i - 2 X^+ X_-  \, , $$
dove l'indice $i$ si riferisce alle $(D-2)$ coordinate spazio-temporali 
trasverse.
La simmetria residua cui si \`e accennato permette di imporre il gauge di 
cono di luce 
\beq  
X^+(\s,\t) = x^+ + 2 \a' p^+ \t \, ,
\eeq
che classicamente corrisponde a porre a zero i coefficienti dei modi $\a_n^+ $
 per $ n \neq 0 $, ovvero a scegliere sul \emph{world-sheet} una direzione del 
sistema di coordinate lungo la quale la stringa non oscilla.

Avendo fissato $ X^+ $ , i vincoli di Virasoro $ {(\dot{X} \pm X' )}^2 = 0 $ 
possono essere risolti esprimendo le coordinate $ X^- $ in termini di quelle 
trasverse $ X^i $, e quindi si ottiene 
\beq
{( \dot{X} + X' )}^- = \frac{1}{2 (2 \a' p^+)} {( \dot{X}^i + {X'}^i )}^2 \, ,
\eeq
da questa si deduce che gli oscillatori nella direzione ``-'' possono essere 
espressi in termini di quelli trasversi 
\beq
\a_n^- = \frac{1}{\sqrt{2 \a'} p^+} \sum _{k = 1}^{\infty} \a_{n - k}^i \a_{k , i} = \sqrt{\frac{2}{\a'}} \frac{1}{p^+} (L_n^{\perp} + a \d_{n , 0}),
\label{am}
\eeq
ed analogamente per gli $ {\bar{\a}}_n^- $.
Gli $ L_n^{\perp} \,  ({\bar{L}}_n^{\perp})$  sono gli operatori di 
Virasoro \emph{trasversi}, ovvero costruiti dalle sole $X^i$. Anche in 
questo caso, a causa delle relazioni di commutazione (\ref{com1})-(\ref{com3}),
 essi risultano essere mal 
definiti per $ n = 0 $; di qui la presenza in (\ref{am}) del termine 
proporzionale a $ \d_{n ,0} $, indice appunto della necessit\`a di ordinare 
normalmente $ L_0^{\perp} $ o, se vogliamo, della necessit\`a di una 
ridefinizione dell'energia di punto zero. 
Dunque, il calcolo esplicito di $ L_0^{\perp} $ evidenzia la presenza di una 
somma infinita, la cui regola\-rizzazione, effettuata ad esempio mediante l'uso 
della funzione $  \zeta $  di Riemann 
\beq
\zeta (s) = \sum_{k = 1}^{\infty} k^{- s} \, ,
\eeq
fornisce il valore della costante $ a $ in funzione della dimensione 
spazio-temporale $D$ \cite{Brink:1973ja}, ovvero 
$$ a = \frac{(D-2)}{24} , $$
determinata dal valore notevole della funzione $\zeta$ di Riemann,  
$ \zeta (-1) = - \frac{1}{12} $.
Dalla (\ref{am}) \`e possibile ottenere la condizione di \emph{mass-shell} 
della teoria.

Per la stringa bosonica chiusa si ha 
$$ \a_0^- = \sqrt{\frac{\a'}{2}} p^- = {\bar{\a}}_0^-   \, , $$
e quindi sfruttando la relazione  $ - m^2 = p^{\m} p_{\m} = 2 p^+ p^- - 
p^i p_i $
si ricava 
\beq
m^2 = \frac{2}{\a'} ( N + \bar{N} - 2 a ) \, ,
\eeq
con  $ N = \bar{N} $,   da cui \`e possibile leggere lo spettro.
Lo stato fondamentale $ (N = 0) $, \, $ | 0 ; p^{\m} \bra $,  \`e un tachione 
con $ m^2 = - \frac{4 a}{\a'}$, mentre gli stati corrispondenti ad $ N = 
\bar{N} = 1 $, \,$ \e_{ij} \a_{-1}^i  {\bar{\a}}_{-1}^i |0 ; p^{\m} \bra $, 
con  $ m^2 = 0 $, se $ a=1$, ovvero se $D=26$, possono essere decomposti in 
rappresentazioni irriducibili di $SO(D-2)$ 
\beq
\a_{-1}^i {\bar{\a}}_{-1}^j | 0 ; p^{\m}\bra = 
 \a_{-1}^{[i} {\bar{\a}}_{-1}^{j]} | 0 ; p^{\m} \bra + [ \a_{-1}^{(i} {\bar{\a}}_{-1}^{j)} |0 ; p^{\m} \bra - \d^{ij} \a_{-1}^i \a_{-1}^j 
|0 ; p^{\m} \bra ] & + & \nonumber \\
\d^{ij} \a_{-1}^i {\bar{\a}}_{-1}^j |0 ; p^\m \bra \nonumber
\eeq
ed associati, rispettivamente, alle componenti fisiche di un tensore 
antisimmetrico $ B_{ij} $ e di un tensore simmetrico a traccia nulla, 
$ h_{ij} $. Queste ultime possono essere interpretate come i gradi di 
libert\`a trasversi di una particella di spin due, il {\it gravitone}, il cui 
campo $ h_{\m\n} $ \`e associato alle fluttuazioni della metrica $ g_{\m\n} 
= \eta_{\m\n} + h_{\m\n} $. Si noti come la condizione $D=26$ sia 
necessaria per avere uno spettro  di stati fisici invariante di Lorentz.
Infatti, solo per stati di massa nulla le polarizzazioni sono puramente 
trasverse e descritte da $SO(D-2)$.

Infine, la traccia  di  $ g_{\m\n} $ ha un'interpretazione naturale come un 
campo scalare detto \emph{\bf dilatone}.

Per la stringa aperta \, $ \a_0^- = \sqrt{2 \a'} p^- $, e dunque la (\ref{am}) 
d\`a 
\beq
m^2 = \frac{1}{\a'} ( N - a ).
\eeq
Anche in questo caso lo stato fondamentale  \`e un tachione, ma con massa pari 
ad un quarto della massa del tachione di stringa chiusa ovvero  
$ m^2 = - \frac{a}{\a'} $, perch\'e, come vedremo $ a=1 $.

Per $ N = 1 $, infatti, gli stati hanno $ m^2 = 0 $, se $D=26$, 
e corrispondono a $ \a_{-1}^i | 0 ; p^\m \bra $. Essi 
vengono associati alle componenti trasverse di un vettore, privo di massa,  
di $SO(D-2)$. Questa \`e nuovamente l'unica possibilit\`a per avere uno 
spettro invariante di Lorentz, perch\'e non ha la componente longitudinale
necessaria per descrivere un vettore massivo. 

Osserviamo che sia nel caso chiuso che in quello aperto gli operatori numero 
N  $ (\bar{N}) $ sono costruiti a partire dai soli oscillatori trasversi 
$ \a_n^i $  $ ({\bar{\a}}_n^i) $.
A questo punto non ci resta che controllare la covarianza della teoria. 
Per farlo occorre verificare che l'algebra di Poincar\`e, generata dalle 
correnti di N\"other
\be
P_{\a}^{\m} = T \part_{\a} X^{\m},\\
\label{impulso}
\ee
\be
J_{\a}^{\m\n} = T ( X^{\m} \part_{\a} X^{\n} - X^{\n} \part_{\a} X^{\m} )
\label{mom}
\ee
sia chiusa.  $ P_{\a}^{\m} $ e $ J_{\a}^{\m\n} $ sono le correnti  
(nel gauge conforme) associate rispettivamente all'invarianza dell'azione 
(\ref{p2}) sotto  traslazioni e trasformazioni di Lorentz delle coordinate 
$ X^{\m}$.
L'impulso totale $ P^{\m}$ ed il momento angolare totale $ J^{\m\n} $ della 
stringa, si ottengono integrando le correnti (\ref{impulso}) e (\ref{mom}) 
su $\s$ a $ \t = 0 $, da cui risulta 
$$ P^{\m} = p^{\m} \, , $$
ovvero l'impulso totale coincide con il modo-zero della coordinata della 
stringa, e questo \`e vero sia per il caso chiuso che aperto.
Inoltre 
$$ J^{\m\n} = T \int_0^{\pi} d \s ( X^{\m} \frac{d X^{\m}}{d \t } - X^{\n} 
\frac{d X^{\m}}{d \t} ), $$
che, inserendo l'espansione in modi, assume la forma seguente:

\quad  $ J^{\m\n} = x^{\m} p^{\n} - x^{\n} p^{\m} + E^{\m\n} + 
{\bar{E}}^{\m\n} $ 
\quad per la stringa chiusa ,

\quad  $ J^{\m\n} = x^{\m} p^{\n} - x^{\n} p^{\m} + E^{\m\n} $ \qquad \qquad  
per la stringa  aperta ,

dove 
$$ E^{\m\n} = - i \sum_{n = 1}^{\infty} \frac{1}{n} 
( \a_{-n}^{\m} \a_{n}^{\n} - \a_{-n}^{\n} \a_{n}^{\m} ), $$

$$ {\bar{E}}^{\m\n} = - i \sum_{n = 1}^{\infty} \frac{1}{n} 
( {\bar{\a}}_{-n}^{\m} {\bar{\a}}_{n}^{\n} - 
{\bar{\a}}_{-n}^{\n} {\bar{\a}}_{n}^{\m} ). $$

Nell'ambito della OCQ, utilizzando le relazioni di commutazione 
(\ref{com1})-(\ref{com3}), non \`e difficile dimostrare che l'algebra 
di Poincar\`e 
$$ [ p^{\m} ,p^{\n} ] = 0 \, ,   $$
$$ [ J^{\m\n} , p^{\r} ] = i \eta^{\m\r} p^{\n} - i \eta^{\n\r} p^{\m}  \, , $$
$$ [ J^{\m\n} , J^{\r\s} ] = i \eta^{\m\r} J^{\n\s} + i \eta^{\n\s} J^{\m\r} 
- i \eta^{\n\s} J^{\m\s} - i \eta^{\m\s} J^{\n\r}   \, , $$
\`e soddisfatta. Inoltre, poich\`e $ [ L_n , J^{\m\n} ] = 0 $, le condizioni 
di stato fisico sono invarianti sotto trasformazioni di Lorentz e gli stati 
fisici formano multipletti di Lorentz.
Nella quantizzazione nel gauge del cono di luce occorre prestare particolare 
attenzione al commutatore $ [ J^{i-} , J^{j-} ]$, perch\'e gli $\a^-$ sono
funzioni non lineari degli $\a^i$, e dal calcolo esplicito si evince 
che la teoria \`e invariante se e solo se \ $ D = 26 $  e \ $ a = 1 $.
In questo modo si trova nuovamente la condizione precedente.
\chapter{La superstringa di RNS}
La stringa bosonica discussa nel capitolo precedente presenta due 
importanti problematiche:
\begin{enumerate}
\item l'assenza di fermioni, che quindi non consente una descrizione fisica 
diretta  della natura;
\item la presenza di tachioni, indice di una sbagliata identificazione del 
vuoto.  Un vuoto stabile (come per il modello di Higgs), avr\`a in generale
caratteristiche diverse.
\end{enumerate}
Questo conduce in modo naturale alla formulazione di una diversa teoria di 
stringa, ottenuta come generalizzazione della precedente.
Un primo tentativo in questa direzione che permette,  la soluzione 
del primo problema, si basa sull'introduzione di ``partners'' fermionici delle 
coordinate spazio-temporali $ X^{\m}$: i campi anticommutanti $ \p^{\m}(\xi)$.
La \emph{\bf superstringa} \cite{Schwarz:1982jn} 
(o {\it stringa fermionica}) \cite{Neveu:1971rx} 
\`e quindi ottenuta estendendo l'azione invariante sotto riparametrizzazioni 
(\ref{p}) mediante l'aggiunta di campi fermionici in modo consistente con la 
supersimmetria locale in due dimensioni. L'azione risultante \`e 
\cite{Brink:1976sc}
\be
S = - \frac{T}{2} \int d^2 \xi \sqrt{-g} [  g^{\a\b} \part_{\a} X^{\m} 
\part_{\b} X^{\m} + i {\bar{\p}}^{\m} \g^{\a} \btd_{\a} \p_{\m} + 
i \c^{\a} \g^{\b} \g^{\a} \p^{\m} ( \part_{\b} X_{\m} + 
\frac{i}{4} \c_{\b} \p_{\m} ) ] \, ,
\label{sf}
\ee
e descrive la supergravit\`a bidimensionale accoppiata ai campi 
$ X^{\m} $ e $\p^{\m}$. $ \c^{\a} $ \`e il campo di gauge della
supersimmetria, un gravitino di 
Majorana, e come $ \sqrt{-g} g^{\a\b} $, \`e un moltiplicatore di Lagrange 
non dinamico, mentre  le $\g^{\a} $ sono le due  matrici gamma-di Dirac 
bidimensionali.
L'introduzione dei gradi di libert\`a fermionici aumenta le simmetrie 
dell'azione. L'azione (\ref{sf}) infatti risulta invariante sotto le 
trasformazioni di supersimmetria locali 
\begin{center}
$\d g_{\a\b} = 2 i \g_{\a} \c_{\b}$ \ ,   \\
 $ \d \c_{\a} =  2 \btd_{\a} \epsilon $ \ ,  \\
$ \d{\p}^{\m} = \g^{\a} ( \part_{\a} X^{\m} - \frac{i}{2} \c_{\a} \p^{\m} ) 
\epsilon $  \, ,  \\
$ \d X^{\m} = i  \epsilon \  \p^{\m} \ , $
\end{center}
dove $ \epsilon $, il parametro della trasformazione di supersimmetria, \`e 
uno spinore di Majorana.
Il gauge conforme della stringa bosonica ha un analogo in questo caso, 
definito dalle condizioni 
\begin{center}
$ g_{\a\b} = e ^{2 \Lambda} \eta_{\a\b} \, , $

$ \c_{\a} = \g_{\a} \zeta  \, ,  $
\end{center} 
e noto come gauge \emph{superconforme}. Con questa scelta, l'azione (\ref{sf}) 
si riduce a 
\be
S = - \frac{T}{2} \int d^2 \xi ( \part_{\a} X^{\m} \part^{\a} X_{\m} + 
i {\bar{\p}}^{\m} \g^{\a} \part_{\a} \p_{\m} ) \, ,
\label{sfc}
\ee
che dal punto di vista della teoria bidimensionale \`e l'azione di $D$ campi 
scalari e di $D$ campi fermionici liberi.

Le equazioni del moto per i campi $ X^{\m}$ e $\p^{\m}$ sono 
\beq
 \frac{\part^2 X^{\m}}{\part \t^2} - \frac{\part^2 X^{\m}}{\part \s^2} & = & 0
 \ , \nonumber \\
 i \, \g^{\a} \part_{\a} \p^{\m} & = & 0 \ ,
\label{eqmf}
\eeq 
ovvero l'equazione delle onde (o di Klein-Gordon) e l'equazione di Dirac a 
massa nulla.

Le matrici $ \g^{\a} $ sono i generatori dell'algebra di Clifford 
$\{ \g^{\a} , \g^{\b} \} = 2 \eta^{\a\b} $. Scegliendo per esse la base 
puramente immaginaria
\begin{center}
$ \g^0 = \s_2 =  \left(
\ba{cc}
0 & {-i}\\
i & 0
\ea
\right) \ , 
\qquad 
\g^1 = i \s_1 =  \left(
\ba{cc} 
0 & i\\
i & 0 
\ea
\right) \, ,
$
\end{center}
con operatore di chiralit\`a
$$ 
\g^3 = \g^0 \g^1 = \s_3 =  \left(
\ba{cc}
1 & 0\\
0 & {-1}
\ea
\right) \ ,
$$
l'operatore di Dirac \`e reale, e quindi le componenti dello spinore sul 
\emph{world-sheet}, $ \p^{\m}$, possono essere scelte reali. Di 
conseguenza $ \p^{\m} $ pu\`o essere consistentemente uno spinore di 
Majorana con due componenti ma di chiralit\`a  opposta
$$
\p^{\m} = \left(
\ba{c}
\p_1^{\m}\\
\p_2^{\m}
\ea
\right),
$$
le quali descrivono due spinori di Majorana-Weyl.

Applicando il teorema di N\"other, per la superstringa \`e possibile 
costruire due  correnti conservate, il tensore energia-impulso 
$$ 
T_{\a\b} = \part_{\a} X^{\m} \part_{\b} X_{\m} + i {\bar{\p}}^{\m} 
\g_{\a} \part_{\b} \p_{\m} - \frac{1}{2} g_{\a\b} ( \part^{\s} X_{\m} 
\part_{\s} X^{\m} + i {\bar{\p}}^{\m} \g^{\s} \part_{\s} \p_{\m} ) \, ,
$$
e la supercorrente (ovvero la corrente di supersimmetria)
$$ J^{\a} = \frac{1}{2} \g^{\a} \g^{\b} \p^{\m} \part_{\b} X^{\m}  
\quad (\a = 0 , 1)  ,  $$
che possiamo riscrivere in termini delle componenti $J_{+,-}$ come 
\begin{center}
$ J_+ = ( \part_+ X_{\m}) \p_1^{\m} \quad,\quad  J_- = ( \part_- X_{\m} ) \p_2^{\m} $
\end{center}
$$  \part_{\pm} = \frac{1}{2} ( \part_{\t} \pm \part_{\s} ) \, . $$
In questo caso i vincoli sul sistema, detti di \emph{super-Virasoro}, sono 
$$ T_{\a\b} = 0 \ , \quad  J^{\a} = 0 , $$
e seguendo l'analogia con la stringa bosonica, possono essere espressi 
in una forma che generalizza quella di Fubini e Veneziano 
\be
T_{++} = \frac{1}{4} {[ \dot{X} + X' ]}^2 + \frac{i}{4} [ \p_2 {\p_2}' 
+ \p_2 \dot{\p}_2 ] = 0 \ ,
\label{vin5}
\ee
\be
T_{--} = \frac{1}{4} {[ \dot{X} - X' ]}^2 + \frac{i}{4} [ \p_1 {\p_1}' + 
\p_1 \dot{\p}_1 ] = 0 \ ,
\label{vin6}
\ee
\be
J_+ = J_- = 0 \, .
\label{vin7}
\ee
Osserviamo che, come  nel caso puramente bosonico, $ T_{\a\b} $ ha traccia 
nulla, e quindi 
$$  T_{+-} = T_{-+} = 0 . $$

\section{Superstringhe chiuse e aperte}
Come per la teoria bosonica,  la stazionariet\`a dell'azione (\ref{sfc}) 
\`e garantita dalle equazioni del moto (\ref{eqmf}) e dall'annullarsi del 
termine di bordo 
\be
[ \d S ]_{boundary} = \int _{-\infty}^{\infty} d \t [ X'_{\m} \d X^{\m} 
+ i ( \p_1 \d \p_1 - \p_2 \d \p_2 )]_0^{\pi} \, .
\ee
Quindi si definiscono \emph{chiuse} le superstringhe che soddisfano le 
seguenti condizioni di periodicit\`a o di antiperiodicit\`a 
$$ X^{\m}(\t,\s) = X^{\m}(\t, \s +\pi)  \, , $$
\be
\p_i(\pi) = + \p_i(0) \, ,
\label{cR}
\ee
\be
\p_i(\pi) = - \p_i(0) \, ,
\label{cNS}
\ee
con $ i= 1,2.$
Le (\ref{cR}) vengono dette condizioni di {\bf{Ramond}} ({\bf{R}}), e le 
(\ref{cNS}) condizioni di {\bf{Neveu-Schwarz}}, 
({\bf{NS}}) \cite{Neveu:1971rx}.
Entrambe preservano i generatori del gruppo di Lorentz, quadratici nei 
fermioni. Da queste segue che la superstringa chiusa, in corrispondenza 
delle possibili combinazioni di $ \p_1 $ e  $ \p_2 $, \`e caratterizzata da 
quattro settori distinti: {\bf{ NS-NS, NS-R, R-NS, R-R }}.

Le soluzioni dell'equazione di Dirac per le quattro condizioni al bordo, 
espresse in termini dei modi di Fourier, divengono 
$$
\emph{R-destro} : \quad   \p_1^{\m}(\t,\s) = \sqrt{2 \a'} 
\sum_{n \in Z} d_n^{\m} e^{-2in(\t-\s)} \, , $$
$$ \emph{R-sinistro} : \quad  \bar{\p}_2^{\m}(\t,\s) = \sqrt{2 \a'} 
\sum_{n \in Z} \bar{d}_n^{\m} e^{-2in(\t+\s)} \, , $$
$$ \emph{NS-destro} : \quad   \p_1^{\m}(\t,\s) = \sqrt{2 \a'} 
\sum_{r \in Z +\frac{1}{2}} b_r^{\m} e^{-2ir(\t-\s)} \, , $$
$$\emph{NS-sinistro} : \quad   \bar{\p}_2^{\m}(\t,\s) = \sqrt{2 \a'} 
\sum_{r \in Z +\frac{1}{2}} \bar{b}_r^{\m} e^{-2ir(\t+\s)} \, . $$
I generatori di super-Virasoro sono dati dai modi di $ T_{\a\b} $ e 
$ J_{\a} $. Ad esempio per il settore di NS destro si ha 
\be
L_n^{NS} = \frac{T}{2} \int_0^{\pi} d \s T_{--} e ^{2in\s} = L_n^B +
 \frac{1}{2} \sum_{r \in Z+\frac{1}{2}} (r - \frac{n}{2}) :b_{n-r}^{\m} 
 b_{r,{\m}} : \, ,
\label{gSV1}
\ee
\be
G_n = T \int _0^{\pi} d \s J_{-} e^{2ir\s} = 
\sum_{r \in Z + \frac{1}{2}} \a_{n-r}^{\m} b_{r,{\m}} \, ,
\label{gSV2}
\ee
dove $ L_n^B $ sono gli operatori di Virasoro della stringa bosonica. 
Per il settore di NS sinistro sussistono analoghe relazioni , con $ T_{--}$ 
sostituito da $ T_{++} $ e $ J_- $ sostituito da $ J_+ $, mentre per il 
settore di R-destro 
\be
L_n^R = L_n^B + \frac{1}{2} \sum_{k \in Z} ( k - \frac{1}{2}) 
: d_{n-k}^{\m} d_{k ,{\m}} : \ ,
\label{gSV3}
\ee
\be
F_n = \sum_{k \in Z} \a_{n-k}^{\m} d_{k ,{\m}} \ .
\label{gSV4}
\ee
Anche la quantizzazione della superstringa  pu\`o essere eseguita utilizzando 
le tecniche gi\`a descritte per la teoria di stringa bosonica 
\cite{Cohn:1986bn}.
La quantizzazione delle coordinate $ X^{\m} $ resta inalterata rispetto al 
caso non supersimmetrico, mentre la quantizzazione delle coordinate 
fermioniche coinvolge le relazioni canoniche di anticommutazione. Infatti 
definendo il momento coniugato a $ \p^{\m} $ come 
$$ P_F^{\m} = 2 \frac{\d S}{\d {\dot{\p}}^{\m}} = i T \p^{\m,T},$$
si ottiene :
$$ \{ \p_i^{\m}(\t,\s) , P_{Fj}^{\n}(\t,\s')\} = i \eta^{\m\n} 
\d_{ij} \d (\s - \s') . $$
Queste relazioni implicano che i modi $ b_r^{\m} $ e $ d_n^{\m} $ soddisfano 
\be
\{ b_r^{\m}, b_s^{\n} \} = \eta^{\m\n} \d_{r+s,0} \, ,
\ee
\be
\{ d_n^{\m} , d_m^{\n} \} = \eta^{\m\n} \d_{m+n,0} \, ,
\label{aC}
\ee
ed analogamente per i modi delle coordinate $ \bar{\p}_i $.
Ne segue che gli operatori $ L_n^{NS} $ e $G_n$ 
($ \bar{L}_n^{NS} ,\bar{G}_n  $) soddisfano 
\emph{l'algebra di NS} (\emph{o di super Virasoro}) 
\be
[ L_n^{NS} , L_m^{NS} ] = ( n - m ) L_{m+n}^{NS} + 
\frac{\frac{3}{2} D}{12} m ( m^2 - 1 ) \d_{m+n,0} \, ,
\ee
\be
\{ G_n , G_m \} = 2 L_{n+m}^{NS} + \frac{\frac{3}{2} D}{12} 
(4 r^2 -1 )
 \d_{m+n,0} \, ,
\ee
\be
[ L_n^{NS} , G_m ] = ( \frac{n}{2} - m ) G_{m+n} \, ,
\ee
con carica centrale $ c = \frac{3}{2} D $; 
mentre gli operatori $ L_n^R $ e $ F_n $ 
( $ \bar{L}_n^R , \bar{F}_n $ ) 
soddisfano \emph{l'algebra di Ramond} 
\be
[ L_n^R , L_m^R ] = (n - m ) L_{n+m} + 
\frac{\frac{3}{2} D}{12} m^2 \d_{m+n ,0}\, ,
\ee
\be
\{ F_n , F_m \} = 2 L_{n+m} + \frac{6D}{12} m^2 \d_{m+n,0} \, ,
\ee
\be
[ L_n^R , F_m ] = ( \frac{n}{2} - m) F_{n+m} \, ,
\ee
ma con diverse estensioni centrali.
Lo spazio di Fock della teoria complessiva si ottiene dal prodotto tensoriale 
degli spazi di Fock relativi ai quattro settori, costruiti a partire dallo 
stato di vuoto definito come 

$ \a_n^{\m} | 0 ; p^{\m} \bra = 0 = b_r^{\m} 
| 0 ; p^{\m} \bra = 0 $ \quad 
per $ n,r >0 $ \quad    nel settore di NS;

$ \a_n^{\m} | 0; p^{\m} \bra = 0 = d_n^{\m} 
| 0; p^{\m} \bra$  \quad  per 
$ n > 0 $ \quad nel settore di Ramond;

ovvero come  lo stato annullato da tutti gli operatori di abbassamento.
Anche la teoria di superstringa presenta, nei livelli eccitati, 
stati di norma negativa. Il modo pi\`u semplice di eliminarli, 
come nel caso bosonico, consiste nello sfruttare le 
simmetrie della teoria per imporre il gauge nel 
cono di luce, definito da 
$$ X^+(\t,\s) = x^+ + 2 \a' p^+ \t ,$$
$$ \p^+ = 0 .$$
Focalizzando l'attenzione sul settore di NS-destro ad esempio,  nel cono di 
luce i vincoli (\ref{vin6}) e (\ref{vin7}) divengono, rispettivamente 
\be
[ \dot{X} - X' ] = \frac{1}{2 (2 \a') p^+} [ ( \dot{X}^i - X'^i )^2 
+ 2i \p_1^i \part_- \p_1^i ] \, ,
\ee
\be
\frac{1}{2} ( \dot{X} - X' )^+ \p_1^- = \frac{1}{2} ( \dot{X} - X')^i 
\p_1^i \, ,
\ee
da cui 
\be
\a_n^- = \frac{1}{\sqrt{2 \a'} p^+} \sum_{n \in Z} : \a_{n-k}^i \a_k^i : 
+ \sum_{k \in Z +\frac{1}{2}} ( k - n) : b_{n - k}^i b_k^i : = 
\sqrt{2}{\a'} \frac{1}{\p^+} L_n^{NS \perp} \, ,
\ee
\be
b_n^- = \sqrt{\frac{2}{\a'}}  \frac{1}{\p^+} \sum_{k \in Z +
\frac{1}{2}} \a_{n-k}^i b_k^i = \sqrt{\frac{2}{\a'}} 
\frac{1}{p^+} G_n^{\perp} \, ,
\ee
dove $ L_n^{NS \perp} $ e $G_n^{\perp} $ sono gli operatori di Virasoro e di 
supersimmetria trasversi, e soddisfano l'algebra di NS in $(D-2)$ dimensioni.
La condizione di \emph{mass-shell} per gli stati fisici si determina, 
al solito, dai modi-zero 
\be
\a_0^- = \sqrt{\frac{2}{\a'}} p^- = \sqrt{\frac{2}{\a'}} \frac{1}{p^+} L_0 
= \sqrt{\frac{2}{\a'}} \frac{1}{p^+} [ \frac{\a_0^2}{2} + N^B + N^F - a ] \, ,
\ee
\be
\bar{\a}_0^- = \sqrt{\frac{2}{\a'}} p^- = \sqrt{\frac{2}{\a'}} 
\frac{1}{p^+} \bar{L}_0^{\perp} = \sqrt{\frac{2}{\a'}} \frac{1}{p^+}
[ \frac{\bar{\a}_0^2}{2} +\bar{N^B} +\bar{N^F} - a ] \, ,
\ee
dove\, $ N_B = \sum_{n = 1}^{\infty} \a_{-n}^i \a_n^i \ , \  
N_F = \sum_{r = \frac{1}{2}}^{\infty} r b_{-r}^i b_r^i \, . $

La costante $ a $ pu\`o essere determinata sia nel settore di NS 
(destro o si\-nistro) che di R (destro o sinistro), con l'ausilio della 
regolarizzazione mediante la $\zeta$ di Riemann, e non \`e difficile 
dimostrare che 

per il settore di NS : \qquad  $ a_{NS} = \frac{D-2}{16} ;$

per il settore di R :   \qquad   $ a_R = 0 ; $

o meglio, che i contributi allo shift $ a $ dell'energia di vuoto, 
in $(D-2)$ dimensioni sono
\begin{itemize}
\item $- \frac{1}{24}$ \, per ogni coordinata bosonica periodica ,
\item $+ \frac{1}{24} $ \, per ogni coordinata fermionica periodica \, 
(settore R),
\item $ - \frac{1}{48} $ \, per ogni coordinata fermionica antiperiodica \, 
(settore NS).
\end{itemize}
Ricordando che \ $ m^2 = - p^{\m} p_{\m} = 2 p^+ p^- - p^i p_i $ \ 
si ottiene 
\beq
 m^2 & = & \frac{2}{\a'} [ N_B + \bar{N}_B + N_F + \bar{N}_F - \frac{D-2}{8}] 
\quad \mbox{ nel settore di NS } ,\nonumber \\
 m^2 & = & \frac{2}{\a'} [ N_B + \bar{N}_B + N_F + \bar{N}_F ]  
\qquad \mbox{nel settore di R} , \nonumber 
\eeq
con: $ N_B +\bar{N}_B = N_F + \bar{N}_F . $

Sebbene molto efficace, il formalismo nel gauge del cono di luce offusca, 
anche in questo caso, l'invarianza di Lorentz. Richiedendo la chiusura 
dell'algebra di Lorentz, la dimensione della teoria viene fissata al valore 
critico   $ D = 10 , $ consistentemente con la presenza delle sole 
polarizzazioni trasverse per i modi con $ m^2 = 0 . $

Passiamo ora ad esaminare lo spettro della teoria. I primi livelli 
del settore di NS sono   lo stato fondamentale  \, $ | 0 ; p^{\m} \bra $ , 
con $ m^2 = -\frac{2}{\a'} $ , quindi un tachione scalare, ed  
il primo stato eccitato  
$ b_{-\frac{1}{2}}^i | 0 ; p^{\m}\bra , \, \a_{-1}^i | 0 ; p^{\m}\bra $ ,  
con $ m^2 = 0$, ovvero vettori di $SO(8)$.
Osserviamo quindi che lo spettro di NS \`e abbastanza semplice da costruire, 
in quanto i fermioni trasversi non hanno modi-zero, di conseguenza il vuoto 
di NS pu\`o essere scelto in maniera unica. Esso \`e uno scalare, uno stato di 
spin zero, e quindi un bosone,  come tutte le eccitazioni costruite con 
gli operatori di innalzamento a partire da esso.
I primi livelli del settore di R  hanno una natura ben diversa. Infatti 
lo stato fondamentale  $ d_0^i |0 ; p^{\m} \bra $  ha massa nulla ed \`e 
sedici volte degenere, in quanto $ d_0$, come si pu\`o verificare dalla eq. 
(\ref{aC}), soddisfa l'algebra di Clifford
$$ \{ d_0^i , d_0^j \} = \d_{ij} , $$
ovvero \`e una matrice $\g$-di Dirac in 8 dimensioni.
Ne consegue che il vuoto di R \`e uno spinore di $SO(8)$ e questo implica, in 
particolare, che tutti gli stati costruiti dal vuoto agendo con gli operatori 
di innalzamento $ \a_{-n}^i , d_{-n}^j $ sono anch'essi fermionici.
Possiamo  concludere che nella superstringa chiusa i settori 
di NS-NS e R-R danno origine ad ulteriori stati bosonici che vanno ad 
aggiungersi a quelli ottenuti con gli usuali operatori 
$ \a_{-n}^i \bar{\a}_{-n}^i $ , mentre 
i settori misti NS-R e R-NS  forniscono gli stati fermionici.
Il modello descritto fin qui, noto come \emph{\bf modello di RNS}
(Ramond-Neveu-Schwarz), non costituisce una teoria di campo consistente. La 
teoria, infatti, continua a contenere un tachione (quindi non risolve il 
problema 2.) ed inoltre esiste un conflitto tra spin e statistica a causa 
della presenza di operatori anticommutanti $\p^{\m} $ che 
mappano bosoni in bosoni.

Nel 1977 Gliozzi, Scherk ed Olive (GSO) \cite{Gliozzi:1977qd} , 
proposero una particolare tecnica di troncamento dello spettro, nota come 
\emph{\bf proiezione GSO}, in grado 
di risolvere le ``patologie'' della superstringa nel formalismo di RNS.

La proiezione GSO  consiste nell'imposizione di ulteriori 
condizioni sugli stati della stringa fermionica. Per far questo si introduce 
l'operatore $ ( - )^F $, dove F \`e il \emph{numero fermionico sul world-sheet}, ovvero 
\begin{center}
$  F_{NS} = \sum_{n = \frac{1}{2}}^{\infty} b_{-n}^i b_{n , i} $ \, , \, $ F_R = \sum_{n = 1}^{\infty} d_{-n}^i d_{n,i} \, , $
\end{center}
e si costruisce  \emph{l'operatore di proiezione GSO} definito come 
\be
 P_{GSO} = \frac{1 - ( - )^{F_{NS}}}{2} \, ,
\ee
per mezzo del quale tutti gli stati di NS ottenuti agendo sul vuoto con un 
numero \emph{pari} di operatori fermionici vengono proiettati via. In questo 
modo il tachione, ottenuto senza agire con operatori fermionici, viene 
eliminato dallo spettro.
Per quanto riguarda il settore di R, l'operatore di proiezione GSO \`e 
definito come 
\be
P_{GSO}^R = \frac{1 - \Gamma_{11} ( - )^{F_R}}{2} \, ,
\ee
dove $ \Gamma_{11} $ \`e l'operatore di chiralit\`a a 10 dimensioni. In 
questo caso nessuno stato viene completamente proiettato via, 
ma a  ciascuno viene 
assegnata una definita chiralit\`a. Il vuoto di R diviene quindi uno spinore 
di Weyl.
O meglio, se inizialmente si aveva  \  $  | 0 {\bra} _R =  | 0 {\bra}_+  +  
| 0 {\bra}_-  \, ,$ \,dopo l'azione di $ P_{GSO}^R $ si ottiene il solo
$ | 0 {\bra}_+$ o il solo $ | 0 {\bra}_-  \,. $
La scelta di quale spinore mantenere nello spettro del settore di R destro 
\`e arbitraria, poich\`e ci\`o che conta \`e la chiralit\`a relativa 
dei due vuoti di 
R-R:
se i due vuoti di R  sinistro e destro vengono proiettati alla stessa maniera, 
ovvero se hanno la stessa chiralit\`a, si ottiene la \emph{superstringa di 
Tipo IIB}; se invece essi hanno chiralit\`a opposta si ottiene la  
\emph{superstringa di Tipo IIA}.

Il contenuto dei corrispondenti spettri di bassa energia \`e quello delle 
supergravit\`a in 10 dimensioni. In particolare, la superstringa di Tipo IIA 
ha come il limite di bassa energia la supergravit\`a non chirale di Tipo IIA, 
i cui campi bosonici, oltre alla metrica, sono uno scalare, un vettore, una 
due-forma ed una tre-forma.
Viceversa la superstringa di Tipo IIB ha come limite di bassa energia la 
supergravit\`a chirale di tipo IIB che contiene due scalari, due due-forme 
antisimmetriche ed una quattro-forma con $field-strength$ autoduale.
Vedremo nel prossimo capitolo come arrivare a dedurre questi risultati. 

Passiamo ora a considerare le superstringhe \emph{aperte}. Esse sono definite 
dalle seguenti condizioni ai bordi
\be
 X'_{\m}(\t,0) = 0 = X'_{\m}(\t,\pi) \ ,
\ee
\be
\p_1(0) =  \p_2(0) \ ,
\ee
\be
\p_1(\pi) = \pm \p_2(\pi) \ .
\ee
Dall'analisi si deduce come in questo caso le componenti spinoriali destre e 
si\-nistre non siano pi\`u indipendenti.
In particolare vengono dette superstringhe aperte nel \emph{settore di R} 
quelle tali che 
\begin{center}
$  \p_1(0) = \p_2(0) $  \, , \,   $ \p_1(\pi) = \p_2(\pi) \, . $
\end{center}
In questo caso l'espansione in modi delle soluzioni dell'equazione di Dirac 
(\ref{eqmf}) \`e 
\beq
\p_1^{\m}(\t,\s) = \sqrt{\a'} \sum_{n \in Z} d_n^{\m} e ^{-in(\t + \s)} \ , \\
\p_2^{\m}(\t,\s) = \sqrt{\a'} \sum_{n \in Z} d_n^{\m} e^{-in(\t-\s)} \ .
\eeq
Viceversa, nel \emph{settore di NS} si ha 
\begin{center}
$ \p_1(0) = \p_2(0) $   \, , \,  $ \p_1(\pi) = - \p_2(\pi) \, , $
\end{center}
e quindi i campi fermionici hanno la seguente espansione in modi 
\beq
\p_1^{\m}(\t,\s) = \sqrt{\a'} \sum_{r \in Z + \frac{1}{2}} b_r^{\m} 
e^{-ir(\t+\s)} \, ,  \\
\p_2^{\m}(\t,\s) = \sqrt{\a'} \sum_{r \in Z +\frac{1}{2}} b_r^{\m} 
e^{-ir(\t-\s)} \, .
\eeq
Osserviamo cos\`\i \  che per la stringa fermionica aperta esistono due soli 
settori, quello di NS e quello di R. Da questo punto di vista la superstringa 
aperta pu\`o essere pensata come ``met\`a'' di quella chiusa, nel senso che 
per essa valgono  le relazioni precedentemente ricavate per la stringa 
fermionica chiusa, ma per un singolo settore (destro o sinistro).
\`E possibile esplicitare e risolvere i vincoli definendo i generatori di 
super Virasoro nel settore di R ed NS ($ L_n^R$ e $F_n$ ; $ L_n^{NS}$ e 
$ G_n $) tramite espressioni identiche alle eq. (\ref{gSV1})-(\ref{gSV4}), 
avendo solo l'accortezza di estendere l'intervallo di variabilit\`a di $ \s $ 
a $ [-\pi,\pi]. $ Questi danno luogo alle algebre di NS ed R gi\`a incontrate 
in precedenza, e anche lo studio della quantizzazione nel gauge  del cono di 
luce risulta lo stesso di quello di un singolo settore della superstringa 
chiusa. Esprimendo quindi $ X^-(\t,\s) $ e $ \p^-(\t,\s) $ in termini delle 
componenti trasverse  $ X^i $ e $ \p^i$,
per i corrispondenti oscillatori si ricava 
\be
\a_n^- = \frac{1}{\sqrt{2 \a'} p^+} L_n^{R \perp} \, ,
\label{oscR}
\ee
\be
d_n^- = \frac{1}{\sqrt{2\a'} p^+} F_n^{\perp} \, ,
\ee
nel settore di R, e 
\be
\a_n^- = \frac{1}{\sqrt{2\a'} p^+} L_n^{NS \perp} \, ,
\ee
\be
b_n^- = \frac{1}{\sqrt{2\a'} p^+} G_n^{\perp} \, ,
\label{oscNS}
\ee
nel settore di NS. Inoltre gli operatori trasversi che compaiono nelle eq. 
(\ref{oscR})-(\ref{oscNS})  soddisfano le usuali superalgebre in $(D-2)$ 
dimensioni, e la covarianza della teoria fissa la dimensione dello 
spazio-tempo a quella cri\-tica, $ D = 10 $.
Per ricavare lo spettro, al solito,  si prende in considerazione l'operatore 
di Virasoro trasverso per $ n = 0$  
\be
\a_0^- = \frac{1}{\sqrt{2 \a'} p^+} L_0^{\perp} = \sqrt{2 \a'} p^- ,
\ee
e ricordando che $ L_0^{\perp}$  \`e sensibile all'ordinamento normale, si 
ottiene la condizione di \emph{mass-shell} 
$$ m^2 = \frac{1}{\a'} [ N_B + N_F ] , $$
nel settore di R, dove \, $ N_F = \sum_{n = 1}^{\infty} n d_{-n}^i d_{n,i}$ \,.
In questo settore gli stati sono fermionici, e in particolare il vuoto \`e 
uno spinore di Majorana di $SO(8)$ con degenerazione $ 2^{\frac{D}{2}} = 16 . $

Il settore di NS da' invece origine a stati bosonici, ed infatti per esso 
si ha 
$$ m^2 = \frac{1}{\a'} [ N_B + N_F - \frac{1}{2} ] \, . $$
Per quanto riguarda i primi livelli, lo stato fondamentale $ | 0 ; p^{\m} \bra
 $ \`e  tachionico, mentre il primo stato eccitato $ \p_{-1/2}^i |0 \bra $, 
\`e un bosone vettore di $SO(8)$ con $ m^2 = 0 .$ 
Anche in questo caso il tachione pu\`o essere eliminato mediante 
la proiezione GSO, che d\`a luogo ad una teoria completamente consistente
 nota come 
\emph{Superstringa di Tipo I}. Il limite di bassa energia della teoria di 
stringhe di Tipo I \`e descritto da una supergravit\`a dieci-dimensionale con 
 supersimmetria  $ N = (1,0) $, accoppiata ad un sistema di super 
Yang-Mills, che come vedremo, \`e basato sul gruppo di gauge $SO(32)$.

Da quanto descritto fin qui si evince un'altra importantissima propriet\`a 
della proiezione GSO originale: essa fornisce, sia nel modello chiuso che in 
quello aperto, una teoria supersimmetrica anche dal punto di vista spazio-temporale, 
ovvero fornisce uno spettro in cui, ad ogni livello di massa, il numero 
di stati bosonici uguaglia quello degli stati fermionici (molteplicit\`a e 
massa uguali per stati bosonici e fermionici).
\chapter{Espansione di Polyakov e Funzioni di Partizione}
L'integrale sui cammini di Feynman fornisce un metodo molto utile ed 
in\-tuitivo di rappresentare una teoria quantistica, in cui il calcolo 
delle ampiezze \`e ottenuto sommando su tutte le possibili 
``traiettorie'' che collegano gli stati iniziali a quelli finali.
Nella teoria quantistica dei campi, infatti, l'interazione tra le particelle 
\`e descritta, perturbativamente, da una somma su tutti i diagrammi di
Feynman, 
che possono essere costruiti a partire dai vertici elementari (ottenuti dai 
termini di grado superiore a due che compaiono nell'azione), caratterizzati 
dall'avere topologie inequivalenti.

Un argomento molto simile venne suggerito da Polyakov nel 1981 per la teoria 
delle (super)stringhe \cite{Polyakov:1981re,Friedan:1982is}; 
in questo contesto la somma sui cammini viene 
sostituita da una somma sui \emph{world-sheets} che connettono delle date 
curve  (chiuse e/o  aperte) iniziali e finali.

Sia nella teoria di stringa chiusa che in quella di stringa aperta, il 
{\it world-sheet} \`e una superficie bidimensionale, che pu\`o essere 
opportunamente mappata (localmente)  nel piano complesso. 
Di conseguenza, dopo una rotazione di Wick, 
esso globalmente ha la struttura di una superficie di Riemann, e quindi nella 
serie perturbativa di Polyakov la somma sui \emph{world-sheets} si traduce 
in somma su superfici di Riemann ``conformemente inequivalenti'' 
(condizione, quest'ultima, richiesta dall'invarianza sotto riparametrizzazioni
 e sotto trasformazioni di Weyl). In altri termini, la somma sulle superfici 
diventa la regola di Feynman basilare per la teoria di stringa.

Sebbene l'espansione perturbativa di Polyakov appaia come una 
genera\-liz\-zazione
dell'integrazione sui cammini di Feynman della teoria quantistica dei campi, 
in realt\`a i due casi si differenziano per aspetti molto profondi.
La prima fondamentale distinzione sta nel fatto che la descrizione delle 
stringhe mediante integrale ``sui cammini'' conduce ad una teoria 
primo-quantizzata (nella formulazione che descriviamo in questa tesi non si 
quantizzano campi di stringa, ma il moto di singola stringa), mentre, la serie
 perturbativa della teoria dei campi \`e dedotta da 
un'azione secondo-quantizzata. Quindi, questa descrizione \`e l'equivalente
dell'integrale sui cammini per singola particella.

\begin{figure}[htbp]\unitlength1cm
\begin{picture}(13,3)
\put(0,0){\epsfig{file=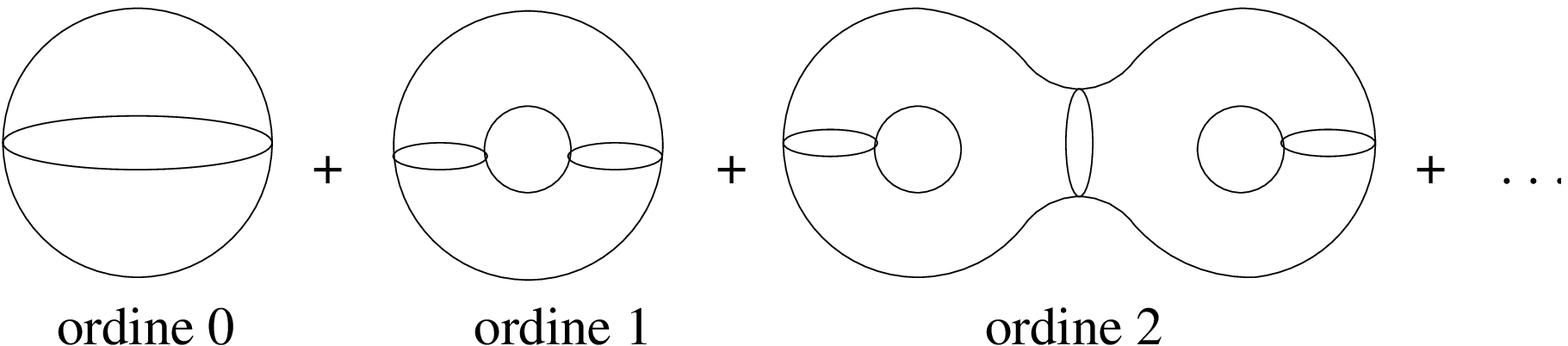,width=13cm,height=3cm}}
\end{picture}
\caption{\footnotesize{Serie perturbativa}}
\label{fig:2} 
\end{figure}

In secondo luogo, i diagrammi della teoria di stringa sono molto 
minori in numero, di  quelli che possono essere costruiti in teoria dei campi.
Per ogni diagramma della teoria di campo c'\`e un corrispondente diagramma in 
teoria di stringa, che pu\`o essere ottenuto semplicemente trasformando le 
linee di universo delle particelle in ``tubi di universo'', come  nel caso 
seguente:
\begin{figure}[htb]\unitlength1cm
\begin{picture}(9,3)
\put(2,0){\epsfig{file=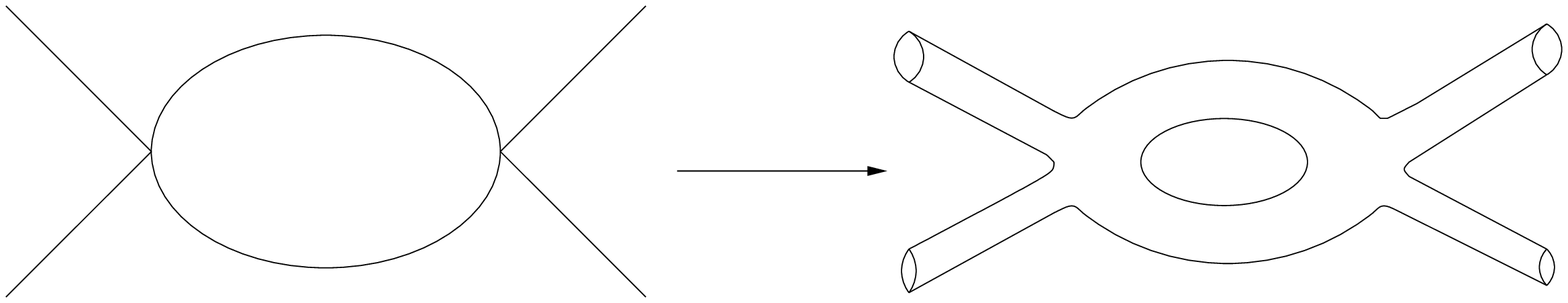,width=9cm,height=3cm}}
\end{picture}
\caption{\footnotesize{Tubi d'universo}}
\end{figure}
Differenti diagrammi di Feynman possono per\`o dar luogo, in 
questo modo, allo stesso diagramma di stringa. Ad esempio i modelli di 
stringhe chiuse o\-rientate bosoniche hanno la caratteristica di ricevere uno 
ed un solo contri\-buto (diagramma) a ciascun ordine della serie perturbativa 
(vedi fig.(\ref{fig:2})). 

Tutto questo corrisponde al fatto che la topologia delle superfici di Riemann 
chiuse ed orientate \`e completamente specificata dal loro numero di ``manici''
{\bf {h}} \cite{Alvarez:1983zi}. $h$ definisce il ``genere'' della 
corrispondente superficie di Riemann e sostituisce, in teoria di stringa, il 
numero di loops della teoria di campo. Di conseguenza esso conta l'ordine 
della serie perturbativa.

La classificazione dei diagrammi di stringa diviene  pi\`u complessa 
nelle teorie con stringhe aperte e/o di stringhe chiuse non orientate, ma 
anche in questo caso il loro numero \`e molto inferiore rispetto ai diagrammi 
di Feynman; questa situazione verr\`a approfondita  nel capitolo successivo.

\begin{figure}[htb]\unitlength1cm
\begin{picture}(10,6)
\put(2.5,0){\epsfig{file=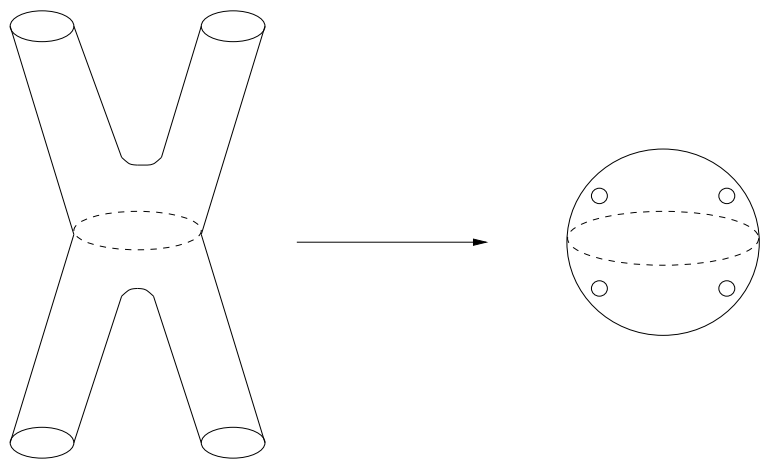,width=10cm,height=6cm}}
\end{picture}
\caption{\footnotesize{Esempio di scattering di stringhe chiuse}}
\label{fig:3}
\end{figure}

L'idea di sommare su tutte le superfici d'universo limitate da certe curve 
iniziali e finali \`e sicuramente la pi\`u naturale, ma conduce ad ampiezze 
piut\-tosto complicate da calcolare. Il calcolo delle ampiezze diviene invece 
particolarmente semplice nel limite in cui le sorgenti di stringa 
vengono poste all'infinito, ovvero nel limite che corrisponde ad ampiezze di 
scattering con specificati stati entranti ed uscenti. Questa semplificazione 
\`e dovuta all'inva\-rianza della teoria sotto riscalamento conforme della 
metrica sul \emph{world-sheet}. Infatti un'opportuna mappa conforme rende 
equivalenti world-sheets con $n$ stringhe esterne, e \emph{world-sheets} 
compatti con un uguale numero di piccoli fori o ``punture'', in 
corrispondenza dei quali andranno inseriti degli operatori locali che 
contengano i numeri quantici degli stati di stringa mappati in quel punto. 
Tali operatori, detti \emph{vertici di interazione}, descrivono appunto lo 
scattering di stringhe (vedi fig.(\ref{fig:3})).

\section{ Ampiezze di vuoto con caratteristica di Eulero nulla}
Le ampiezze di vuoto rivestono un ruolo di particolare rilievo in teoria di 
stringa. Esse descrivono ampiezze di processi in cui non viene inserito alcun 
operatore di vertice e la loro importanza risiede nella possibilit\`a  di 
descrivere in modo compatto il contenuto completo dello spettro perturbativo.
Si \`e precedentemente accennato che il genere di una superficie di Riemann 
conta l'ordine della teoria delle perturbazioni, ovvero il numero di potenze 
della costante di accoppiamento  $ g_s$ del modello in considerazione 
associate ai diagrammi. La serie perturbativa \`e pesata dal fattore  
$ g_s^{- \c}  $, dove $ {\chi} $ \`e la \emph{\bf caratteristica di Eulero} 
della superficie,  che infatti per superfici chiuse ed orientate \`e definita 
come 
$$ \c  = 2 - 2h   .$$    
$ g_s $ \`e determinata dal valore di aspettazione nel vuoto del
\emph{dilatone}:
$$  g_s = e ^{< \f >} .$$
In questo capitolo l'attenzione verr\`a focalizzata sul calcolo 
delle ampiezze di vuoto con $ \c = 0 $ , dette anche 
\emph{ \bf funzioni di partizione a 
genere uno}, in quanto gi\`a in esse risiedono le propriet\`a pi\`u 
interessanti per l'analisi dei vari modelli.

Sia nelle teorie di stringhe chiuse che aperte, il percorso pi\`u semplice da 
seguire per questo tipo di calcolo fa  riferimento alla teoria dei campi. 
A questo scopo andiamo  a considerare, per semplicit\`a, la teoria di un
campo 
scalare $ \phi $ di massa m in $D$-dimensioni spazio-temporali.
Il funzionale generatore da cui dedurre l'energia di vuoto $ \G $ \`e 
\be
Z  = \int [ D \f ] e^{-S_E}  =   
\int [ D \f ] e^{- \int d^D x \frac{1}{2} {(\part_{\m} \f  \part^{\m} \f 
+ m^2 \f^2 )}} 
\sim {[ det {( - \part_{\m}\part^{\m}  + m^2 )}]}^{-1/2} \ ,
\label{Z}
\ee
da cui si ricava 
\be
\G =  \frac{1}{2} log [ det (- \part_{\m}\part^{\m} + m^2 )] = \frac{1}{2}  
tr [ log ( - \part_{\m}\part^{\m} + m^2 ) ] \, ,
\label{env}
\ee

Facendo uso della parametrizzazione di Schwinger per un generico 
op\-eratore A 
\be
log{A} = - \int_{\e}^{\infty} \frac{dt}{t} e^{- t A} \, ,
\label{Schw}
\ee
dove t \`e il parametro di Schwinger, ed $\e$ \`e un cut-off ultravioletto 
(data la divergenza della (\ref{Schw}) in $ t = 0 $), si ottiene l'identit\`a 
\be
log ( det A ) = tr ( log A ) = - \int_{\e}^{\infty} \frac{d t}{t} tr 
( e^{-tA} ) \ ,
\ee
da utilizzare nell'espressione (\ref{env}).
Per calcolare la traccia in (\ref{env}) basta inserire un set completo di 
autostati dell'impulso p,  $| p \bra $, in modo da rendere diagonale 
l'operatore cinetico 
$ \part_{\m} \part^{\m}.$ La (\ref{env}) diviene quindi 
\be
\G_B = - \frac{V}{2} \int_{\e}^{\infty} \frac{dt}{t} e^{-tm^2} 
\int \frac{d^D p}{{(2 \pi)}^D} e^{-tp^2} \, ,
\ee
da cui, effettuando l'integrale gaussiano, viene estratta la dipendenza da 
$ m^2$
\be
\G_B = - \frac{V}{2 {(4 \pi)}^{\frac{D}{2}}} \int_{\e}^{\infty} 
\frac{dt}{t^{\frac{D}{2} + 1}} e^{-tm^2} \, .
\label{fpb}
\ee
Un procedimento del tutto analogo pu\`o essere utilizzato per un campo di 
Dirac $\p $ di massa $ m $ in D dimensioni, e porta al risultato 
\be
\G_F = V \,  \frac{2^\frac{D}{2}}{{(4 \pi)}^\frac{D}{2}}  
\int_{\e}^{\infty} \frac{dt}{t^{\frac{D}{2}+1}} e^{-tm^2} \, ,
\label{fpf}
\ee
che ha segno opposto rispetto alla (\ref{fpb}), data la natura fermionica 
dei campi nel {\it path-integral}. Osserviamo inoltre che le espressioni 
(\ref{fpb}) e (\ref{fpf}), dipendono solo da  $ m^2 $, e possono essere 
facilmente estese a qualunque campo boso\-nico e fermionico.
Quindi l'espressione pi\`u generale che le contempla entrambe \`e data da 
\be
\G = \frac{V}{ 2 {( 4 \pi)}^{\frac{D}{2}}} \int_\e^{\infty} 
\frac{dt}{t^{\frac{D}{2}+1}} Str ( e^{-t m^2} ) \, ,
\label{fp}
\ee
dove $Str$ tiene conto delle molteplicit\`a e dei rispettivi segni degli 
stati fermio\-nici e bosonici.

Applichiamo la (\ref{fp}) alla stringa bosonica chiusa.
Abbiamo visto che la dimensione critica per essa \`e $ D = 26 $ e che lo 
spettro \`e dato da 
$$
m^2 = \frac{2}{\a'} ( N + \bar{N} - 2 ) \quad    \mbox{con} \quad  N = \bar{N} \ ,
$$
\be
\G_{closed} =  \frac{V}{ 2 {( 4 \pi )}^{13}} \int_\e^{\infty} 
\frac{dt}{t^{14}} tr ( e^{- \frac{2}{\a'} ( N + \bar{N} - 2)} ) \ .
\ee
Per tener conto della condizione di raccordo dei livelli per gli stati fisici, basta introdurre ``a mano'' la seguente {\it funzione delta} 
\be
\d ( N - \bar{N} ) = \int_{- \infty}^{+\infty} ds \  
e ^{2 \pi i s ( N - \bar{N} )} \  ,
\ee
ottenendo quindi 
\be
\G_{closed} = \frac{V}{2 {( 4 \pi)}^{13}}  \int_{-\infty}^{+\infty} ds 
\int_\e^{\infty} \frac{dt}{t^{14}} tr ( e^{- \frac{2}{\a'} 
( N + \bar{N} - 2 )t}  e^{2 \pi is ( N - \bar{N})} ) \ .
\label{fpc}
\ee
La (\ref{fpc}) pu\`o essere posta in forma pi\`u elegante definendo  il 
parametro di Schwinger \emph{complesso} 
\be
\t = \t_1 + i \t_2 = s + i \frac{t}{\pi \a'} \, ,
\label{pSchw}
\ee
e ponendo 
\be
q = e^{2 \pi i \t }  \qquad , \quad \bar{q} = e^{- 2 \pi i \bar{\t}} \, ,
\ee
per mezzo delle quali si ottiene 
\be
\G = \frac{V}{2 {( 4 \pi^2 \a' )}^{13}} \int_{- \infty}^{+\infty} 
d \t_1 \int_\e^{\infty} \frac{d \t_2}{{\t_2}^{14}} tr ( q^{N - 1} 
\bar{q}^{\bar{N} - 1} )\, .
\label{fp1}
\ee
Il parametro complesso $ \t $ che compare in $ \G $ \`e strettamente 
legato al \emph{world-sheet} dell' ampiezza che si sta calcolando. 
Nel caso in esame infatti, viene identificato con il \emph{\bf modulo} o  
\emph{parametro di Teichm\"uller}  del toro, ovvero della superficie 
spazzata, ad un loop, da una stringa chiusa.

\begin{figure}[htb]\unitlength1cm
\begin{picture}(7,3)
\put(4,0){\epsfig{file=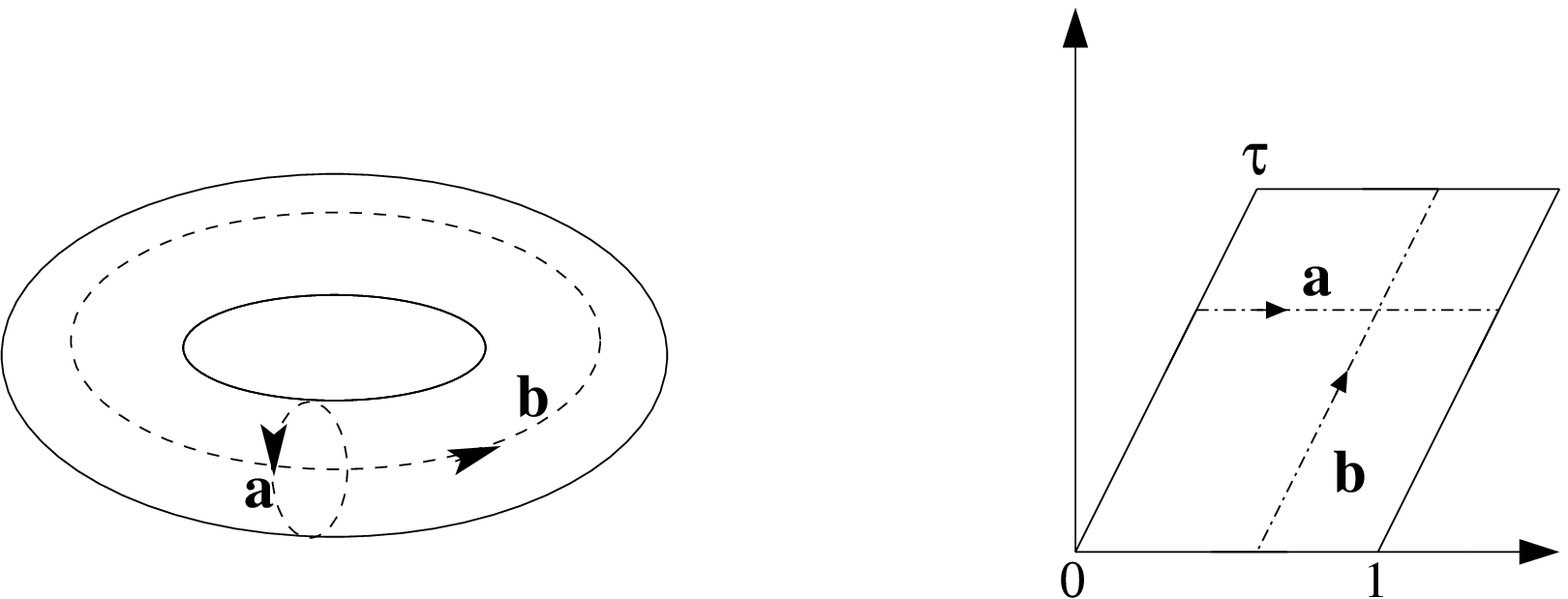,width=7cm,height=3cm}}
\end{picture}
\caption{\footnotesize{Toro}}
\label{fig:4} 
\end{figure}

Il toro \`e una superficie di Riemann chiusa ed orientata, che riportiamo nella figura 
(\ref{fig:4}) evidenziando i due cicli non contraibili $a$ e $b$.

Mediante due tagli opportuni, un toro pu\`o essere trasformato in un 
pa\-rallelogramma nel piano Euclideo con lati opposti identificati.
 In questo caso $ \t $ definisce \emph{la struttura complessa del toro}. Un 
lato pu\`o essere scelto orizzontale e di lunghezza unitaria, mentre la sua 
parte immaginaria,  $  \t_2 $, pu\`o essere scelta positiva.
Il parallelogramma descrive cos\`\i \  una cella fondamentale di un reticolo 
bidimensionale specificato da $ \t $, ma in realt\`a la scelta del parametro 
$ \t $ non \`e unica. Infatti lo stesso reticolo \`e specificato da una cella 
qualsiasi, tra le infinite di area $ \t_2 $.
Ogni cella reticolare \`e legata ad ogni altra dal gruppo degli automorfismi 
del reticolo, noto come \emph{\bf Gruppo Modulare} 
$  PSL(2,Z) = SL(2,Z )/ Z_2 $  
definito dalle trasformazioni:
$$ 
\t  \longrightarrow \frac{a\t + b}{c\t + d} \quad \mbox{con} \, ad - bc = 1 
\quad ( a,b,c,d, \in Z ).
$$
PSL(2,Z) pu\`o essere generato dalle due trasformazioni modulari 
$$ T : \, \t \longrightarrow \t + 1$$
$$ S : \, \t \longrightarrow - \frac{1}{\t} $$
che soddisfano la relazione
$$  S^2 = {(ST)}^3 . $$
Ogni valore di $\t $, ottenuto applicando le trasformazioni del gruppo 
modulare ad un valore iniziale $ \t_0 $, definisce un toro 
``conformemente equivalente'' a quello di partenza. Di qui l'esigenza di 
definire una regione del semipiano complesso entro la quale valori di  
$ \t $ diano origine a tori inequivalenti. Un esempio \`e fornito dalla 
\emph{regione fondamentale standard} $  \mathcal{F} $  definita come 
$$ \mathcal{F} = \{ \t \in C : -\frac{1}{2} \leq \t_1 \leq \frac{1}{2} \, 
\mbox{e} \,  | \t | \geq 1 \} \, .  $$

\begin{figure}[htb]\unitlength1cm
\begin{picture}(4,3.5)
\put(5,0){\epsfig{file=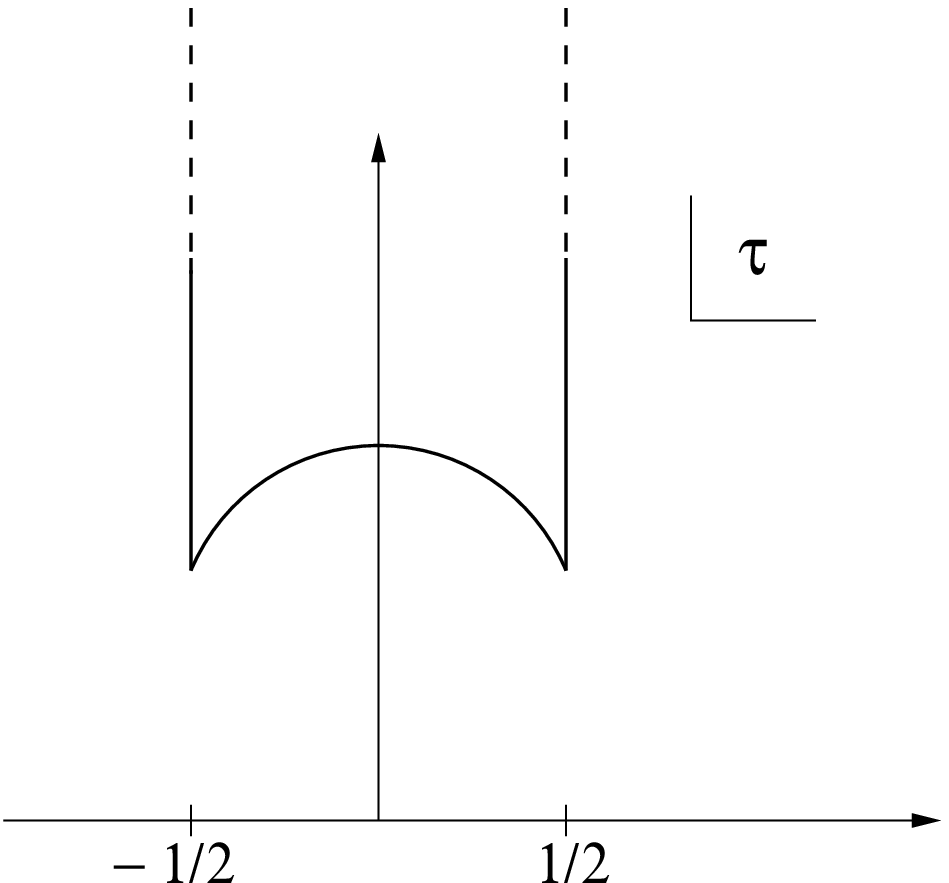,width=4cm,height=3.5cm}}
\end{picture}
\caption{\footnotesize{Regione Fondamentale}}
\label{fig:5} 
\end{figure}

La geometria del toro suggerisce quindi di restringere la regione 
d'integra\-zione della (\ref{fp}) alla regione fondamentale $ \mathcal{F},$ in modo 
da evitare il conteggio multiplo di tori equivalenti.
Effettuando un riscalamento, la funzione di partizione della stringa bosonica 
chiusa (\ref{fp1}), detta anche \emph{\bf ampiezza di toro}, assume la 
seguente forma finale:
\be
\mathcal{T} = \int_{\mathcal{F}} \frac{d^2 \t}{{\t_2}^{14}} tr ( q^{N - 1} 
\bar{q}^{\bar{N} - 1} ) \, ,
\label{T}
\ee
dove \  $ d^2 \t = d\t_1 d\t_2  .$ 

La (\ref{T}) \`e in realt\`a  un'espressione del tutto generale, che pu\`o 
essere applicata a tutti i modelli di stringhe chiuse orientate distinte dai 
soli operatori $ N $ ed $ \bar{N} $.

La restrizione dell'integrazione ad $ \mc{F} $ nella (\ref{T}) introduce un 
naturale cut-off ultravioletto per i modi di stringa, in quanto elimina 
il punto $\t_2 = 0 $ nel quale esiste una divergenza ultravioleta della 
teoria. $ \t_2 $ ha infatti un preciso significato fisico: esso rappresenta 
il {\it tempo proprio} impiegato dalla stringa chiusa per spazzare il toro. 
Ne segue che queste teorie non hanno divergenze ultraviolette.

Per ricavare l'espressione esplicita di $ \mathcal{T} $  occorre calcolare 
$ tr ( q^N ). $ 
Ricordiamo anzitutto che 
\be
N = \sum_{k = 1}^{\infty} \a_{-k}^i \a_{k , i} = \sum_{k = 1}^{\infty} 
k {a}^{\dag i}_k  a_{k , i} \, ,
\ee
ovvero che $ N $ \`e l'operatore numero di un set infinito di oscillatori 
armonici, e che lavoriamo nella base degli autostati dell'operatore numero tali 
che 
$$
{a}^{\dag }_k a_k | k\bra = k | k\bra.
$$
Allora, per ogni stato  $ | k\bra $ e per ogni direzione trasversa dello 
spazio-tempo, si avr\`a 
$$
tr ( q ^{k a^{\dag}_k a_k}  ) = \sum_{n = 0}^{\infty} \ket n | 
q^{k a^{\dag}_k a_k } | n \bra = \sum_{n = 0}^{\infty} {( q^k)}^n 
= 1 + q^k + q^{2k} + \dots = \frac{1}{1 - q^k}  ,
$$
e quindi 
$$
tr ( q^N ) =  \prod_{k = 1}^{\infty} tr ( q^{k a^{\dag}_k a_k} ) 
= \prod_{k = 1}^{\infty} \frac{1}{1 - q^k}  .
$$
Inoltre,  essendo il numero delle dimensioni trasverse pari a 24, 
la (\ref{T}) assumer\`a la forma
\be
\mathcal{T} = \int_{\mathcal{F}} \frac{d^2 \t}{\t_2^{14}} 
\frac{1}{q \bar{q}} \Big | \frac{1}{\prod_{k = 1}^{\infty} 
{( 1 - q^k)}^{24}} \Big |^2 = \int_{\mathcal{F}} \frac{d^2}{\t_2^{14}} 
\frac{1}{| \eta  (\t) |^{48}} \, .
\ee
In quest'ultima espressione \`e stata introdotta la funzione $ \eta$ di 
Dedekind
\be
\eta  ( \t ) = q^{\frac{1}{24}} \prod_{k = 1}^{\infty} ( 1 - q^k ) \, ,
\ee
che, sotto le trasformazioni modulari $T$ ed $S$ si comporta  
nel modo seguente
\beq
\eta  ( \t + 1 ) = e^{\frac{i \pi}{12}} \eta  ( \t ) \ , \nonumber \\
\eta  (- \frac{1}{\t} ) = \sqrt{ -i \t} \, \eta  ( \t) \ .
\label{eta}
\eeq
Osservando che la misura d'integrazione $ \frac{d^2 \t}{\t_2^2} $ \`e 
invariante modulare, si pu\`o verificare che l'intero integrando di $ 
\mathcal{T}$ \`e invariante sotto il gruppo modulare. Infatti utilizzando le 
(\ref{eta}) si pu\`o dimostrare che la combinazione  
$ \sqrt{\t_2} | \eta ( \t) |^2 $, corrispondente al contributo di una singola
direzione trasversa, \`e  essa stessa invariante sotto 
l'azione di $ PSL(2,Z)$.
L'invarianza modulare dell'ampiezza di toro \`e una diretta conseguenza 
dell'unitariet\`a della teoria, che a sua volta viene assicurata 
dall'introduzione della condizione di raccordo dei livelli. Ci\`o nonostante 
sappiamo che la teoria non \`e del tutto consistente a causa della presenza 
del tachione.  Questa particella di massa 
immaginaria introduce  in $ \mathcal{T} $ delle divergenze 
infrarosse $ ( \t_2 \rightarrow \infty ) $. Per vederlo basta considerare 
l'espansione in serie di potenze di $ ( q \bar{q} ) = e^{ - 4 \pi \t_2} $ 
dell'integrando di $ \mathcal{T} $, che corrispondono agli stati fisici 
che soddisfano la condizione di raccordo dei livelli.
A meno delle potenze di $ \t_2 $ e della misura d'integrazione, si ha 
$$
\frac{1}{(q \bar{q})} \, ( 1 + (24)^2 \, (q \bar{q} ) + \dots ) 
= ( q \bar{q} )^{-1} + (24)^2 + \dots = e^{4 \pi \t_2} + (24)^2 + \dots \ ,
$$
e quindi, poich\'e le potenze di $ (q \bar{q} ) $ indicano i livelli di massa 
degli stati 
dello spettro (in unit\`a di $ \frac{\a'}{2} $), ed i coefficienti la 
loro molteplicit\`a, si deduce immediatamente che il primo termine 
della serie, che \`e quello che produce la divergenza infrarossa nel 
limite $ \t_2 \ra \infty $, corrisponde proprio al tachione.

Passiamo ora al calcolo della funzione di partizione per le 
supersrtinghe chiuse.
L'ampiezza di vuoto per le stringhe fermioniche si ottiene seguendo la 
trattazione sviluppata per le stringhe bosoniche chiuse, e quindi partendo 
dall'espressione generale (\ref{fp}) con la sola differenza che in questo caso 
la traccia va calcolata anche sullo spazio di Fock degli oscillatori 
fermionici (ovvero anticommutanti). Ricordando che lo spettro di superstringa 
chiusa \`e dato da 
\be
m^2 = \frac{2}{\a'} ( N_B + N_F + \bar{N_B} + \bar{N_F} - 
( a_R + a_L)) \ ,
\ee
dove 
$$
N_B = \sum_{k = 1}^{\infty} \a_{-k}^i \a_{k,i} \ ,
$$
$$
N_F^{NS} = \sum_{r = 1/2}^{\infty} r b_{-r}^i b_{r,i} \ ,
$$
$$
N_F^R =  \sum_{r = 1}^{\infty} r d_{-r}^i d_{r,i} \ ,
$$
mentre $a_L $ ed $ a_R $ sono gli shifts all'energia di vuoto, che valgono 
$$ a_L = a_R = 1/2   \qquad  \mbox{ nel settore di NS,} $$
$$ a_L = a_R = 0     \qquad  \mbox{ nel settore di R.} $$

Utilizzando la (\ref{T}) si ottiene
\be
\G = \frac{V}{2  {( 4\pi^2 \a' )}^3} \int \frac{d^2 \t}{\t_2^2} 
\frac{d \t_2}{\t_2^4} tr ( q^{N_B}  \bar{q}^{\bar{N_B}} q^{N_F} 
\bar{q}^{\bar{N_F}} q^{-a_R} \bar{q}^{-a_L} ) \ ,
\label{fp2}
\ee
dove al solito $q = e^{2\pi i \t} $  ,  $ \bar{q} = e^{-2 \pi i \bar{\t}}.$
Il calcolo di $ tr ( q^{N_F} ) $ pu\`o essere effettuato facendo riferimento 
ai risultati standard per un gas di Fermi. Indicando con $ \l_r $ un generico 
oscillatore fermionico, si ha 
\be
tr ( q^{N_F} ) = tr ( q^{\sum_r r \l_{-r}^i \l_{r,i}} ) = 
\prod_r tr ( q ^{r \l_{-r}^i \l_{r,i}} ) = \prod_r {( 1 + q^r )}^{D-2} \ ,
\label{trf}
\ee
in quanto il principio di esclusione di Pauli consente due sole scelte di 
stati per ogni fermione, e dove $(D-2)$ si riferisce al numero di coordinate 
trasverse. La (\ref{trf}) pu\`o essere applicata tanto al settore di NS 
quanto a quello di R, purch\`e $ r \in Z+1/2 $ nel primo caso e $ r \in  Z $ 
nel secondo. Dunque, ricordando che per la superstringa $D = 10$ si 
ottiene per i due settori 
\be
tr_{NS} ( q^{N_F^{NS}} ) = \prod_{r = 1/2}^{\infty} {( 1 + q^r )}^8 = 
\prod _{k = 1}^{\infty}  {( 1 + q^{k - 1/2} )}^8 \, ,
\ee
\be
tr_R ( q^{N_F^R} ) = 16  \prod_{k =1}^{\infty} {( 1 + q^k )}^8 \, ,
\ee
dove il prefattore 16 conta la degenerazione del vuoto di R (uno spinore di 
Majorana-Weyl). Per tener conto dell'espressione completa della $ tr $
che compare nella (\ref{fp2}) si pone
\be
L_0^{\perp} = N_B + N_F - a_L \ ,
\ee
e quindi si ottiene 
\be
tr_{NS} ( q^{L_0^{\perp}} ) = \frac{\prod_{k=1}^{\infty} 
{( 1 + q^{k - 1/2} )}^8}{q^{1/2} \prod_{k = 1}^{\infty}  {( 1 - q^k )}^8}
\label{trNS}
\ee
nel settore di NS, e 
\be
tr_R ( q^{L_0^{\perp}} ) = 16 \frac{\prod_{k=1}^{\infty} 
{( 1 + q^k )}^8}{\prod_{k=1}^{\infty} {( 1 - q^k )}^8} 
\label{trR}
\ee
nel settore di R.

In realt\`a la (\ref{trNS}) e la (\ref{trR}) non conducono alla corretta 
ampiezza di vuoto, poich\'e come si \`e visto nel capitolo precedente, solo 
tramite la proiezione GSO si riesce ad ottenere uno spettro consistente di 
stati fisici. Questo si\-gnifica che nelle tracce di eq. (\ref{trNS}) e
(\ref{trR}) andranno inseriti i rispettivi proiettori GSO per i settori
 di NS e di R.
Quindi il settore di NS proiettato \`e descritto da 
\be
tr ( P_{GSO}^{NS}   q^{L_0^{\perp}} ) = tr \Big(  \frac{1 - 
( - )^{F^{NS}}}{2} q^{L_0^{\perp}} \Big) =  
\frac{  \prod_{k=1}^{\infty} {( 1 + q^{k - 1/2} )}^8 - 
\prod_{k=1}^{\infty} {( 1 - q^{k - 1/2} )}^8}{ 2  q^{1/2} 
\prod_{k=1}^{\infty} {( 1 - q^k )}^8} \ ,
\ee
dove l'operatore $ G = (-)^{F^{NS}} $ inverte i segni di 
tutti gli oscillatori $ q^k$ associati ai fermioni.
Il settore di R proiettato \`e descritto da 
\be
tr ( P_{GSO}^R \, q^{L_0^{\perp}} ) = tr \Big(  \frac{1 - \G_{11} 
( - )^{F^R}}{2} q^{L_0^{\perp}} \Big) = \frac{16}{2}  
\frac{\prod_{k=1}^{\infty}{( 1 + q^k )}^8}{\prod_{k=1}^{\infty} 
( 1 - q^k )^8} \ .
\ee
L'operatore $ \G_{11} ( - )^{F^R} $ agisce sullo spettro semplicemente 
rimuovendo met\`a degli stati, assegnando loro una definita chiralit\`a.
La funzione di partizione completa per le superstringhe chiuse dovr\`a 
ovviamente  contenere tutte le eccitazioni trasverse, ovvero sia quelle 
fermioniche che bosoniche. Quindi $\G $ sar\`a data dalla somma (con segni)
delle ampiezze di vuoto per i quattro settori di superstringa, ovvero 
$$
\G = \G_{NS-NS} + \G_{R-R} + \G_{NS-R} + \G_{R-NS} \ ,
$$
oppure, effettuando un opportuno riscalamento 
(che elimini il fattore $ \frac{V}{2 \, {( 4 \pi^2 \a' )}^4}$)
\be
\mathcal{T} = \mathcal{T}_{NS-NS} + \mathcal{T}_{R-R} + \mathcal{T}_{NR-R} + 
\mathcal{T}_{R-NS} \ ,
\label{Ttot}
\ee
dove 
\beq
\mathcal{T}_{NS-NS} = \int \frac{d^2 \t}{\t_2^6} \frac{1}{4} 
\Big| \frac{\prod_{k=1}^{\infty} {( 1 + q^{k - 1/2})}^8 - 
\prod_{k=1}^{\infty} {( 1 - q^{k - 1/2})}^8}{q^{1/2} 
\prod_{k=1}^{\infty} {( 1 - q^k )}^8} \Big|^2  ,
\eeq
\beq
\mc{T}_{R-R}  & = & \int \frac{d^ \t}{\t_2^6} \, \frac{1}{4} 
\Big| \frac{2^4 \, \prod_{k=1}^{\infty} ( 1 + q^k )^8}{\prod_{k = 1}^{\infty} 
( 1 - q^k)^8} \Big |^2  \ , \\
\mc{T}_{NS-R} & = & - \int \frac{d^2 \t}{\t_2^6} \, \frac{1}{4} 
\frac{1}{q^{1/2}} \frac{[ \prod_{k = 1}^{\infty} {( 1 + q^{k - 1/2} )}^8 - 
\prod_{k = 1}^{\infty} {( 1 - q^{k - 1/2} )}^8]}{\Big| 
\prod_{k=1}^{\infty} {( 1 - q^k )}^8 \Big|^2} \nonumber \\ 
             & \times  &[-2^4 \prod_{k = 1}^{\infty} {( 1 + \bar{q}^k )}^8]
 \ , \\
\mc{T}_{R-NS} & = & - \int \frac{d^2 \t}{\t_2^6} \frac{1}{4} 
\frac{1}{\bar{q}^{1/2}} \frac{[ 2^4 \prod_{k=1}^{\infty} {( 1 + q^k)}^8 ]}
{ \Big | \prod_{k=1}^{\infty} {( 1 - q^k )}^8 \Big|^2} \nonumber \\ 
& \times & [ \prod_{k=1}^{\infty} {( 1 + \bar{q}^{k - 1/2} )}^8 - 
\prod_{k=1}^{\infty} {( 1 - \bar{q}^{k-1/2} )}^8 ] \ .
\eeq
I quattro contributi di (\ref{Ttot}) possono essere riscritti in forma 
piuttosto compatta introducendo le \emph{\bf funzioni} $ \theta $ 
\emph{ \bf di Jacobi}, definite per mezzo di somme gaussiane o di prodotti 
infiniti 
\beq
&\th & \left[ \ba{c}
\a\\
\b
\ea \right]
( z | \t ) = \sum_{n \in Z} q^{1/2 ( n+\a)^2} e^{2 \pi i ( n+ \a ) 
( z + \b )} \\ 
           & = &  e^{2 \pi i \a ( z + \b)} q^{\frac{\a^2}{2}} \prod_{n=1}^{\infty} ( 1 - q ^n ) ( 1 + q^{n + \a - 1/2} 
e^{2\pi i (z+\b)})(1 + q^{n - \a - 1/2} e^{-2\pi i (z+\b)}), \ \nonumber
\eeq
dove $\a $ e $  \b $ sono parametri reali e $ q = e^{2\pi i \t} $.
Sotto le trasformazioni modulari $T$ ed $S$ queste hanno rispettivamente il 
seguente comportamento:
\be
T : \quad \th \left[ \ba{c}
\a\\
\b
\ea \right]
( z | \t + 1 ) = e^{-i \pi \a(\a-1)} \th \left[ \ba{c}
\a\\
\b+ \a - 1/2
\ea \right] ( z| \t ) ,
\label{Tsuteta}
\ee
\be
S : \quad \th \left[  \ba{c}
\a\\
\b
\ea \right] ( z | -\frac{1}{\t} ) = \sqrt{-i\t} \,  \th \left[ \ba{c}
\b\\
-\a
\ea \right] ( z | \t ) .
\label{Ssuteta}
\ee
Di particolare interesse per il nostro scopo sono le espressioni delle 
$ \th $ in $z=0$ e con caratteristiche $\a $ e $ \b $ uguali a 0 e 1/2 
\be
\th \left[  \ba{c}
1/2\\
1/2
\ea \right] = \th_1 ( 0| \t) = 0 ,
\label{teta1}
\ee
\be
\th \left[ \ba{c}
1/2\\
0
\ea \right] = \th_2 ( 0 | \t ) = 2 q^{1/8} \prod_{n = 1}^{\infty} ( 1 - q^n){(1 + q^n)}^2 ,
\ee
\be
\th \left[ \ba{c}
0\\
0
\ea \right] = \th_3 ( 0 | \t ) = \prod_{n=1}^{\infty} ( 1- q^n){( 1 + q^{n-1/2})}^2 ,
\ee
\be
\th \left[ \ba{c}
0\\
1/2
\ea \right] = \th_4 ( 0 | \t ) = \prod_{n=1}^{\infty} ( 1- q^n){( 1- q^{n-1/2} )}^2 ,
\label{teta4}
\ee
dalle (\ref{teta1})-(\ref{teta4}) segue che 
\beq
\frac{\th_2^4 ( 0 | \t )}{\eta^{12} ( \t )} & = &
16 \frac{\prod_{k=1}^{\infty} {( 1 + q^k )}^8}{\prod_{k=1}^{\infty} {( 1 - q^k )}^8} 
= tr_R ( q^{L_0^{\perp}} ) \ , \nonumber \\
\frac{\th_3^4 ( 0 |\t )}{\eta^{12} (\t)}    & = & 
\frac{\prod_{k=1}^{\infty} {( 1 + q^{k-1/2})}^8}{q^{1/2} 
\prod_{k=1}{\infty} {( 1 - q^k)}=8} = tr_{NS} ( q^{L_0^{\perp}}) \  ,
\nonumber \\
\frac{\th_4^4 ( 0 | \t )}{\eta^{12} (\t)}    & = & 
\frac{\prod_{k=1}^{\infty} {( 1 - q^{k-1/2} )}^8}{q^{1/2} 
\prod_{k=1}^{\infty} {( 1 - q^k )}^8}  \ , \nonumber \\
\eeq
che sono direttamente legate alle ampiezze di vuoto delle superstringhe. 
Infatti la (\ref{Ttot}) pu\`o essere riscritta come 
\be
\mc{T} = \int \frac{d^2 \t}{\t_2^6} \Big| \frac{\th_3^4 ( 0 | \t ) - \th_4^4 ( 0| \t) - \th_2^4 ( 0| \t)}{\eta^{12} (\t)} \Big|^2.
\label{fpsf}
\ee
Questa espressione racchiude in s\`e due importantissime propriet\`a 
direttamente legate alla proiezione GSO. 
\begin{enumerate}
  \item Dalla famosa identit\`a di Jacobi
\be
\th_3^4 - \th_4^4 - \th_2^4 = 0 \ ,
\ee
nota come \emph{AEQUATIO IDENTICA SATIS ABSTRUSA}, si deduce che la 
funzione di partizione del toro per la superstringa chiusa (\ref{fpsf}) \`e 
identicamente nulla.
Questo \`e un risultato  notevole, in quanto implica che  l'intero 
spettro perturbativo delle eccitazioni di stringa contiene, ad ogni livello 
di massa, un ugual numero di stati bosonici e fermionici, ovvero che nello 
spazio-tempo lo spettro \`e supersimmetrico.
 \item  Dal comportamento delle funzioni $\th$ di Jacobi sotto le 
trasformazioni modulari $S$ e $T$ si pu\`o verificare che la funzione di 
partizione di toro  (\ref{fpsf}) \`e invariante modulare 
\cite{Alvarez-Gaume:1986es}, sebbene i singoli termini che la compongono, 
relativi ai quattro settori, non lo siano.
\end{enumerate}
Vogliamo ora mostrare come l'ampiezza (\ref{fpsf}) possa essere riscritta in 
maniera ancora pi\`u chiara e compatta, e come, imponendo i vincoli di 
invarianza modulare, si riescano a costruire tutti  i modelli consistenti di 
stringhe chiuse orientate a 10 dimensioni.
A questo scopo, si introducono i cosiddetti \emph{\bf caratteri di $so(8)$ a 
livello 1} indicati come $ O_8, V_8 , S_8 $ e $ C_8 $ \cite{Lerche:1989np}
\cite{Goddard:1983at}, 
particolari combinazioni delle funzioni $ \th $ di Jacobi e della $ \eta $ di Dedekind, 
 definiti da
\beq
O_8 = \frac{ \th_3^4 + \th_4^4}{2 \eta^4} \, ,  \qquad  V_8 = \frac{\th_3^4 - \th_4^4}{2 \eta^4} \, , \nonumber \\
S_8 = \frac{\th_2^4 + \th_1^4}{2 \eta^4} \, ,   \qquad  C_8 = \frac{\th_2^4 - \th_1^4}{2 \eta^4} \, .
\eeq
$ O_8 $ e $ V_8 $ corrispondono ad una decomposizione ortogonale dello spettro di NS, dove sopravvivono rispettivamente solo un numero pari o dispari di 
eccitazioni fermioniche. Tutto ci\`o appare ancora pi\`u evidente riscrivendo 
esplicitamente $O_8 $ e $ V_8 $ come traccia sugli oscillatori GSO proiettati 
\beq
\frac{O_8}{\eta^8} & = & tr_{NS} \Big ( \frac{ 1 + ( - )^F_{NS}}{2} \ 
q^{L_0^{\perp}} \Big) \  , \nonumber \\ 
\frac{V_8}{\eta^8}  & = & tr_{NS} \Big( \frac{1 - ( - )^F_{NS}}{2} \ 
q^{L_0^{\perp}} \Big) = tr _{NS} ( P_{GSO}^{NS} \, q^{L_0^{\perp}} ) \ . 
\nonumber
\eeq
Come \`e stato osservato in precedenza, l'operatore $ P_{GSO}^{NS} $ elimina 
dallo spettro gli stati ottenuti dal vuoto con un numero pari di oscillatori 
fermionici, ed in particolare elimina il tachione. Quindi $ V_8 $, al livello 
pi\`u basso in massa contiene  un vettore di massa nulla. $ O_8 $, invece,  
contenendo un operatore che proietta via gli stati ottenuti dal vuoto agendo 
con un numero dispari di modi fermionici, parte al livello pi\`u basso con un 
tachione.
I caratteri $ S_8 $ e $ C_8 $ sono associati alle due classi spinoriali 
appartenenti al settore di R, e anche essi descrivono porzioni ortogonali 
dello spettro di R, che al livello di massa nulla inizia con due spinori di 
chiralit\`a opposta. In questo caso l'operatore di proiezione GSO anticommuta 
con i campi $  \p^{\m}$, e quindi le eccitazioni massive sia di $ S_8 $ 
che di $ C_8 $ hanno chiralit\`a alternate. 
Sottoliniamo per\`o che nel settore di R 
\`e presente un'ambiguit\`a, legata all'annullarsi di $ \th_1 $ all'origine, 
che implica l'uguaglianza numerica di $ S_8 $ e $ C_8 $.

Infine osserviamo che le trasformazioni modulari (\ref{Tsuteta}) e 
(\ref{Ssuteta}) determinano le corrispondenti matrici $T$ ed $S$ per i quattro 
caratteri di $so(8)$.  Non \`e difficile mostrare che
\be
T = diag ( -1, 1, 1, 1 ) \, ,
\ee
\be
S = \frac{1}{2}  \left( \ba{cccc}
1 & 1 & 1 & 1\\
1 & 1 & -1 & -1\\
1 & -1 & 1 & -1\\
1 & -1 & -1 & 1
\ea \right).
\ee
Sotto l'azione di queste matrici, le combinazioni  $ \frac{O_8}{\t_2^4 \eta^8}, \frac{V_8}{\t_2^4 \eta^8}, \frac{S_8}{\t_2^4 \eta^8}, \frac{C_8}{\t_2^4 \eta^8} $ si trasformano le une nelle altre, e l'ambiguit\`a del settore di R viene risolta.
La funzione di partizione (\ref{fpsf}) assume quindi una delle due 
forme seguenti 
\beq
\mc{T}_1 & = & \int \frac{d^2 \t }{\t_2^6} 
( \bar{V}_8 - \bar{S}_8 )( V_8 - S_8 ) 
= \int \frac{d^2 \t}{\t_2^6} | V_8 - S_8 |^2  \ , \nonumber \\
\mc{T}_2 & = & \int \frac{d^2 \t }{\t_2^6} 
( \bar{V}_8 - \bar{S}_8 )( V_8 - C_8 ) \ .
\label{Tau}
\eeq
In realt\`a le (\ref{Tau}) non sono le uniche ampiezze invarianti modulari 
che si possano determinare a 10 dimensioni.
Infatti l'integrando dell'ampiezza di toro si pu\`o porre nella forma  
$$
\mc{T} = \c^{\dag}  M  \c \ ,
$$
dove $ \c $ descrive i caratteri di $so(8)$, e la matrice $M$, che 
definisce la proiezione GSO,  soddisfa i vincoli di invarianza modulare  
$$ M = S^{\dag} M S  \ , \  M = T^{\dag} M T  \ , $$
ed \`e tale da soddisfare la relazione tra spin e statistica.

Sotto queste ipotesi  non \`e difficile vedere che esistono solo quattro 
ampiezze di toro associate alla superstringa di RNS.
Due di queste corrispondono alle \emph{ \bf superstringhe di Tipo IIA} e di 
\emph{ \bf Tipo IIB} 
\beq
\mc{T}_{IIA} & = & \frac{1}{{( \sqrt{\t_2} \eta \bar{\eta} )}^8} 
( \bar{V}_8 - \bar{S}_8 ) ( V_8 - C_8 ) \ , \nonumber \\
\mc{T}_{IIB} & = & \frac{1}{{( \sqrt{\t_2} \eta \bar{\eta} )}^8}
| V_8 - S_8 |^2  ,
\label{IIAB}
\eeq
(osserviamo che le (\ref{IIAB}) coincidono con le (\ref{Tau})).

Le altre due ampiezze corrispondono a due teorie \emph{non} 
supersimmetriche, la \emph{ \bf 0A} e la \emph{ \bf 0B} \cite{Seiberg:1986by},
  descritte da 
\be
\mc{T}_{0A} = \frac{1}{{( \sqrt{\t_2} \eta \bar{\eta} )}^8} 
( | O_8 |^2 + | V_8 |^2 + \bar{S}_8 C_8 + \bar{C}_8 S_8 )  ,
\label{0A}
\ee
\be
\mc{T}_{0B} = \frac{1}{{( \sqrt{\t_2} \eta \bar{\eta} )}^8}
 ( | O_8 |^2 + | V_8 |^2 + | S_8 |^2 + | C_8 |^2 ).
\label{0B}
\ee
Sottolineiamo come l'introduzione dei caratteri O, V, S e C conduca ad una 
descrizione molto elegante delle funzioni di partizione e dei relativi 
spettri, in cui ogni settore della teoria di superstringa resta 
associato ad un carattere indipendente.
Per apprezzare meglio questo risultato, 
\`e istruttivo riassumere il contenuto dello spettro di bassa 
energia di questi modelli.

Iniziamo dalle (\ref{IIAB}), e notiamo anzitutto che, come 
anticipato nel capitolo precedente, nella superstringa di Tipo IIA i vuoti 
del settore di R hanno chiralit\`a opposta ($ \bar{S}_8 $ e $ C_8 $), mentre 
nella Tipo IIB essi hanno la stessa chiralit\`a ($\bar{S}_8 $ e $ S_8 $).
Esaminiamo i campi bosonici che derivano dai settori NS-NS ed R-R.
Entrambe le teorie includono, nel settore di NS-NS (=$ V_8 \bar{V}_8 $), un 
dilatone $ \f $, un tensore antisimmetrico $ B_{\m\n} $ ed un gravitone.
Per la Tipo IIA, il settore di R-R ($=\bar{S}_8 C_8 $) contiene un vettore 
$ A_{\m} $ ed una tre-forma $ C_{\m\n\r} $, mentre per la IIB, 
($=\bar{S}_8 S_8 $) contiene un ulteriore scalare $\f' $, 
un ulteriore tensore antisimmetrico 
$ B'_{\m\n}$ ed una quattro-forma $ D_{\m \n \r \s}^{+} $ con 
\emph{field-strength} autoduale.
Per quanto riguarda i campi fermionici, derivanti dai settori misti, entrambe 
le teorie contengono una coppia di gravitini  $ \p_{\a}^{\m} $ ed una coppia 
di spinori  $ \c_{\a} $, ma, mentre per la Tipo IIB ($\bar{V}_8 S_8 $ e $ 
\bar{S}_8 V_8 $) entrambi i gravitini hanno la stessa chiralit\`a (\emph{left}) 
cos\`\i \  come gli spinori (entrambi \emph{right}), 
nella IIA ($\bar{V}_8 C_8 $ e $ \bar{S}_8 V_8 $) entrambe le coppie di campi 
hanno chiralit\`a opposta.
Da tutto questo e dal fatto che le due teorie sono prive di tachioni, segue 
che gli stati di massa nulla della Tipo IIA completano il multipletto non 
chirale della supergravit\`a (1,1) a 10 dimensioni, mentre gli stati a massa 
nulla della Tipo IIB completano il multipletto della supergravit\`a chirale 
(2,0) a 10 dimensioni.

Lo spettro di bassa energia delle teorie di Tipo IIA e IIB \`e riassunto nella
 tabella seguente:
\vskip .2in
\begin{tabular}{ccc}
\quad               & TIPO IIA     &      TIPO IIB\\
campi bosonici &  &  \\
NS-NS &     $ \f $, $ B_{\m\n} $, $ g_{\m\n}$   &      $ \f, B_{\m\n}^N, g_{\m\n}  $\\
R-R   &     $ A_{\m}, C_{\m\n\r} $   &  $ \f', B_{\m\n}^R, D_{\m\n\r\s}^{\dag} $\\
campi fermionici\\
NS-R  & $ \c_{\a}, \p_{\dot{\a}}^{\m} $   & $ \c_{\dot{\a}}, \p_{\a}^{\m}$  \\
R-NS & $ \c_{\dot{\a}}, \p_{\a}^{\m} $   & $   \c_{\dot{\a}}, \p_{\a}^{\m}$
\end{tabular}
\vskip .2in
Passiamo ora alle teorie 0A e 0B.
Queste non contengono fermioni spazio-temporali e sono non chirali.
In questo caso i settori di NS-NS di entrambe le teorie contengono, oltre ai 
modi di bassa energia dei modelli precedenti, anche un tachione T. Inoltre gli
 stati di R-R della 0A sono due copie di quelle della IIA, mentre per la 
teoria 0B  essi sono una coppia di scalari, una coppia di due-forme 
$ B'$, $ B''$ ed una quattro-forma $ D_{\m\n\r\s} $ completa. 
\vskip .2in
\begin{tabular}{ccc}
\quad     & 0A  &      0B \\
NS-NS   & $ T, \f, g_{\m\n}, B_{\m\n} $  &  $ T, \f, g_{\m\n}, B_{\m\n} $\\
R-R     & $ A_\m, A'_\m, C_{\m\n\r}, C'_{\m\n\r}  $ & $ \f', \f'', B'_{\m\n}, 
B''_{\m\n}, D_{\m\n\r\s} $
\end{tabular}

\section{Compattificazioni}
Le teorie di superstringa chiusa sviluppate fin qui, sebbene rappresentino un 
passo in avanti verso la descrizione della natura, risultano essere non 
completamente soddisfacenti. Infatti 

1) \ la dimensione critica dello spazio-tempo in cui le superstringhe sono 
immerse, $D = 10$, \`e  troppo elevata rispetto al ``nostro'' spazio-tempo
 Minkow\-skiano tetradimensionale.

2) \ Questi modelli, pur includendo la gravit\`a, non introducono simmetrie 
di gauge, indispensabili per la descrizione delle interazioni 
fondamentali.

Di qui l'esigenza di un meccanismo di riduzione dimensionale \`a la 
Kaluza-Klein, capace di rendere compatte e ``microscopiche'' le dimensioni 
in eccesso e, contemporaneamente, di render conto dei gruppi di gauge interni 
(ed eventualmente della loro rottura).
In altri termini, l'idea \`e quella di ripensare allo spazio-tempo 
Minkowskiano $ M^D $ come al seguente prodotto:
$$ M^D = M^d \times K  ,$$
dove $K$ \`e una variet\`a compatta, la cui scala caratteristica 
\`e cos\'\i \ piccola da non poter essere rivelata alle energie accessibili.
$D$, invece, \`e il numero di dimensioni che si vogliono mantenere estese.

Vedremo come, in relazione all'idea della compattificazione, sia possibile 
costruire un nuovo modello di superstringa chiusa, nota come 
\emph{\bf Superstringa Eterotica} \cite{Gross:1985fr}, 
con gruppi di gauge $SO(32)$ o $ E_8 \times E_8 $.

\section{Compattificazioni toroidali}
L'esempio pi\`u semplice di compattificazione si ottiene esaminando la 
stringa bosonica chiusa con un'unica coordinata compatta e scegliendo $K$
coincidente con un toro unidimensionale $ T^1 $ (ovvero un cerchio di raggio
 R). L'espansione in modi di Fourier della coordinata compatta \`e 
\be
X (\t, \s) = x + 2 \a' p \t + 2 n R \s + i \sqrt{\frac{\a'}{2}} 
\sum_{k \neq 0} \{ \frac{\bar{\a}_k}{k} e^{-2ik(\t-\s)} + 
\frac{\a_k}{k} e^{-2ik(\t+\s)} \}.
\label{coordc}
\ee
La (\ref{coordc}) segue dal fatto che la stringa, oltre ad avere un impulso 
quantizzato lungo la direzione circolare 
\be
p = \frac{m}{R}  \qquad  \mbox{con} \  m \in Z \ ,
\ee
ha la possibilit\`a di avvolgersi $ n $ volte attorno ad essa. A questo 
proposito si definisce il vettore di avvolgimento (\emph{winding number}) 
$w$ come 
\be
w = n R   \qquad     \mbox{con} \   n \in Z.
\label{imp}
\ee
Quando $ w \neq 0 $, la (\ref{coordc}) descrive configurazioni solitoniche di 
stringa, la cui energia \`e divergente per $ R \rightarrow \infty . $
Decomponendo l'espansione (\ref{coordc}) in termini di onde destre e sinistre, 
la (\ref{imp})  pu\`o essere riscritta in termini di
\be
p_L = \frac{1}{\sqrt{2}} \Big[ \sqrt{\frac{\a'}{R^2}} m + 
\sqrt{\frac{R^2}{\a'}} n \Big] \ , 
\label{pL}
\ee
e
\be
p_R = \frac{1}{\sqrt{2}} \Big[ \sqrt{\frac{\a'}{R^2}} m - 
\sqrt{\frac{R^2}{\a'}} n \Big] \ .
\label{pR}
\ee
In questo caso lo spettro, ottenibile al solito dai modi zero trasversi del 
tensore energia-impulso, \`e dato da 
\be
m^2 = \frac{2}{\a'} ( N + \bar{N} - 2 + \frac{1}{2} p_L^2 + 
\frac{1}{2} p_R^2 )\ ,
\label{msc}
\ee
\be
\mbox{con}  \qquad    N - \bar{N} = \frac{1}{2} ( p_L^2 - p_R^2 ) = m n \ .
\label{cond}
\ee
La condizione (\ref{cond}) garantisce l'invarianza sotto riparametrizzazioni, 
e gli operatori numero hanno la seguente espressione:
$$ N = \sum_{k=1}^{\infty} ( \a_{-k}^i \a_{i,k} + \a_{-k} \a_k ) \ , $$
$$\bar{N} = \sum_{k=1}^{\infty} ( \bar{\a}_{-k}^i \bar{\a}_{i,k} + 
\bar{\a}_k \bar{\a}_k ) \ . $$
L'introduzione dei nuovi numeri quantici $m$ ed $n$ fa s\`\i \ che lo spettro 
in consi\-derazione sia pi\`u ricco di quello ottenuto nel  caso non campatto. 
In particolare, lo 
spettro a massa nulla, per $m = n = 0$ contiene, oltre agli usuali campi 
bosonici, due vettori di gauge relativi al gruppo $ U(1)_L \times U(1)_R . $
Infatti, identificando ogni stato con $ | m, n \bra $, per $ p_R = p_L = 0 $ 
ed $ N = \bar{N} = 1 $, lo spettro contiene gli stati :
$$ \a_{-1}^i \bar{\a}_{-1}^j |0,0\bra , \a_{-1}^i \bar{\a}_{-1} |0,0\bra , 
\a_{-1} \bar{\a}_{-1}^i |0,0\bra , \a_{-1} \bar{\a}_{-1} |0,0\bra \ , $$
dove si pu\`o notare che alcuni indici spazio-temporali sono stati trasformati
 in indici relativi alla variet\`a interna.
Inoltre, sempre per lo spettro a massa nulla, ma per $(p_R,p_L) \neq (0,0) $, 
si pu\`o dimostrare che  se il raggio di compattificazione soddisfa 
$ R^2 = \a' $, il gruppo di simmetria interna viene esteso ad 
$ SU(2)_L \times SU(2)_R $. In questo caso infatti, la condizione 
(\ref{cond}) restringe la scelta di $ ( N, \bar{N} )$ e di $ ( p_L,p_R ) $
 e fornisce altri valori compatibili con $ R^2 = \a' $, ossia \\
a) \ $ ( N = 1 , \bar{N} = 0 ) $  e  $ ( p_L^2 = 0 , p_R^2 = 2 ) $ \ 
che implica l'esistenza degli ulteriori  stati di norma nulla
$$ \a_{-1}^i |1,1\bra , \a_{-1} | 1,1\bra , \a_{-1}^i |-1,-1\bra , \a_{-1} |-1,-1\bra . $$ 
b) \ $ ( N = 0 , \bar{N} = 1 ) $  e  $ ( p_L^2 = 2 , p_R^2 = 0 ) $ \  
che implica l'esistenza degli ulteriori stati di norma nulla 
$$ \bar{\a}_{-1}^i |1,-1\bra , \bar{\a}_{-1} |1,-1\bra , \bar{\a}_{-1}^i 
|-1,1\bra , \bar{\a}_{-1} |-1,1\bra . $$
In conclusione, quando $ R^2 = \a' $, nello spettro a massa nulla compaiono 
degli stati che, relativamente ai numeri quantici $m$ ed $n$, formano i sei 
vettori della rappresentazione $ (3,1) \oplus (1,3) $ di 
$ SU(2)_L \times SU(2)_R $.

Avendo a disposizione la condizione di {\it mass-shell}, 
\`e possibile ricavare la 
funzione di partizione per la stringa bosonica compattificata su $ T^1 $. 
Inserendo nell'espressione generale (\ref{fp}) la (\ref{msc}) e la 
(\ref{cond}) e ricordando che la dimensione dello spazio-tempo \`e ora 
$D=26-1$, si ottiene 
\be
\G_{T^1} = {(\frac{1}{4 \pi^2 \a'})}^{13-1/2} \int_{\mc{F}}  
\frac{d^2 \t}{\t^{14-1/2}} \frac{1}{{( \eta(\t) \eta(\bar{\t}) )}^{23}} 
\frac{\sum_{(m,n) \in Z} q^{\frac{p_L^2}{2}} 
\bar{q}^{\frac{p_R^2}{2}}}{( \eta(\t) \eta(\bar{\t}))} ,
\label{Ttoro}
\ee
dove $ q = e^{2 \pi i \t}. $
La (\ref{Ttoro}) \`e invariante modulare, ed inoltre, definendo 
\be
Z_{T^1} ( R ) = \sum_{m,n} \frac{q^{\frac{p_L^2}{2}} 
\bar{q}^{\frac{p_R^2}{2}}}{( \eta(\t) \eta(\bar{\t}))} ,
\label{fptoro}
\ee
si pu\`o verificare che 
\be
\lim_{R \to \infty} \frac{Z_{T^1}(R)}{R} = 
\frac{1}{\sqrt{\t_2} ( \eta(\t) \bar{\eta}(\bar{\t}))} ,
\ee
ovvero nel limite di decompattificazione si riottiene la funzione di 
partizione di un bosone non compatto.

L'altra propriet\`a di notevole interesse della (\ref{fptoro}), \`e 
l'invarianza sotto la trasformazione 
$$ R \longleftrightarrow \frac{\a'}{R} , \quad m \lra n , $$ 
nota come \emph{\bf T-dualit\`a}. La T-dualit\`a \`e una simmetria esatta 
della teoria delle perturbazioni della stringa bosonica chiusa. Essa lega le 
compattificazioni su cerchi di raggio R a quelle di su cerchi di raggio 
$R ' = \frac{\a'}{R}$, scambiando impulsi e avvolgimenti, rendendole quindi 
fisicamente indistinguibili. Nelle superstringhe, questa scambia la chiralit\`a
relativa dei vuoti di Ramond, e di conseguenza scambia la teoria $0A$ con la
$0B$ e la teoria $IIA$ con la $IIB$.

Occorre per\`o   prestare attenzione al dilatone, sul quale la T-dualit\`a  
agisce in  modo non banale, come si evince dall'azione effettiva
$$ R e^{-2 \f} = R' e^{-2 \f'}  , $$
e quindi \`e necessario ridefinirlo nella maniera seguente
\be
\f' = \f + \frac{1}{2} \log \frac{R'}{R} = \f + \frac{1}{2} 
\log \frac{\a'}{R^2}  .
\ee

L'analisi precedente pu\`o essere  generalizzata al caso di tori di 
dimensione arbitraria, considerando il fatto che un toro unidimensionale 
pu\`o essere pensato come un reticolo unidimensionale di passo $ 2 \pi R $ 
in cui punti traslati vengono identificati 
$$ T^1 = {\mathbf R} / 2 \pi R . $$
Un toro $k$-dimensionale si potr\`a quindi esprimere come il quoziente 
$$ T^k = {\mathbf R} ^k/\L , $$
dove $ \L $ \`e un reticolo quadrato che possiamo immaginare come il prodotto 
di $k$ cerchi.
Le coordinate compattificate hanno allora espressioni simili a quelle del 
caso unidimensionale, con la differenza che in questo caso le (\ref{pL}) e 
(\ref{pR}) divengono 
\beq
p_L^i = m_a \tld{e}^{a i} + \frac{n^a}{\a'} e_a^i - \frac{B_{ab}}{\a'} n^b 
\tld{e}^{a i}   ,    \nonumber \\
p_R^i = m_a \tld{e}^{a i} - \frac{n^a}{\a'} e_a^i - \frac{B_{ab}}{\a'} n^b 
\tld{e}^{a i}   ,
\eeq
dove $g_{ab}$ \`e la metrica del toro interno, e $B_{ab}$ un tensore 
antisimmetrico. Inoltre $\{ e^i_a \}$ \`e un'opportuna base di vettori  
del reticolo  $\L$ ($e^i_a e_{i b}=g_{ab}$), mentre $\{ \tld{e}^{a i} \}$ 
\`e la corrispondente base di vettori del reticolo duale $ \L^* $ 
($\tld{e}^{a i}=g^{ab} e^i_b$). 

La funzione di partizione per la stringa bosonica compattificata su un toro 
$ T^k $ \`e 
\be
Z_{T^k} = \frac{1}{{( \eta(\t) \bar{\eta}(\bar{\t}) )}^k}  \sum_{(m,n) \in Z} 
q^{\frac{p_L^2}{2}} \bar{q}^{\frac{p_R^2}{2}} ,
\ee
 la diretta generalizzazione della (\ref{fptoro}).
Applicando a questa espressione una trasformazione modulare $T$, l'invarianza 
di $Z_{T^k}$ richiede $ e^{i \pi (p_L^2-p_R^2)}=1$, ovvero che i vettori 
$ p_L^2 $ e $  p_R^2 $ soddisfino la seguente relazione:
\be
p_L^2 -  p_R^2  \in 2Z .
\ee
Quindi, $ ( \vec p_L , \vec p_R ) $ pu\`o essere considerato un vettore di un 
reticolo $ W^{k,k} $, costruito a partire da $ \L $ e $ \L^* $, con metrica 
lorentziana di segnatura $ ( + ^ k , - ^ k ) $, e con prodotto scalare 
\be
{ | ( p_L, p_R ) |}^2 = p_L^2 - p_R^2 .
\label{pLpR}
\ee
L'invarianza sotto $T$ richiede che il reticolo sia pari, e allo stesso modo, 
dopo una risommazione di Poisson l'invarianza sotto $S$ richiede che sia 
autoduale. Quindi la stringa bosonica pu\`o essere compattificata su un 
toro qualsiasi purch\`e il reticolo che lo definisce sia lorentziano, pari ed 
autoduale, condizione queste ultime, dettate dall'invarinza modulare della 
teoria.

Dalla discussione svolta fin qui si deduce che gi\`a la compattificazione su 
variet\`a semplici quali il toro, consente l'introduzione di gruppi di gauge 
nella teorie delle stringhe bosoniche chiuse.

\section{ La stringa Eterotica}
Dal capitolo precedente \`e emerso come le uniche teorie di superstringa 
consi\-stenti a 10 dimensioni siano la Tipo IIA, la Tipo IIB e la Tipo I.
In realt\`a esiste un ultimo tipo di teoria di stringa supersimmetrica e 
consistente a 10 dimensioni, noto come \emph{\bf  stringa eterotica}, 
ottenuta combinando i modi destri della superstringa con i modi sinistri 
della stringa bosonica \cite{Gross:1985fr}.
L'osservazione alla base della costruzione dei modelli eterotici, che tra 
l'altro ne giustifica il nome, \`e la seguente:
nelle teorie di stringhe chiuse i modi destri e sinistri sono indipendenti; 
\`e quindi possibile immaginare una teoria di stringhe chiuse in cui i modi 
sinistri siano di un certo tipo e quelli destri di un altro. Allora, per 
incorporare la supersimmetria spazio-temporale, che consente la presenza di 
fermioni e l'assenza di tachioni, si considerano i modi destri della consueta 
superstringa a dieci dimensioni, ai quali vanno ad aggiungersi dieci dei 
ventisei modi sinistri della stringa bosonica.
Indicando rispettivamente con $ X^{\m}(\t,\s)$ e $\p^{\m}(\t,\s)$ le 
coordinate bosoniche e fermioniche, con $ \small{\m = 0, \dots,9}$, 
restano fuori le 
ulteriori 16  coordinate bosoniche sinistre, che quindi dovranno essere 
compattificate su una qualche variet\`a interna. In questo modo si riescono 
ad introdurre anche i gradi di libert\`a di gauge.

L'invarianza modulare richiede che la variet\`a compatta sia descritta da un 
reticolo di rango 16, euclideo, pari ed autoduale.
In sedici dimensioni esistono due soli reticoli di questo tipo:\\
$-  \G_8 \times \G_8 $, ovvero il reticolo delle radici dell'algebra 
$ E_8 \times E_8 $;\\
$-   \G_{16} $, il reticolo dei pesi di $ Spin(32)/Z_2 $.\\
Queste danno origine a due distinti modelli eterotici supersimmetrici,
\emph{la stringa} $ E_8 \times E_8 $ e \emph{la stringa} $ Spin(32)/Z_2 $.
Le funzioni di partizione per i due modelli, effettuando nel settore destro 
le appropriate proiezioni GSO, sono
\be
Z_{E_8 \times E_8} = \frac{1}{{(\sqrt{\t_2} \eta \bar{\eta})}^8} (V_8 - O_8 ) 
( \bar{O}_{16} + \bar{S}_{16} ) ( \bar{O}_{16} + \bar{S}_{16} ) ,
\label{HE}
\ee
\be
Z_{Spin(32)/Z_2} = \frac{1}{{(\sqrt{\t_2} \eta \bar{\eta})}^8} (V_8 - S_8 ) 
( \bar{O}_{32} + \bar{S}_{32} ) ,
\label{HO}
\ee
Dalle eq. (\ref{HE}) e (\ref{HO}) \`e possibile ricavare lo spettro delle due 
teorie, ed osserviamo che il settore destro \`e lo stesso della superstringa 
di Tipo II.
Imponendo le condizioni di stato fisico, ottenibili al solito dai modi-zero 
trasversi del tensore-energia impulso, $ L_0 - a_R = 0 $ \  e  \ 
$ \bar{L}_0 - a_L = 0 , $ 
e ricordando che il contributo allo shift dell'energia di vuoto \`e 
$$ - \frac{1}{24} \,  \mbox{per ogni bosone periodico}, \quad \frac{1}{24} \,   
\mbox{per ogni bosone antiperiodico}, $$ 
$$ \,\frac{1}{24} \, \mbox{per ogni fermione periodico},  \quad  - \frac{1}{48} \, 
\mbox{per ogni fermione antiperiodico}, $$
si ricava che a massa nulla, nel settore di NS, oltre al solito multipletto
del gravitone $( \f , g_{\m\n} , B_{\m\n} )$, compaiono gli stati che 
corrispondono ai vettori di gauge, nella rappresentazione aggiunta, del 
gruppo $SO(32)$ e del gruppo  $ E_8 \times E_8 $.
Nel settore di R sono invece presenti i superpartners degli stati precedenti, 
necessari per completare lo spettro  della 
supergravit\`a (1,0) accoppiata alla teoria di super-Yang-Mills 
(SYM); questo \`e infatti il modello di bassa energia per le due stringhe 
eterotiche appena descritte.

In realt\`a \`e possibile descrivere ulteriori modelli di stringa eterotica
a dieci dimensioni, oltre alle $SO(32)$ ed $ E_8 \times E_8 $. Questi 
corrispondono ad altre versioni della stessa teoria ma con diverse
proiezioni dello spettro. Ad esempio, la stringa $O(16) \times O(16)$
\`e una teoria eterotica non supersimmetrica \cite{Alvarez-Gaume:1986jb,
Ginsparg:1987bx} .

Negli ultimi dieci anni sono stati fatti notevoli sforzi nel tentativo di 
dimostrare che i vari  modelli supersimmetrici possono essere 
pensati come aspetti diversi di un'unica e pi\`u fondamentale teoria ad 
11 dimensioni (della quale non conosciamo i gradi di libert\`a microscopici), 
la \emph{\bf M-teoria} \cite{Witten:1995ex, Hull:1995mz}.

Tale congettura \`e suggerita  essenzialmente dal fatto che la 
supergravit\`a di Tipo IIA pu\`o essere ottenuta come riduzione dimensionale 
della supergravit\`a a 11 dimensioni (compattificando la seconda su un
cerchio $ S^1 $ di raggio R).
Alla fine degli anni '70 Cremmer, Julia e Scherk costruirono l'unica teoria 
di supergravit\`a in 11 dimensioni \cite{Cremmer:1978km}.
Poich\'e tutte le teorie di supergravit\`a  a 10 dimensioni sono  limiti di 
bassa energia delle corrispondenti teorie di stringhe, \`e naturale pensare 
che esista una teoria, la M-teoria appunto, che contienga la supergravit\`a a 
11 dimensioni come suo limite di bassa energia \cite{Witten:1995ex}.

Postulando l'esistenza della M-teoria, esistono due possibilit\`a di 
riduzione dimensionale che danno origine ad altrettante teorie
 di superstringa a 10 dimensioni:\\
a) se la M-teoria viene compattificata su un cerchio $ S^1 $, si ottiene la 
superstringa di Tipo IIA (o equivalentemente possiamo dire che la stringa di 
Tipo IIA, nel limite di forte accoppiamento, che coincide con il limite di 
decompattificazione, sia descritta dalla M-teoria );\\
b) compattificando invece la M-teoria sull'intervallo $ S^1/Z_2 $, ottenuto 
identificando punti opposti sul cerchio, si riesce a descrivere la stringa 
eterotica $ E_8 \times E_8 $.
\vskip .1in
Si pu\`o inoltre mostrare come da queste due ultime teorie di superstringa 
(la Tipo IIA e la $ E_8 \times E_8 $) sia possibile ottenere le restanti tre 
mediante trasformazioni di dualit\`a \cite{Sen:1997yy}.
La situazione pu\`o essere riassunta con lo schema seguente:
\begin{figure}[htb]\unitlength1cm
\begin{picture}(9,6)
\put(3,0){\epsfig{file=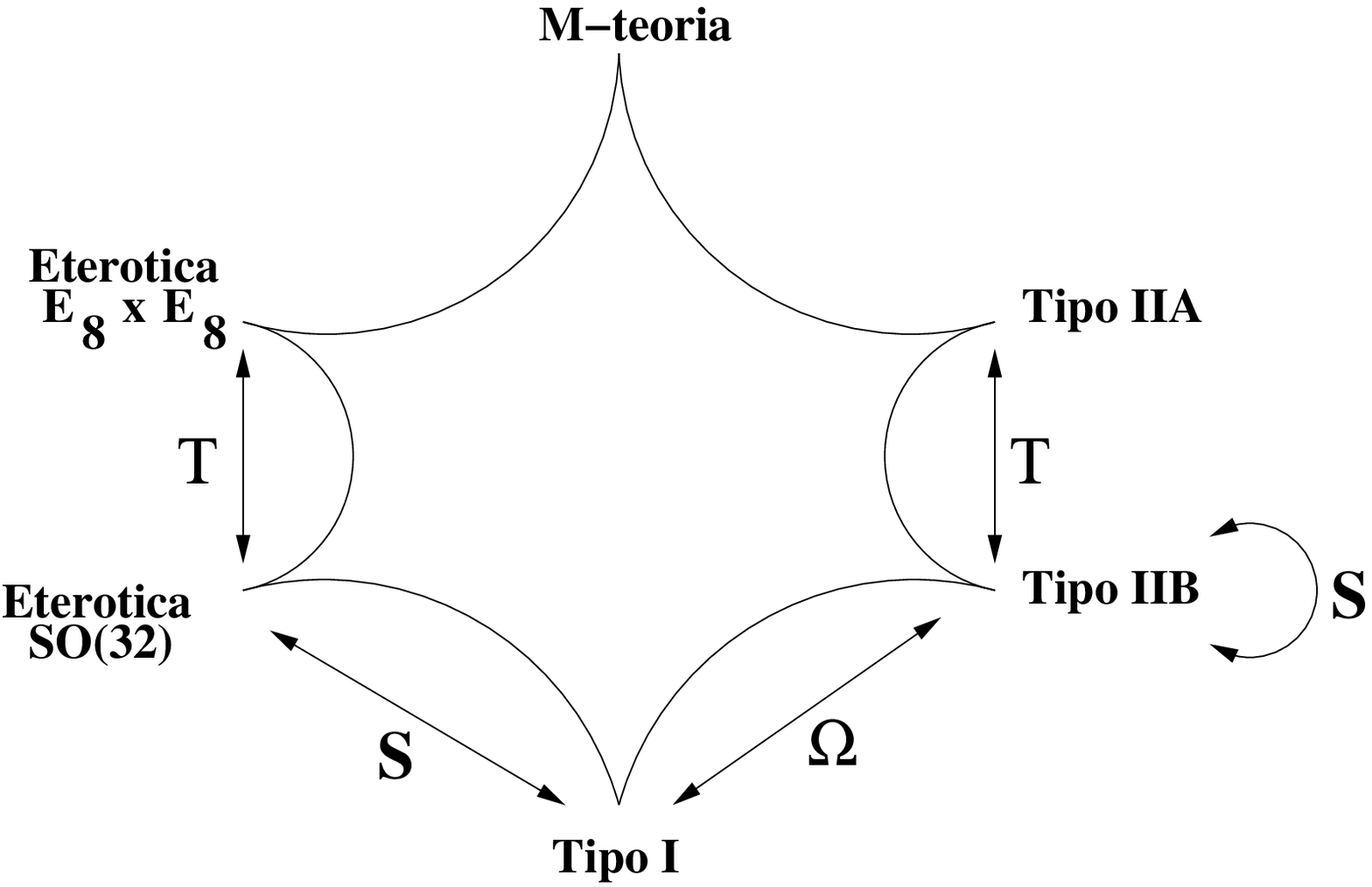,width=9cm,height=6cm}}
\end{picture}
\caption{\footnotesize{Sito delle Dualit\`a}}
\label{fig:6} 
\end{figure}

dove
\vskip.2in
\begin{description}
\item[$ S : g \rightarrow \frac{1}{g} $] \quad \`e l'operazione che identifica
 il limite di accoppiamento forte di una teoria con il limite di accoppiamento
 debole dell'altra (quale la Tipo I-SO(32) e l'eterotica SO(32) ; la Tipo IIB
 \`e autoduale) \cite{Sen:1994fa};
\item[$ T : R \rightarrow \frac{1}{R} $] \quad identifica teorie compattificate su cerchi di raggio R con quelle compattificate su cerchi di raggio 1/R;
\item[$ \O : x + i y \rightarrow x - i y $] \quad   \`e un'operazione di 
parit\`a sul world-sheet, che nelle teorie di stringhe chiuse, quali la 
Tipo IIB, scambia modi destri e sinistri.
\end{description}
 
\section{Compattificazioni su orbifols}
Nel paragrafo precedente abbiamo visto come con la compattificazione 
toroi\-dale si riesce ad introdurre i gruppi di gauge nei modelli eterotici. 
In generale  questo metodo presenta per\`o uno svantaggio: conduce a teorie 
con supersimmetria estesa in $D<10$, e dunque non chirali.

In effetti il toro \`e una variet\`a molto semplice, e quindi si \`e portati 
in modo naturale a considerare variet\`a pi\'u complesse sulle quali 
compattificare la teoria.
La compattificazione su \emph{ \bf  orbifolds} consente di raggiungere un 
buon compromesso tra (relativa) semplicit\`a dell'analisi e ricchezza dei 
risultati.

Un orbifold $ \mc{O} $ \`e definito come  uno spazio che localmente, ma non 
ne\-cessariamente globalmente, \`e il quoziente di una variet\`a M per 
l'azione 
di qualche gruppo discreto  $G$ \cite{Dixon:1985jw, Dixon:1987qv}: 
$$ \mc{O} = M/G \ . $$
Ovvero, dato un punto $ x $ di M, questo viene ad essere identificato con 
tutti i punti $ g\cdot x $, per $ g \in G $. In generale $G$ \`e un 
sottogruppo del gruppo delle isometrie di M 
(ovvero $G$ preserva la metrica su M);
se per esempio M \`e lo spazio Euclideo d-dimensionale $ {\bf R}^d $, gli 
elementi di $G$ sono le rotazioni e le traslazioni.

Un orbifold cessa di essere una variet\`a liscia nei punti fissi rispetto 
all'azio\-ne di $G$, poich\'e questi ultimi diventano delle singolarit\`a 
coniche nello spazio quoziente. \`E sempre possibile per\`o ``riparare'' tali 
singolarit\`a sostituendo un intorno del punto singolare con una variet\`a 
liscia non compatta che abbia il corretto comportamento asintotico (tecnica 
di ``blow-up''), ma in generale solo nel limite di orbifold \`e possibile 
costruire  esplicitamente lo spettro di stringa.
\vskip .1in
La costruzione di un orbifold comporta inoltre due modifiche per lo spazio di 
Hilbert degli stati di stringa \cite{Dixon:1987qv}. Anzitutto, poich\'e punti 
dello spazio-tempo vengono identificati mediante l'azione degli elementi di
$ G$, \`e necessario 
tener conto anche delle stringhe che si chiudono (sul ricoprimento) a meno 
dell'azione di $G$, definite dalle seguenti condizioni di (quasi) 
periodicit\`a:
$$ X ( \s + 2 \pi ) = g X ( \s ) \ . $$   
Queste danno origine ai cosiddetti \emph{\bf settori twisted} dello spazio di 
Hilbert, che verr\`a quindi ampliato e suddiviso in vari settori $ H_g $, uno 
per ogni classe di coniugio di $G$, ed in ciascuno dei quali le condizioni al 
bordo dei campi bidimensionali X saranno distinte.

Inoltre lo spazio di Hilbert degli stati fisici, che descrive stringhe chiuse 
sull'orbifold, dovr\`a essere invariante sotto l'azione di tutti gli elementi 
di $G$; ogni settore $ H_g $ dovr\`a quindi essere proiettato su un
sottospazio 
$G$-invariante. In generale sappiamo che per ottenere uno stato fisico di 
stringa chiusa, si tensorizza, allo stesso livello di massa, un modo destro 
con uno sinistro. Nel caso degli orbifolds, bisogna richiedere che gli 
autovalori dei modi destri siano complessi coniugati di quelli dei modi 
destri.
 
In definitiva possiamo dire che l'effetto dell'identificazione 
sotto il gruppo 
di simmetria discreto $G$ \`e quello di aggiungere allo spazio di Hilbert 
stringhe chiuse ``twistate'' e di proiettare lo spettro su stati 
$G$-invarianti \cite{Hamidi:1987vh}.

Gli esempi pi\'u semplici di orbifolds si ottengono identificando $M$ con un 
toro, e $G$ con un gruppo di rotazioni di $ {\bf R}^d $, 
i cui elementi agiscano 
compatibilmente con il reticolo che definisce il toro.
Allora, quando $ M = T^k $ e $ G \equiv Z_N $ si parla di 
\emph{ \bf orbifolds abeliani di tori}.\\
I due orbifolds pi\`u semplici sono : $ I = S^1 / Z_2 $  \, e \, $ T = T^2 / Z_2 . $ 
\vskip .1in
L'orbifold unidimensionale $ S^1 / Z_2 $ pu\`o essere pensato come il 
risultato dell'azione combinata di due operazioni \\
- l'identificazione periodica:  \  $ X \sim X + 2 \pi R $ \, che da' 
    origine al cerchio $ S^1 $\\
- l'inversione delle coordinate (o riflessione):  \  $ X \sim - X $ \, 
    dovuta a $ Z_2 $, che identifica punti opposti sul cerchio.\\
Lo spazio compatto che ne risulta \`e un segmento I, con $ 0 \leq X \leq
  \pi R $, i cui estremi, X = 0 e $  X = \frac{1}{2} ( 2 \pi R ) $, 
sono i punti fissi  di $ Z_2 $.

La stessa procedura pu\`o essere applicata a tori $k$-dimensionali,
 utilizzando la riflessione per l'identificazione delle coordinate.

L'orbifold corrispondente sar\`a caratterizzato in questo caso da $ 2^k $ 
punti fissi. 
In particolare l'orbifold $ T^2 / Z_2 $ \`e un tetraedro, che pu\`o esere 
ottenuto a partire dal toro bidimensionale $ T^2 $ identificandone i punti 
sotto 
$ X \rightarrow - X $ (ovvero sotto una rotazione di $ \pi $ attorno 
all'origine D).
\begin{figure}[htb]\unitlength1cm
\begin{picture}(8,3)
\put(3,0){\epsfig{file=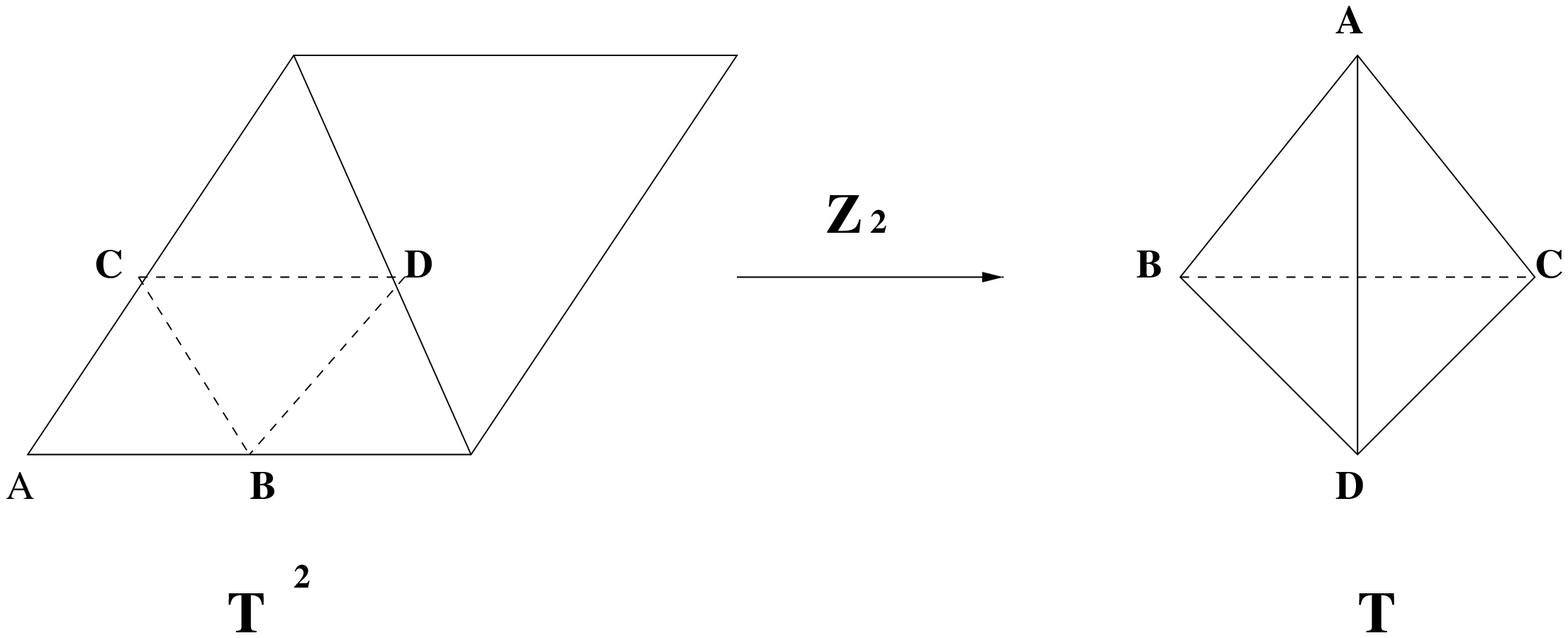,width=8cm,height=3cm}}
\end{picture}
\caption{\footnotesize{Orbifold $T^2/Z_2$}}
\label{fig:7} 
\end{figure}

A,B,C e D sono i 4 punti fissi della rotazione e coincidono con i 
vertici del tetraedro.

Ciascun punto fisso fornisce una possibile locazione dello stato fondamentale 
di un settore $twisted$
della stringa; gli stati eccitati nei settori $twisted$ dello spazio di 
Hilbert corrispondono a fluttuazioni della stringa attorno ad ognuno dei 
punti fissi.

Per comprendere meglio il ruolo di queste costruzioni, passiamo al calcolo 
della funzione di partizione per un singolo campo bosonico compattificato 
sull'intervallo $ S^1 / Z_2 $.

L'espressione generale della funzione di partizione, nel caso di 
compattificazioni su orbifolds $ T^k / Z_N $, \`e data da 
\be
\G = \frac{1}{{( 4 \pi)}^{\frac{D-2}{2}}} \int_0^{\infty} 
\frac{d t}{t^{\frac{D-k}{2}+1}} \sum_k tr_k (\mc{P} e^{-tm^2} ) ,
\ee
dove $ \mc{P} $ \`e il proiettore che seleziona il corretto spazio di Hilbert 
degli stati $G$-invarianti, e dove il pedice $k$ indica che la traccia 
\`e sul settore $k-twisted.$

L'azione della riflessione $ \mc{R} $  su un generico stato di stringa \`e 
indotta dalle seguenti trasformazioni:
\beq
\part X^i \rightarrow - \part X^i  \quad \bar{\part} X^i \rightarrow 
- \bar{\part} X^i , \nonumber \\
\part X^{\m} \rightarrow \part X^{\m} . 
\label{riflex} 
\eeq  

Iniziamo dal calcolo del contributo alla funzione di partizione del 
settore \emph{untwisted}. In questo settore lo spettro viene proiettato sugli 
stati con autovalore di $ \mc{R}$  uguali a 1 , e quindi inserendo nella 
traccia degli stati il proiettore $ \frac{1 + \mc{R}}{2} $, si ottiene 
$$
Z_{untw} = \frac{1}{2} tr ( 1 + \mc{R} ) q^{L_0 - 1/24} 
\bar{q}^{\bar{L}_0 - 1/24} \ .$$
Esplicitamente 
\be
Z_{untw} =  \frac{1}{2} Z (R) + \Big | \frac{\eta}{\th_2} \Big | 
= \frac{1}{2} \sum_{(m,n)} \frac{q^{\frac{p_L^2}{2}} 
\bar{q}^{\frac{p_R^2}{2}}}{\eta (\t) \bar{\eta} (\t) } + 
\frac{1}{2} \frac{1}{{(q \bar{q})}^{24}  {| \prod_{k=1}^{\infty} ( 1 + q^k ) |}^2} \ ,
\ee
dove si vede che $ Z (R) $ \`e l'usuale funzione di partizione sul toro. Si 
noti che, nel termine della traccia contenente $ \mc{R} $, gli elementi diagonali 
di $ \mc{R} $ devono avere $ m = n = 0 $  
(sono gli stati con $p_L = p_R = 0$), mentre le (\ref{riflex})  cambiano 
$ q \ra -q $ e $ \bar{q} \ra - \bar{q}$ in corrispondenza di 
$ \part X \rightarrow - \part X $  \ e 
$ \bar{\part} X \rightarrow - \bar{\part} X $ .

Nel settore \emph{twisted} il campo X \`e invece antiperiodico, e quindi deve 
avere un'espansione in modi seminteri del tipo
\be
X (\t, \s) = x + i \sqrt{\frac{\a'}{2}} \sum_{k \in Z} 
\Big ( \frac{\a_{k+1/2}}{k + \frac{1}{2}} e^ {- i ( k +1/2 ) ( \s + \t )} 
+ \frac{\bar{\a}_{k + 1/2}}{k + \frac{1}{2}} e^{- i ( k + 1/2 )( \s - \t )} \Big ) ,
\ee
dove $ x $ corrisponde ad uno dei due punti fissi  $ x = 0 $ o $ x = \pi R $.\\
In questo settore la funzione di partizione ha la forma seguente:
$$
Z_{tw} = \frac{1}{2} tr ( 1 + \mc{R} ) q^{L_0 + 1/48} \bar{q}^{\bar{L}_0 + 1/48} , 
$$
ovvero 
\be
Z_{tw} = \Big | \frac{\eta}{\th_4} \Big| + \Big| \frac{\eta}{\th_3} \Big| =  
\frac{{( q \bar{q})}^{1/48}}{{| \prod_{k=1}^{\infty} ( 1 - q^{k - 1/2} )|}^2} 
+ \frac{{( q \bar{q} )}^{1/48}}{{| \prod_{k=1}{\infty} ( 1 + q^{k + 1/2} ) |}^2} .\ee
Ne segue che complessivamente si avr\`a 
\be
Z_{orb} = Z_{untw} + Z_{tw} = \frac{1}{2} Z (R) + \Big | \frac{\eta}{\th_2} 
\Big | + \Big | \frac{\eta}{\th_4} \Big | + \Big | \frac{\eta}{\th_3} \Big | .
\label{Zorb}
\ee
In precedenza abbiamo gi\`a discusso il fatto che  il primo termine della 
(\ref{Zorb}) \`e invariante modulare.
Sfruttando le propriet\`a delle funzioni $ \th$-di Jacobi e della $\eta$ di 
Dedekind sotto le trasformazioni modulari $S$ e $T$, non \`e difficile 
verificare che anche la somma degli ultimi tre termini  di $ Z_{orb} $ 
\`e invariante modulare, bench\'e essi non lo siano singolarmente.

A tale proposito \`e utile  evidenziare come la richiesta di invarianza 
mo\-dulare della funzione di partizione ad un loop delle stringhe chiuse 
fornisca uno strumento per la costruzione dei modelli, in quanto la sua 
violazione implicherebbe l'inconsistenza della teoria. La (\ref{Zorb}) 
infatti  pu\`o essere ottenuta  seguendo un percorso alternativo, e per 
vederlo  cominciamo a  riscriverla nel modo seguente:
\be
\mc{T}_{orb} = \mc{T}_{untw} + \mc{T}_{tw} = ( \mc{T}_{++} + \mc{T}_{+-} ) 
+ ( \mc{T}_{-+} + \mc{T}_{--} ).
\ee
Partendo dall'ampiezza, nota, $ \mc{T}_{++} = \frac{1}{2} Z(R) $ per le 
stringhe chiuse compattificate sul toro, si vanno ad aggiungere tutti i 
contributi che servono a renderla invariante sotto l'azione di $G$, ovvero di 
$ Z_2 $ nel caso in esame; questo significa simmetrizzare rispetto a  
$ X \rightarrow - X $.

Il completamento della simmetrizzazione si ottiene da 
\be
\mc{T}_{++} = \frac{1}{2} \frac{\sum_{(m,n)} q^{\frac{p_L^2}{2}} 
\bar{q}^{\frac{p_R^2}{2}}}{ {( q \bar{q} )}^{1/24} {| \prod_{k=1}^{\infty} 
( 1 - q^k )|}^2} ,
\label{T++}
\ee
eliminando i modi-zero ed invertendo il segno degli oscillatori
$ q^k \rightarrow - q^k $ e $\bar{q}^k \ra - \bar{q}^k$

Ci\`o da' luogo all'ampiezza 
\be
\mc{T}_{+-} = \frac{1}{{(q \bar{q} )}^{1/24}  { | \prod_{k=1}^{\infty} 
( 1 + q^k ) | }^2} \ ,
\ee
ed in questo modo si ricava il settore $untwisted$.
La (\ref{T++}) \`e palesemente non invariante modulare, e quindi occorre 
sommare i termini ottenuti da $ \mc{T}_{+-} $ mediante le trasformazioni $S$ 
e $ T$.
Ad esempio $ \mc{T}_{-+} $ si ottiene da $ \mc{T}_{+-} $ operando 
una trasformazione 
$S (\t \ra - 1 / \t) $. Questo equivale a rendere i bosoni  di 
$ \mc{T}_{++} $ 
antiperiodici e ad introdurre la degenerazione dei punti fissi $( 2^2 )$ 
dell'orbifold e quindi 
\be
\mc{T}_{-+} = 2^2 \frac{ {( q \bar{q} )}^{1/48}}{{| \prod_{k=1}^{\infty} 
( 1 - q^{k - 1/2}) |}^2} \ .
\ee
Infine, la simmetrizzazione di $ \mc{T}_{-+} $ si ottiene applicando ad essa 
la trasformazione $ T : \t \rightarrow \t + 1 $ che conduce a 
\be
\mc{T}_{--} = 2^2 \frac{ {( q \bar{q} ) }^{1/48}}{{| \prod_{k=1}^{\infty} 
( 1 + q^{k - 1/2} ) |}^2} \ .
\ee
Dal punto di vista del $path-integral$ sul toro, $ \mc{T}_{++} $ deriva dai 
campi di stringa quantizzati con condizioni al bordo di tipo 
``periodico-periodico'' lungo le due direzioni, ``spaziale'' e ``temporale''. 
\begin{figure}[htb]\unitlength1cm
\begin{picture}(5,3)
\put(4,0){\epsfig{file=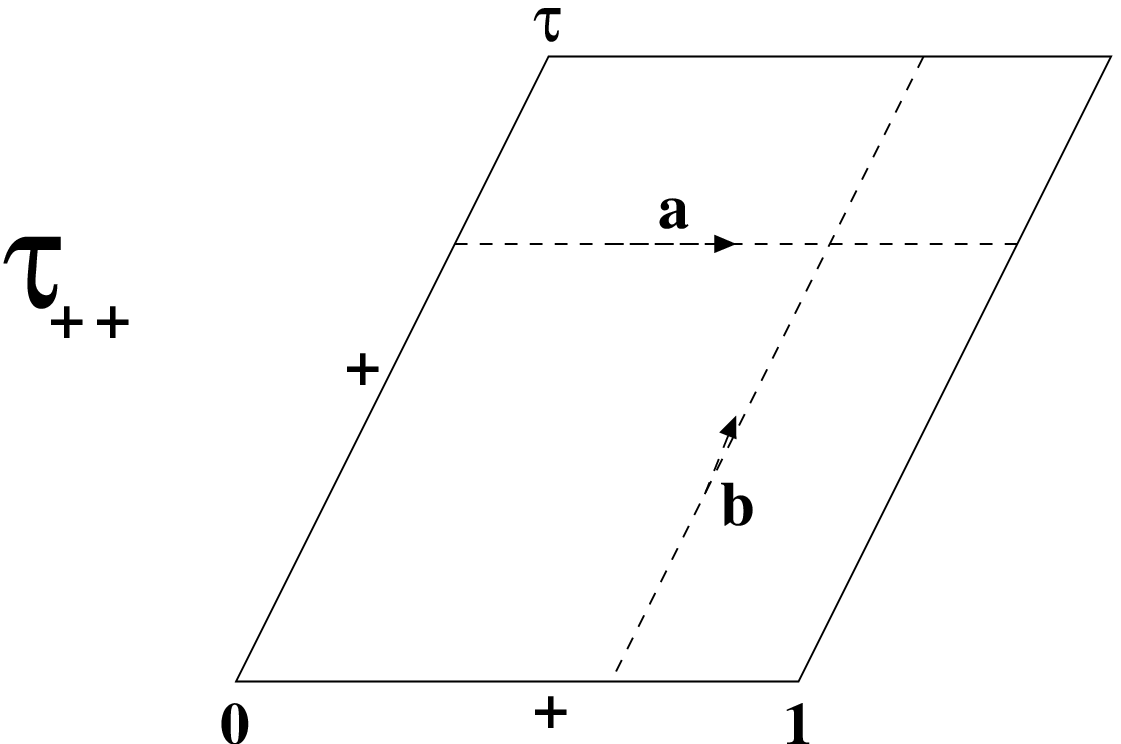,width=5cm,height=3cm}}
\end{picture}
\caption{\footnotesize{$\mc{T_{++}}$}}
\label{fig:8} 
\end{figure}

Viceversa, il termine $ \mc{T}_{+-} $, che deriva dal settore $untwisted$ con 
l'operatore $ \mc{R} $ nella traccia, \`e caratterizzato da campi periodici 
nella direzione spaziale ed antiperiodici in quella temporale.
\begin{figure}[h]\unitlength1cm
\begin{picture}(5,3)
\put(4,0){\epsfig{file=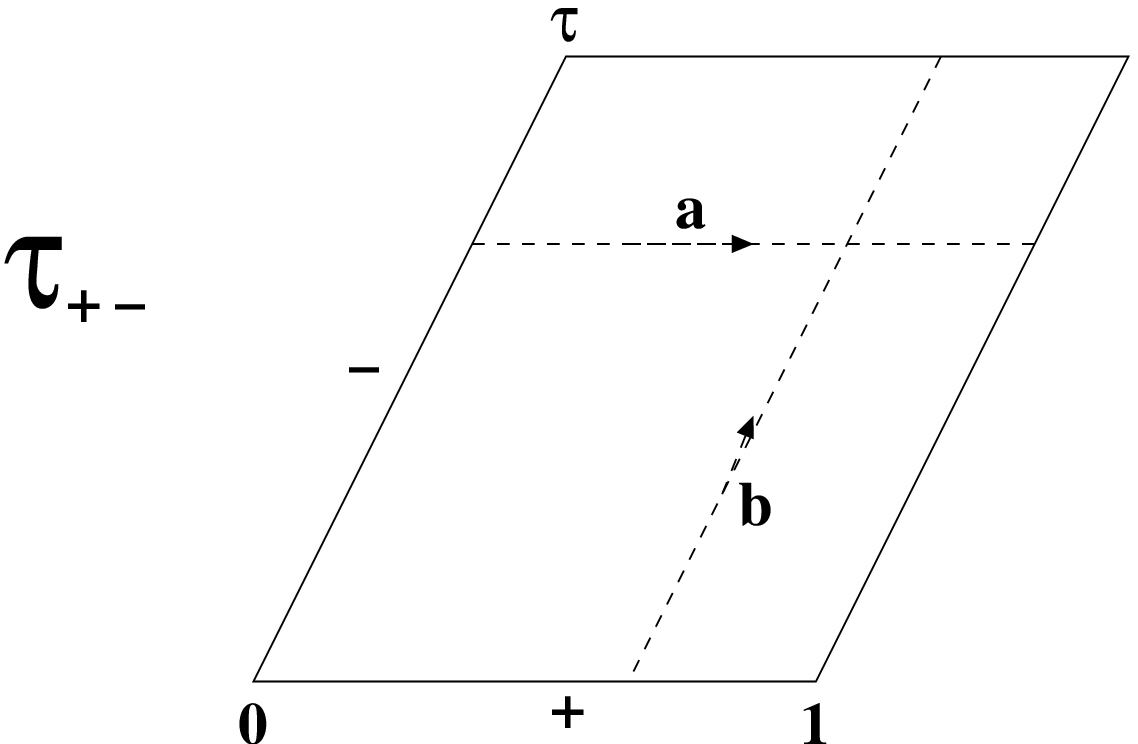,width=5cm,height=3cm}}
\end{picture}
\caption{\footnotesize{$\mc{T_{+-}}$}}
\label{fig:9} 
\end{figure}

$ \mc{T}_{-+} $ , che deriva dal settore $twisted$ con 1 nella traccia, ha 
condizioni al bordo opposte rispetto a $ \mc{T}_{+-} $ e la trasformazione 
modulare $S$ connette questi due termini.

Infine,  il contributo  $ \mc{T}_{--} $ del settore $twisted$ con $ \mc{R} $ 
nella traccia, che deriva dai campi con condizioni al bordo di tipo 
``antiperiodico-antiperiodico'' sul toro, pu\`o essere generato operando una
trasformazione $T (\t \ra \t+1)$ su $ \mc{T}_{-+}. $

\begin{figure}[h]\unitlength1cm
\begin{picture}(5,3)
\put(4,0){\epsfig{file=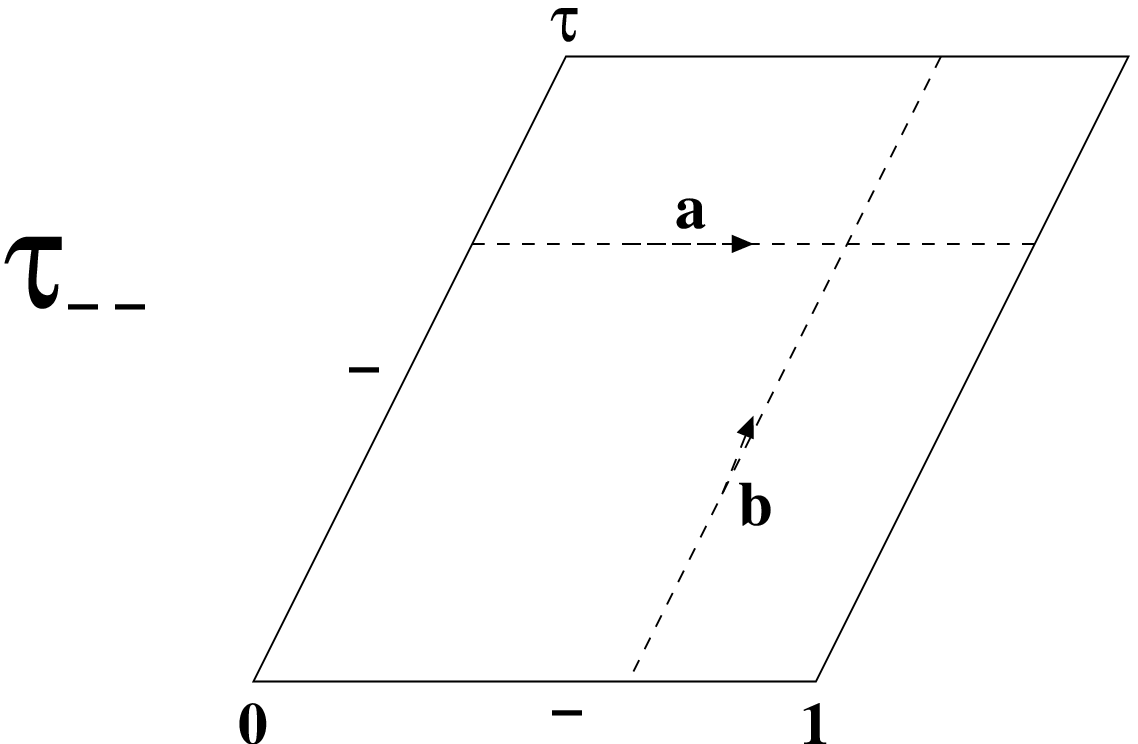,width=5cm,height=3cm}}
\end{picture}
\caption{\footnotesize{$\mc{T_{--}}$}}
\label{fig:10} 
\end{figure}

\begin{figure}[h]\unitlength1cm
\begin{picture}(10,7)
\put(2,0){\epsfig{file=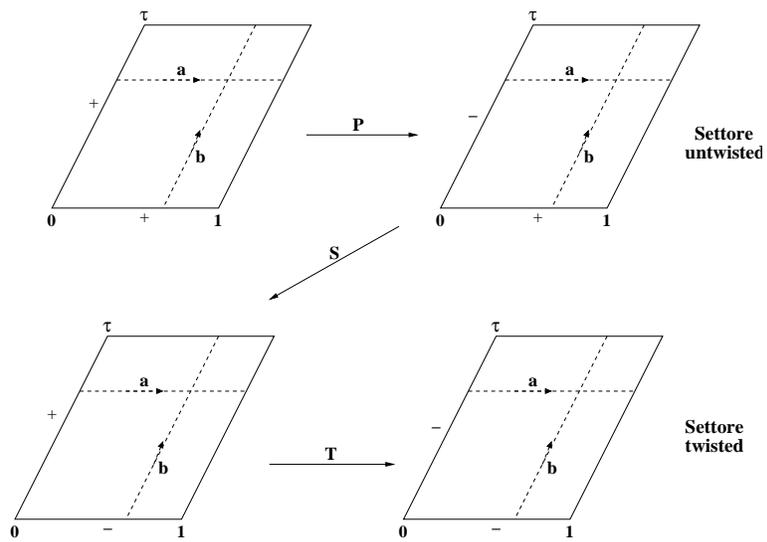,width=10cm,height=7cm}}
\end{picture}
\caption{\footnotesize{Settore Twisted + Settore Untwisted}}
\label{fig:11} 
\end{figure}

In definitiva la situazione pu\`o essere schematizata come in figura
(\ref{fig:11}), che illustra come i settori $twisted$ siano necessari 
per l'invarianza modulare.
\chapter{Stringhe aperte}
\section{ Gruppi di simmetria nelle stringhe aperte (Metodo di Chan-Paton)}
Nel capitolo precedente abbiamo illustrato come sia possibile introdurre, 
nelle teorie di stringhe chiuse, dei gruppi di simmetria interna mediante 
compattificazioni su variet\`a pi\`u o meno complesse, anche se le simmetrie 
di gauge non sono affatto presenti nella stringa bosonica a $D=26$. 
In realt\`a esiste una diversa e pi\`u naturale modalit\`a di introduzione 
dei gruppi di gauge, fornita dalle stringhe aperte, basata sull'assunzione che le 
simmetrie interne siano gi\`a presenti a 10 dimensioni.
Sappiamo che le stringhe aperte, a differenza di quelle chiuse, sono 
caratterizzate da due punti molto speciali, i loro estremi. 
\`E quindi naturale assumere che esse presentino, oltre ai campi 
che si propagano nello spazio 
Minkowskiano ($bulk$), degli ulteriori gradi di libert\`a, non dinamici, 
associati ai loro estremi: questi possono essere associati alle cariche di 
un gruppo di simmetria interna, note in letteratura come 
\emph{\bf cariche di Chan-Paton}.
In base ad alcune propriet\`a delle ampiezze di scatte\-ring ad albero, 
nel 1969 Chan e Paton proposero una tecnica per associare i gruppi
di gauge $U(2)$ ed $U(3)$
al settore aperto della teoria di stringa boso\-nica \cite{Paton:1969je}. 
Una caratteristica fondamentale delle ampiezze di stringa aperta, per N bosoni
esterni identici, \`e la loro simmetria ciclica (``nelle gambe e\-sterne'').
Chan e Paton osservarono che tale propriet\`a rimaneva inalterata se le 
ampiezze venivano moltiplicate per tracce di prodotti di matrici a valori 
nell'algebra $u(2)$ e $u(3)$.
Successivamente Schwarz not\`o che tali ampiezze dovevano soddisfare
una serie infinita di vincoli, per poter essere compatibili con
l'unitariet\`a della teoria \cite{Schwarz:1982md}.
Questi vincoli (di unitariet\`a), risolti da Marcus e Sagnotti nel 1982,
hanno portato a concludere che con il metodo di Chan-Paton possono essere
incorporate tutte le algebre di Lie classiche $ so(n)$, $sp(2n)$ ed $u(n)$, 
ma non le algebre eccezionali \cite{Marcus:1982fr}.
Tali risultati possono essere riassunti dicendo che i due estremi delle 
stringhe aperte devono avere valori nelle rappresentazioni fondamentali dei 
gruppi classici. Quindi per $SO(n)$ e $Sp(2n)$ si avranno stringhe aperte non 
orientate, in quanto le corrispondenti rappresentazioni sono reali e 
pseudo-reali; viceversa per $U(n)$, la cui rappresentazione fondamentale \`e 
complessa ($ n \neq \bar{n} $), le stringhe aperte saranno genericamente 
orientate.
I gruppi di Chan-Paton classici e le corrispondenti rappresentazioni possono 
essere ottenuti anche dalla dinamica di ulteriori gradi di libert\`a posti 
sui bordi della stringa aperta \cite{Marcus:1987cm}. 
Per far questo si pu\`o aggiungere per esempio un 
numero (pari) $n$ di fermioni liberi unidimensionali $ \p^I $, con azione di 
Dirac 
\be
S = \frac{i}{4} \int_{\part \S} ds \p^I (s) \frac{\part \p_I(s)}{\part s} \ ,
\ee
dove $ \part \S $ \`e il bordo del \emph{world-sheet}.

In particolare, $n$ fermioni liberi danno origine a rappresentazioni di 
dimensione $ 2^{n/2} $.
Associare delle cariche agli estremi delle stringhe aperte \`e rilevante 
ovviamente solo nei modelli che li contengono. A tale proposito, ricordiamo 
che tra le teorie supersimmetriche, l'unica teoria contenente stringhe 
aperte \`e la superstringa di Tipo I. A livello ad albero, una qualsiasi 
scelta del gruppo di simmetria \`e consistente per una teoria di stringhe 
aperte interagenti.
Al livello quantistico, le condizioni di consistenza richiedono invece
$SO(32)$ come unica possibile scelta del gruppo di gauge. Quindi la 
Tipo I-$SO(32)$\`e l'unica teoria di superstringa  consistente 
con la quantizzazione, e 
contiene  stringhe chiuse ed aperte non orientate in interazione.
La presenza del settore chiuso \`e conseguenza della richiesta di
unitariet\`a della teoria, in base alla quale teorie di sole stringhe
aperte non sono possibili
(intuitivamente \`e sempre possibile ottenere una stringa chiusa dall'unione
di una o due stringhe aperte, e quindi la propagazione di stringhe chiuse 
deve essere inclusa per consistenza). Se ne evince che nel modello di Tipo I, 
includere numeri quantici di gauge con il metodo di Chan-Paton significa 
avere stringhe chiuse neutre sotto il gruppo di Chan-Paton.
\section {Discendenti Aperti}
La formulazione perturbativa dei modelli aperti risulta molto pi\`u complessa 
rispetto a quella dei modelli chiusi. Questo \`e dovuto alla presenza, 
nel loro spettro, di settori chiusi ed aperti non orientati. 
Come conseguenza, la loro espansione di Polyakov include una somma su 
superfici di Riemann con bordi e 
non orientabili, ovvero su superfici caratterizzate da un numero variabile di 
due nuove strutture: i \emph{\bf bordi} ed i \emph{\bf crosscaps}.
Questi ultimi possono essere visualizzati, a partire da una sfera, nella 
maniera seguente:\\
(a) \ identificando coppie di punti di latitudine opposta si
ottiene un disco, e l'equatore, ovvero la linea di punti fissi della 
costruzione, \`e  il suo \emph{bordo};\\
(b) \ identificando punti antipodali,  non si ottengono punti 
fissi, ma i punti equatoriali opposti devono essere identificati. A tale 
linea di punti si d\`a il nome di \emph{crosscap} e la superficie 
risultante, chiusa e non orientabile, \`e il piano reale proiettivo $ RP^2 $. 
La caratteristica di Eulero per una superficie Riemann con $ h $ manici, 
$ b $ bordi e $ c $ crosspacs \`e
$$ \c = 2 - 2 h - b - c  , $$
e quindi, a differenza dei modelli di stringhe chiuse orientate, in questo 
caso la serie perturbativa contiene potenze di $ g_s = e^{\ket\f\bra} $ sia 
pari che dispari. A genere 1/2 $(\c = 1)$, per esempio, abbiamo il disco ed 
il piano reale proiettivo. A genere 1 $ (\c = 0) $ invece, si hanno quattro 
possibilit\`a, poich\'e $ \c = 0 $ pu\`o essere ottenuto, oltre che dal toro, 
da altre tre superfici: 
la \emph{\bf bottiglia di Klein} $ (h = 0 = b ; \,  c = 2) $, 
l'\emph{\bf anello} $(h = 0 = c ; \, b = 2)$, e la 
\emph{\bf striscia di M\"obius} $ (h = 0 ; \, b = 1 = c)$. 
Come abbiamo fatto vedere per il toro, queste superfici possono essere 
associate ad un reticolo del piano complesso, mediante un numero opportuno di 
tagli.

Iniziamo dalla bottiglia di Klein. 
Questa pu\`o essere descritta dalla cella rettangolare $ (1 , i \t_2)$ con 
lati orizzontali identificati dopo un'inversione.

\begin{figure}[htb]\unitlength1cm
\begin{picture}(7,3)
\put(4,0){\epsfig{file=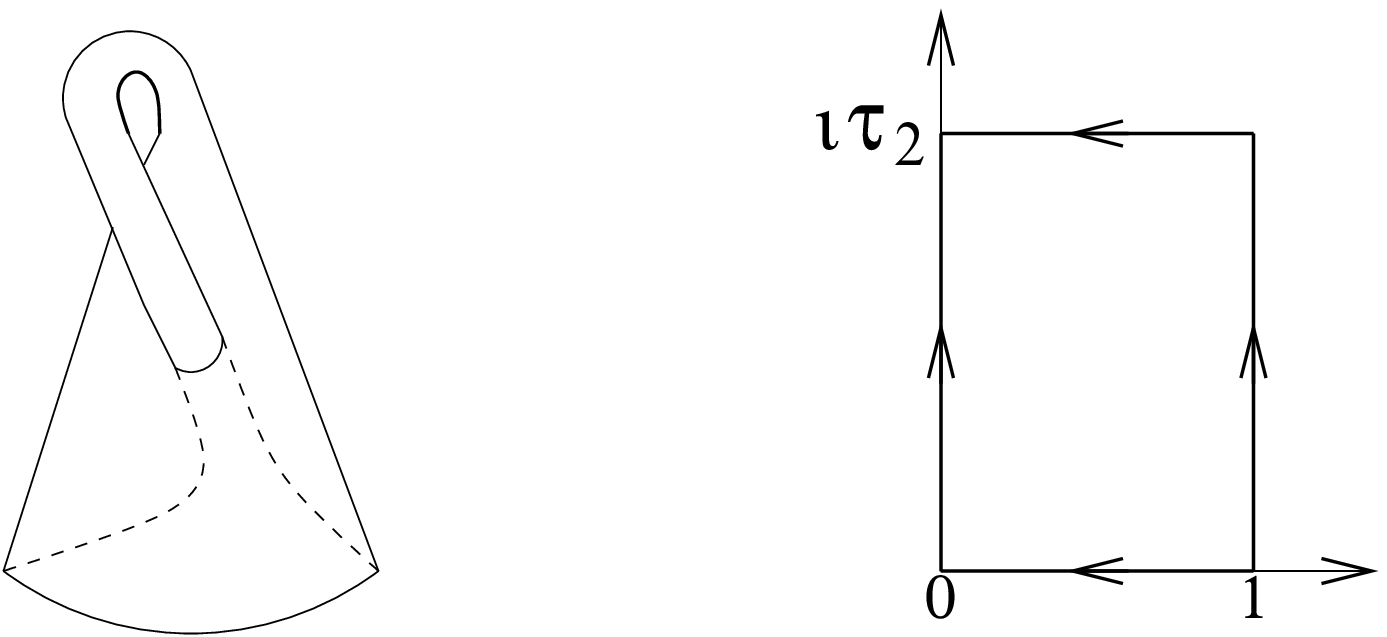,width=7cm,height=3cm}}
\end{picture}
\caption{\footnotesize{Bottiglia di Klein}}
\label{fig:12} 
\end{figure}

Osserviamo che, come per il toro, $ \t_2 $ ha un preciso significato fisico, 
in quanto rappresenta il ``tempo proprio'' impiegato dalla stringa chiusa per 
spazzare la bottiglia di Klein. In realt\`a esiste una seconda scelta per il
poligono fondamentale, che definisce una scelta distinta del ``tempo proprio''.
Raddoppiando verticalmente la bottiglia di Klein si ottiene un toro di mo\-dulo
$ \t = 2 i \t_2 $. Dimezzando i lati orizzontali in modo da mantenere l'area 
inalterata, si ottiene una rappresentazione equivalente della superficie, 
come un tubo terminante su due crosscaps.

\begin{figure}[htb]\unitlength1cm
\begin{picture}(6,3)
\put(4,0){\epsfig{file=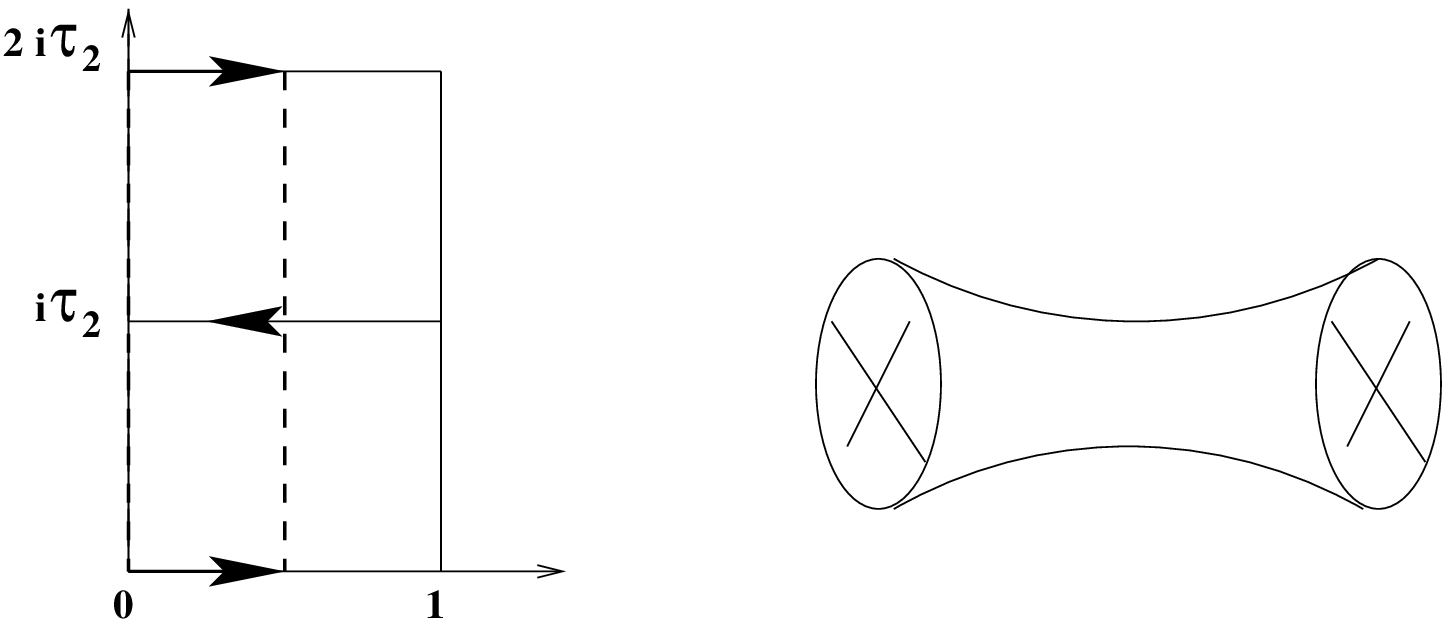,width=6cm,height=3cm}}
\end{picture}
\end{figure}

In questo caso il tempo orizzontale rappresenta il ``tempo proprio'' 
im\-piegato
dalla stringa chiusa per propagarsi tra i due crosscaps.

Passiamo al settore aperto, iniziando dall'anello.
Il suo poligono fondamentale \`e il rettangolo $(1 , i \t_2 )$, in cui i lati 
orizzontali sono identificati, mentre quelli verticali corrispondono ai due 
bordi.

\begin{figure}[htb]\unitlength1cm
\begin{picture}(7,3)
\put(4,0){\epsfig{file=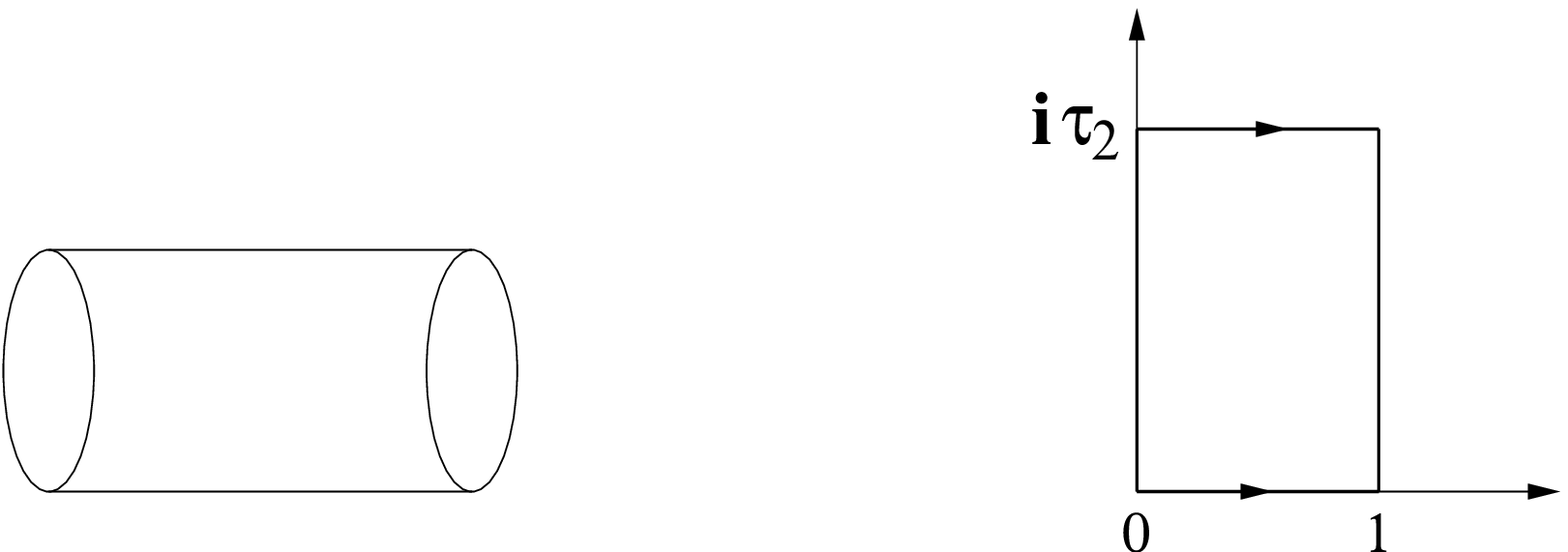,width=7cm,height=3cm}}
\end{picture}
\caption{\footnotesize{Anello}}
\label{fig:13} 
\end{figure}

In questo caso $ \t_2 $ rappresenta il tempo proprio impiegato dalla stringa 
aperta per spazzare l'anello, e come per la bottiglia di Klein, esiste una 
seconda scelta per la cella fondamentale: basta raddoppiare orizzontalmente 
l'anello, ottenendo cos\`\i \ un toro di modulo $ \t = i \t_2 /2 $, per poi 
dimezzarlo verticalmente. La superficie che ne risulta \`e un tubo terminante 
su due bordi, e quindi il tempo orizzontale definisce il tempo proprio 
impiegato dalla stringa chiusa per propagarsi tra due bordi.

\begin{figure}[htb]\unitlength1cm
\begin{picture}(6,3)
\put(4,0){\epsfig{file=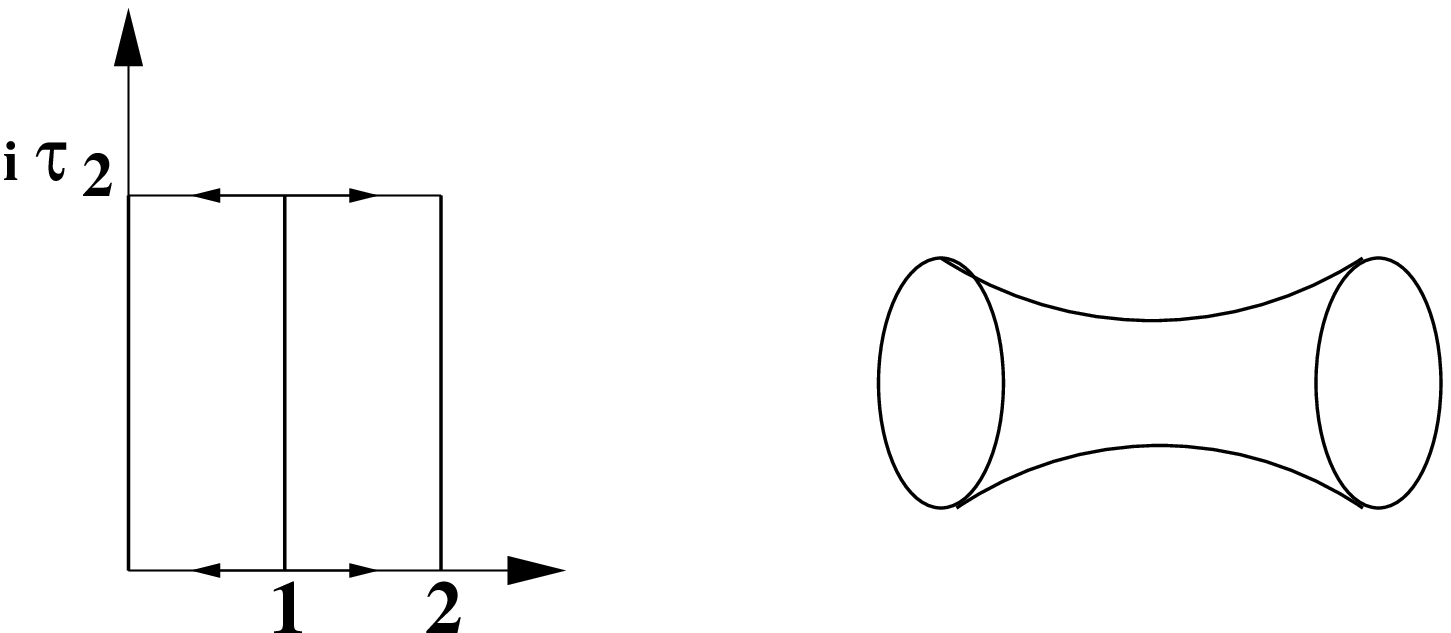,width=6cm,height=3cm}}
\end{picture}
\end{figure}

Infine, la striscia di M\"obius ha come poligono fondamentale 
il rettangolo $(1 , i \t_2)$, in cui i lati orizzontali hanno orientazione 
opposta, mentre quelli verticali sono le due met\`a di un singolo bordo.

\begin{figure}[htb]\unitlength1cm
\begin{picture}(8,3)
\put(3,0){\epsfig{file=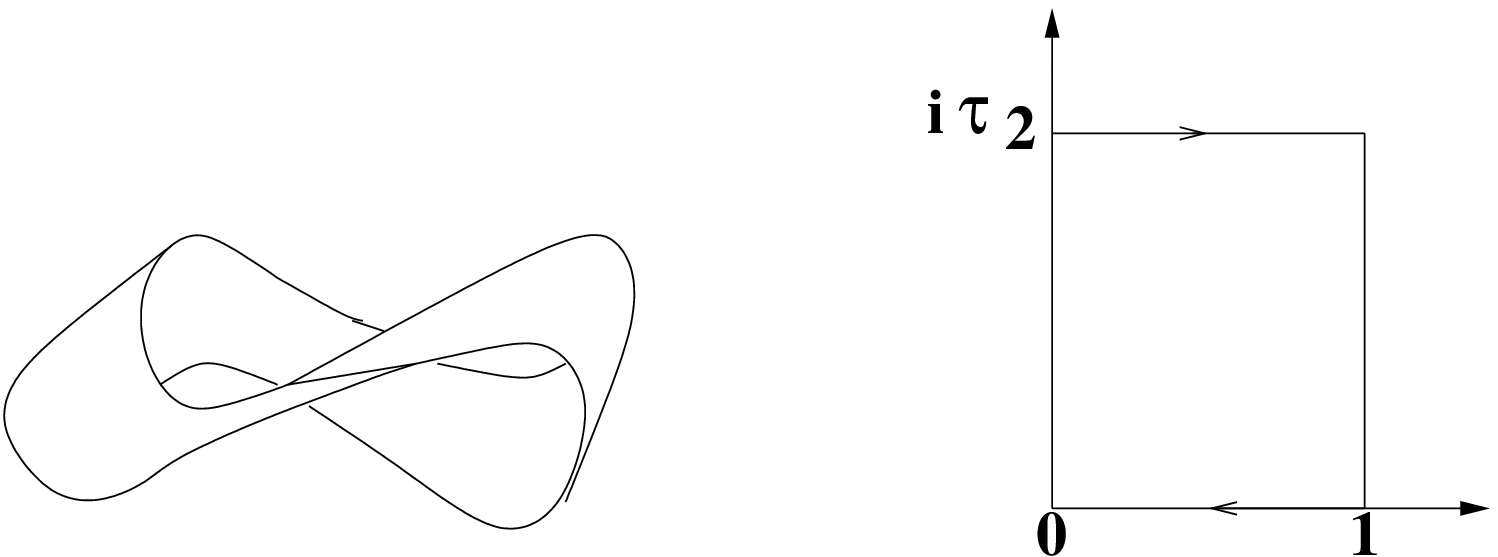,width=8cm,height=3cm}}
\end{picture}
\caption{\footnotesize{Striscia di M\"obius}}
\label{fig:14} 
\end{figure}

Anche in questo caso $\t_2$ descrive il tempo proprio impiegato dalla stringa 
aperta per spazzare la striscia di M\"obius, e di nuovo si pu\`o scegliere
un differente poligono fondamentale, in grado di fornire una rappresentazione 
equivalente della superficie come un cilindro terminante su un bordo e su un
crosscap. Questo si ottiene semplicemente raddoppiando in verticale e 
dimezzando in orizzontale, ed  in questo modo uno dei due lati verticali \`e 
l'intero bordo della striscia di M\"obius, mentre l'altro \`e un 
crosscap. Il corrispondente tempo orizzontale definisce il tempo proprio di
propagazione di una stringa chiusa  tra un bordo ed un crosscap. 
\`E interessante 
osservare che in questo caso il raddoppiamento verticale della striscia di 
M\"obius non definisce un toro doppiamente ricoprente. Tuttavia 
quest'ultimo esiste, ed \`e caratterizzato da un modulo che non \`e puramente 
immaginario, ma dato da $ \t = \frac{1}{2} + i \frac{\t_2}{2} $, 
ovvero \`e descritto  da una cella reticolare obliqua con vertici a $0,2 $
e a $ (i \t_2 + 1)$.

\begin{figure}[htb]\unitlength1cm
\begin{picture}(7,4)
\put(3,0){\epsfig{file=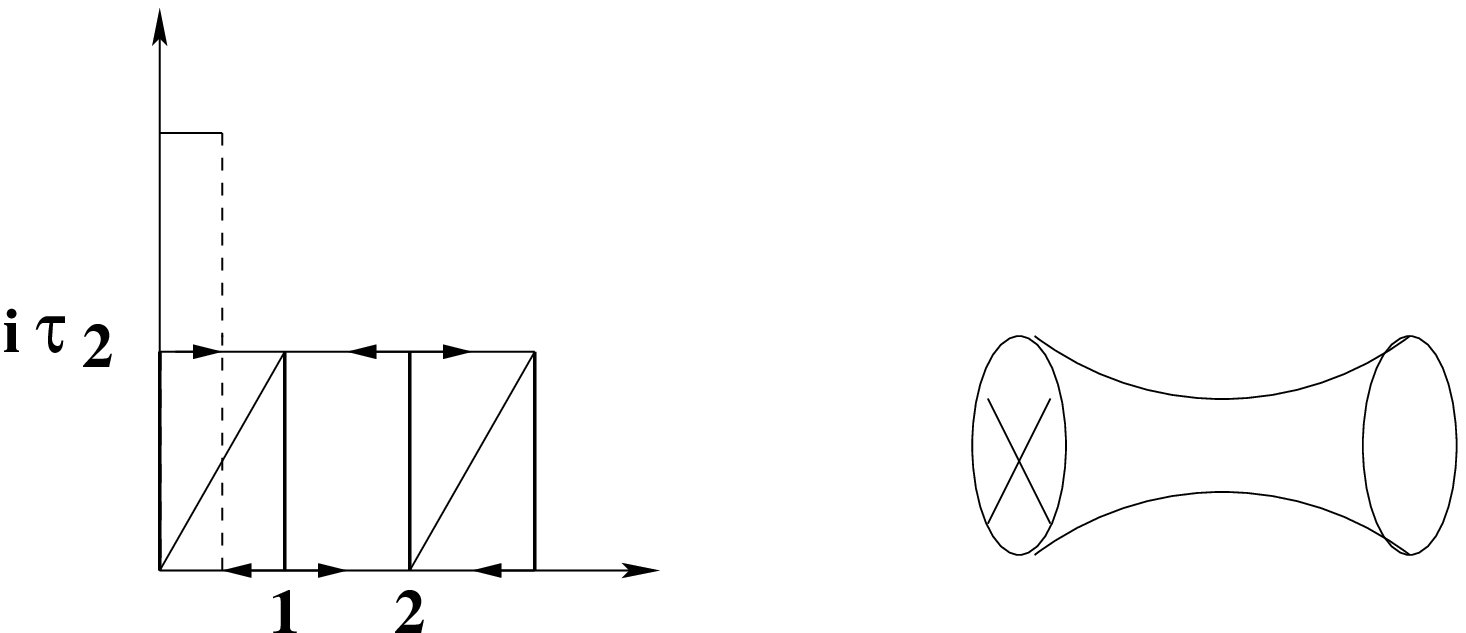,width=7cm,height=4cm}}
\end{picture}
\end{figure}

Il passaggio dal tempo verticale a quello orizzontale, nella striscia di 
M\"obius, viene effettuato mediante la cosiddetta trasformazione 
modulare P
$$ P : \quad \frac{1}{2} + i \frac{\t_2}{2} \rightarrow \frac{1}{2} + 
i \frac{1}{2 \t_2}. $$
$P$ pu\`o essere ottenuta tramite una sequenza di trasformazioni modulari 
$S$ e $T$, ossia $ P = (T S) T (T S) $, che, come conseguenza 
della relazione  $ S^2 = {(S T )}^3 $, soddisfa  $ P^2 = S^2 = {(S T)}^3 $.
Dall'analisi svolta fin qui si pu\`o dedurre che superfici con bordi e 
crosscaps ammettono come ricoprimenti doppi superfici chiuse orientate. Si
pu\`o mostrare infatti che la bottiglia di Klein, l'anello e la striscia di 
M\"obius possono essere definite come quozienti ({\it cosets}), 
a partire dai loro ricoprimenti doppi, mediante opportune involuzioni 
\cite{Alessandrini:1971cz}.
La bottiglia di Klein, per esempio, pu\`o essere descritta partendo da un
toro, con modulo $ 2 i \t_2 $ e definito dai generatori 
$ K_1 (\s_1 , \s_2) = ( \s_1 + 1 , \s_2 ) $ e 
$ K_2 (\s_1 , \s_2) = (\s_1 , \s_2 + 2 \t_2) $, mediante 
l'involuzione $ K_3 (\s_1 , \s_2) = ( 1 - \s_1 , \s_2 + \t_2) $ che non ha 
punti fissi e squadra alla traslazione verticale $ K_2 
$ (dove $(\s_1 , \s_2) = \s_1 + i \s_2 $).\\
L'anello si pu\`o ottenere dal toro di modulo $ \frac{i \t_2}{2} $, definito 
dalle traslazioni $ A_1 (\s_1 , \s_2) = (\s_1 + 1 , \s_2)$ e 
$ A_2 (\s_1 , \s_2) = (\s_1 , \s_2 + \t_2)$, tramite l'identificazione 
$ A_3 (\s_1 , \s_2) = 
 (2 - \s_1 , \s_2)$, i cui punti fissi sono i bordi dell'anello $ \s_1 = 0 $
e $ \s_1 = 1 $. \\
Infine la striscia di M\"obius si pu\`o ottenere dal toro di modulo 
$ \t = \frac{1}{2} + i \frac{\t_2}{2}$, definito dai generatori delle 
traslazioni $ M_1 (\s_1 , \s_2) = (\s_1 + 1 , \s_2 + \t_2) $ e 
$ M_2 (\s_1 , \s_2) = (\s_1 , \s_2 + 2 \t_2) $, mediante l'involuzione 
$ M_3 (\s_1 ,\s_2) = 
(1 - \s_1 , \s_2 + \t_2) $.
Ne segue che queste superfici di Riemann a genere 1 sono diversi 
orbifolds $Z_2$ del toro, dove lo $Z_2$ \`e antinconforme.

Come abbiamo avuto modo di apprezzare in precedenza, le regole di base che 
determinano gli spettri perturbativi delle stringhe chiuse orientate sono 
dovute essenzialmente ai due vincoli sulle loro ampiezze di vuoto:\\
(a) \, l'invarianza modulare, che rende irrilevante 
la scelta del ``tempo'' sul \emph{world-sheet}; \\
(b) \, la relazione tra spin e statistica, che fissa i segni relativi 
dei contributi bosonici e fermionici.

In particolare, l'invarianza modulare della funzione di partizione 
costituisce un principio costruttivo per i modelli di stringhe chiuse 
orientate, assicurandone anche l'unitariet\`a. 
La situazione per i modelli di stringhe aperte \`e assai pi\`u complessa. 
La presenza di superfici con bordi e crosscaps, infatti, non consente di 
ravvisare immediatamente un concetto equivalente all'invarianza modulare. 
Vedremo in seguito che anche  cancellazioni di anomalie o di divergenze, 
se presenti, sono uno dei criteri di consistenza per i modelli aperti. 
D'altra parte, la descrizione delle superfici di Riemann con bordi e crosscaps
 in termini di orbifolds di tori ha ispirato lo sviluppo di una tecnica ben 
precisa per la costruzione perturbativa dei modelli di stringhe aperte non 
orientate: \emph{\bf il metodo dei discendenti aperti} (noti anche con il 
nome di \emph{orientifolds}).

Esso si fonda sull'ipotesi, suggerita da Sagnotti nel 1987, che i modelli di 
stringhe aperte siano \emph{orbifolds nello spazio dei parametri} (o 
\emph{discendenti}) dei corrispondenti modelli chiusi con spettro simmetrico 
sotto lo scambio dei modi destri e sinistri \cite{Sagnotti:1987tw}. 
L'operazione di simmetria rispetto alla quale la teoria chiusa viene 
proiettata \`e la parit\`a sul \emph{world-sheet} 
$ \O  :\,  \s \rightarrow \pi - \s $, che  nelle
 teorie di stringhe chiuse scambia i modi sinistri e destri  
($ \a_m^{\m} \leftrightarrow \bar{\a}_m^{\m} $), mentre nelle stringhe 
aperte  inverte gli estremi ($ \a_m^{\m} \leftrightarrow (-)^m \a_m^{\m}$).
Il metodo dei discendenti aperti, oltre ad aver permesso di reinterpretare i 
modelli noti a 26 e 10 dimensioni come orbifolds dei corrispondenti modelli 
chiusi orientati, si \`e evoluto in un  algoritmo capace di 
supplementare le numerose teorie di stringhe chiuse orientate, 
 fornendo anche nuove descrizioni per alcune  classi di modelli di 
(super)gravit\`a.
In una serie di esempi vedremo  come questo algoritmo sia in grado di 
associare, ad opportuni spettri chiusi, addizionali spettri aperti con 
definiti esempi di rottura della simmetria interna di Chan-Paton. 

La costruzione dei modelli mediante la proiezione di orientifold segue una 
strategia analoga a quella descritta per gli $orbifolds$. Il punto di partenza 
\`e la definizione del settore ``untwisted'', che in questo caso consiste del
settore chiuso non orientato. La proiezione dello spettro chiuso viene 
effettuata dalla bottiglia di Klein. Successivamente occorre introdurre i 
settori ``twisted'', identificati con i settori aperti, determinati 
dall'ampiezza di anello, e quindi proiettati dalla striscia di M\"obius.

\section{La stringa bosonica in D=26}
Allo scopo di rendere pi\`u chiara la procedura, passiamo ad illustrare gli 
esempi pi\`u semplici di discendenti aperti, iniziando dalla stringa 
bosonica \cite{Bianchi:1990yu}. 
\`E opportuno sottolineare che l'ingrediente di base per la 
costruzione degli $orientifolds$ \`e una teoria di strighe chiuse orientate 
\emph{simmetrica} sotto l'operazione di parit\`a sul \emph{world-sheet} $\O$,
o sotto una sua opportuna genera\-lizzazione.
La stringa bosonica rispetta questa propriet\`a, ed in
particolare le eccitazioni di energia pi\`u bassa del suo spettro 
$(T ; \f , g_{\m\n} , B_{\m\n})$ hanno
simmetria definita sotto lo scambio $ \a_m^i \leftrightarrow \bar{\a}_m^i $.
Ricordiamo che l'ampiezza di toro per la stringa bosonica \`e
$$ \mc{T} = \int_{\mc{F}} \frac{d^2 \t}{\t_2^2} \frac{1}{\t_2^{12}} \frac{1}
{{| \eta (\t) |}^{48}}  \  .$$
Considerare stringhe chiuse non orientate significa proiettare lo spazio di 
Fock nel sottoinsieme di stati invarianti sotto $ \O $. Questo \`e quanto
si ottiene con la bottiglia di Klein che, completando la simmetrizzazione 
dell'ampiezza di toro, elimina dallo spettro a massa 
nulla il tensore antisimmetrico di rango due $B_{\m\n} $. 
Il livello a massa nulla, contenente originariamente ${(24)}^2$ stati, dopo 
la proiezione, dovr\`a contenere solamente $ \frac{24 (24 + 1)}{2} $ stati, 
mentre il tachione  sar\`a ancora presente nello spettro. 

L'ampiezza di bottiglia di Klein descrive il diagramma di vuoto di una stringa
 chiusa che subisce un'inversione nell'orientazione. Operatorialmente equivale
a calcolare 
\be
\mc{K} = \frac{1}{2} \int_{\mc{F}_k} \frac{d^2 \t}{\t_2^2} \frac{1}{\t_2^{12}}
 tr ( q^{N - 1} \bar{q}^{\bar{N} - 1} \O ) \, ,
\label{K}
\ee
dove $\O$ \`e l'operatore che agisce sugli stati nel modo seguente:
\be
\O | L , R \bra = | R , L \bra.
\ee  
Quindi nella (\ref{K}) 
\be
tr ( q^{N - 1} \bar{q}^{\bar{N} - 1} \O ) = \sum_{L,R} \ket L,R| q^{N -1} 
\bar{q}^{\bar{N} - 1} \O |L,R \bra = \sum_{L} \ket L,L| 
{( q \bar{q} )}^{N - 1} |L,L \bra  \ .
\label{tr}
\ee
Si noti che il prodotto scalare restringe $\sum_{L,R}$ al sottoinsieme 
diagonale $ |L,L\bra $, conducendo cos\'\i \ all'identificazione dei modi 
destri e sinistri $(N $ ed $ \bar{N})$.
Calcolando la traccia (\ref{tr}) in maniera analoga a quanto fatto per 
l'ampiezza del toro,  si ottiene quindi
\be
\mc{K} = \frac{1}{2} \int_0^{\infty} \frac{d \t_2}{\t_2^{14}} \frac{1}{{\eta 
( 2 i \t_2 )}^{24}}  \, .
\label{Kb}
\ee

Osserviamo che l'ampiezza risultante dipende da $ 2 i \t_2 $ che, come abbiamo
visto, \`e il modulo (puramente immaginario) del toro che ricopre doppiamente
la bottiglia di Klein. Inoltre la presenza di $\O$ rompe l'invarianza 
modulare, e quindi la regione d'integrazione di $\mc{K}$ diviene l'intero asse
immaginario del piano $\t$.
Per controllare che $\mc{K}$ descriva la corretta simmetrizzazione degli 
stati,
\`e sufficiente controllare, nelle espansioni in $q$ degli integrandi di 
$\mc{T}$ e $\mc{K}$,  i termini che compaiono nella combinazione 
$ ( q \bar{q} )$ (questi corrispondono agli stati fisici \emph{on-shell} 
che soddisfano le condizioni di raccordo dei livelli) 
$$ \mc{T} \rightarrow \, ( {(q \bar{q})}^{-1} + {(24)}^2 + \cdots ) \ , $$
$$ \mc{K} \rightarrow \, \frac{1}{2} ( {(q \bar{q})}^{-1} + (24) + 
\cdots)  \ .$$
Si vede che, per tener conto delle molteplicit\`a degli stati nello spettro
proiettato, occorre dimezzare il contributo del toro (che conta lo spettro 
chiuso di partenza) sommando ad esso l'ampiezza della bottiglia di Klein 
(che conta tutti gli stati fissi sotto $\O$). Quindi 
$$ \mc{T} \rightarrow \, \frac{1}{2} \mc{T} + \mc{K} \  ,$$
da' il giusto conteggio degli stati del settore ``untwisted''.
\`E possibile riferire l'ampiezza (\ref{Kb}) al suo ricoprimento doppio, 
effettuando la seguente ridefini\-zione della variabile d'integrazione: 
$ t = 2 \t_2 $. In questo modo la (\ref{Kb}) diviene 
\be
\mc{K} = \frac{2^{13}}{2} \int_0^{\infty} \frac{d t}{t^{14}} \frac{1}
{{\eta ( i \t )}^{24}} \ .
\label{Kbose}
\ee
Ma abbiamo visto che la  bottiglia di Klein consente
due scelte naturali di ``tempo proprio'':\\
1) \ il tempo ``verticale'', $\t_2$, che porta alla definizione operatoriale 
della traccia e quindi all'ampiezza di vuoto ad un loop (\ref{Kb}), detta 
anche ampiezza  nel \emph{canale diretto};\\
2) \ il tempo ``orizzontale'', $ l $ , che rende la bottiglia di Klein 
equivalente ad un tubo che termina con due crosscaps, e che quindi descrive
la propagazione di una stringa chiusa tra due crosscaps. L'ampiezza 
corrispondente, al livello ad albero in questo caso, \`e detta ampiezza nel 
\emph{canale trasverso}.

Il passaggio dal canale diretto al canale trasverso si ottiene applicando 
alla (\ref{Kbose}) la trasformazione modulare $ S : \, t \rightarrow 1/l $. 
L'espressione risultante, che indichiamo con $ \tld{\mc{K}}$, \`e 
\be
\tld{\mc{K}} = \frac{2^{13}}{2} \int_0^{\infty} d l 
\frac{1}{{\eta ( i l )}^{24}} \, .
\label{Kbtil}
\ee
Il passo successivo consiste nella costruzione del settore aperto 
(``twisted''). In questo caso la funzione di partizione a genere 1 
si ottiene dal calcolo
del determinante dell'operatore D'Alambertiano  su un \emph{world-sheet} 
che ha la topologia di un anello. Quindi l'ampiezza ad un loop nel canale 
diretto, definita in termini della traccia su stati di stringa aperta, 
\`e data da 
\be
\mc{A} = \frac{\mc{N}^2}{2} \int_0^{\infty} \frac{d \t_2}{\t_2^{14}} tr ( q^{\frac{1}{2} ( N - 1 )} ) \, ,
\label{A}
\ee
dove il prefattore $\mc{N}^2$, la molteplicit\`a associata agli estremi della 
stringa,  tiene conto della simmetria interna di Chan-Paton. Inoltre 
l'esponente in (\ref{A}) risulta dimezzato rispetto al caso chiuso, poich\'e 
la condizione di \emph{mass-shell} per la stringa bosonica aperta \`e data da 
$$ m^2 = \frac{1}{\a'} ( N - 1 ). $$
L'usuale calcolo della traccia conduce quindi all'espressione 
\be
\mc{A} = \frac{\mc{N}^2}{2} \int_0^{\infty} \frac{d \t_2}{\t_2^{14}} 
\frac{1}{{\eta ( i \frac{\t_2}{2} )}^{24}} \ ,
\label{Ab}
\ee
e, di nuovo, osserviamo che l'ampiezza  \`e espressa in termini del 
modulo (puramente immaginario) del toro che ricopre doppiamente la superficie,
 $ \frac{i \t_2}{2} $. Come per la bottiglia di Klein, \`e conveniente usare 
la ridefinizione della variabile d'integrazione  $ t = \frac{\t_2}{2} $ per
riscrivere la (\ref{Ab}) come 
\be
\mc{A} = \frac{2^{-13} \mc{N}^2}{2} \int_0^{\infty} \frac{d t}{t^{14}} \frac{1}{{\eta ( i t )}^{24}} \, .
\label{Abose}
\ee
Il passaggio all'ampiezza nel canale trasverso, $\tld{\mc{A}}$, si ottiene
quindi applicando alla (\ref{Abose}) la trasformazione modulare 
$ S: t \ \rightarrow 1/l$.  L'espressione risultante,
\be 
\tilde{\mc{A}} = \frac{2^{-13} \mc{N}^2}{2} \int_0^{\infty} d l \frac{1}{{\eta ( i l )^{24}}} \, ,
\label{Abtil}
\ee
\`e funzione del tempo ``orizzontale'' $ l $ che descrive l'anello come un 
tubo  che termina su due bordi, e quindi definisce l'ampiezza ad albero di una 
stringa chiusa che si propaga tra due bordi. A questo proposito \`e essenziale 
sottolineare che, nell'ampiezza nel canale trasverso (\ref{Abtil}), 
la molteplicit\`a $\mc{N}$ delle cariche di Chan-Paton associate agli estremi della 
stringa aperta determina i coefficienti di riflessione per lo spettro chiuso 
di fronte ai bordi. 

La proiezione dello spettro aperto non orientato viene effettuata 
dall'am\-piezza
di striscia di M\"obius, che  rispetto ai casi precedenti
presenta  ulteriori sottigliezze. Queste sono legate al modulo del toro 
doppiamente ricoprente che, come abbiamo visto, non \`e puramente 
immaginario, 
ma \`e caratterizzato da una parte reale fissata ad 1/2. Il fatto che $\t_1$
sia pari ad 1/2 ha due importanti conseguenze:\\
i) \  le eccitazioni di oscillatore ai vari livelli di massa
compaiono in $\mc{M}$ con segni alterni, poich\'e 
$$ q^k = {( e^{2 \pi i \t} )}^k = e ^{2 \pi i k ( \frac{1}{2} + i 
\frac{i \t_2}{2})} = e^{\pi i k} \, e^{- \pi \t_2 k} = ( - )^k q_M^{k/2} , $$
ii) \ l'integrando dell'ampiezza di M\"obius, a differenza di quelli
per $\mc{K}$ e $\mc{A}$, contiene a priori una fase moltiplicativa.

L'espressione per l'ampiezza di M\"obius pu\`o essere ricavata come segue:\\
1) \ si determina dapprima $\tilde{\mc{M}}$, l'ampiezza (ad albero) per una 
stringa che si propaga tra un bordo ed un crosscap, prendendo la ``media 
geometrica'' delle due ampiezze $\tilde{\mc{K}}$ ed  $\tilde{\mc{A}}$, che
nel caso della stringa bosonica \`e
\be
\tilde{\mc{M}} = 2 \cdot \frac{N}{2} \int_0^{\infty} d l \frac{1}{{\hat{\eta}
( \frac{1}{2} + i l )}^{24}} \ ,
\label{Mbtil}
\ee
dove l'ultimo fattore 2 nella (\ref{Mbtil}) 
riflette la possibilit\`a di interpretare $\tilde{\mc{M}}$ come il diagramma 
di una stringa chiusa che si propaga tra un bordo ed un crosscap 
o tra un crosscap ed un bordo.

\begin{figure}[htb]\unitlength1cm
\begin{picture}(8,1.5)
\put(2.5,0){\epsfig{file=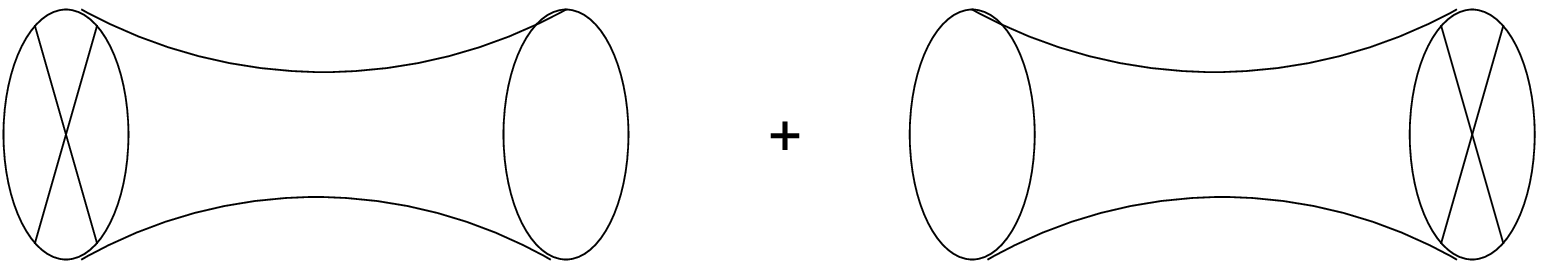,width=8cm,height=1.5cm}}
\end{picture}
\end{figure}

$ N$ determina il coefficiente di riflessione della stringa chiusa di 
fronte ad un singolo bordo. Inoltre l'integrando, ridefinito per un fattore
di fase, pu\`o essere espresso in termini di 
\be
\hat{\eta} ( \frac{1}{2} + i l ) = q^{1/24} \prod_{n=1}^{\infty} 
( 1 - ( - )^n q^n ) \, ,
\ee
che \`e manifestamente reale;\\
2) \ si effettua una ridefinizione della variabile d'integrazione, ponendo
$ l = \frac{t}{2} $, e quindi una trasformazione modulare $P^{-1}$ 
\be
P^{-1} : \quad \frac{1}{2} + \frac{i t}{2} \longrightarrow \frac{1}{2} 
+ \frac{i}{2 \t_2}
\ee
sotto l'azione della quale 
\be
\hat{\eta}\, ( \frac{1}{2} + \frac{i}{2 t} ) = \sqrt{t} \, \hat{\eta} \ 
( \frac{1}{2} + \frac{i t}{2} ).
\ee
L'ampiezza diretta di M\"obius per la stringa bosonica aperta assume 
cos\`\i \  la forma
\be
\mc{M} = \frac{N}{2} \int_0^{\infty} \frac{d \t_2}{\t_2^{14}} 
\frac{1}{{\hat{\eta} ( \frac{1}{2} + \frac{i \t_2}{2} )}^{24}} \ .
\label{Mbose}
\ee
Le ampiezze $\mc{A}$ ed $\mc{M}$ determinano lo spettro aperto proiettato 
(non orientato). Come fatto in precedenza, per assicurarsi della correttezza 
della simmetrizzazione degli stati, espandiamo in potenze di $ \sqrt{q} 
= e^{- \pi \t_2} $ gli integrandi di $\mc{A}$ ed $ \mc{M} $ delle equazioni 
(\ref{Ab}) e (\ref{Mbose}), ottenendo 
$$ \mc{A} \longrightarrow \frac{N^2}{2} 
( {(\sqrt{q})}^{-1} + (24) + \cdots ) \ ,$$
$$ \mc{M} \longrightarrow \frac{N}{2} 
( {(\sqrt{q})}^{-1} - (24) + \cdots ) \ . 
$$ 
Per i  due livelli di massa pi\`u bassa si ottiengono quindi 
$$ \frac{N (N + 1)}{2} \quad  \mbox{tachioni, \ e } \qquad 
\frac{N (N - 1)}{2} 
\quad \mbox{ vettori a massa nulla}. $$
Pertanto  $ \frac{1}{2} ( \mc{A} + \mc{M} ) $ fornisce il corretto conteggio 
degli stati del settore ``twisted''. Ricapitolando, lasciando implicite le 
misure d'integrazione, i di\-scendenti aperti della stringa bosonica a 26 
dimensioni sono descritti da 
$$ \mc{T} = \frac{1}{2} \, \frac{1}{\t_2^{12} \, {| \eta ( \t ) |}^{48}} \quad
, \quad  \mc{K} = \frac{1}{2} \, \frac{1}{\t_2^{12} \, {\eta ( 2 i \t_2 )}^{24}}\  , $$
che definiscono lo spettro chiuso non orientato,  e da 
$$ \mc{A} = \frac{N^2}{2} \, \frac{1}{\t_2^{12} \ 
{\eta ( \frac{i \t_2}{2} )}^{24}} \quad , \quad 
\mc{M} = \frac{N}{2} \, \frac{1}{\t_2^{12} \ {\hat{\eta} 
( \frac{1}{2} + \frac{i \t_2}{2} )}^{24}} \  , $$
che definiscono lo spettro aperto non orientato.

In realt\`a, mentre $\mc{K}$ ha un segno ben definito, fissato dall'ampiezza
di toro, $\mc{M}$ ha un segno indeterminato. Quindi, un valore positivo di $N$
corrisponde ad un gruppo di gauge ortogonale $SO(N)$, 
mentre un valore negativo (e pari) di $N$ corrisponde ad un gruppo 
di gauge simplettico $USp(2N)$. 
\`E allora 
naturale chiedersi come fissare il segno dell'intero $N$, ovvero come fissare 
il gruppo di gauge della teoria. La risposta \`e fornita dall'analisi del 
comportamento delle quattro ampiezze con $\c = 0$ nel limite di piccolo tempo
verticale, $ \t_2 \rightarrow 0 $. Per i modelli chiusi, abbiamo visto che 
l'origine ($ \t_2 = 0 $) \`e esclusa dalla regione d'integrazione, poich\'e
\`e fuori dalla regione fondamentale del gruppo modulare. Quindi $ \mc{T} $
\`e protetta dall'invarianza modulare. Viceversa, le altre tre superfici sono 
integrate su una regione che interseca l'asse reale, e questo introduce nelle
ampiezze $\mc{K}$, $\mc{A}$ ed $\mc{M}$ delle
corrispondenti divergenze ultraviolette. Il passaggio al canale trasverso 
consente una migliore interpretazione della situazione. In questo caso infatti
le divergenze si palesano nel limite infrarosso delle equazioni (\ref{Kbtil}), 
(\ref{Abtil}) e (\ref{Mbtil}), ovvero nel limite di tempo orizzontale molto 
grande $ ( l \rightarrow \infty )$. Queste divergenze infrarosse 
sono associate agli stati non massivi e tachionici dello spettro chiuso.

In generale, stati di massa $ m $ danno un contributo proporzionale a 
\be
\int_0^{\infty} d l e^{- l m^2} = \frac{1}{m^2} = \frac{1}{( p^2 + m^2 )_
{p=0}} \, .
\label{int}
\ee
La divergenza dominante in queste ampiezze di vuoto, dovuta al tachione,
pu\`o formalmente essere regolata. Vicerversa non c'\`e alcuna 
possibilit\`a di regolare le  divergenze dovute agli scambi di stati di 
massa nulla.
In questo modello, l'invarianza di Lorentz associa il modo singolare non 
massivo all'unico modo scalare a massa nulla della stringa chiusa, 
il dilatone.
L'origine di tale divergenza pu\`o essere compresa intuitivamente osservando
che nel limite $ l \rightarrow \infty $ le tre ampiezze ad albero, $\tilde
{\mc{K}}$, $\tilde{\mc{A}}$ e $\tilde{\mc{M}}$, sono assimilabili a tubi molto
lunghi (che terminano su bordi e/o crosscaps). Esse sono quindi  simili
a diagrammi di processi in cui uno stato di stringa chiusa, di impulso nullo,
viene creato nel vuoto, si propaga per una certa distanza, per poi sparire 
nuovamente nel vuoto.
\begin{figure}[htb]\unitlength1cm
\begin{picture}(10,1.2)
\put(1.5,0){\epsfig{file=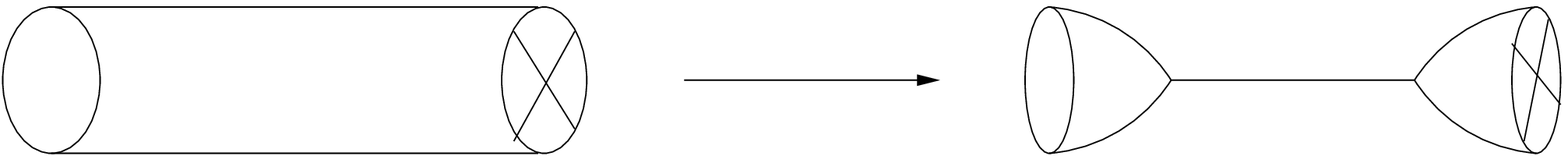,width=10cm,height=1.2cm}}
\end{picture}
\end{figure}

Come suggerito dalla (\ref{int}), il propagatore di uno stato con massa ed 
impulso nulli, \`e divergente, ma  i corrispondenti 
residui sono finiti e defini\-scono due elementi di base della teoria: le 
funzioni ad un punto per i campi di stringa chiusa (\emph{\bf tadpoles}) in 
presenza di un bordo o di un crosscap. Nella teoria non orientata le 
singolarit\`a delle ampiezze (\ref{Kbtil}), (\ref{Abtil}) e (\ref{Mbtil}) 
si combinano in un contributo proporzionale a 
\be
\tilde{\mc{K}} + \tilde{\mc{A}} + \tilde{\mc{M}} \sim \frac{1}{2} ( 2^{13} + 
2^{-13}N^2 + 2 \e N ) = \frac{2^{-13}}{2} {( 2^{13} + \e N )}^2 \, ,
\label{tadb}
\ee
dove $ \e = \pm 1 $ \`e legato alla scelta del segno di $\mc{M}$.

L'equazione (\ref{tadb}) \`e una \emph{tadpole condition}, in questo caso  
la condizione di cancellazione delle divergenze dovute al dilatone. Essa 
conduce ad un'am\-piezza finita solo se $ \e =-1 $ ed $ N = 2^{13} $, ovvero 
se il gruppo di Chan-Paton \`e  $SO( 2^{13}) = SO(8192)$ 
\cite{Douglas:1987eu}. 
Si noti inoltre come la scelta di $ N = 2^{13} $ riproduce la potenza di 2 
$(= 2^{D/2})$ incontrata nella discussione dei gruppi di simmetria per le 
stringhe aperte. 
Le condizioni di cancellazione dei {\it tadpole} forniscono uno strumento di 
controllo cruciale della consistenza dei modelli aperti. 
Infatti, se per la stringa bosonica la particolare scelta di $N$ elimina una 
correzione della teoria effettiva di bassa energia, in modelli pi\`u complessi
 il suo analogo elimina 
contributi legati alla presenza di anomalie di gauge e gravitazionali 
\cite{Weinberg:1987ie}.  
Vedremo in seguito come, in teorie con fermioni chirali, alcuni 
{\it tadpoles} siano connessi alle corrispondenti anomalie.

\section{Discendenti aperti a 10 dimensioni}
Gli esempi pi\`u semplici di modelli aperti a 10 dimensioni sono i discendenti
della superstringa di Tipo IIB e delle teorie 0A e 0B, i due modelli non 
supersimmetrici e tachionici discussi nel capitolo precedente. Come la stringa
bosonica, queste teorie hanno la caratteristica di essere simmetriche sotto lo
scambio dei modi destri e sinistri. Ricordiamo infatti che le loro ampiezze di
toro sono 
$$\mc{T}_{IIB} = \int_{\mc{F}} \frac{d^2 \t}{\t_2^2} \, \frac{1}{{ (\sqrt{\t_2}\, \eta ( \t ) \bar{\eta} ( \bar{\t} ) )}^8} \, {| V_8 - S_8 |}^2, $$
$$ \mc{T}_{0A} = \int_{\mc{F}} \frac{d^2 \t}{\t_2^2} \, \frac{1}{{ (\sqrt{\t_2} \, \eta( \t ) \bar{\eta} ( \bar{\t} ) )}^8} \, ( | O_8 |^2 + | V_8 |^2 + S_8
\bar{C}_8 + C_8 \bar{S}_8 ), $$
$$ \mc{T}_{0B} = \int_{\mc{F}} \frac{d^2 \t}{\t_2^2} \, \frac{1}{{ (\sqrt{\t_2}\, \eta ( \t ) \bar{\eta} ( \bar{\t} ) )}^8} \, ( | O_8 |^2 + | V_8 |^2 + | S_8 |^2 + | C_8 |^2 ). $$ 
Tutti gli altri modelli a 10 dimensioni, compresa la superstringa di Tipo IIA, 
hanno uno spettro non simmetrico sotto $ \O $, e quindi non ammettono
 discendenti aperti. Abbiamo visto come il punto di partenza della costruzione di un discendente aperto sia infatti una proiezione dello 
spettro chiuso che mescola modi destri e sinistri. Questo si ottiene sommando 
alla funzione di partizione di toro, opportunamente dimezzata, il contributo 
di bottiglia di Klein nel canale diretto. $\mc{K}$ deve contenere tutti i 
caratteri che in $ \mc{T} $ compaiono come moduli quadri, ed in generale non 
\`e unica.
In linea di principio,  la teoria di Tipo IIB  ammette quattro possibilit\`a 
per $ \mc{K} $, una per ogni scelta di segno dei due caratteri $ V_8 $ ed 
$ S_8 $
che compaiono in $ \mc{T} $. In realt\`a questi segni sono soggetti a forti 
restrizioni dettate dall'algebra di fusione \cite{Fioravanti:1994hf,
Pradisi:1995qy}, ovvero dall'algebra che definisce le regole di superselezione
 per i vari settori dello spettro.
Per ogni coppia di famiglie di operatori $ [ \f_i ] $ e $ [ \f_j ]$, gli 
interi $ N_{ij}^k, $ definiti da 
\be
[ \f_i ] \times [ \f_j ] = \sum_k N_{ij}^k [ \f_k ]
\ee
determinano quali famiglie di stati possono essere ottenute 
dal loro prodotto. 
Nel caso in considerazione, $[ \f_i ]$ 
viene identificata con una delle quattro fami\-glie $ [ O_8 ] , [ V_8 ] ,
 [S_8 ] $ e $ [ C_8 ] $. Il cosiddetto \emph{coefficiente di fusione}, 
$ N_{ij}^k $, conta il numero di accoppiamenti indipendenti tra le 
tre famiglie $ [ \f_i ] , [ \f_j ] $ e $ [ \f_k ] $, ha valori non negativi,
ed \`e determinato dalla formula di Verlinde 
\be
N_{ij}^k = \sum_m \frac{S_{im}  S_{jm}  S_{km}^{\dag}}{S_{im}} \ ,
\ee
dove $S$ \`e la matrice modulare associata alla trasformazione 
$ \t \rightarrow - 1/ \t $. Dall'unitariet\`a di
$S$ segue che  $ N_{1j}^k = \d_j^k $ , e quindi l'identit\`a \`e l'elemento 
neutro dell'algebra di fusione.
 Il risultato generale cui si giunge \`e che le possibili 
scelte di segno per i caratteri corrispondono agli isomorfismi di $ Z_2 $
dell'algebra di fusione compatibili con la proiezione GSO del toro. Nel nostro
caso, scegliendo come base di caratteri $( O_8 , V_8 ,- S_8 ,- C_8 )$, ovvero 
associando un naturale segno negativo ai caratteri che descrivono fermioni, 
l'identit\`a \`e  $ [ V_8 ] $,  ed appare nel quadrato di 
tutte le famiglie \cite{Englert:1986na, Nicolai:1986aq} . 
Questa scelta \`e la pi\`u naturale per il gruppo di Lorentz $SO(1,9)$,
e conduce alle seguenti ampiezze di bottiglia di Klein:
$$ \mc{K}_{IIB} = \frac{1}{2} ( V_8 - S_8 ) , $$
\be
\mc{K}_{0A} = \frac{1}{2} ( O_8 + V_8 ) \ ,   
\label{Ks}
\ee  
$$ \mc{K}_{0B} = \frac{1}{2} ( O_8 + V_8 - S_8 - C_8 ) , $$
dove per semplicit\`a sono stati omessi i contributi dei bosoni trasversi e 
della misura d'integrazione. L'argomento delle espressioni delle eq. 
(\ref{Ks})
\`e $ 2 i \t_2 $, con $\t_2$ l'usuale ``tempo proprio'' della stringa chiusa.
La proiezione di bottiglia di Klein determina l'azione di
$\O$ sui vari settori dello spettro chiuso non orientato, che
pu\`o essere sintetizzata come segue: un segno positivo (negativo) nei 
caratteri implica la simmetrizzazione (antisimmetrizzazione) dei settori 
sotto lo scambio dei modi destri e sinistri.

Concentriamoci per il momento sul caso della superstringa di Tipo IIB. 
$ \mc{K}_{IIB} $ simmetrizza il settore di NS-NS, proiettando via dallo 
spettro a massa
nulla la due-forma $B_{\m\n}$, ed antisimmetrizza il settore di R-R, 
eliminando
cos\`\i \  il secondo scalare $\f' $ e la quattro-forma autoduale 
$ D_{\m\n\r\s}^+ $.
I modi fermio\-nici vengono dimezzati, di conseguenza lo spettro a
 massa nulla risultante dal settore chiuso contiene il gravitone, il dilatone,
 la due-forma di R,  un gravitino {\it left} ed uno spinore {\it right}. 
Questo \`e esattamente lo spettro corrispondente alla supergravit\`a N = (1,0)
 a dieci dimensioni, ovvero lo spettro chiuso della teoria di Tipo I.
Nel canale trasverso le ampiezze delle eq. (\ref{Ks}) divengono 
$$ \tld{\mc{K}}_{IIB} = \frac{2^5}{2} ( V_8 - S_8 ) , $$
\be
\tld{\mc{K}}_{OA} = \frac{2^5}{2} ( O_8 + V_8 )  \ ,  
\ee  
$$ \tld{\mc{K}}_{OB} = \frac{2^6}{2} V_8 , $$
dove le potenze di due derivano dalla misura d'integrazione, una volta che
tali ampiezze vengono espresse in termini dei moduli dei loro ricoprimenti 
doppi.

Il passo successivo nella costruzione coinvolge il settore ``twisted'', 
determinato dalla proiezione  $ \frac{1}{2} ( \mc{A} + \mc{M} ) $.
Per la deduzione di $ \mc{A} $ \`e conveniente ini\-ziare dal canale 
trasverso.
Quindi, anche in questo caso il punto di partenza \`e il settore chiuso, o 
meglio la porzione dello spettro chiuso che pu\`o fluire nell'anello trasverso
$\tld{\mc{A}} $. Le ampiezze $\tld{\mc{A}}$ ed $\tld{\mc{M}}$
hanno rispettivamente due bordi, e un bordo ed un crosscap; quindi sono 
polinomi di secondo e primo grado nelle molteplicit\`a dei vari settori di 
carica. La struttura di questi polinomi  \`e fortemente vincolata dalle 
condizioni di fattorizzazione dell'ampiezza sul disco. In particolare, 
poich\'e in $ \tld{\mc{A}}$ lo spettro chiuso si propaga tra una coppia di 
bordi, i suoi coefficienti possono essere interpretati come i quadrati della 
normalizzazione della funzione ad un punto sul disco.
Quindi, data la funzione di partizione (quasi) diagonale per lo spettro 
chiuso 
\be
\mc{T} = \sum_{i,j} X_{ij} \c_i {\bar{\c}}_j \ ,
\ee
dove $X_{ij}$ \`e una matrice di interi non negativi,
i caratteri $\c_i $ che possono fluire nell'anello trasverso $ \tld{\mc{A}} $
sono quelli accoppiati con i loro coniugati dalla proiezione GSO del modello
chiuso; se il bordo deve rispettare una data simmetria, la riflessione pu\`o 
avvenire solo se il coniugato GSO \`e coniugato anche rispetto alla simmetria
in considerazione. Tutto ci\`o fornisce quindi una regola per la 
determinazione dei settori permessi in $\tld{\mc{A}}$ e, conseguentemente, per
il numero  dei coefficienti di riflessione.
Osserviamo che, per i modelli in questione, la simmetria locale che i bordi 
devono rispettare \`e legata al gruppo di Lorentz trasverso $SO(8)$, le cui 
rappresentazioni sono tutte autoconiugate. Questo implica che nel modello 0A
in $\tld{\mc{A}}$ potranno fluire i soli caratteri $O_8$ e $ V_8$,  gli unici 
accoppiati ai 
loro coniugati dalla proiezione GSO delle ampiezze di toro $ \mc{T}_{0A} $; 
viceversa, nel modello 0B, tutti i caratteri sono accoppiati con i loro 
coniugati, e quindi tutti i caratteri presenti in $\mc{T}_{0B} $ potranno 
fluire in $\tld{\mc{A}}$.
Nel caso della teoria di Tipo IIB c'\`e un singolo settore permesso, se 
vogliamo preservare la supersimmetria $ ( V_8 - S_8 ) $, e quindi 
\be
\tld{\mc{A}}_{IIB} = \frac{2^5}{2} \, n^2  \, ( V_8 - S_8 ) \ ,
\ee
dove l'argomento \`e $ i l $, il modulo del ricoprimento doppio.
L'ampiezza di M\"obius nel canale trasverso, $\tld{\mc{M}}$, \`e 
determinata da $ \tld{\mc{A}}_{IIB} $ e da $ \tld{\mc{K}}_{IIB} $, ed \`e 
\be
\tld{\mc{M}}_{IIB} =  - \frac{2}{2} \, n \, ( \hat{V}_8 - \hat{S}_8 ) \ ,
\ee
dove il coefficiente di riflessione \`e dato dalla media geometrica di quelli
di $ \tld{\mc{A}}$ e di $\tld{\mc{K}}$. L'ulteriore fattore 2 riflette la 
combinatorica di $\tld{\mc{M}}$, mentre l'argomento \`e in questo caso 
$ ( i l + 1/2 ) $, il modulo del ricoprimento doppio. 
A questo punto \`e possibile trasformare $\tld{\mc{A}}_{IIB}$ nell'ampiezza 
di anello del canale diretto usando la matrice modulare 
$S$. Si ottiene cos\`\i \ l'espressione 
\be
\mc{A}_{IIB} = \frac{n^2}{2} ( V_8 - S_8 ) \ ,
\ee
e  argomento \`e $ \frac{i \t_2}{2} $, dove $\t_2$ \`e il ``tempo proprio'' 
della stringa aperta.

Per quanto riguarda $\tld{\mc{M}}_{IIB}$, osserviamo che nella sua 
espressione 
compaiono i caratteri reali $\hat{\c}$; il passaggio nel canale diretto 
richiede quindi l'uso di una differente matrice P, legata alle matrici S e T 
da 
\be
P = T^{1/2} ( S T^2 S ) T^{1/2} = T = diag ( -1 , 1 , 1 , 1 ),
\ee
dove il coniugio corrisponde alla ridefinizione della fase.
L'ampiezza di M\"obius assume quindi la forma 
\be
\mc{M}_{IIB} = - \frac{n}{2} ( \hat{V}_8 - \hat{S}_8 )
\ee
e l'argomento \`e  $ \frac{i \t_2}{2} + \frac{1}{2} $. Si noti che un valore
positivo di $ n $ implica  un gruppo ortogonale $SO(n)$ con 
$ \frac{n ( n - 1 )}{2} $ vettori di gauge, 
mentre un valore negativo di $ n $ implica un gruppo 
simplettico $USp(|n|)$, con $ \frac{|n| ( |n| + 1 )}{2}$ bosoni di gauge.
Lo spettro aperto a massa nulla \`e dunque un multipletto di super 
Yang-Mills $ N = 1$ con gruppo di gauge $ SO(n)$ o  $USp(|n|)$.
Come per la stringa bosonica, il gruppo di gauge \`e in realt\`a 
determinato dalla 
condizione di cancellazione dei $tadpole$, che si ottiene annullando  
la funzione ad un punto totale associata ai due settori $V_8 \bar{V}_8$ ed
$S_8 \bar{S}_8$ :
\be
\frac{2^5}{2} + \frac{2^{-5}}{2} \,n^2 - \frac{2}{2} \, n = \frac{2^5}{2} 
{( n - 2^5 )}^2 = 0 \, ,
\label{tadIIB}
\ee
Questa equazione equivale quindi alla cancellazione di due {\it tadpoles}, 
uno per il settore di NS-NS ed uno per il settore di R-R. 
La sua unica soluzione, $ n = 32 $, seleziona quindi  il gruppo di 
gauge $ SO(32)$. In questo modo  la superstringa di Tipo I-$SO(32)$, priva di
anomalie, ottenuta per la prima volta da Green e Schwarz nel 1984 
\cite{Green:1984sg}, emerge come discendente aperto della Tipo IIB.

La relazione tra condizione di $tadpole$ e cancellazione delle anomalie venne 
messa in evidenza per la prima volta proprio per la superstringa di Tipo I, da
Polchinski e Cai, i quali collegarono  le anomalie ai $tadpole$ di stati a 
massa nulla non fisici \cite{Polchinski:1988tu}. Un'analisi pi\`u generale 
della relazione \`e contenuta in \cite{Aldazabal:1999nu, Bianchi:2000de}   
Uno stato di questo genere \`e presente nel settore  $S_8 \bar{S}_8$, uno 
scalare di R-R a massa nulla eliminato dallo spettro 
chiuso mediante la proiezione di bottiglia di Klein.
I $tadpole$ di R-R sono un segnale di una inconsistenza interna della 
teoria e quindi devono essere sempre annullati. I $tadpole$ di NS-NS, 
viceversa,
segnalano la necessit\`a di una ridefinizione del $background$, poich\'e 
producono dei potenziali per i campi corrispondenti.

Torniamo ora ad esaminare i discendenti aperti dei modelli non supersimmetrici,
iniziando dalla teoria 0A \cite{Bianchi:1990yu}. 
L'ampiezza di bottiglia di Klein $\mc{K}_{0A}$
di eq.(\ref{Ks}) simmetrizza il settore di NS-NS, eliminando dai livelli di 
massa pi\`u bassa dello spettro chiuso il tensore antisimmetrico $B_{\m\n}$, ma
lasciando il tachione, il gravitone ed il dilatone. Il settore di R-R viene 
``dimezzato'', e quindi nello spettro a massa nulla compaiono anche un vettore
ed una tre-forma, combinazioni lineari di quelli presenti nel modello 0A
di partenza.

Per quanto riguarda la costruzione dello spettro aperto a partire da quello
chiuso nel canale trasverso, abbiamo gi\`a accennato al fatto che in 
$\tld{\mc{A}}_{0A}$ possono fluire solo i caratteri $O_8$ e $V_8$, in 
quanto $S_8$ e $C_8$ sono autoconiugati in $SO(8)$. Quindi 
\be
\tld{\mc{A}}_{0A} = \frac{2^{-5}}{2} ( \a^2 O_8 + \b^2 V_8 ) ,
\label{Atld0A}
\ee
e di conseguenza  $ \tld{\mc{A}}_{0A}$ contiene due  coefficienti di 
riflessione indipendenti, che possono essere associati ai due caratteri 
permessi. Questo dato \`e  di particolare rilievo nella costruzione, 
poich\'e determina il numero 
di settori di carica del modello. Effettuando la parametrizzazione 
$$ \frac{\a^2 + \b^2}{2} = n_B^2 + n_F^2 , $$
$$ \frac{\b^2 - \a^2}{2} = 2 n_B n_F , $$
per i coefficienti di riflessione, che, come vedremo, consente 
un'interpretazione di $\mc{A}_0$ in termini di cariche di Chan-Paton, l'ampiezza di 
eq. (\ref{Atld0A}) assume la forma 
\be
\tld{\mc{A}}_{0A} =  \frac{2^{-5}}{2} ( {( n_B - n_F )}^2 \, O_8 + 
{( n_B + n_F )}^2 \, V_8  ) \, ,
\ee
dove l'argomento \`e  $ i l $.
L'espressione per l'ampiezza di M\"obius $\tld{\mc{M}}$, nel canale 
trasverso si 
ottiene accomodando tutti i caratteri comuni ad $\tld{\mc{A}}_{0A}$  e 
$\tld{\mc{K}_{0A}}$, con coefficienti di riflessione che, a meno dei loro 
segni, sono medie geometriche di quelli in $\tld{\mc{A}}$ e $\tld{\mc{K}}$. 
In conclusione
\be
\tld{\mc{M}}_{0A} =  - \frac{2}{2} ( ( n_B - n_F ) \,\hat{O}_8 + ( n_B + n_F ) 
\ \hat{V}_8  ) \ ,
\label{Mtld0A}
\ee
dove l'argomento \`e, al solito, $ i l + \frac{1}{2} $.
La funzione di partizione di stringa aperta $\mc{A}$ e l'ampiezza di striscia 
di M\"obius nel canale diretto $\mc{M}$ si determinano quindi mediante  
trasformazioni $S$ e $P$ rispettivamente, seguite dalle appropriate 
ridefinizioni delle variabili, e il risultato finale \`e 
\be
\mc{A}_{0A} = \frac{ ( n_B^2 + n_F^2 )}{2} ( O_8 + V_8 ) - n_B n_f 
( S_8 + C_8 )  \ ,
\label{A0A}
\ee
\be
\mc{M}_{0A} = - \frac{1}{2} \Big ( ( n_B - n_F ) \,\hat{O}_8 + ( n_B + n_F ) \ 
\hat{V}_8 \Big) \, .
\ee
L'espressione (\ref{A0A}) mette in luce un aspetto particolarmente 
interessante di questi pur semplici modelli, ovvero la presenza di tre tipi 
di stringhe aperte risultanti dai due settori di carica di Chan-Paton. 
Infatti, mentre gli stati di NS descrivono stringhe non orientate con cariche 
di Chan-Paton dello stesso tipo agli estremi ($ n_B $ o $ n_F $), gli stati di 
R restano associati a stringhe con cariche diverse di Chan-Paton agli estremi.

\begin{figure}[hb]\unitlength1cm
\begin{picture}(10,3)
\put(2,0){\epsfig{file=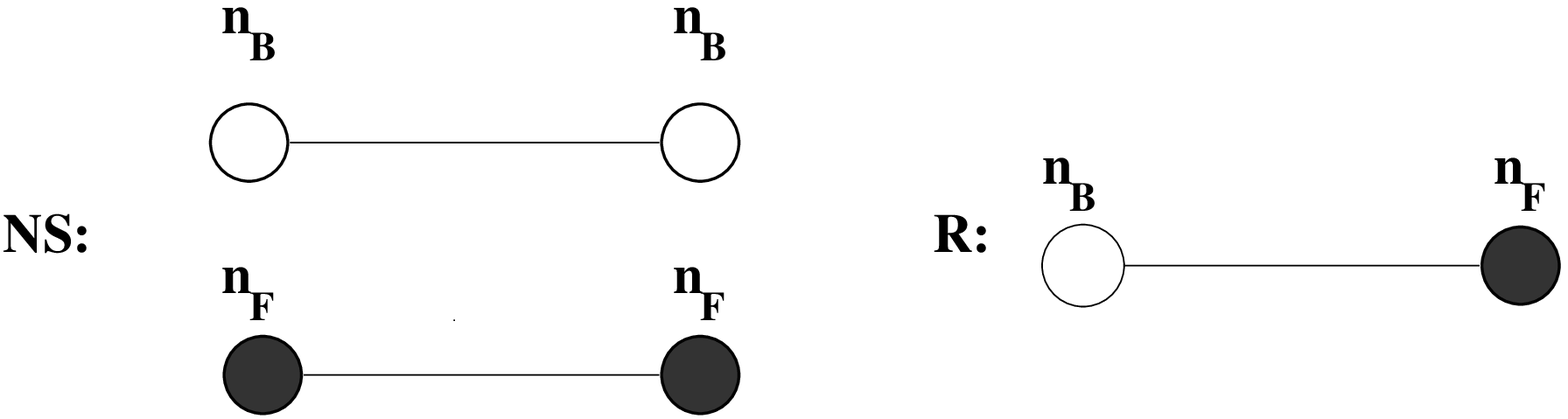,width=10cm,height=3cm}}
\end{picture}
\caption{\footnotesize{Cariche di Chan-Paton}}
\end{figure}

Questa struttura riflette la rottura spontanea della simmetria di Chan-Paton.
Osserviamo infatti che, se $ n_F = 0 $, esiste un solo settore di stringa 
aperta, corrispondente a $ ( O_8 + V_8 )$, e tutte le stringhe aperte portano 
una coppia di cariche identiche; se $ n_F \neq 0 $, sono presenti due tipi
di srtinghe aperte nel settore di $ ( O_8 + V_8 )$ e ulteriori 
stringhe aperte, corrispondenti al settore $ ( S_8 + C_8 ) $, con agli 
estremi due differenti tipi di carica di molteplicit\`a $ n_B $ ed $ n_F $.
La condizione di $tadpole$  
$$ n_B + n_F = 32  , $$
deriva in questo modello dal solo settore $ V_8 $, e quindi, a differenza del 
caso della IIB, non 
\`e legata ad anomalie di gauge  e gravitazionali.  Si ottiene cos\`\i \ una 
classe di modelli con gruppo di gauge $ SO(n_B) \times SO(32-n_B)$.
Lo spettro aperto di bassa energia, che pu\`o essere dedotto dalle ampiezze 
di eq.(\ref{Mtld0A}) e (\ref{A0A}), contiene quindi bosoni di gauge 
(corrispondenti a $V_8 $) 
nella rap\-presentazione aggiunta di  $ SO(n_B) \times SO(32-n_B) $, tachioni
(corrispondenti ad $O_8 $) nella rappresentazione $\frac{n_B (n_B+1)}{2}$ e
$\frac{n_B (n_B+1)}{2}$, e fermioni di 
Majorana (corrispondenti ai caratteri $ S_8$ e $ C_8 $ del settore di R)
nella rappresentazione bifondamentale $ (n_B, n_F) $ dello stesso gruppo.
Lo spettro risultante \`e chiaramente non chirale sia nel settore chiuso 
proiettato che in quello aperto.

L'ultima teoria che rimane da trattare \`e la Tipo 0B che, come vedremo, 
conduce a diversi tipi di discendenti aperti, in corrispondenza delle 
possibili scelte di proiezione di bottiglia di Klein.
Iniziamo dall'ampiezza di Klein di equazione (\ref{Ks}), per la quale 
il settore chiuso di NS-NS viene simmetrizzato mentre quello di R-R viene
antisimmetrizzato, e l'argomento \`e il \cite{Bianchi:1990yu}. 
Lo spettro che ne risulta contiene un tachione, il 
gravitone ed una coppia di tensori antisimmetrici. Mediante una 
trasformazione $S$ ed una ridefinizione della  variabile 
($ t = 2 \t_2 $ e t = 1/l) l'ampiezza di Klein nel canale trasverso assume 
quindi la forma
\be
\tld{\mc{K}}_{0B} = \frac{2^6}{2} V_8 \, .
\label{kappatld0B}
\ee
Come in $ {\mc{K}}_{0B}$, anche nella corrispondente ampiezza di anello nel 
canale trasver\-so $ {\mc{A}}_{0B}$  possono fluire tutti i caratteri 
$O_8, V_8, C_8$ ed $ S_8$ e quindi 
\be
\tld{\mc{A}}_{0B} = \frac{2^{-6}}{2} ( \a^2 V_8 + \b^2 O_8 - \g^2 S_8 - 
\d^2 C_8 ).
\label{Antld0B}
\ee
Prima di scrivere in maniera pi\'u esplicita i coefficienti di riflessione 
$ \a ,\b , \g, \d,$ passiamo a considerare la funzione di partizione di 
stringa aperta nel canale diretto, ${\mc{A}}_{0B}$, per la quale esiste in 
questo caso  un ansatz per 
l'assegnazione delle cariche di Chan-Paton, noto come\emph{ ansatz di Cardy}
\cite{Cardy:1989ir}. 
Questo si applica in generale ai modelli le cui ampiezze di toro siano 
costruite mediante la matrice di coniugio di carica $\mc{C}$:
\be
\mc{T} = \sum_{ij} \mc{C}_{ij} \c_i \bar{\c}_j \, .
\ee
In tali condizioni, Cardy  stabil\`\i \ una corrispondenza uno ad uno tra 
tipi 
di bordi e settori dello spettro, fornendo in tal modo  una nuova 
interpretazione 
dei coefficienti dell'algebra di fusione $N_{ij}^k$ : essi determinano il 
contenuto  $(k)$ nello spettro di anello corrispondente alle condizioni al 
bordo $(i)$ e $(j)$. Data questa corrispondenza, l'ampiezza di anello  pu\`o 
essere espressa in questo caso come
\be
\mc{A} = \frac{1}{2} \sum_{ijk} N_{ij}^k n^i n^j \c_k \,  .
\ee
Passando al canale traverso, si ottiene
\be
\tld{\mc{A}} = \frac{2^{-D/2}}{2} \sum_{ijkl} N_{ij}^k n^i n^j S_k^l \c_l  \, ,
\ee
ed utilizzando infine la formula di Verlinde, 
\be
\tld{\mc{A}} = \frac{2^{-D/2}}{2} \sum_{m} \c_m { \Big( \sum_k 
\frac{S_{mk} n^k}{\sqrt{S_{m1}}}\Big)}^2  \, .
\ee
Quindi tutti i coefficienti di $ \tld{\mc{A}} $ sono quadrati perfetti, 
esattamente come \`e richiesto per la costruzione di $ \tld{\mc{M}} $.
Per i gruppi $SO(4n)$, ed in particolare per $SO(8)$, la matrice di coniugio 
$ \mc{C} $ coincide con la matrice identit\`a. 
L'ansatz di Cardy pu\`o essere cos\`\i \ applicato direttamente 
al modello 0B, e quindi l'ampiezza di anello nel canale trasverso 
assume la forma 
\beq
\tld{\mc{A}}_{0B} & = & \frac{2^{-6}}{2} [ {( n_o + n_v + n_s + n_c )}^2 V_8 + 
{( n_o + n_v - n_s - n_c )}^2 O_8  \nonumber \\
                  & + & {( n_o - n_v + n_s - n_c )}^2 S_8  
+ {( n_o - n_v - n_s + n_c )}^2 C_8 ] \, ,
\label{Atld0B}
\eeq
che collega i quattro coefficienti $(\a,\b,\d,\g)$ di eq.(\ref{Antld0B}) alle
quattro molteplici\-t\`a di Chan-Paton.

L'ampiezza di M\"obius nel canale trasverso, determinata completamente da 
$ \tld{\mc{K}}_{0B} $ e $ \tld{\mc{A}}_{0B}$, \`e
\be
\tld{\mc{M}}_{0B} = - \frac{2}{2} ( n_o + n_v + n_s + n_c ) \hat{V}_8 \ ,
\label{Mtld0B}
\ee
e quindi il settore aperto del modello 0B \`e descritto da 
\beq
\mc{A}_{0B} & = & \frac{n_o^2 + n_v^2 + n_s^2 + n_c^2}{2} \ V_8 
+ ( n_o n_v + n_s n_c ) \ O_8 \nonumber \\
            & - & ( n_v n_s + n_o n_c ) \ S_8 
- ( n_v n_c + n_o n_s ) \  C_8 \ , \\
\mc{M}_{0B} & = & - \frac{n_o + n_v + n_s + n_c}{2} \  \hat{V}_8 \ , 
\eeq
legato come al solito, a $ \tld{\mc{A}}_{0B}$ e $ \tld{\mc{M}}_{0B} $ da
trasformazioni $S$ e $P$, rispettivamente, e da corrispondenti 
ridefinizioni delle variabili. 
Dalle espressioni (\ref{kappatld0B}), (\ref{Atld0B}) e (\ref{Mtld0B}) vediamo 
che questo modello richiede tre condizioni di $tadpole$, quelle relative ai 
tre settori $V_8,S_8,$ e $ C_8$ che comprendono eccitazioni a massa nulla: 
\be 
V_8 : \quad    n_o + n_v + n_s + n_c  = 2^6 = 64 \ ,
\ee
\be
S_8 : \qquad   n_o - n_v + n_s - n_c  = 0  \, ,
\label{tadS}
\ee
\be
C_8 : \qquad   n_o - n_v - n_s + n_c  = 0  \, .
\label{tadC}
\ee
Le soluzioni sono $ n_o = n_v $, $n_s = n_c $, $ n_o+n_s=32 $ ed il risultante 
gruppo di gauge \`e $ SO(n_o) \times SO(n_v) \times SO(n_s) \times SO(n_c) $,
 o meglio $ SO(n_o) \times SO(n_v) \times SO(n_o - 32) \times SO(n_v - 32) $.
La presenza di $tadpole$ derivanti dal settore di R-R implica che lo spettro 
aperto per questa classe di modelli \`e chirale e potenzialmente anomalo, 
ma le condizioni (\ref{tadS}) e (\ref{tadC}) assicurano la cancellazione di 
tutte le anomalie di gauge e gravitazionali. 
Oltre ai vettori di gauge, lo spettro comprende tachioni, fermioni 
\emph{left} e \emph{right} tutti in rappresentazioni bi-fondamentali.

Come accennato in precedenza, il modello $0B$ \`e quello che possiede la 
struttura pi\`u ricca. Esistono infatti altri due classi di discendenti, 
associate alle due ulteriori scelte, inequivalenti, di bottiglia di Klein:
\be
\mc{K}'_{0B} = \frac{1}{2} ( O_8 + V_8 + S_8 + C_8 ) \, ,
\ee
\be
\mc{K}''_{0B} = \frac{1}{2} ( - O_8 + V_8 + S_8 - C_8 ) \, .
\ee
Queste nuove ampiezze hanno origine dalle altre possibili scelte di base dei 
caratteri compatibili con la proiezione GSO del toro, ovvero
$$ ( O_8 , V_8 , S_8 , C_8 ) , \, ( - O_8 , V_8 , - S_8 , C_8 ) , \, 
( - O_8 , V_8 , S_8 , - C_8 ) $$
dove si vede che le ultime due sono equivalenti a meno di una trasformazione 
di parit\`a. Queste ampiezze comportano diverse scelte per la simmetrizzazione
o antisimmetrizzazione dei vari settori. In particolare  $\mc{K}''_{0B}$  
rimuove dallo spettro chiuso il tachione e la due-forma $ B_{\m\n} $ di NS-NS,
e da' luogo ad uno spettro chiuso \emph{chirale} che comprende una   
quattro-forma autoduale $D_{\m\n\r\s}^+$ \cite{Sagnotti:1995ga, Angelantonj:1998gj, 
Sagnotti:1997qj}.

Le corrispondenti ampiezze di bottiglia di Klein nel canale trasverso
\be
\tld{\mc{K}}'_{0B} = \frac{2^6}{2} O_8 \ ,
\ee
\be
\tld{\mc{K}}''_{0B} = - \frac{2^6}{2} C_8 \ ,
\label{Ksec}
\ee
sono caratterizzate da espressioni particolarmente semplici, e la presenza di 
$C_8$ in (\ref{Ksec}) implica  per questo modello la presenza di un $tadpole$ 
di R-R. Le corrispondenti ampiezze di anello $ \mc{A} $ e di 
M\"obius  $ \mc{M} $ nel canale diretto possono essere ottenute semplicemente
``fondendo'' i vari termini di $ \mc{A}_{0B} $ ed $ \mc{M}_{0B} $ con $ O_8 $
e $ - C_8 $ rispettivamente. Per il primo modello si ottiene quindi 
\beq
\mc{A}'_{0B} & = & \frac{n_o^2 + n_v^2 + n_s^2 + n_c^2}{2} \ O_8 +
 ( n_o n_v + n_s n_c ) \ V_8   \nonumber \\
             & - & ( n_v n_s + n_o n_c ) \ C_8 - 
( n_v n_c + n_o n_s ) \ S_8 \ ,
\label{A0Bprime}
\eeq
e 
\be
\mc{M}'_{0B} = \pm \frac{1}{2} ( n_o + n_v - n_s - n_c ) \hat{O}_8 \ .
\label{M0Bprime}
\ee
Le corrispondenti ampiezze nel canale trasverso sono :
\beq
\tld{\mc{A}}'_{0B} & = & \frac{2^{-6}}{2} [ {( n_o + n_v + n_s + n_c )}^2 \ 
V_8 + {( n_o + n_v - n_s - n_c )}^2 O_8  \nonumber \\
                   & + & {( n_o - n_v + n_s - n_c )}^2 \ C_8 + 
{( n_o - n_v - n_s + n_c )}^2 S_8 ] \ ,
\eeq
e
\be
\tld{\mc{M}}'_{0B} = \pm  \frac{2}{2} ( n_o + n_v - n_s - n_c ) \hat{O}_8 \ ,
\ee
dove l'indeterminazione nel segno di $ \mc{M} $ \`e dovuta all'assenza di una 
condizione di $tadpole$ per il tachione.

Dalla forma delle ampiezze nel canale diretto (ed in particolare 
dall'espres\-sione dei coefficienti di $V_8$) si pu\`o dedurre che 
il gruppo di gauge per questa 
classe di modelli deve essere unitario, e  le cariche (``complesse'') 
andranno reinterpretate come segue:
$ n_o = n $ , $ n_v = \bar{n}$, $ n_s = m$, $ n_c = \bar{m} . $
Quindi la (\ref{A0Bprime}) e la (\ref{M0Bprime}) assumono la forma
\beq
\mc{A}'_{0B} & = & \frac{1}{2} ( n^2 + \bar{n}^2 + m^2 + \bar{m}^2 ) \ O_8 +
( n \bar{n} + m \bar{m} ) \ V_8 \nonumber \\
             & - & ( \bar{n} m + n \bar{m} ) \ C_8 - 
( n m + \bar{n} \bar{m} )  \ S_8 \ ,
\eeq
e
\be
\mc{M}'_{0B}  = \pm  \frac{1}{2}  ( n + \bar{n} - m -\bar{m} ) \ \hat{O}_8 \ .
\ee

Le condizioni di $tadpole$, da $V_8$, $ C_8$ ed $ S_8$, richiedono che 
$ n = m $, ma 
lasciano indeterminata la dimensione del gruppo di gauge.
Lo spettro aperto che si ottiene \`e chirale e, a parte i bosoni di gauge di
$ U(n) \times U(m) $, le prime eccitazioni  comprendono 
tachioni nella rappresentazione (anti)simmetrica e fermioni chirali in 
diverse rappresentazioni bi-fondamentali.

In conclusione passiamo al modello descritto da $ \tld{\mc{K}}''_{0B} $,
le cui ampiezze di anello e di M\"obius nel canale diretto sono
\beq
\mc{A}''_{0B} & = & - \frac{n_o^2 + n_v^2 + n_s^2 + n_c^2}{2} \ C_8
- ( n_o n_v + n_s n_c ) \ S_8   \nonumber \\
              & + & (  n_v n_s +  n_o  n_c ) \ V_8 + ( n_v n_c + n_o n_s ) 
\ O_8 \ ,
\eeq
e
\be
\mc{M}''_{0B} =  \frac{1}{2} ( n_o - n_v - n_s + n_c ) \  \hat{C}_8 \ .
\ee
Le corrispondenti ampiezze nel canale trasverso sono quindi
\beq
\tld{\mc{A}}''_{0B} & = & \frac{2^{-6}}{2} [ {( n_o + n_v + n_s + n_c )}^2 \ 
V_8 - {( n_o + n_v - n_s - n_c )}^2 \ O_8   \nonumber \\
                    & + & {( n_o - n_v - n_s + n_c )}^2 \ C_8 
+ {( n_o - n_v + n_s - n_c )}^2 \ S_8 ] \, ,
\eeq
\be
\tld{\mc{M}}''_{0B} = \frac{2}{2} ( n_o - n_v - n_s + n_c ) \hat{C}_8 \, .
\ee
Osservando l'espressione per $ \tld{\mc{A}}''_{0B} $, si pu\`o notare che i 
settori $ O_8 $ ed $ S_8 $ compaiono in essa con segni che non rispettano 
la relazione tra spin e statistica e che rendono $ \tld{\mc{A}}''_{0B} $  
non compatibile 
con la condizione di positivit\`a dei coefficienti di riflessione 
dell'ampiezza nel canale trasverso. La consistenza di $ \tld{\mc{A}}''_{0B} $
pu\`o essere ripristinata imponendo che i coefficienti dei settori $ O_8 $ ed 
$ S_8 $ siano numericamente nulli, ovvero:
$$ n_o + n_v = n_s + n_c \ ,  $$
$$ n_o - n_v = n_c - n_s \ ,  $$
o, equivalentemente :
\beq
n_o = n_c  \, , \nonumber \\ 
n_v = n_s  \, .
\label{charge} 
\eeq
D'altra parte, come per la classe di modelli precedenti, dalla forma delle 
ampiezze nel canale diretto si deduce che le cariche $ n_i (i = o,v,s,c) $ 
devono essere reinterpretate in termini dei gruppi unitari, consistentemente 
con la condizione (\ref{charge}). Quindi ponendo 
$ n_v = n$, $n_s = \bar{n}$, $n_o = m$, $n_c = \bar{m}$, la (\ref{charge}) 
diviene :
$$ m = \bar{m} \ ,  $$
$$ n = \bar{n} \ , $$
ed \`e quindi identicamente soddisfatta.

Lo spettro aperto, descritto da 
\beq
\mc{A}''_{0B} & = & - \frac{n^2 + \bar{n}^2 + m^2 + \bar{m}^2}{2} \ C_8 + 
( n \bar{n} + m \bar{m} ) \ V_8 \nonumber \\
              & + & ( n \bar{m} + m \bar{n} ) \ O_8 - ( m n + \bar{m} \bar{n} )
\ S_8 \, ,
\eeq
\be
\mc{M}''_{0B} = \frac{m + \bar{m} - n - \bar{n}}{2} \hat{C}_8 \, ,
\ee
\`e soggetto alla condizione di $tadpole$ relativa al settore di R-R $ C_8 $ 
$$ m - n = 32  \  , $$
che assicura la cancellazione di anomalie di gauge e gravitazionali;
tuttavia non \`e possibile annullare il $tadpole$ del dilatone, e 
pertanto la dimensione del gruppo di gauge risulta indeterminata.
La scelta $ n = \bar{n} = 0 $  \`e per\`o l'unica  che d\`a luogo ad uno 
spettro privo di tachioni sia nel settore chiuso che in quello aperto,
e seleziona il gruppo di gauge $U(32)$.

Prima di addentrarci nello studio dei discendenti aperti di modelli pi\`u 
complessi, \`e istruttivo iniziare da un semplice esempio, 
legato alla compattificazione a nove dimensioni della superstringa di Tipo I.
Questo fornir\`a anche l'occasione per introdurre i concetti di 
\emph{D-brana} e di \emph{O(rientifold)-plane} 
\cite{Polchinski:1995mt, Pradisi:1989xd}.

La funzione di partizione di toro per la superstringa di Tipo IIB 
compattificata su un toro $ T^k $  si scrive
\be
\mc{T}_{IIB}^{T^K} = \int_{\mc{F}} \frac{d^2 \t}{ \t_2^{\frac{D-2}{2}+1}}\,
\frac{{| V_8 - S_8 |}^2 (\t)}{{( \eta(\t) \bar{\eta}(\bar{\t}))}^{D-k-2}}\,
\frac{\sum_{(p_L,p_R)} q^{\frac{\a'}{2} p_L^2} \bar{q}^{\frac{\a'}{2} p_R^2}}
{{( \eta(\t) \bar{\eta} (\bar{\t}))}^k}
\ee
dove ricordiamo che:
$$ \vec{p}_L = \frac{1}{\sqrt{\a'}} ( \a' \vec{p} + \vec{w} ) , $$
$$ \vec{p}_R = \frac{1}{\sqrt{\a'}} ( \a' \vec{p} - \vec{w} ) , $$
e $ q = e^{2 \pi i \t} $.
Lo spettro chiuso proiettato si ottiene al solito dall'ampiezza di 
bottiglia di Klein nel canale diretto. La struttura degli 
operatori di vertice su un toro mostra che in $\mc{K} $ possono 
fluire solamente gli stati caratterizzati da $ p_L = p_R $, ovvero gli stati 
con vettore di $winding$ nullo $(n = 0)$, gli unici mappati da $\O$ in s\'e 
stessi.
Considerando quindi il caso unidimensionale, $ k = 1 $, le ampiezza di toro e 
di bottiglia di Klein assumono la forma
\be
\mc{T} = {| V_8 - S_8 |}^2 ( \t ) \frac{\sum_{(m,n)}  q^{\frac{\a'}{2} p_L^2}
 \bar{q}^{\frac{\a'}{2} p_R^2}}{( \eta(q) \bar{\eta} (\bar{q}))} \, ,
\ee
\be
\mc{K} = \frac{1}{2} ( V_8 - S_8 )(2 i \t_2)  \frac{\sum_{m \in Z} q^{\frac{\a' m^2}{2 R^2}}}{\eta (q^2)}  \, , 
\label{Kcomp}
\ee
dove abbiamo lasciato impliciti sia i contributi bosonici che quelli della 
misura d'integrazione. Inoltre, al solito, in eq. (\ref{Kcomp}) 
$ q = e^{- 2 \pi \t_2} $.

Il passaggio al canale trasverso richiede una ridefinizione della variabile
($t=2 \t_2$), una trasformazione S ($ t \ra 1/l $) ed una risommazione di 
Poisson:
\be
\sum_{m={-\infty}}^{\infty} e^{- \pi \a^2 m^2} = \frac{1}{\a'} 
\sum_{k={-\infty}}^{\infty} e^{- \pi k^2/\a^2} \ ,
\ee
e pertanto 
\be
\tld{\mc{K}} = \frac{2^5}{2} \frac{R}{\sqrt{\a'}} ( V_8 - S_8 )(i l) 
\sum_{n \in Z} \frac{\sum_{n \in Z} q^{\frac{n^2 R^2}{\a'}}}{\eta (i l)} \ .
\ee
Da quest'ultima espressione emerge, per confronto con la successiva 
espressione per $ \tld{\mc{A}}$, che attraverso un crosscap possono fluire in
questo caso, solo i settori chiusi  con \emph{windings} pari, in quanto  
$ q^{\frac{n^2 R^2}{\a'}} $ pu\`o essere interpretato come 
$ e^{- \frac{\pi l}{2 \a'} {(2n)}^2 R^2} $. 
Si noti inoltre che in $ \tld{\mc{K}} $ compare il 
volume del toro unidimensionale  $ \frac{R}{\sqrt{\a'}}. $

Per ottenere il settore aperto, come al solito \`e conveniente partire 
dall'am\-piezza di anello nel canale trasverso, $ \tld{\mc{A}} $. 
In questo caso la condizione al bordo di Neumann assicura che non ci 
sia impulso nel tubo, ovvero: $ \vec{p_L} + \vec{p_R} = 0 $.
Quindi 
\be
\tld{\mc{A}} = \frac{2^{-5}}{2} N^2  \frac{R}{\sqrt{\a'}} ( V_8 - S_8 )(i l)
\frac{\sum_{n \in Z} {( e^{-2 \pi l})}^{\frac{n^2 R^2}{4 \a'}}}{\eta (i l)} 
\ ,
\ee
da cui con procedimento analogo al precedente, si ricava 
\be
\mc{A} = \frac{N^2}{2} ( V_8 - S_8 ) (i \frac{\t_2}{2}) 
\frac{\sum_{m \in Z} q^{\frac{\a' m^2}{2 R^2}}}{\eta (\sqrt{q})} \ .
\label{Acomp}
\ee
Si noti come eq. (\ref{Acomp}) sia esattamente consistente con la scala delle 
eccitazioni di momento nel settore aperto.
Infine, l'ampiezza di M\"obius nel canale trasverso \`e 
\be
\tld{\mc{M}} = - 2 \frac{N}{2} \frac{R}{\sqrt{\a'}} ( \hat{V}_8 - \hat{S}_8 )
(i l + i \frac{\t_2}{2}) \frac{\sum_{n \in Z} {( e^{-2 \pi l})}^{\frac{n^2 R^2}{ \a'}}}{\eta (i l + 1/2)} \ ,
\ee
e nel canale diretto diventa
\be
\mc{M} = - \frac{N}{2} ( \hat{V}_8 - \hat{S}_8 )(\frac{1}{2} + 
i \frac{\t_2}{2}) \frac{\sum_{m \in Z} q^{\frac{\a' m^2}{2 R^2}}}{\hat{\eta} 
(-\sqrt{q})} \ .
\label{Mcomp}
\ee
Anche per questo modello la condizione di {\it tadpole} fissa $N = 32$.  

Si osservi come, a differenza della stringa eterotica, lo spettro di questa
teoria non sia invariante sotto T-dualit\`a. Ricordiamo a tal fine che per
$ R \lra \frac{\a'}{R} $ gli impulsi quantizzati $ \frac{m}{R} $ si 
trasformano in {\it windings} $ \frac{n R}{\a'} $, e quindi 
$ T p_L = p_L $ \ e  
$ T p_R = - p_R $ ; \ o anche : $ T X_L = X_L $e $ T X_R = - X_R $ , da cui 
si deduce che T \`e un'operazione di parit\`a ``asimmetrica''
sul {\it world-sheet}, nel senso che essa agisce  solo sui modi destri.
Il comportamento ``non simme\-trico'' delle coordinate interne di stringa 
sotto T-dualit\`a ha delle implicazioni anche sulle condizioni
 ai bordi \cite{Dai:1989ua, Horava:1989vt} . 
Infatti l' usuale condizione al bordo di Neumann, 
che essenzialmente associa il settore di stringa aperta ad $ X_L + X_R $, 
viene trasformata dalla T-dualit\`a, in condizione di Dirichlet per il 
settore aperto corrispondente ad $ X_L - X_R $. In altri termini, la 
T-dualit\`a scambia le variabili del {\it world-sheet} $\s$ e $\t$, 
e quindi scambia
le derivate normali $\part_n = \part_\s$ e tangenziali $\part_t = \part_\t$ 
(ovvero condizioniai bordi di Neumann (N) con condizioni ai bordi di Dirichlet
 (D)). Di conseguenza, una generica stringa aperta con condizioni al 
contorno di Neumann, viene collegata dalla T-dualit\`a ad un nuovo tipo di
stringa aperta, con condizioni ai bordi di Dirichlet. 
Questa situazione ha fornito la base intuitiva per l'introduzione della 
nozione (geometrica) delle \emph{\bf D-brane}, ovvero di iperpiani sui quali 
gli estremi (di Dirichlet) delle stringhe aperte possono terminare 
\cite{Polchinski:1996fm} . 
La T-dualit\`a fornisce cos\`\i $\:$ l'interpretazione spazio-temporale dei 
due ingredienti di base per la descrizione della teoria non orientata sul 
{\it world-sheet}, ossia i bordi ed i crosscaps, mettendoli in
corrispondenza con due tipi di oggetti estesi, le \emph{\bf D-brane} e gli 
\emph{\bf O(rientifold)-planes}, rispettivamente. Alla luce di quanto appena 
detto, la bottiglia di Klein pu\`o essere interpretata come l'ampiezza che 
descrive la propagazione di 
una stringa chiusa che parte e termina su due O(rientifold)-planes, cos\`\i 
\ come l'ampiezza di anello  descrive la propagazione tre due D-brane. 
In realt\`a un'analisi pi\`u accurata mostra come 
le D-brane siano soluzioni solitoniche di stringa, e quindi oggetti dinamici, 
parte integrante delle configurazioni di vuoto di stringa aperta. In questo
senso gli stati delle stringhe aperte, ed in particolare gli stati privi di
 massa, sono interpretati come fluttuazioni delle D-brane. 
Viceversa nella teoria perturbativa di stringa gli O-planes non sono dinamici,
 cio\`e non esistono modi di stringa legati alle loro fluttuazioni.

Si definiscono $D_p$ - brane le ipersuperfici su cui terminano le 
stringhe aperte tali che $(p+1)$ delle coordinate hanno condizioni al 
bordo di Neumann, $ X^\a $ con 
$ \a = 0,\dots,p $  (tangenti al volume d'universo) e con $ (d-9-p)$ 
coordinate con condizioni al bordo di Dirichlet , $ X^m$, con $m=p, \dots, d$  
(normali al volume d'universo). 
Poich\'e la T-dualit\`a scambia condizioni ai bordi di 
Neumann con condizioni ai bordi di Dirichlet, essa agisce sulle D-brane  
alterandone la dimensionalit\`a; in particolare, pensando alle stringhe di 
Neumann come a  stringhe aperte che terminano su D9-brane, cio\`e su 
oggetti che invadono l'intero spazio-tempo a 10 dimensioni, una T-dualit\`a 
lungo una qualunque delle coordinate le trasforma in D8-brane.
Quindi, in generale, la T-dualit\`a fa aumentare o diminuire la dimensione di
una D-brana a se\-conda che essa agisca lungo una direzione ortogonale o 
parallela ad essa. In definitiva: 
$$\mbox{ agendo con T lungo una direzione tangente (N) alla $D_p$-brana} :  
D_p \ra D_{p-1}, $$
$$\mbox{ agendo con T lungo una direzione normale (D) alla $D_p$-brana} :  
D_p \ra D_{p+1}. $$
Una delle caratteristiche chiave delle D-brane e degli O-planes \`e che essi
sono le cariche elementari (le sorgenti elettriche) dei campi di
R-R \cite{Polchinski:1995mt}. 
Quindi, pi\'u precisamente, una D-brana (D) \`e tale se 
possiede carica di R-R positiva, mentre se la sua carica di R-R \`e negativa
\`e detta \emph{\bf anti}-D-brana ($\bar{D}$). Ovviamente un discorso analogo
pu\`o essere fatto per gli O-planes. Non \`e difficile dimostrare come
l'accoppiamento di una D-brana al corrispondente campo di Neveu-Schwarz sia 
proporzionale alla sua tensione, ma, mentre per le D-brane la tensione
\`e sempre positiva, per gli O-planes essa pu\`o essere anche negativa. Di
conseguenza gli orientifold-planes si presentano in due ulteriori variet\`a
rispetto alle D-brane, come riportato nella tabella (\ref{tab:1}).
\begin{table}[htbp]
\begin{center}
\begin{tabular}{|p{.7in}|*{6}{c|}}
\hline
                           & $ O_+$ & $O_-$ & $\bar{O}_+$ & $\bar{O}_-$ & $D$ & 
$\bar{D}$ \\
\hline
\bfseries carica           &    -   &    +  &      +      &     -       & + &   -     \\
\hline
\bfseries tensione         &    -   &    +  &      -      &     +       & + &   +     \\
\hline
\end{tabular}
\end{center}
\caption{Carica e tensione delle D-brane e degli O-piani}
\label{tab:1}
\end{table}
Abbiamo visto che tutte le teorie di (super)stringa consistenti a 10
dimensioni, eccetto l'eterotica, contengono nel loro spettro tensori 
antisimmetrici provenienti dal settore di R-R. Per la 
teoria di Tipo IIA sono consentite solo $D_p$ - brane supersimmetriche (BPS)
con $p$ pari ($p=0,2,4,6,8$), in quanto i corrispondenti tensori hanno
un numero dispari di indici, mentre per la Tipo IIB, i cui tensori hanno
numeri pari di indici, sono consentite solo $D_p$ - brane con $p$ dispari 
($p=-1,1,3,5,7,9$). Infine la teoria di Tipo I ammette solo D9 e D5 brane,
a meno di T-dualit\`a, le uniche consistenti con la proiezione $\O$.
Un altro aspetto particolarmente interessante delle D-brane, sul quale 
vogliamo soffermarci, \`e come esse forniscano un'interpretazione alternativa
della rottura della simmetria di Chan-Paton.
Per cominciare possiamo considerare il caso di stringhe aperte orientate, 
che corrispondono al gruppo di Chan-Paton $U(n)$. 
Compattificando una dimensione \`e possibile includere una linea di Wilson 
$ W = e^{\int_{\part \S} A_i d x^i} , $
con un campo di gauge interno $A_i$ abeliano e costante sul bordo. 
Un primo effetto
della linea di Wilson \`e quello di produrre una ridefinizione dell'impulso, 
a causa dell'accoppiamento minimale che altera i livelli
energetici (e questo \`e avvertibile solo se $p$ \`e quantizzato, ovvero in una 
variet\`a interna compatta). 
Ad esempio, per la stringa bosonica con un alinea di Wilson a $\s=0$, si ha:
$$ S = - \frac{T}{2} \int_{\S} d^2 \xi ( \dot{X}^2 - {X'}^2 ) + q \int  d \t 
A_i \dot{x}^i |_{\s=0}  . $$
Quindi, l'introduzione di una linea di Wilson costante sui bordi
\`e equivalente ad uno \emph{shift} negli impulsi degli stati di stringa 
aperta, perch\'e il momento coniugato diventa
$$ \Pi_i = T \dot{X}_i + q A_i \d (\s) . $$
Questo ha come conseguenza che nella teoria T-duale, dove il $winding$
sostituisce l'impulso, gli estremi dei campi non apparranno pi\'u allo 
stesso iperpiano, quindi, data una stringa ai cui estremi siano
associati i gradi di libert\`a di Chan-Paton $ | i,j \bra $ 
($ i,j=1,\cdots,N$),
l'estremo nello stato $i$, a meno di una costante additiva, si trover\`a 
nella posizione della corrispondente D-brana (le cui coordinate sono 
gli autovalori del campo di gauge $A_i$ moltiplicati per $2 \pi \a'$). 
Di conseguenza si avranno,  in generale, 
ipersuperfici in differenti posizioni, ovvero  differenti
D-brane. In conclusione, nella teoria duale, se le $N$ D-brane non coincidono,
$U(N)$ viene rotto ad $ U(1)^N $ (esiste un vettore di gauge a massa nulla 
per ogni $U(1)$), con stringhe aperte che si distendono tra queste superfici; 
se $k$ D-brane coincidono, allora nella teoria di partenza la linea di Wilson
lascer\`a un sottogruppo $U(k)$ non rotto (la simmetria verr\`a estesa); 
se, infine, le $N$ D-brane coincidono, il gruppo di gauge $ U(N)$ verr\`a 
ripristinato. In presenza di $\O$ questi gruppi divengono tipicamente $O(N)$,
come vedremo dalla discussione che segue.

\`E istruttivo considerare un semplice esempio, che \`e una 
generalizzazione del modello toroidale descritto dalle equazioni (\ref{Acomp})
ed (\ref{Mcomp}), dove l'inclusio\-ne di linee di Wilson rompe $SO(32)$ a 
$ U(M) \times SO(N) $ con $ N + 2M = 32 $ \cite{Bianchi:1992eu}. 
Le linee di Wilson alterano i livelli energetici a causa dell'accoppiamento 
mi\-nimale e della quantizzazione degli impulsi, e quindi in questo caso 
la ampiezza di anello e di striscia di M\"obius diventano
\beq
\mc{A} & = & (V_8 - S_8)(\sqrt{q}) \sum_{m \in Z} \Big[  
(M \bar{M} + \frac{N^2}{2}) 
\frac{q^{\a' m^2/ 2 R^2}}{\eta(\sqrt{q})}  \nonumber \\
       & + &  M N \frac{q^{\a' {(m+a)}^2 /2 R^2}}{\eta(\sqrt{q})} + \bar{M} N \frac{q^{\a' {(m-a)}^2/2 R^2}}{\eta(\sqrt{q})}  \nonumber \\
       & + & \frac{1}{2} M^2 \frac{q^{\a' {(m+2a)}^2 /2 R^2}}{\eta(\sqrt{q})} 
+ \frac{1}{2} {\bar{M}^2} \frac{q^{\a' {(m-2a)}^2 /2 R^2}}{\eta(\sqrt{q})} 
\Big]  \ ,
\label{Abreak}
\eeq
\beq
\mc{M} & = & - (\hat{V}_8 - \hat{S}_8)(- \sqrt{q}) \sum_{m \in Z} \Big[
\frac{1}{2} N \frac{q^{\a' m^2 / 2 R^2}}{\hat{\eta}(- \sqrt{q})}  \nonumber \\
       & + & \frac{1}{2} M \frac{q^{\a' {(m+2a)}^2 / 2 R^2}}{\hat{\eta}(-\sqrt{q})} + \frac{1}{2} \bar{M} \frac{q^{\a' {(m-2a)}^2 / 2 R^2}}{\hat{\eta}(-\sqrt{q})} \Big] \ .
\label{Mbreak}
\eeq
Le 32 cariche nella rappresentazione fondamentale di $SO(32)$ vengono
scisse in tre insiemi: i primi due comprendono $M$ cariche ciascuno, e nel
primo insieme il numero quantico dell'impulso $m$ viene traslato a $m+a$,
mentre nel secondo \`e traslato a $m-a$. Infine, nell'ultimo insieme l'impulso
rimane inalterato. Si noti come il contributo di $\mc{M}$ sia associato 
solamente ai termini di $\mc{A}$ con cariche identiche, ma i corrispondenti
$shifts$ negli impulsi appaiono raddoppiati, poich\'e il bordo della striscia
di M\"obius ha lunghezza doppia rispetto ai due bordi dell'anello.

Il passaggio al canale trasverso rende pi\'u limpida la struttura di questa
deformazione. Le ampiezze corrispondenti sono 
\be
\tld{\mc{A}} = \frac{2^{-5}}{2} \frac{R}{\sqrt{\a'}} (V_8 - S_8)(i l) \Big[
\sum_{n \in Z} \frac{q^{\frac{n^2 R^2}{4 \a'}}}{\eta(i l)} {(N + M e^{2 \pi i a n} + \bar{M} e^{- 2 \pi i a n})}^2 \Big] \, ,
\ee
\be
\tld{\mc{M}} = - \frac{2}{2} \frac{R}{\sqrt{\a'}} ({\hat{V}}_8 - {\hat{S}}_8 )(i l + \frac{1}{2}) \Big[ \sum_{n \in Z} \frac{q^{\frac{n^2 R^2}{\a'}}}{\hat{\eta}(i l + \frac{1}{2})} (N + M e^{4 \pi i a n} + \bar{M} e^{- 4 \pi i a n}) \Big] \, ,
\ee
dove le fasi che moltiplicano le cariche $M$ ed $\bar{M}$ sono associate alle
linee di Wilson delle componenti interne dei vettori di gauge. Osserviamo che 
se $a$ \`e un intero si ottiene il gruppo $SO(32)$, ma se $a = \frac{1}{2}$ le
cariche $M$ ed $\bar{M}$ hanno gli stessi coefficienti di riflessione in
$\tld{\mc{A}}$ ed $\tld{\mc{M}}$ e quindi $\mc{A}$ ed $\mc{M}$ contengono la
loro somma. Di conseguenza anche in questo caso si ha un allargamento 
(o un parziale ripristino)
del gruppo di simmetria da $U(M)$ ad $SO(2M) \times SO(32-M)$.

\begin{figure}[h]\unitlength1cm
\begin{picture}(3,3)
\put(5.5,0){\epsfig{file=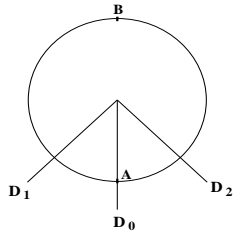,width=3cm,height=3cm}}
\end{picture}
\caption{\footnotesize{``Spostamenti'' di D-brane}}
\label{fig:displ}
\end{figure}

Effettuando un'operazione di T-dualit\`a, che trasforma le traslazioni nello
spazio degli impulsi in traslazioni nello spazio delle coordinate (duali),
la rottura della simmetria pu\`o essere pensata in termini di ``spostamenti''
delle D-brane \cite{Polchinski:1996fm}. 
A tal fine consideriamo la figura (\ref{fig:displ}).

Dopo una trasformazione di T-dualit\`a nella dimensione compatta, la stringa 
aperta vive su un segmento $S^1/Z_2$, con ai due estremi una coppia di 
O8-planes (che giacciono quindi sui due punti fissi di $S^1/Z_2$). Nell'ambito
di questa descrizione, la rottura della simmetria in (\ref{Abreak}) e 
(\ref{Mbreak}) coinvolge stringhe aperte che si allungano tra una D8-brana 
($D_0$) posta nel punto A del cerchio, ed un'ulteriore coppia di D8-brane 
($D_1,D_2$) situate in due punti immagine simmetrici rispetto alla 
precedente.  Il risultato appena illustrato pu\`o quindi
essere riprodotto in questo contesto,
``muovendo'' le brane da un punto fisso ad un punto nella regione interna della
circonferenza, e da qui al secondo punto fisso. In altri termini, 
l'allargamento 
del gruppo di gauge ha luogo quando $D_1$ e $D_2$ si incontrano in B, mentre 
il completo ripristino di $SO(32)$ si ottiene quando $D_0$, $D_1$ e $D_2$ si
trovano tutte in A. 

Passiamo ora a descrivere la compattificazione della teoria
di Tipo IIB su tori di dimensione maggiore di uno ed i
corrispondenti vuoti di tipo I. 
Questo tipo di analisi  permette di descrivere una particolare 
deformazione discreta, legata alla due-forma $B_{ab}$ del settore di NS-NS
\cite{Bianchi:1992eu}.
Ricordiamo che lo spettro della superstringa di Tipo I non contiene il 
tensore antisimmetrico $B_{\m\n}$, che viene rimosso dalla proiezione
di bottiglia di Klein. Ciononostante \`e possibile introdurre dei campi di
\emph{background} $B_{ab}$ (quantizzati), compatibilmente con la simmetria 
della teoria di Tipo IIB sotto lo scambio dei modi destri e sinistri, e 
studiare cos\`\i \ le propriet\`a dei rispettivi discendenti aperti.
Vedremo che la presenza di $backgrounds$ di questo genere implica una riduzione
del rango del corrispondente gruppo di Chan-Paton.

Per tori $k$-dimensionali, che non siano il prodotto di $k$ cerchi, i momenti
destri e sinistri della teoria chiusa di partenza possono essere espressi nel
modo seguente:
\be
p_{L,a} = m_a + \frac{1}{\a'} (g_{ab} - B_{ab}) n^b \, ,
\label{pLa}
\ee
\be
p_{R,a} = m_a - \frac{1}{\a'} (g_{ab} + B_{ab}) n^b \, ,
\label{pRa}
\ee
dove la metrica $g_{ab}$ descrive la dimensione e la forma del toro interno. 
Riscalando gli impulsi 
(\ref{pLa}) e (\ref{pRa}) con un vielbein inverso, l'ampiezza di toro pu\`o
essere espressa in modo da contenere solo la metrica interna, ovvero
$$ \mc{T} = {| V_8 - S_8 |}^2 \sum_{(m,n)} \frac{q^{\frac{\a'}{4} p_L^T g^{-1} p_L} \bar{q}^{\frac{\a'}{4} p_R^T g^{-1} p_R}}{{| \eta(q) |}^{2k}} \ . $$
Abbiamo visto  che l'ingrediente alla base della costruzione dei 
discendenti aperti \`e la simmetria, della teoria di partenza, sotto lo scambio dei modi destri e sinistri. Per questo tipo di compattificazione, la simmetria
del \emph{world-sheet} si manifesta per valori generici della metrica 
$g_{ab}$ solo su quei tori per i quali, dato un certo $p_{L,a}$, esista un 
corrispondente $p'_{R,a}$ tale che
$$ p_{L,a} = p'_{R,a} \ .$$
Questa condizione si traduce in un vincolo sul tensore $B_{ab}$, in quanto
implica che  $ \frac{2}{\a'} B_{ab}   \, \in Z ,$ ovvero che $B_{ab}$ sia 
\emph{quantizzato} nelle unit\`a appropriate.

Una volta imposta la simmetria richiesta per la teoria di stringa chiusa di 
partenza, si pu\`o procedere al calcolo dei discendenti aperti, cominciando,
al solito, dall'ampiezza di bottiglia di Klein (che in questo caso non viene
alterata) 
\be
\mc{K} = \frac{1}{2} ( V_8 - S_8 )(q^2) \sum_m \frac{q^{\frac{\a'}{2} m^T g^{-1} m}}{\eta^k (q^2)} \ ,
\ee
per cui la corrispondente ampiezza nel canale trasverso \`e 
\be
\tld{\mc{K}} = \frac{2^5}{2} \sqrt{det(\frac{g}{\a'})} ( V_8 - S_8 )(i l)
\sum_n \frac{q^{\frac{1}{\a'} n^T g n}}{\eta^k (i l)} \ .
\ee
L'ampiezza di anello nel canale trasverso $\mc{A}$ pu\`o contenere solo gli 
stati accoppiati ai  loro coniugati, ossia gli stati per i quali 
$$ p_{L,a} = - p_{R,a} .  $$
Questa condizione richiede che 
\be
\frac{2}{\a'} B_{ab} n^b = 2 m_a \ .
\label{Bab}
\ee
Data la condizione di quantizzazione sul campo $B_{ab}$, la (\ref{Bab}) 
diviene un vincolo per $n^b$, $ \frac{2}{\a'} B_{ab} n^b \, \in 2  Z \ , $
del quale si tiene conto introducendo un opportuno proiettore nell'ampiezza
$\tld{\mc{A}}$, che  assume cos\`\i \ la forma 
\be
\tld{\mc{A}} = \frac{2^{r-k-5}}{2} N^2 \sqrt{det(\frac{g}{\a'})} (V_8 - S_8)(i l) \sum_{\e=0,1} \sum_m \frac{q^{\frac{1}{4 \a'} n^T g n} e^{\frac{2 \pi i}{\a'} n^T B \e}}{\eta^k(i l)} \ .
\ee
 $r$ \`e il rango di B e la sua presenza in $\tld{\mc{A}}$ assicura la 
corretta normalizzazione di $\mc{A}$.
Come al solito l'ampiezza di anello nel canale diretto si ottiene mediante la 
trasformazione modulare $S$ 
\be
\mc{A} = \frac{2^{r-k}}{2} N^2 (V_8 - S_8)(\sqrt{q}) \sum_{\e=0,1} \sum_m 
\frac{q^{\frac{\a'}{2} {(m + \frac{1}{\a'} B \e)}^T g^{-1} (m + \frac{1}{\a'} B \e)}}{\eta^k(\sqrt{q})}  \ ,
\ee
e di nuovo la presenza di $r$ garantisce che i vettori a massa nulla abbiano 
precisamente molteplicit\`a di Chan-Paton pari ad $N$.
La proiezione del settore aperto viene completata dall'ampiezza 
di striscia di M\"obius, che nel canale trasverso assume la forma 
\be
\tld{\mc{M}} = - \frac{2}{2} 2^{r/2-k/2} N \sqrt{det(\frac{g}{\a'})} (\hat{V}_8 - \hat{S}_8)(i l + \frac{1}{2}) \sum_{\e=0,1} \sum_n 
\frac{q^{\frac{1}{\a'} n^T g n} e^{\frac{2 \pi i}{\a'} n^T B \e}}{{\hat{\eta}}^k (i l + \frac{1}{2})} \g_{\e}  \  ,
\ee
e mediante una trasformazione modulare $P$ :
\be
\mc{M} = - \frac{2^{r/2-k/2}}{2} N (\hat{V}_8 - \hat{S}_8)(-\sqrt{q}) 
\sum_{\e=0,1}  \sum_m  \frac{q^{\frac{\a'}{2} {(m + \frac{1}{\a'} B \e)}^T g^{-1} (m + \frac{1}{\a'} B \e)}}{{\hat{\eta}}^k(-\sqrt{q})} \g_{\e}  \ .
\ee
Osserviamo subito come l'ampiezza di striscia di M\"obius presenti una nuova 
sottigliezza legata alla presenza di ulteriori segni $\g_\e$, necessari per 
garantire la corretta normalizzazione nel canale diretto.
L'ultimo passo nella costruzione dei discendenti aperti consiste nell'imporre
la cancellazione dei {\it tadpole} degli stati a massa nulla non fisici che 
fluiscono nel canale trasverso. In questo caso la condizione di {\it tadpole}
 \`e 
$$ \frac{2^5}{2} + \frac{2^{r-5}}{2} N^2 + 2 \frac{2^{r/2-k/2}}{2} N \sum_{\e=0,1} \g_{\e} = 0  ,  $$
che si pu\`o scrivere come un quadrato perfetto solo se 
$$ \sum_{\e=0,1} \g_{\e} = 2^{k/2} ,  $$
e  conduce al risultato  
$$ N = 2^{5-r/2}  .  $$
Si noti che i segni $\g_{\e}$ non sono tutti positivi. Questo fatto ha 
importanti conseguenze, ed in particolare consente di passare con continuit\`a da gruppi ortogonali $O(n)$ a gruppi $USp(n)$.

Come anticipato, il gruppo di gauge risultante ha in generale un rango minore 
rispetto al caso senza $B_{ab}$. Si vede inoltre come, se la matrice 
antisimmetrica $B_{ab}$ ha rango $r$ (necessariamente pari), 
esistono $2^{r/2}$ ``immagini'' dello spettro aperto che condividono la 
carica di Chan-Paton.

\section{Modelli in sei dimensioni}
Vogliamo ora trattare la costruzione di due classi di modelli in sei 
dimensioni, discendenti dell'orbifold $ T^4/Z_2$ della teoria di Tipo IIB.
Per far questo \`e opportuno introdurre le  combinazioni di 
supercaratteri di $SO(4)$
\beq
Q_o = V_4 O_4 - C_4 C_4  \quad , \quad  Q_v = O_4 V_4 - S_4 S_4  , \nonumber \\
Q_s = O_4 C_4 - S_4 O_4  \quad , \quad  Q_c = V_4 S_4 - C_4 V_4  \  ,
\label{scar}
\eeq
definiti da 
\beq
O_4 = \frac{\vth_3^2 + \vth_4^2}{2 \eta^2} \ , \ V_4 = \frac{\vth_3^2 - \vth_4^2}{2 \eta^2} \ , \ S_4 =  \frac{\vth_2^2 - \vth_1^2}{2 \eta^2} \ , \ 
C_4 = \frac{\vth_2^2 + \vth_1^2}{2 \eta^2} \ .
\eeq
Questi ci consentono di porre la funzione di partizione di toro (comune ai due
modelli) nella forma
\beq
\mc{T} & = & \frac{1}{2} {|Q_o + Q_v|}^2 \L_4 + \frac{1}{2} {|Q_o - Q_v|}^2 
{\Big|\frac{2 \eta}{\th_2}\Big|}^4 \nonumber \\
       & + & \frac{1}{2} {|Q_s + Q_c|}^2 
{\Big|\frac{2 \eta}{\th_4}\Big|}^4 + \frac{1}{2} {|Q_s - Q_c|}^2
{\Big|\frac{2 \eta}{\th_3}\Big|}^4 \ ,
\label{TT4}
\eeq
dove al solito sono stati omessi i contributi dei bosoni trasversi e la misura
d'integrazione, mentre  $\L_4$ rappresenta il reticolo di compattificazione.

Come abbiamo visto, la formulazione perturbativa della teoria di
Tipo I si fonda sull'invarianza modulare dello spettro di stringa chiusa, e su
alcune condizioni che legano la bottiglia di Klein al settore aperto e  non 
orientato. A queste occorre  aggiungere le condizioni di \emph{tadpole}
di R-R, connesse  con la cancellazione delle anomalie,  che possono essere 
pensate come condizioni di neutralit\`a globale per le cariche di R-R nello
spazio compatto interno. La cancellazione dei {\it tadpole} deriva da 
contributi 
opposti di bordi e \emph{crosscaps}, o, nel linguaggio spazio-temporale, di
brane ed \emph{orientifolds}. \`E possibile costruire modelli dove alcuni dei
contributi di \emph{orientifold} sono invertiti, e quindi la cancellazione 
dei $tadpole$ di R-R non pu\`o essere imposta nella maniera usuale. 
Per tener conto di questa possibilit\`a scriviamo l'ampiezza di bottiglia di 
Klein 
\be
\mc{K} = \frac{1}{4} \Big[ (Q_o + Q_v)(P_4 + W_4) + 2 \e \times 16 (Q_s + Q_c)
{\Big(\frac{\eta}{\th_4}\Big)}^2 \Big] \ ,
\ee 
dove $\e = \pm 1$ \`e il fattore che discrimina tra i due tipi di modelli, 
mentre $P_4$ e $W_4$ sono le somme reticolari ristrette rispettivamente a 
\emph{windings} ed impulsi nulli.

Per entrambe le scelte di $\e$, lo spettro di stringa chiusa ha supersimmetria
$N = (1,0)$, ma le proiezioni che ne derivano sono profondamente diverse:
\begin{itemize}
\item per $\e =+1$, lo spettro chiuso consiste di un multipletto 
gravitazionale $(1 G = g_{\m\n}, 2 \p_{\m,L}, 5 B_{\m\n}^+)$, 
un multipletto tensoriale $(1 T = B_{\m\n}^-, \c_R, \f)$, 
e 20 ipermultipletti $(1H = \p_R,4\f)$. Il modello corrispondente \`e noto 
come \emph{\bf modello  U(16) $\times$ U(16)} 
\cite{Pradisi:1989xd,Gimon:1996rq}.
\item per $\e =-1$, tutti i contributi del settore \emph{twisted} vengono
invertiti, e lo spettro chiuso non orientato che ne risulta contiene, oltre al
multipletto gravitazionale, 17 multipletti tensoriali e  4 ipermultipletti.
Il modello corrispondente esibisce un nuovo fenomeno, noto come 
\emph{\bf Brane Supersymmetry Breaking (BSB)} \cite{Antoniadis:1999xk}.
\end{itemize}
Dopo  una trasformazione modulare $S$ si ottiene l'ampiezza di bottiglia di 
Klein nel canale trasverso 
\be
\tld{\mc{K}} = \frac{2^5}{4} \Big[ (Q_o + Q_v)(v_4 W_4^e + \frac{P_4^e}{v_4})
+ 2 \e (Q_o - Q_v) \Big(\frac{2 \eta}{\th_2}\Big)^2 \Big] \ ,
\ee
dove $v_4$ \`e il volume dello spazio compatto e $P_4^e$ $(W_4^e)$ indica la 
somma reticolare ristretta ad impulsi (\emph{windings}) pari.\\
\`E istruttivo controllare la forma assunta da $\tld{\mc{K}}$ all'origine del reticolo di $T^4$ (dove $P_4^e = W_4^e = 1$), in quanto si verifica che i 
coefficienti di $Q_o$ e $Q_v$ si arrangiano in quadrati perfetti, anche per 
$\e =-1$ :
\be
\tld{\mc{K}}_0 = \frac{2^5}{4} \Big[ Q_o (\sqrt{v_4} + 
\frac{\e}{\sqrt{v_4}})^2 + Q_v (\sqrt{v_4} - \frac{\e}{\sqrt{v_4}})^2 
\Big] \ .
\label{K0tiluntw}
\ee
Dobbiamo ora costruire il settore aperto dei due modelli, ma per il momento 
concentriamoci sul caso $\e =+1$. Dalla (\ref{TT4}) si vede che in 
$\tld{\mc{A}}$ possono fluire tutti i settori (sia $twisted$ che $untwisted$),
 ma ci\`o che possiamo scrivere facilmente \`e la parte \emph{untwisted} di 
$\tld{\mc{A}}_0$, poich\'e i coefficienti di $Q_o$ e $Q_v$  devono essere 
quadrati perfetti (descrivono la propagazione di stringhe chiuse tra due 
bordi). Quindi :
\be
{\tld{\mc{A}}}_0^{untw} = \frac{2^{-5}}{4} \Big[  Q_o (N \sqrt{v_4} + 
\frac{D}{\sqrt{v_4}})^2 + Q_v (N \sqrt{v_4} - \frac{D}{\sqrt{v_4}})^2 
\Big]  , ,
\label{A0tiluntw}
\ee
dalla quale si possono desumere immediatamente i contributi \emph{untwisted}
di $\tld{\mc{A}}$ includendo le somme reticolari complete, ovvero :
\be
{\tld{\mc{A}}}^{untw} = \frac{2^{-5}}{4} \Big[ (Q_o + Q_v) ( N^2 v_4 W_4 + 
\frac{D^2 P_4}{v_4} ) + 2 N D (Q_o - Q_v ) \Big] \ .
\label{Auntw}
\ee
Dalla (\ref{Auntw}) si deduce che, nel canale trasverso, in presenza di 
D9-brane, stringhe che si stendono tra due bordi di tipo 
Neumann, corrispondono ai soli stati con impulso nullo (come si pu\`o leggere
dal termine di (\ref{Auntw}) proporzionale ad $ N^2 v_4 W_4 $), mentre in 
presenza di D5-brane,  stringhe che si allungano tra due bordi di tipo 
Dirichlet, corrispondono ai soli stati con avvolgimento nullo (come si 
pu\`o leggere dal termine di (\ref{Auntw}) proporzionale ad 
$\frac{D^2 P_4}{v_4}$). Infine, in presenza di bordi ``misti'' 
Neumann-Dirichlet, i modi di stringa non posseggono zero modi 
n\'e di  impulso n\'e di avvolgimento.

Notiamo inoltre che l'assenza di settori $twisted$ in $\tld{\mc{K}}_0$ ed
$\tld{\mc{A}}_0$, fa s\`\i \quad che anche $\tld{\mc{M}}_0$ ne sia priva. 
$\tld{\mc{M}}_0$ \`e quindi caratterizzata da un'espressione del tipo
\be
\tld{\mc{M}}_0 = - \frac{2}{4} \Big[ \hat{Q}_o  (\sqrt{v_4} + 
\frac{\e}{\sqrt{v_4}}) (N \sqrt{v_4} + \frac{D}{\sqrt{v_4}})+
\hat{Q}_v  (\sqrt{v_4} - \frac{\e}{\sqrt{v_4}}) (N \sqrt{v_4} - 
\frac{D}{\sqrt{v_4}})  \Big] \ ,
\label{M0tiluntw}
\ee
da questa, introducendo le somme reticolari, si ottiene l'ampiezza completa
di striscia di M\"obius nel canale trasverso 
\be
\tld{\mc{M}} = - \frac{2}{4} \Big[ (\hat{Q}_o + \hat{Q}_v) (N v_4 W_4^e + 
D \frac{P_4^e}{v_4}) + (\hat{Q}_o - \hat{Q}_v) (N + D) \Big] \ .
\label{Mtil}
\ee
Le eq. (\ref{K0tiluntw}), (\ref{A0tiluntw}) e (\ref{M0tiluntw})  
permettono di ricavare la condizione di $tadpole$ per questo modello, 
\be
\Big[ \sqrt{v_4} (2^5 - N) + \frac{1}{\sqrt{v_4}} (2^5 - D) \Big]^2 = 0 \ ,
\ee
che conduce alle soluzioni
\beq
N = 32  \ ,  \nonumber \\
D = 32  \ .
\eeq
Il modello \`e quindi consistente se sono presenti 32 D9-brane e 32 D5-brane.\\
Per poter leggere lo spettro aperto a  massa nulla della teoria, \`e necessario
passare al canale diretto. Per quanto riguarda $\mc{M}$ basta applicare alla
(\ref{Mtil}) una trasformazione modulare $P^{-1}$, ottenendo cos\`\i
\be
\mc{M} = - \frac{1}{4} [ (\hat{Q}_o + \hat{Q}_v) (N P_4 + D W_4) - (\hat{Q}_o -
\hat{Q}_v) (N + D) ] \ .
\ee
All'origine dei reticoli di $T^4$ 
\be
\mc{M}_0 = - \frac{1}{4} \times 2  \hat{Q}_v (N + D) \ ,
\label{M0}
\ee
e il coefficiente di $\hat{Q}_o$ si annulla, che implica la presenza di  
gruppi di gauge unitari.

La deduzione di $\mc{A}$ non \`e diretta come quella di $\mc{M}$. Applicando 
la trasformazione modulare $S^{-1}$ ad $\tld{\mc{A}}^{untw}$ si ottiene 
\be
\mc{A} = \frac{1}{4} [ (Q_o + Q_v) (N^2 P_4 + D^2 W_4) + 2 N D (Q_s + Q_c) ]
 \ ,
\ee
che, pur includendo le somme reticolari, non fornisce l'espressione completa
di $\mc{A}$. Mancano infatti i termini che simmetrizzano lo spettro 
rispetto all'involuzione che definisce l'orbifold. Quindi, se $N^2$,
$D^2$ ed $ND$ sono i fattori di Chan-Paton standard, occorre introdurre dei
 nuovi fattori di Chan-Paton che tengano conto dei bordi, che 
sotto involuzione, presentano un comportamento differente rispetto ai 
precedenti. Di conseguenza l'espressione corretta per l'ampiezza di 
anello \`e 
\beq
\mc{A} & = & \frac{1}{4}\, [\, (Q_o + Q_v) (N^2 P_4 + D^2 W_4) 
+ 2 N D (Q_s + Q_c) \,] \nonumber \\
       & + &  \frac{1}{4}\, [\, (Q_o - Q_v) (R_N^2 + R_D^2) + 2 R_N R_D 
(Q_s - Q_c )\,] \ ,
\eeq
che all'origine dei reticoli diviene
\beq
\mc{A}_0 & = & \frac{1}{4}\, [\, (Q_o + Q_v) (N^2 + D^2 ) 
+ 2 N D (Q_s + Q_c) \,] \nonumber \\
       & + &  \frac{1}{4}\, [\, (Q_o - Q_v) (R_N^2 + R_D^2) + 2 R_N R_D 
(Q_s - Q_c )\,] \ .
\label{A0compl}
\eeq

I contributi $twisted$ di $\tld{\mc{A}}$ (\emph{termini di rottura})
sono quindi dati da 
\be
{\tld{\mc{A}}}^{tw} =  \frac{2^{-5}}{4} \, [\, 16\, (R_N^2 + R_D^2) (Q_s + Q_c) - 8 R_N R_D (Q_s - Q_c) \,] \ .
\ee
Questi termini rivestono un ruolo particolarmente importante, poich\'e  
dipendono in modo molto chiaro dalla disposizione delle brane, e quindi 
consentono  di dare un'interpretazione geometrica ai modelli. 
Consideriamo a tal fine il coefficiente di $Q_s$.
L'unica maniera di scriverlo in termini di quadrati perfetti \`e 
\be
Q_s : \qquad  \frac{1}{4} \ [\, (R_N - 4 R_D)^2 + 15 R_N^2 \,] \ .
\label{branegeom}
\ee
Ricordiamo che l'orbifold $T^4/Z_2$ \`e un tetraedro con 16 punti fissi.
Dalla (\ref{branegeom}) si evince correttamente che, 
mentre le D9-brane invadono tutto lo
spazio-tempo, le D5-brane giacciono tutte su un unico punto fisso del toro
interno, essendo $ 4 = \sqrt{16} $ .

Infine, dal coefficiente di $Q_v$ delle eq. (\ref{M0}) e (\ref{A0compl})
leggiamo che il gruppo di gauge \`e $ U(N) \times U(D); $ ma usando la
seguente parametrizzazione delle cariche (complesse) di Chan-Paton :
$$ N = n + \bar{n}  \quad , \quad D = d + \bar{d}  \, , $$
$$ R_N = i (n - \bar{n})  \quad , \quad R_D = i (d - \bar{d})  \, , $$
con la condizione $ n = \bar{n} $ e $ d = \bar{d} $, la teoria viene ad essere
caratterizzata dal gruppo di simmetria $ U(16)_9 \times U(16)_5 $,
come di pu\`o leggere dall'espressione per lo spettro aperto
\beq
\mc{A}_0 + \mc{M}_0 & = & Q_o  \ ( n \bar{n} + d \bar{d} )  + 
Q_s \ \frac{1}{2} ( n \bar{d} + \bar{n} d ) \nonumber \\
                    & + & Q_v \ \Big[ \  \frac{n (n-1)}{2} +  
\frac{\bar{n} (\bar{n}-1)}{2} +  \frac{d (d-1)}{2} + 
\frac{\bar{d} (\bar{d}-1)}{2} \ \Big] \ . \nonumber \\
\eeq
Passiamo ora a discutere la costruzione del settore aperto del modello con
$\e =-1$. In questo caso la modifica della proiezione di bottiglia di Klein
e la cancellazione del {\it tadpole} di R-R, sono compatibili con la 
positivit\`a
dello spettro aperto in $\mc{A}$ a patto di introdurre delle 
\emph{anti}-D5-brane (D$\bar{5}$), ovvero delle D5-brane con carica di R-R
negativa, ma con accoppiamenti di NS-NS identici a quelli delle brane
\beq
\mc{A}_{BSB} & = & \frac{1}{4} \Big[ \ ( Q_o + Q_v ) \ ( N^2 P_4 + D^2 W_4 )
+ 2 N D  ( Q'_s + Q'_c ) \Big( \frac{\eta}{\vth_4} \Big)^2  \nonumber \\
             & + &  ( R^2_N + R^2_D ) ( Q_o - Q_v )  
\Big( \frac{ 2 \eta}{\vth_2} \Big)^2
\nonumber \\
             & + & 2 R_N R_D ( V_4 C_4 + S_4 V_4 - O_4 S_4 - C_4 O_4 ) 
\Big( \frac{\eta}{\vth_3} \Big)^2 
\ \Big]   \ ,
\eeq
dove sono stati introdotti i supercaratteri primati
\beq
Q '_o = V_4 O_4 - S_4 S_4  \quad , \quad  Q '_v = O_4 V_4 - C_4 C_4  , \nonumber \\
Q '_s = O_4 S_4 - C_4 O_4  \quad , \quad  Q '_c = V_4 C_4 - S_4 V_4  \  ,
\eeq
legati a quelli di eq. (\ref{scar}) dallo scambio (della chiralit\`a) 
$S_4 \lra C_4$.

Nel canale trasverso, questa scelta inverte i segni di tutti i contributi
di R-R nella parte N(eumann)-D(irichlet), lasciando inalterati tutti gli 
altri termini $untwisted$
\beq
\tld{\mc{A}}_{BSB} & = & \frac{2^{-5}}{4} \Big[ \ ( Q_o + Q_v ) \ 
( N^2 v_4 W_4 + D^2 \frac{P_4}{v_4} ) 
+ 2 N D  ( Q'_o - Q'_v ) \Big( \frac{2 \eta}{\vth_2} \Big)^2  
\nonumber \\
                   & + & 16  ( R^2_N + R^2_D ) ( Q_s + Q_c )  
\Big( \frac{\eta}{\vth_4} \Big)^2
\nonumber \\
                   & + & 8 R_N R_D ( V_4 S_4 + C_4 V_4 - O_4 C_4 - S_4 O_4 ) 
\Big( \frac{\eta}{\vth_3} \Big)^2 
\ \Big]   \ .
\eeq
Questo ha come conseguenza la rottura della 
supersimmetria, a livello ad albero, nei settori di D$\bar{5}$-brana, mentre
gli stati dei settori D9-D9 o D9-O9 rimangono supersimmetrici.
Infatti ricordando che $\tld{\mc{M}}$ descrive la propagazione tra
brane ed {\it orientifold planes}, dalla sua espressione
\beq
\tld{\mc{M}}_{BSB} & = & - \frac{1}{2} \Big[ \ N v_4 W_4^e 
( \hat{Q}_o + \hat{Q}_v ) - N ( \hat{Q}_o - \hat{Q}_v ) 
\Big( \frac{2 \hat{\eta}}{\hat{\vth}_2} \Big)^2 \nonumber \\
                   & - & D  \ \frac{P_4^e}{v_4} \
( \hat{V}_4 \hat{O}_4 +  \hat{O}_4 \hat{V}_4 + \hat{C}_4 \hat{C}_4 
+ \hat{S}_4 \hat{S}_4 ) \nonumber \\
                  & + & D \  ( \hat{V}_4 \hat{O}_4 -  \hat{O}_4 \hat{V}_4 
+ \hat{C}_4 \hat{C}_4 - \hat{S}_4 \hat{S}_4 ) 
\Big( \frac{2 \hat{\eta}}{\hat{\vth}_2} \Big)^2
\  \Big] \ ,
\eeq
si pu\`o vedere che tutti i segni dei termini D9-O5, D$\bar{5}$-O5 
nel settore NS-NS e dei termini D$\bar{5}$-O9 nel settore R-R, 
sono invertiti rispetto ai segni della (\ref{Mtil}).
Come al solito, una trasformazione modulare $P$ determina 
l'ampiezza nel canale diretto 
\beq
\mc{M}_{BSB} & = & - \frac{1}{4} \Big[ \ N P_4 ( \hat{Q}_o + \hat{Q}_v ) - 
N ( \hat{Q}_o - \hat{Q}_v ) \Big( \frac{2 \hat{\eta}}{\hat{\vth}_2} \Big)^2 
\nonumber \\
       & - & D W_4 ( \hat{V}_4 \hat{O}_4 +  \hat{O}_4 \hat{V}_4 
+ \hat{C}_4 \hat{C}_4 + \hat{S}_4 \hat{S}_4 ) \nonumber \\
       & + & D (  \hat{O}_4 \hat{V}_4 - \hat{V}_4 \hat{O}_4 + \hat{S}_4 \hat{S}_4 
 - \hat{C}_4 \hat{C}_4 )  \Big( \frac{2 \hat{\eta}}{\hat{\vth}_2} \Big)^2   
\  \Big]  \ .
\eeq
Le condizioni di {\it tadpole} possono essere cos\`\i $\:$ risolte 
disaccoppiando i contributi di NS-NS e di R-R. 
Mentre si riescono a risolvere tutti
i {\it tadpoles} di R-R, che fissano
$$ N = D = 32 , \quad \mbox{e} \quad   R_N = R_D = 0 , $$
lo stesso non \`e vero per quelli di NS-NS, che 
danno un contributo all'energia di vuoto localizzato 
sulle antibrane.
Parametrizzando le cariche come
$$ N = n_1 + n_2 \ , \qquad D = d_1 + d_2  \  , $$
$$ R_N = n_1 - n_2 \ , \qquad R_D = d_1 - d_2 \ , $$
si ricava che questo modello, contenente 32 D9-brane e 32 D$\bar{5}$-brane,
\`e caratterizzato da uno spettro aperto, descritto da
\beq
\mc{A}_0 + \mc{M}_0 & = & \frac{1}{2} [ \ n_1 (n_1 + 1) + n_2 (n_2 + 1) +
d_1 (d_1 + 1) + d_2 (d_2 + 1) \ ] V_4 O_4 \nonumber \\
                   & - & \frac{1}{2} [ \ n_1 (n_1 - 1) + n_2 (n_2 - 1) +
d_1 (d_1 - 1) + d_2 (d_2 - 1) \ ] C_4 C_4 \nonumber \\
                   & + & ( n_1 n_2 + d_1 d_2 ) ( O_4 V_4 - S_4 S_4 ) 
\nonumber \\
                   & + & ( n_1 d_2 + n_2 d_1 )  O_4 S_4 
\nonumber \\
                   & - & ( n_1 d_1 + n_2 d_2 )  C_4 O_4  \ ,
\eeq
che  \`e non supersimmetrico, chirale, privo di 
tachioni e caratterizzato dal gruppo di gauge  
$[ SO(16) \times SO(16) ]_9 \times [ USp(16) \times USp(16) ]_5 $. 
Inoltre viene generata una costante cosmologica al livello $ \c=1$.
\chapter{Deformazioni magnetiche e shift-orbifolds}
Dai capitoli precedenti dovrebbe essere emerso 
come le stringhe aperte possono essere considerate una generalizzazione
 dei campi di Yang-Mills. Pertanto 
lo studio della loro dinamica in campi elettrici e magnetici di $background$ 
risulta particolarmente interessante \cite{Fradkin:1985qd}.

Anzitutto bisogna distinguere tra campi elettromagnetici 
spazio-temporali ed interni. 
Nel primo caso sono essenzialmente due le motivazioni che spingono ad
una loro trattazione nel contesto della teoria di stringhe aperte. Una \`e di 
valenza concettuale, perch\'e nell'espansione perturbativa delle teorie di 
stringa aperta la gravit\`a  compare a genere zero, mentre 
l'elettromagnetismo compare a genere $\frac{1}{2}$,  e si 
presenta  come una correzione quantistica dello spazio-tempo piatto. La 
seconda, di natura pi\`u tecnica, si fonda sul fatto che la presenza di un 
particolare campo elettromagnetico permette di cambiare, in
modo continuo, le condizioni al bordo dei campi di stringa aperta.

La presenza di campi magnetici interni  ha varie importanti conseguenze,
ed in particolare pu\`o dar luogo alla rottura della supersimmetria 
\cite{Witten:1984dg, Bachas:1995ik}. 
Gli estremi carichi delle
stringhe aperte possono infatti accoppiarsi al campo, e di conseguenza
 particelle di diversi spin avvertono i suoi effetti in ragione dei loro 
momenti di dipolo, acquistando cos\`\i \ masse diverse. 
Ci occuperemo in maniera approfondita di questa seconda 
situazione nei paragrafi successivi. Iniziamo dal problema della
 stringa bosonica aperta in un campo di gauge abeliano, che nel limite di 
campo lentamente variabile risulta essere esattamente risolubile.
Nel seguito descriveremo come sia possibile combinare una magnetizzazione
interna con costruzioni di {\it shift-orbifolds} in quattro dimensioni.

\section{La stringa bosonica aperta in un campo magnetico uniforme}   
Nel gauge conforme, l'azione per la stringa bosonica aperta in un campo 
elettromagnetico  \`e 
\beq
S & = & \frac{1}{4 \pi \a'} \int_{- \infty}^{+\infty}d \t \int_0^{\pi} d \s 
\part_\a X^\m \part^\a X_\m \nonumber \\
  & + & \frac{1}{2 \pi \a'} \int_{-\infty}^{+\infty} d \t
[ q_1 A_\m \part_\t X^\m (0) + q_2 A_\m \part_\t X^\m (\pi) ] \ .
\label{Sboseem}
\eeq
Per un campo elettromagnetico uniforme, descritto da un tensore $F_{\m\n}$, 
si pu\`o scegliere un potenziale 
\be
A_\m = - \frac{1}{2} F_{\m\n} X^\n  \, ,
\label{gaugecond}
\ee
e la (\ref{Sboseem}) diviene 
\beq
S & = & \frac{1}{4 \pi \a'} \int_{- \infty}^{+\infty}d \t \int_0^{\pi} d \s 
\part_\a X^\m \part^\a X_\m \nonumber \\
  & - & \frac{1}{ 4 \pi \a'} \int_{-\infty}^{+\infty} d \t
[ q_1 F_{\m\n} X^\n \part_\t X^\m (0) + q_2 F_{\m\n} X^\n \part_\t X^\m (\pi)
 ]
\ .
\eeq
In generale, il campo abeliano e costante, \`e  immerso nel 
gruppo di gauge della stringa aperta, e  $q_1$ e $q_2$ sono le cariche, 
associate
ai due estremi della stringa, che si accoppiano ad esso.

Poich\'e il campo elettromagnetico esterno si accoppia solo ai $bordi$ della 
stringa, l'equazione del moto dei campi $X^\m$ \`e l'usuale equazione d'onda 
\be
( \part_\t^2 - \part_\s^2 )  X^\m = 0   \, ,
\ee
ma le condizioni al bordo divengono 
\beq
\part_\s X_\m - q_1 F_{\m\n} \part_\t X^\n = 0  \qquad \mbox{per}\, \s = 0 
\ , 
\nonumber \\
\part_\s X_\m + q_2 F_{\m\n} \part_\t X^\n = 0  \qquad \mbox{per}\, \s = \pi 
\ .
\label{bcem}
\eeq
Se ora consideriamo un campo con componenti non nulle solo nelle direzioni 
1 e 2, ovvero $ F_{\m\n} = H \e_{\m\n} $,  con $ {\small{\m,\n = 1,2}}$, e 
$ \e_{12}=1=-\e_{21}$, le (\ref{bcem}) divengono
\beq
\part_\s X^1 (\pi) + q_2 H \part_\t X^2 (\pi) = 0 \, , \nonumber \\
\part_\s X^2 (\pi) - q_2 H \part_\t X^1 (\pi) = 0 \, ,
\label{bcem1}
\eeq
e
\beq
\part_\s X^1 (0) - q_1 H \part_\t X^2 (0) = 0 \, , \nonumber \\
\part_\s X^2 (0) + q_1 H \part_\t X^1 (0) = 0 \, .
\label{bcem2}
\eeq
A questo punto \`e conveniente introdurre le coordinate di stringa 
in notazione complessa 
\be
X_{\pm} = \frac{1}{\sqrt{2}} ( X^1 \pm i X^2 ) , \quad \mbox{con} \, 
(X_+)^\dag = X_- \ ,
\ee
poich\'e in questo modo le condizioni ai bordi (\ref{bcem1}) e (\ref{bcem2}) 
si separano ed assumono la forma 
\beq
\part_\s X_+ + i q_1 H \part_\t X_+ = 0 \quad,\quad \part_\s X_- - i q_1 H 
\part_\t X_- = 0   \quad \mbox{a} \,   \s = 0 \, , \nonumber \\
\part_\s X_+ - i q_2 H \part_\t X_+ = 0 \quad,\quad \part_\s X_- + i q_2 H 
\part_\t X_- = 0   \quad \mbox{a} \,   \s = \pi \, .
\label{bcnotcplx}
\eeq

Quando la carica netta della stringa \`e non nulla ($q=q_1+q_2\neq0$),
l'espressione in modi normali per $X_\pm$ \`e data da 
\be
X_+ (\t,\s) = x_+ +  i ( \sum_{n=1}^\infty a_n \f_n (\t,\s) - 
\sum_{m=0}^\infty b_m^\dag \f_{-m} (\t,\s) )  \, , 
\label{Xpiu}
\ee
\be
X_- (\t,\s) = x_- +  i ( \sum_{m=0}^\infty b_m \bar{\f}_{-m} (\t,\s) - 
\sum_{n=1}^\infty a_n^\dag \bar{\f}_n (\t,\s) )  \, ,
\label{Xmeno}
\ee
con  $(X_+)^\dag = X_-$ .
Inoltre, nelle (\ref{Xpiu}) e (\ref{Xmeno}) 
$$ \f_n = \frac{1}{\sqrt{|n - \e|}} e^{- i ( n - \e ) \t} 
\cos [ ( n - \e ) \s + \g ] $$
con 
$$ \e = \frac{1}{\pi} ( \g + \g' ) \quad \mbox{e}  \quad \g = \arctan q_1 H
 \ , \g' = \arctan q_2 H \ . $$
La quantizzazione della teoria pu\`o essere effettuata in modo canonico,
notando che, nel gauge (\ref{gaugecond}), il momento coniugato ad $X_+$ \`e 
\be
 P_+ (\t,\s) = \frac{\part \mc{L}}{\part \dot{X}_+} = \frac{1}{2 \pi \a'} 
\dot{X}_- + \frac{i}{4 \pi \a'} A_- ( q_1 \d (\s) + q_2 \d (\pi-\s)) \ ,
\ee 
che nel caso bidimensionale  diviene
\be
 P_+ (\t,\s)  = \frac{1}{2 \pi \a'} \dot{X}_- + \frac{i}{4 \pi \a'} 
X_- ( q_1 H \d (\s) + q_2 H \d (\pi-\s)) \ .
\ee 
Le relazioni di commutazione per gli oscillatori sono quindi
\begin{center}
$ [ a_n , a_m^\dag ] = \d_{n,m} \qquad \mbox{per} \quad  m \geq 0 \, ,$\\
$ [ b_n , b_m^\dag ] = \d_{n,m} \qquad \mbox{per} \quad  m \geq 1 \, ,$\\
$ [ a_n , a_m ] = [ a_n^\dag , a_m^\dag ] =  [ b_n , b_m ] =
 [ b_n^\dag , b_m^\dag ] = 0 \, , $\\
\be
 [ x_+ , x_- ] = \frac{1}{2 (q_1 + q_2) H} \qquad \mbox{con } 
\quad (x_+)^\dag = x_- \ .
\label{comm.rel.}
\ee
\end{center}
Si noti che $x_+$ ed $x_-$, che nel limite classico corrispondono alle coordinate 
del centro di un'orbita di Landau, non commutano. Costruendo gli operatori di Virasoro, 
\`e possibile verificare che questi 
soddisfano l'algebra
$ [ L_n , L_m ] = ( n - m ) L_{n+m} + \d_{n+m,0} [ \frac{1}{12} n ( n^2 - 1 ) +
n \e ( 1 - \e ) ] \, ,$
ovvero portano all' usuale risultato, a meno del termine centrale che, 
essendo lineare in $n$,  pu\`o essere ricondotto a quello standard mediante 
la ridefinizione  $ L_0 \ra L_0 + \frac{\e}{2} ( 1 - \e ) $. Questo 
equivale ad una modifica del valore originale della costante di ordinamento
normale $a$, o dell'energia di vuoto, che passa da 1 a 
$ 1 - \frac{\e}{2} ( 1 - \e ) $. 

In conclusione, sui settori con carica totale non nulla, un campo magnetico di 
$background$ costante altera la frequenza degli oscillatori $a$ di $ -\e$, 
e degli oscillatori
$b$ di $\e$, modificando al contempo l'energia di vuoto.

Per stringhe con carica totale nulla ($q=q_1+q_2=0$), $\e = 0$,  e quindi i modi non
sono alterati dalla presenza di $F_{\m\n}$ (sebbene $\g$ e $\g'$ non siano 
nulle), ma soprattutto si ha una modifica nella struttura dei modi zero,
in quanto il commutatore (\ref{comm.rel.}) non \`e ben definito per 
$(q_1+q_2)  = 0$. Pertanto l'espansione di $X_+$ diviene
\be 
X_+ = \frac{x_+ + p_- ( \t - i q_1 H ( \s - \frac{\pi}{2} ))}
{\sqrt{1 + q_1^2 H^2}} + i \sum_{n=1}^\infty [ a_n \f_n (\t,\s) - b_n^\dag 
\f_{-n} (\t,\s) ] \ .
\label{Xpiu1}
\ee
Si osservi che in questo caso $X_+$ contiene un modo zero proporzionale a 
$\t-iq_1H$ che soddisfa la condizione ai bordi quando $(q_1+q_2)H=0$.
La presenza del momento totale $p_-$ nella (\ref{Xpiu1}) pu\`o essere spiegata 
in maniera semplice ed intuitiva pensando che gli estremi della stringa
aperta (con cariche opposte), soddisfino la legge di forza
\be
m \frac{d \vec v}{d t} = q \vec v \times \vec B = q \frac{d}{d t} ( \vec r 
\times \vec B) \ ,
\ee
dalla quale si ottengono le due costanti del moto
\beq
\vec U_1 = \vec p_1 - \frac{q}{m} \vec r_1 \times \vec B \ , \nonumber \\
\vec U_2 = \vec p_2 - \frac{q}{m} \vec r_2 \times \vec B \ .
\eeq
Definendo $\vec p = \vec p_1 + \vec p_2$, ed essendo $q_1=q=-q_2$, si
deduce che la grandezza conservata \`e $\vec p-q(\vec r_1-\vec r_2) \times 
\vec B$. Pertanto in questa situazione la coordinata relativa 
$\vec r_1-\vec r_2$ commuta con $\vec p$, e quindi le componenti di 
$\vec U_1$ e $\vec U_2$ commutano tra loro.

Non \`e difficile modificare i risultati precedenti per studiare 
il comportamento di una stringa in presenza di un campo elettrico 
\cite{Bachas:1992bh}. 
Basta scegliere  $ F_{01} = E $ ed utilizzare le coordinate nel 
gauge del cono di luce $ X^{\pm} = \frac{1}{\sqrt{2}} ( X^0 \pm X^1 ) . $
Questo consente, per esempio, di calcolare esattamente la probabilit\`a di
produzione di coppie di stringhe aperte (bosoniche o fermioniche) in un
campo elettrico costante, ovvero l'analogo, in teoria di stringa, dell'effetto
Schwinger della teoria dei campi. Il risultato \`e in accordo con quello 
classico, nel limite di campo debole, ma quando il campo elettrico tende ad
un valore critico (dell'ordine della tensione di stringa), la probabilit\`a
di produzione di tali coppie diverge \cite{Bachas:1992bh}. Questo comportamento 
limite ammette una semplice spiegazione intuitiva, legata al fatto che le forze 
elettriche introdotte dai campi possono superare la forza di richiamo dovuta 
alla tensione della stringa. 

\section{Stringhe aperte su un toro magnetizzato}
Nel caso della stringa bosonica aperta con carica netta non nulla,  
la presenza di un campo magnetico altera drasticamente la struttura dei modi 
zero della teoria.
Non compaiono pi\`u, infatti, gli operatori di momento totale della stringa
$p_\pm$, ma sono presenti degli ulteriori operatori di Fourier, $ b_0^\dag $ e
 $ b_0$, che creano e distruggono quanti di frequenza $\e$. Inoltre, per 
$F \neq 0 $, $x_+$ ed $x_-$ commutano con  $L_0$, ed \`e 
quindi possibile scegliere gli stati come autostati di $x_+$, per esempio.
Dato che $L_0$ non dipende da $x_\pm$, esiste una degenerazione infinita
se le coordinate $X^1$ ed $X^2$ non sono compatte, e finita se lo sono.
Nel primo caso la situazione \`e analoga a quella di una particella in un
campo magnetico costante sul piano: esistono livelli di Landau con degenerazione 
infinita, ma equamente spaziati, collegati tra loro dagli operatori $b_0^\dag$
e $b_0$. Se  $qH$ \`e sufficientemente piccolo (dove $q=q_1+q_2$), 
la separazione nella frequenza $\e$ \`e proporzionale a $qH$. Viceversa, se
le coordinate $X^1$ ed $X^2$ sono compatte, i modi zero corrispondono
alla situazione di una particella carica che si
muove in un campo magnetico costante su un toro. Dunque \`e di particolare
interesse analizzare in dettaglio il caso in cui $x^1$ ed $x^2$ 
siano coordinate compatte di lunghezza $L = 2 \pi R $.

Un campo magnetico su un toro \`e in realt\`a un campo di monopolo:
esiste un flusso totale non nullo di H all'interno del toro, e 
corrispondentemente non esiste un potenziale vettore continuo ed univocamente 
definito su di esso.  Possiamo allora  separare il toro in due regioni, 
definendo in esse dei potenziali continui legati da una trasformazione di gauge. 

Dunque, se $ a \in [ 0 , L ]$ possiamo scegliere 
$$
A_1 = \Bigg\{ \ba{ll} 
- H x_2 & \textrm{$0 \leq x_2 < a$} , \\
- H ( x_2 - L ) & \textrm{$a \leq x_2 < L$} , \\
   \ea 
\qquad A_2 = 0 .   
$$
Le due scelte per $A_1$ sono compatibili, in quanto connesse da una 
trasformazione di gauge, poich\'e
$$ A_1 \Big|_{x_2+L} = A_2 - i e^{- i \th}  \part_2 e^{i \th} \Big|_{x_2} =
- H x_2 + \part_2 \th , $$
da cui segue che  $\th = H L x_2 .$
La funzione di transizione tra le due regioni in cui \`e stato
suddiviso il toro,  $ e^{\frac{i q \th}{2 \pi \a'}}$, \`e ad un sol valore 
solo se viene soddisfatta \emph{la relazione di quantizzazione di Dirac}
\be
q H L^2 = 4 \pi^2 \a' k
\label{Diraccond}
\ee
per qualche intero $k$. Si osservi che nella (\ref{Diraccond}) $k$ \`e anche la
\emph{degenerazione dei livelli di Landau}. 
A questo proposito risulta particolarmente interessante sottolineare
come l'eq. (\ref{Diraccond}) possa essere dedotta ``geometricamente'' dalla
nozione di D-brane e T-dualit\`a su un toro magnetizzato.
Abbiamo visto nel paragrafo precedente che per una stringa bosonica
neutra i modi zero della coordinata di stringa sono
\be 
X_+^{z.m.} = \frac{x_+ + p_- [ \t - i q_1 H ( \s - \frac{\pi}{2} )]}
{\sqrt{1 + q_1^2 H^2}}  \ ,
\ee
ovvero, in termini delle coordinate $X_1$ ed $X_2$ 
\beq
X_1^{z.m.} & = & \frac{x_1 + (2 \a') [ p_1 \t - q_1 H p_2 
(\s - \pi/2) ]}{\sqrt{1 + q_1^2 H^2}} \ , \nonumber \\
X_2^{z.m.} & = & \frac{x_2 + (2 \a') [ - p_2 \t - q_1 H p_1 
(\s - \pi/2) ]}{\sqrt{1 + q_1^2 H^2}} \ .
\label{X1X2}
\eeq
Supponiamo di effettuare un'operazione di T-dualit\`a, per esempio lungo la 
coordinata $X_2$ ($X_2 \ra Y_2$), ricordando che sul {\it world-sheet} 
un'operazione di T-dualit\`a equivale allo scambio di $\t $ con $\s$.
Questo implica che $(2 \a') p_2$ debba essere reinterpretato come
$2 w_2$, e pertanto le (\ref{X1X2}) divengono
\beq
X_1^{z.m.} & = & \frac{x_1 + (2 \a')  p_1 \t - 2 q_1 H w_2 
(\s - \pi/2)}{\sqrt{1 + q_1^2 H^2}} \ , \nonumber \\
Y_2^{z.m.} & = & \frac{y_2 - (2 \a') q_1 H p_1 \t - 2 w_2 
(\s - \pi/2)}{\sqrt{1 + q_1^2 H^2}} \ .
\label{X1Y2}
\eeq
Definendo 
$$ \cos \th = \frac{1}{\sqrt{1 + q_1^2 H^2}} \ , \quad 
   \sin \th = \frac{q_1 H}{\sqrt{1 + q_1^2 H^2}} \ , $$
$$ \tld{x}_1 = x_1 \cos \th \ , \quad   \tld{y}_2 = y_2 \cos \th \ , $$
le (\ref{X1Y2}) assumono la forma seguente:
\beq
X_1^{z.m.} & = & \tld{x}_1 + (2 \a')  p_1 \cos \th \t - 2 w_2 \sin \th 
(\s - \pi/2) \ , \nonumber \\
Y_2^{z.m.} & = & \tld{y}_2 + (2 \a')  p_1 \sin \th \t - 2 w_2 \cos \th 
(\s - \pi/2) \ .
\label{X1tldY2}
\eeq
Si noti che la particolare combinazione delle (\ref{X1tldY2}),
$X_1 \sin \th  + Y_2 \cos \th $,  contiene solo $windings$, e quindi
 rispetto alla configurazione
iniziale della D-brana, sulla quale terminano gli estremi della stringa 
aperta, questa risulta ruotata di un angolo $\th$, ovvero caratterizzata
da un versore normale $\hat{n} = (\sin \th, \cos \th)$.
In questa situazione, affinch\`e la posizione della 
D-brana ``ruotata'' sia consistente con la cella fondamentale del toro 
\`e necessario imporre che 
\be 
k \tld{R} \cot \th = R  \ ,
\label{intero}
\ee
dove $k$ \`e un intero , ed $\tld{R}$ \`e il raggio T-duale del toro:
$\tld{R}=\frac{\a'}{R}$. Essendo $R=L/2 \pi$, la (\ref{intero}) 
si potr\`a riscrivere come 
\be
\tan \th = \frac{4 \pi^2 \a'}{L^2} k \ ,
\ee
ma $\tan \th = q_1 H $, e quindi si riottiene la (\ref{Diraccond}).

A scopo illustrativo, applichiamo ora questi risultati alla superstringa
di Tipo I, ricavando in particolare la funzione di partizione di stringa aperta.
 L'ampiezza di anello nel canale diretto per la superstringa ad otto dimensioni
\`e
\be
\mc{A} = \frac{1}{2} N^2  (V_8 - S_8)(i \t/2) P_2 \ ,
\label{AT2}
\ee
dove $P_2$ \`e la somma reticolare ristretta a {\it windings} nulli
\be
P_2 = \Big[ \sum_{m_1 \in Z} q^{\frac{\a'}{4} \frac{m_1^2}{R^2}} \Big] 
\Big[ \sum_{m_2 \in Z} q^{\frac{\a'}{4} \frac{m_2^2}{R^2}} \Big] \ .
\ee 
In presenza di un campo magnetico costante su $T^2$, le cariche di Chan-Paton 
$N$ si accoppiano ad esso. A tal fine \`e conveniente scindere $N$ nella 
maniera seguente: $N \ra n + m_+ + m_-$, dove $n$ indica il numero di cariche 
elettricamente neutre, ed $m_{\pm}$ indicano i numeri di cariche con carica 
elettrica $\pm 1$.
L'espressione (\ref{AT2}) viene corrispondentemente deformata, ed assume 
quindi la forma seguente:
\beq
\mc{A} & = & \frac{1}{2} \Big[ \ [ n^2 P_2 +2 m_+ m_- \tld{P}_2 ] 
( V_8 - S_8 )(0) - 2 i n  m_+ (V_8 - S_8)(z \t ; \t) \ 
\frac{k \eta}{\th_1(z \t)}  \nonumber \\
       & - & 2 i n m_- (V_8 - S_8)(-z \t ; \t) \ \frac{k \eta}{\th_1(-z \t)}  ] 
- i [  m_+^2 (V_8 - S_8)(2 z \t ; \t) \ \frac{k \eta}{\th_1(2 z \t)} 
\nonumber \\ 
       & + &  m_-^2 (V_8 - S_8)(-2 z \t ; \t) \ \frac{k \eta}{\th_1(-2 z \t)}  ] 
\ \Big].
\label{AT2magn}
\eeq
Nel passaggio al canale trasverso l'ampiezza (\ref{AT2magn}) diventa
\beq
\tld{\mc{A}} & = & \frac{2^{-5}}{2} \Big[ \ [ n^2 W_2 + 2 m_+ m_- (1 + q^2 H^2) 
\tld{W}_2 ] \ v_2 \ (V_8 - S_8)(0)  \nonumber \\ 
             & + & 2 n m_+ \ (V_8 - S_8 )(z) \  \frac{k \eta}{\th_1(z)} 
+ 2 n m_- \ (V_8 - S_8 )(-z) \  \frac{k \eta}{\th_1(-z)}   \\
             & + & m_+^2 \ (V_8 - S_8 )(2z) \  \frac{2 k \eta}{\th_1(2z)} 
+ m_-^2 \ (V_8 - S_8 )(-2z) \ \frac{2 k \eta}{\th_1(-2z)} \ \Big] \ ,
\nonumber 
\label{AtldT2magn}
\eeq
dove la dipendenza dal modulo $il$ \`e lasciata implicita, e
\be
\tld{W}_2 = \Big[ \sum_{n_1 \in Z} q^{\frac{1}{4 \a'}(n_1^2 R^2 
\sqrt{1 + q^2 H^2})} \Big] 
\Big[ \sum_{n_2 \in Z} q^{\frac{1}{4 \a'}(n_2^2 R^2 
\sqrt{1 + q^2 H^2})} \Big] \ ,
\ee
\be
cotg (\pi z) = \frac{1 - q_1 q_2 H^2}{(q_1 + q_2) H}  \ .
\label{modeshift}
\ee
Dalla (\ref{AtldT2magn}) si pu\`o notare che gli effetti della deformazione 
magnetica sono molteplici. Per quanto riguarda i termini ``carichi''
(proporzionali ad $m_+$, $m_-$, $m_+^2$, $m_-^2$) 
lo $shift$ nei modi delle stringhe
aperte, dato dalla (\ref{modeshift}), si riflette nella presenza 
dei caratteri con argomento non nullo, e nella presenza delle molteplicit\`a
$k$ dei livelli di Landau.
Tra i termini ``scarichi'' invece, appare un nuovo tipo di stringhe 
(quelle proporzionali a $m_+ m_-$). Queste sono caratterizzate dall'avere agli
estremi cariche opposte, e quindi i loro modi restano inalterati,
mentre i modi zero subiscono un $boost$ immaginario. I loro modi di
Kaluza-Klein devono essere riscalati come segue:
$$m_i \ra \frac{m_i}{\sqrt{1 + q^2 H^2}} \ .$$
Ci\`o assicura che i contributi di ordine pi\`u basso delle ampiezze trasverse
si assemblino in quadrati perfetti.

In realt\`a anche questi risultati trovano una spiegazione pi\`u semplice
nella descrizione T-duale, che rivela molto chiaramente a quali restrizioni
sono soggetti gli stati che fluiscono nel tubo, mettendo in risalto il 
significato fisico dei modi riscalati.

In generale i modi zero di stringa bosonica chiusa sul toro sono
\beq
Z_1^{z.m} & = & z_1 + (2 \a') p_1 \t + 2 w_1 \s \ , \nonumber \\
Z_2^{z.m} & = & z_2 + (2 \a') p_2 \t + 2 w_2 \s \ ,
\label{Z1Z2}
\eeq
dove 
$$ p_i = \frac{m_i}{R} = 2 \pi \frac{m_i}{L} \ , \quad
   w_i = n_i R = \frac{n_i L}{2 \pi}  . $$
Seguendo la strategia adottata per la stringa aperta, effettuiamo
un'operazione di T-dualit\`a lungo la coordinata $Z_2$, che trasforma le
(\ref{Z1Z2}) in
\beq
Z_1^{z.m} & = & z_1 + (2 \a') m_1 \frac{2 \pi}{L} \t + n_1 \frac{L}{\pi} \s \ ,
 \nonumber \\
Z_2^{z.m} & = & \tld{z}_2 + m_2 \frac{\tld{L}}{\pi} \s + (2 \a') n_2 
\frac{2 \pi}{\tld{L}} \t \ ,
\eeq
dove $ \frac{\tld{L}}{\pi} \ra \frac{4 \pi \a'}{L}$. Queste, se riespresse
in termini di $L$, assumono la forma seguente:
\beq
Z_1^{z.m} & = & z_1 + (2 \a') m_1 \frac{2 \pi}{L} \t + n_1 \frac{L}{\pi} \s \ ,
 \nonumber \\
Z_2^{z.m} & = & \tld{z}_2 + n_2 \frac{L}{\pi} \t + (2 \a') m_2 
\frac{2 \pi}{L} \s \ .
\eeq
Ma abbiamo visto che l'introduzione di un campo magnetico costante su un
toro equivale ad avere una D-brana ruotata di un angolo $\th$, con 
versori normali e tangenti $\hat{n}=(\sin \th, \cos \th)$ e 
$\hat{t}=(-\cos \th, \sin \th)$, e dove $ \tan \th = 
\frac{4 \pi^2 \a'}{L^2} k $. Quindi, se nel caso della teoria di stringa 
compattificata sul toro, nell'ampiezza $\tld{\mc{A}}$ possono fluire 
solo gli stati con impulso nullo, sul toro ma\-gnetizzato possono fluire 
solo gli stati con impulso nullo nella direzione normale alla D-brana,
e stati con {\it winding} nullo nella direzione tangente.
Tali condizioni, riassunte dalle espressioni
$$ \vec p \cdot \hat{t} = 0 \ , \quad \vec w \cdot \hat{n} =0 \ ,   $$
portano a concludere che 
\beq
m_1 & = & k n_2  \ , \nonumber \\
m_2 & = & -k n_1 \ ,
\label{m1m2}
\eeq
e pertanto implicano una modifica nei modi zero degli stati di stringa
glo\-balmente neutra. Infatti il generico termine della somma reticolare
$$ q^{\frac{\a'}{4}  [ \ ( \frac{m_1^2}{R^2} + \frac{n_1^2 R^2}{\a'^2} )
+ ( \frac{m_2^2}{R^2} + \frac{n_2^2 R^2}{\a'^2} ) \ ]} \ , $$ 
 mediante le (\ref{m1m2}), si potr\`a riscrivere come
$$ q^{\frac{\a'}{4}  [ ( \frac{k^2 n_2^2}{R^2} + \frac{n_1^2 R^2}{\a'^2} )
+ ( \frac{k^2 n_1^2}{R^2} + \frac{n_2^2 R^2}{\a'^2} )  ]}
= q^{\frac{R^2}{4 \a'} (n_1^2+n_2^2)  [ \frac{k^2 \a'^2}{R^2} + 1 ]}
= q^{\frac{R^2}{4 \a'} (n_1^2+n_2^2)  ( 1+ q^2 H^2)} \ .  $$ 
Da cui segue che, come anticipato, i  modi di Kaluza-Klein dei 
modi zero degli stati di stringa neutra devono  essere riscalati come
$$ m_i \ra \frac{m_i}{\sqrt{1 + q^2 H^2}}\ , \quad 
n_i \ra n_i \sqrt{1 + q^2 H^2} \ .  $$ 

\section{Stringhe aperte su orbifols magnetizzati}
Lo scopo del paragrafo che segue \`e la descrizione di esempi pi\`u 
complessi di stringhe aperte in presenza di campi magnetici uniformi.
In particolare, verranno descritti gli effetti di deformazioni magnetiche,
per compattificazioni toroidali o su orbifolds nelle stringhe di Tipo-I,
mostrando ad esempio, come  D9-brane magnetizzate possono emulare delle
(anti)D5-brane (BPS) \cite{Angelantonj:2000hi, Angelantonj:2000rw}.
Una situazione di questo genere pu\`o essere illustrata qualitativamente
nell'ambito della teoria di campo limite.

Si consideri a tal fine l'azione effettiva di bassa energia di una D9-brana 
immersa in un campo abeliano di
$background$, 
\be
S_9 = - T_{(9)} \int_{\mc{M}_{10}} d^{10} x \ e^{-\f} \sum_{a=1}^{32} 
\sqrt{-\det(g_{10} + q_a F)} - \m_{(9)} \sum_{p,a} \int_{\mc{M}_{10}} e^{q_a F}
 \w C_{p+1} + \cdots \ ,
\label{S1}
\ee
dove il primo termine \`e l'azione di Born-Infeld \cite{Leigh:1989jq},
ed il secondo \`e il
termine di Wess-Zumino di accoppiamento con i campi di R-R, e
dove $a$ distingue i tipi di cariche di Chan-Paton che si accoppiano ai campi 
magnetici. $T_{(9)}$ e $ \m_{(9)}$ sono, rispettivamente, la tensione e la
carica di R-R per una $D_p$-brana di Tipo-I, legate dalla relazione
\be
T_{(p)} = \sqrt{\frac{\pi}{2 \k^2}} \ \Big( 2 \pi \sqrt{\a'} \Big)^{3-p} = 
| \m_{(p)} | \ ,
\label{BPS}
\ee
con $\k^2 = 8 \pi G_N^{(10)}$ che definisce la costante di Newton a dieci 
dimensioni.

Compattifichiamo quattro delle dieci dimensioni su due doppi tori, introducendo
in essi due campi magnetici costanti ed abeliani $H_1$ ed $H_2$
$$ \mc{M}_{(10)} = \mc{M}_{(6)} \times T^2 (H_1) \times T^2 (H_2) \ . $$
Come descritto nel paragrafo precedente, i campi magnetici uniformi 
sul toro sono campi di monopolo, che quindi soddisfano le relazioni di 
quantizzazione di Dirac
\be
q H_i v_i = k_i \qquad (i=1,2) \ ,
\label{Diracrelquant}
\ee
dove $ v_i = \frac{R_i^{(1)} R_i^{(2)}}{\a'} $ sono i volumi dei due tori
di raggi $R_i^{(1)}$ ed  $R_i^{(2)}$. $k_i \ (\in Z) $ sono le degenerazioni 
dei corrispondenti livelli di Landau, e $q$ \`e la carica elettrica elementare
del sistema.

In teoria dei campi la cancellazione delle anomalie richiede, per i campi 
magnetici interni, l'annullarsi della carica istantonica $ tr (F \w F)  $ 
come condizione di integrabilit\`a per l'equazione della due-forma $B_{(2)}$.
 Viceversa, in teoria di stringa  un numero istantonico non nullo
pu\`o essere compensato mediante D-brane ed O-piani.
Questo fenomeno pu\`o essere ben descritto in termini delle teorie di
campo limite.  \`E quindi possibile 
introdurre una coppia di campi abeliani allineati con lo stesso sottogruppo
$U(1)$ di $SO(32)$, in maniera tale che la (\ref{S1}) divenga
\beq
S_9 = - T_{(9)} \int_{\mc{M}_{10}} d^{10} x \ e^{-\f} \ \sqrt{-g_6} 
\sum_{a=1}^{32} \sqrt{(1 + q_a^2 H_1^2)(1 + q_a^2 H_2^2)} &+& \nonumber \\
-\  32 \m_{(9)} \int_{\mc{M}_{10}} C_{10} - \Big( 2 \pi \sqrt{\a'} \Big)^4
\m_{(9)} v_1 v_2 H_1 H_2 \sum_{a=1}^{32} q_a^2  \int_{\mc{M}_{10}} C_6 \ ,
\label{S2}
\eeq
dove $g_6$ \`e la metrica spazio-temporale a sei dimensioni e,
per semplicit\`a,
nello spazio interno la metrica \`e stata scelta essere l'identit\`a.
Si noti che, nel caso particolare in cui $H_1$ ed $H_2$ abbiano la stessa 
intensit\`a, il termine di Born-Infeld si semplifica e
l'azione corrispondente alla configurazione (anti)auto-duale
$(H_1=\pm H_2)$ si riduce a 
\beq
S_9 = - 32 \int_{\mc{M}_{10}} \Big(d^{10} x \ \sqrt{-g_6} T_{(9)} e^{-\f} +
\m_{(9)} C_{10}\Big) &+& \nonumber \\
- \sum_{a=1}^{32} \Big(\frac{q_a}{q}\Big)^2 \int_{\mc{M}_6}  \Big(d^6 x \ 
\sqrt{-g_6} |k_1 k_2|  T_{(5)} e^{-\f} + k_1 k_2 \m_{(5)} C_6\Big) \ .
\label{S3}
\eeq
Pertanto una D9-brana magnetizzata mima $|k_1 k_2|$ D5 o anti-D5-brane BPS,
a seconda che $k_1$ e $k_2$ abbiano segno uguale o opposto (ovvero a seconda
delle orientazioni relative di $H_1$ ed $H_2$).

La trasposizione di questo fenomeno alla teoria di stringa richiede, per
ragioni tecniche, che la deformazione magnetica venga applicata a modelli di
{\it orbifold}. Ci\`o \`e dovuto al fatto che una configurazione 
supersimmetrica \`e
realizzata in presenza di sole D9 e D5-brane (e non di antibrane), e
solamente gli { \it orbifolds}, a differenza dei tori, possono contenere degli 
O5-piani in grado di assorbire la carica della D5-brana dovuta alla D9-brana
magnetizzata.

Quindi per ottenere una configurazione supersimmetrica occorre
consi\-derare un {\it orbifold} che richieda l'introduzione di D5-brane. Il caso
pi\`u semplice in cui ci\`o si verifica \`e la compattificazione a sei 
dimensioni della superstringa di Tipo IIB su $ (T^2 \times T^2) / Z_2$.

Analogamente alla trattazione nella teoria di campo limite, \`e possibile 
introdurre i due campi magnetici interni, $H_1$ ed $H_2$ allineati
con lo stesso sottogruppo $U(1)$ di $SO(32)$. In altri termini,
\`e possibile ammettere un numero istantonico non nullo 
($ tr H_i \w H_j \neq 0 $), che  pu\`o  essere compensato da un eccesso di 
(anti)D-brane e/o O5-piani.

Ricordiamo inoltre che un campo magnetico uniforme, con componenti $ H_1$ ed 
$H_2$ nei due tori interni altera le condizioni ai bordi delle stringhe aperte.
I loro modi vengono shiftati di 
\be
z_i^{L,R} = \frac{1}{\pi} [\ arctg (q_L H_i) + arctg (q_R H_i) \ ] \ ,
\ee
dove $q_L$ e $q_R$ sono, rispettivamente, le cariche dell'estremo sinistro e 
destro della stringa aperta che si accoppiano ai campi abeliani $H_i$.
Pertanto il settore chiuso non viene alterato dalle deformazioni magnetiche,
e quindi continua ad assumere la forma usuale
\beq
\mc{T} & = & \frac{1}{2} \Big[ \ |Q_o + Q_v|^2 \L_{(4,4)} + |Q_o - Q_v|^2 
\Big| \frac{2 \eta}{\vth_2 (0)} \Big|^4  \nonumber \\
       & + & |Q_s + Q_c|^2 \Big| \frac{2 \eta}{\vth_4 (0)} \Big|^4 + 
|Q_s - Q_c|^2 \Big| \frac{2 \eta}{\vth_3 (0)} \Big|^4 \ \Big] \ ,
\eeq
con proiezione di bottiglia di Klein
\be
\mc{K} = \frac{1}{4} \Big[ (Q_o + Q_v)(0;0) [ P_1 P_2 + W_1 W_2 ] + 
2 \times 16 (Q_s + Q_c)(0;0) \Big( \frac{\eta}{\vth_4(0)} \Big)^2 \Big] \ .
\label{KT4suZ2}
\ee
Il risultato \`e sempre  lo spettro $(1,0)$ supersimmetrico chiuso non 
orientato  con un multipletto tensoriale e venti ipermultipletti.

Il settore aperto \`e invece  descritto dalla seguente ampiezza di anello:
\beq
\mc{A} & = & \frac{1}{4} \Big[ \ (Q_o + Q_v)(0;0) [ (n + \bar{n})^2 P_1 P_2 + 
2 m \bar{m} \tld{P}_1 \tld{P}_2 + (d + \bar{d})^2 W_1 W_2 ]  \nonumber \\
       & - & 2 (n + \bar{n})(m + \bar{m}) (Q_o + Q_v)(z_1 \t;z_2 \t) 
\frac{k_1 \eta}{\vth_1(z_1 \t)} \frac{k_2 \eta}{\vth_1(z_2\t)}  \nonumber \\
       & - & (m^2 + \bar{m}^2) (Q_o + Q_v)(2z_1 \t;2z_2 \t) 
\frac{2k_1 \eta}{\vth_1(2z_1 \t)} \frac{2k_2 \eta}{\vth_1(2z_2 \t)} \nonumber \\
       & + & 2 (d + \bar{d})  (n + \bar{n}) (Q_s + Q_c)(0;0) 
\Big( \frac{\eta}{\vth_4(0)} \Big)^2 \nonumber \\
       & + & 2 (d + \bar{d})  (m + \bar{m}) (Q_s + Q_c)(z_1 \t;z_2 \t)  
\frac{\eta}{\vth_4(z_1 \t)} \frac{\eta}{\vth_4(z_2 \t)}   \nonumber \\
       & - & [ (n - \bar{n})^2 - 2 m \bar{m} + (d + \bar{d})^2 ] 
(Q_o - Q_v)(0;0) \Big( \frac{2 \eta}{\vth_2(0)} \Big)^2    \nonumber \\
       & - & 2 (n - \bar{n}) (m + \bar{m}) (Q_o - Q_v)(z_1 \t;z_2 \t) 
\frac{2 \eta}{\vth_2(z_1 \t)} \frac{2 \eta}{\vth_2(z_2 \t)} \nonumber \\
       & - & (m^2 + \bar{m}^2) (Q_o - Q_v)(2z_1 \t;2z_2 \t) 
\frac{2 \eta}{\vth_2(2z_1 \t)} \frac{2 \eta}{\vth_2(2z_2 \t)} \nonumber \\
       & - & 2 (d - \bar{d}) (n - \bar{n}) (Q_s - Q_c)(0;0) \Big( \frac{\eta}{\vth_3(0)} \Big)^2 \nonumber \\
       & - & 2 (d - \bar{d}) (m - \bar{m}) (Q_s - Q_c)(z_1 \t;z_2 \t)  
\frac{\eta}{\vth_3(z_1 \t)} \frac{\eta}{\vth_3(z_2 \t)} \ , 
\label{AT4suZ2} 
\eeq
e dalla corrispondente ampiezza di M\"obius
\beq
\mc{M} & = & - \frac{1}{4} \Big[ (\hat{Q}_o + \hat{Q}_v)(0;0) [ (n + \bar{n})
P_1 P_2 + (d + \bar{d}) W_1 W_2 \Big] \nonumber \\
       & - &  (m + \bar{m}) (\hat{Q}_o + \hat{Q}_v)(2 z_1 \t;2 z_2 \t) 
\frac{2 k_1 \hat{\eta}}{\hat{\vth}_1(2 z_1 \t)}  
\frac{2 k_2 \hat{\eta}}{\hat{\vth}_1(2 z_2 \t)} \nonumber \\
       & - &  (n + \bar{n} + m + \bar{m})  (\hat{Q}_o - \hat{Q}_v)(0;0)
\Big( \frac{2 \eta}{\vth_2(0)} \Big)^2  \nonumber \\
       & - &  (m + \bar{m}) (\hat{Q}_o - \hat{Q}_v)(2 z_1 \t;2 z_2 \t) 
\frac{2 \hat{\eta}}{\hat{\vth}_2(2 z_1 \t)}  
\frac{2 \hat{\eta}}{\hat{\vth}_2(2 z_2 \t)} \ .
\label{MT4suZ2}
\eeq
In queste espressioni $n, \bar{n}, m$ ed $\bar{m}$ sono le molteplicit\`a
di Chan-Paton delle D9-brane (rispettivamente, neutre e magnetizzate), mentre
$d$ e $\bar{d}$ sono le molteplicit\`a delle D5-brane. $P_i$ e $W_i$ sono le
usuali somme reticolari, mentre
\be
\tld{P}_i = \Big[ \sum_{m_i=-\infty}^{\infty} q^{\frac{\a'}{2 R^2}
(\frac{m_i}{\sqrt{1+q_a^2 H_i^2}})^2} \Big]^2 \ .
\ee
Si osservi inoltre  come nelle  funzioni di partizione (\ref{KT4suZ2}), 
(\ref{AT4suZ2}) e (\ref{MT4suZ2}),  i supercaratteri a sei 
 dimensioni siano funzioni di una coppia di argomenti, che
descrivono l'effetto delle deformazioni magnetiche sui fermioni interni.
I supercaratteri cos\`\i \ deformati sono
\beq
Q_o(\z_1;\z_2) & = & V_4(0) [O_2(\z_1) O_2(\z_2) + V_2(\z_1) V_2(\z_2)] 
\nonumber \\
               & - & C_4(0) [S_2(\z_1) C_2(\z_2) + C_2(\z_1) S_2(\z_2)]  \ , 
\nonumber \\
Q_v(\z_1;\z_2) & = & O_4(0) [V_2(\z_1) O_2(\z_2) + O_2(\z_1) V_2(\z_2)] 
\nonumber \\
               & - & S_4(0) [S_2(\z_1) S_2(\z_2) + C_2(\z_1) C_2(\z_2)]  \ , 
\nonumber \\
Q_s(\z_1;\z_2) & = & O_4(0) [S_2(\z_1) C_2(\z_2) + C_2(\z_1) S_2(\z_2)] 
\nonumber \\                     
               & - & S_4(0) [O_2(\z_1) O_2(\z_2) + V_2(\z_1) V_2(\z_2)]  \ , 
\nonumber \\
Q_c(\z_1;\z_2) & = & V_4(0) [S_2(\z_1) S_2(\z_2) + C_2(\z_1) C_2(\z_2)] 
\nonumber \\ 
               & - & C_4(0) [V_2(\z_1) O_2(\z_2) + O_2(\z_1) V_2(\z_2)] \ ,
\eeq
dove i quattro caratteri di livello 1 di $O(2n)$ sono legati alle quattro
funzioni theta di Jacobi dalle relazioni
\beq
O_{2n} (\z) & = & \frac{\vth_3^n (\z|\t) + \vth_4^n(\z|\t)}{2 \eta^n (\t)} \ ,
\qquad 
V_{2n} (\z)  =  \frac{\vth_3^n (\z|\t) - \vth_4^n(\z|\t)}{2 \eta^n (\t)} \ ,
\nonumber \\
S_{2n} (\z) & = & \frac{\vth_2^n (\z|\t) + i^n \vth_1^n(\z|\t)}{2 \eta^n (\t)} 
\ ,
\quad 
C_{2n} (\z)  =   \frac{\vth_2^n (\z|\t) -  i^n  \vth_1^n(\z|\t)}{2 \eta^n (\t)} \ . \nonumber
\eeq
Gli argomenti $z_i$ e  $2 z_i$ restano cos\`\i \ associati alle stringhe 
aperte con uno o due estremi carichi. Inoltre, le simmetrie delle funzioni 
theta di Jacobi sotto $z \ra -z$, permettono di assemblare alcuni  termini delle 
(\ref{AT4suZ2}) e (\ref{MT4suZ2}) con cariche opposte (e quindi con argomento 
opposto).

Per valori generici dei campi magnetici, lo spettro aperto non \`e 
supersimmetrico e da' origine ad instabilit\`a tachioniche (instabilit\`a
di Nielsen-Olesen \cite{Nielsen:1978rm}), che possono essere ascritte agli 
accoppiamenti magnetici
delle componenti interne dei campi di gauge. Infatti, per deboli campi
magnetici, la formula di massa per i modi di stringa diviene
\be
\D m^2 = \frac{1}{2 \pi \a'} \sum_{i=1,2} [ \ (2 n_i + 1) | (q_L + q_R) H_i |
+ 2 (q_L + q_R) \S_i H_i  \ ] \ ,
\label{masshift}
\ee
dove il primo termine deriva dalla presenza dei livelli di Landau, mentre il 
secondo deriva dagli accoppiamenti di momento magnetico degli spin $\S_i$.
Da questa relazione si pu\`o facilmente dedurre che le componenti
dei vettori interni, nello stato fondamentale di Landau, possono dare origine
a contributi tachionici a causa dell'accoppiamento di dipolo magnetico.
In realt\`a la (\ref{masshift}) \`e valida per i modi $untwisted$, in quanto
le stringhe di tipo $N D $ non hanno livelli di Landau, e quindi per i modi
$twisted$ il contri\-buto di punto zero non \`e presente.
In questo caso la parte bosonica di $Q_s$, $O_4 C_4$, risente per\`o della 
deformazione magnetica, poich\'e presenta uno $shift$ in massa 
proporzionale a $\pm (H_1 - H_2)$, e quindi nuovamente non sviluppa 
instabilit\`a tachioniche se $H_1=H_2$. Inoltre con questa scelta anche 
la carica di 
supersimmetria, che risiede in $C_4 C_4$, viene preservata.
Infatti, utilizzando l'identit\`a di Jacobi per argomento non nullo, si pu\`o
verificare che per $z_1=z_2$ sia $\mc{A}$ che $\mc{M}$ si annullano
identicamente. Questo significa che l'intero spettro aperto di stringa 
possiede una supersimmetria residua. I modelli supersimmetrici che ne 
risultano sono particolarmente interessanti, e sono caratterizzati da 
condizioni di $tadpole$ ``deformate'' che riflettono gli accoppiamenti 
dell'azione 
(\ref{S1}).  Come al solito, le condizioni di $tadpole$ possono essere 
desunte dalle ampiezze di bottiglia di Klein, di anello e di M\"obius nel
canale trasverso che, nel caso in esame, all'origine dei reticoli assumono
la forma seguente:
\be
\tld{\mc{K}}_0 = \frac{2^5}{4} \Big[ \ Q_o(0;0) (\sqrt{v_1 v_2} + 
\frac{1}{\sqrt{v_1 v_2}})^2 + Q_v(0;0) (\sqrt{v_1 v_2} - 
\frac{1}{\sqrt{v_1 v_2}})^2 \ \Big] \ ,
\label{K0tldT4suZ2}
\ee
\beq
\tld{\mc{A}}_0 & = & \frac{2^{-5}}{4} \Big[ \ (Q_o + Q_v)(0;0) [ (n+\bar{n})^2
v_1 v_2  \nonumber \\      
               & + & 2 m \bar{m} (1 + q^2 H_1^2) (1 + q^2 H_2^2) v_1 v_2 
\nonumber \\
               & + & \frac{(d+\bar{d})^2}{v_1v_2}] + \nonumber \\
               & + & 8 \ (n+\bar{n}) (m+\bar{m}) (Q_o +Q_v)(z_1;z_2) \frac{k_1 
\eta}{\vth_1 (z_1)} \frac{k_2 \eta}{\vth_1 (z_2)} + \nonumber \\
               & + & 4 \ (m^2+\bar{m}^2) (Q_o + Q_v)(2 z_1;2 z_2) \frac{2 k_1 
\eta}{\vth_1 (2 z_1)} \frac{2 k_2 \eta}{\vth_1 (2z_2)} + \nonumber \\
               & + & 2 \ (n+\bar{n}) (d+\bar{d}) (Q_o - Q_v)(0;0) \Big( 
\frac{2 \eta}{\vth_2 (0)} \Big)^2 +  \nonumber \\
               & + &  2 \  (m+\bar{m}) (d+\bar{d}) (Q_o - Q_v)(z_1;z_2) \Big( 
\frac{2 \eta}{\vth_2 (z_1)} \Big)  \Big( \frac{2 \eta}{\vth_2 (z_2)} \Big) +
\nonumber \\
               & - & 16 \  [ (n-\bar{n})^2 - 2 m \bar{m} + (d-\bar{d})^2 ]  
\Big( \frac{\eta}{\vth_4 (0)} \Big)^2 +  \nonumber \\
               & - & 32 \ (n-\bar{n}) (m+\bar{m}) (Q_s + Q_c)(z_1;z_2) \Big( 
\frac{\eta}{\vth_4 (z_1)} \Big)  \Big( \frac{\eta}{\vth_4 (z_2)} \Big) +
\nonumber \\
               & - & 16 \ (m^2 + \bar{m}^2) (Q_s + Q_c)(2 z_1;2 z_2) \Big( 
\frac{\eta}{\vth_4 (2 z_1)} \Big)  \Big( \frac{\eta}{\vth_4 (2 z_2)} \Big) +
\nonumber \\
               & + & 8 \ (n-\bar{n}) (d-\bar{d}) (Q_s - Q_c)(0;0) \Big( 
\frac{\eta}{\vth_3 (0)} \Big)^2 +  \nonumber \\
               & + & 8 \ (m-\bar{m}) (d-\bar{d}) (Q_s - Q_c)(z_1;z_2) \Big( 
\frac{\eta}{\vth_3 (z_1)} \Big)  \Big( \frac{\eta}{\vth_3 (z_2)} \Big) \ \Big]
\ ,
\label{A0tldT4suZ2}
\eeq
\beq
\tld{\mc{M}}_0 & = & -  \frac{2}{4} \  \Big[ \ (\hat{Q}_o + \hat{Q}_v)(0;0) 
[ (n+\bar{n})v_1 v_2 + \frac{(d+\bar{d})^2}{v_1 v_2}] \nonumber \\
               & + & (m+\bar{m}) (\hat{Q}_o + \hat{Q}_v)(z_1;z_2) \frac{2 k_1 
\hat{\eta}}{\hat{\vth}_1 (z_1)} \frac{2 k_2 \hat{\eta}}{\hat{\vth}_1 (z_2)}  
\nonumber \\
               & + & [ (n+\bar{n}) + (d+\bar{d}) ] (\hat{Q}_o - \hat{Q}_v)(0;0)
\Big( \frac{2 \hat{\eta}}{\hat{\vth}_2 (0)} \Big)^2 \nonumber \\
               & + & (m+\bar{m})  (\hat{Q}_o - \hat{Q}_v)(z_1;z_2)
\Big( \frac{2 \hat{\eta}}{\hat{\vth}_2 (z_1)} \Big) 
\Big( \frac{2 \hat{\eta}}{\hat{\vth}_2 (z_2)} \Big) \  \Big] \ .
\label{M0tldT4suZ2}
\eeq

Iniziamo ad esaminare le condizioni di $tadpole$ di R-R $untwisted$ che hanno 
origine dai settori $S_4 S_2 S_2$, $S_4 C_2 C_2$, $C_4 C_2 S_2$ e 
$C_4 S_2 C_2$. Queste conducono a soluzioni tutte tra loro  compatibili. 
Quindi possiamo considerare, per esempio, quella per $C_4 S_2 C_2$
\beq
[ (n & + & \bar{n}) + (m +\bar{m}) + i q (H_1 - H_2) (m -\bar{m}) +
\nonumber \\
     & + & q^2 H_1 H_2 (m +\bar{m}) -32  ] \sqrt{v_1 v_2}  \nonumber \\       
     & + &  [ (d +\bar{d}) - 32 ] \frac{1}{\sqrt{v_1 v_2}}    =  0  \ .
\label{tadpT4suZ21}
\eeq
Utilizzando la relazione di quantizzazione di Dirac (\ref{Diracrelquant}), 
 a meno dei termini che si annullano  identificando le 
molteplicit\`a delle rappresentazioni coniugate $(m,\bar{m})$, la
(\ref{tadpT4suZ21}) pu\`o essere riscritta come
\be
[ (n +\bar{n}) + (m +\bar{m}) - 32 \ ] \sqrt{v_1 v_2}+  
[ k_1 k_2 (m +\bar{m}) + (d +\bar{d}) - 32 \ ] \frac{1}{\sqrt{v_1 v_2}} = 0 \ ,
\label{tadpT4suZ22}
\ee
che implica le condizioni
\beq
n + \bar{n} + m +\bar{m}           & = & 32 \ , \nonumber \\
k_1 k_2 (m + \bar{m}) + d +\bar{d} & = & 32 \ .
\label{tadpT4suZ23}
\eeq
L'espressione (\ref{tadpT4suZ21}) riflette l'accoppiamento di Wess-Zumino 
dell'azione (\ref{S1}). In particolare il termine quadratico nei campi
magnetici descrive come le D9-brane sono cariche rispetto al potenziale di 
sei-forma. Questo effetto \`e messo in maggior risalto dalle eqs. 
(\ref{tadpT4suZ23}), dalle quali si legge che se $k_1 k_2>0$ le D9-brane
acquistano la carica di R-R di $|k_1 k_2|$ D5-brane, mentre se $k_1 k_2<0$ 
esse acquistano la carica di R-R di altrettante antiD5-brane, proprio come
previsto dall'espressione (\ref{S3}).

I $tadpole$  NS-NS $untwisted$, che derivano dai settori $V_4 O_2 O_2$, 
$O_4 V_2 O_2$ e $O_4 O_2 V_2$, mettono in luce la loro relazione con 
l'azione di Born-Infeld di eq. (\ref{S1}) \cite{Leigh:1989jq}. 
Per esempio, il $tadpole$ di $V_4 O_2 O_2$ 
\be
[ n + \bar{n} + (m + \bar{m}) \sqrt{(1 + q^2 H_1^2) (1 + q^2 H_2^2)} - 32 ]
\sqrt{v_1 v_2} + [ d + \bar{d} - 32 ] \frac{1}{\sqrt{v_1 v_2}} \ ,
\label{diltadp}
\ee
\`e legato alla derivata dell'azione (\ref{S2}) rispetto al dilatone $\f$.
Si osservi che se nell'espressione (\ref{diltadp}) si pone $H_1=H_2$ e si
utilizza la relazione di quantizzazione di Dirac, essa assume la stessa forma
di (\ref{tadpT4suZ22}) e quindi si annulla identicamente (come deve essere per 
configurazioni supersimmetriche), quando vengono imposte le condizioni di 
{\it tadpole} R-R.

I $tadpole$ della metrica interna ($O_4 V_2 O_2$ e $O_4 O_2 V_2$) hanno la
particolarit\`a di non assemblarsi in quadrati perfetti a causa del 
comportamento del campo magnetico sotto inversione temporale.
Infatti il contributo dell'ampiezza di anello (\ref{A0tldT4suZ2}) ad 
$O_4 V_2 O_2$, per esempio, \`e
\beq
\Big[ \Big( (n + \bar{n}) & + & (m+\bar{m}) 
\frac{(1-q^2 H_1^2)}{\sqrt{1+q^2 H_1^2}} \sqrt{1+q^2 H_2^2}  +  \nonumber \\
                          & + & (m - \bar{m}) 
\frac{2 q H_1}{\sqrt{1+q^2 H_1^2}} \sqrt{1+q^2 H_2^2} \Big) \sqrt{v_1 v_2} -  
\frac{(d+\bar{d})}{\sqrt{v_1 v_2}} 
 \Big]  \cdot  \nonumber \\
\cdot
\Big[ \Big( (n + \bar{n}) & + & (m+\bar{m}) 
\frac{(1-q^2 H_1^2)}{\sqrt{1+q^2 H_1^2}} \sqrt{1+q^2 H_2^2}  +   \nonumber \\
                          & - & (m - \bar{m}) 
\frac{2 q H_1}{\sqrt{1+q^2 H_1^2}} \sqrt{1+q^2 H_2^2} \Big) \sqrt{v_1 v_2} - 
\frac{(d+\bar{d})}{\sqrt{v_1 v_2}} 
 \Big] . 
\eeq
Le ampiezze nel canale trasverso coinvolgono un'operazione di inversione 
temporale sotto la quale il campo magnetico \`e dispari. Esso introduce 
quindi dei segni che non rendono tali ampiezze delle forme sesquilineari.
Ma la corretta  derivazione per fattorizzazione dell'ampiezza di M\"obius, come
$ \ket \tld{B}|q^{L_0}|C \bra + \ket \tld{C}|q^{L_0}|B \bra$, 
elimina dal canale trasverso i termini  proporzionali ad $(m-\bar{m})$. 
Pertanto il $tadpole$ corrispondente diviene
\be
\Big[ n+\bar{n} + (m+\bar{m}) 
\frac{(1-q^2 H_1^2)}{\sqrt{1+q^2 H_1^2}} \sqrt{1+q^2 H_2^2} - 32 \Big]
\sqrt{v_1 v_2} - \Big[ d+\bar{d} - 32 \Big] \frac{1}{\sqrt{v_1 v_2}} \ ,
\ee
che si annulla, per $H_1 = H_2$, una volta imposta la condizione
(\ref{tadpT4suZ22}).

Per quanto riguarda il settore $twisted$, i $tadpole$  R-R che derivano da
$S_4 O_2 O_2$ non vengono alterati dalle deformazioni magnetiche. 
Essi riflettono la distribuzione delle brane tra i sedici punti fissi.
Infatti
\be
15 \ \Big[ \frac{1}{4} (n-\bar{n} + m-\bar{m}) \Big]^2 +
\Big[ \frac{1}{4} (n-\bar{n} + m-\bar{m}) - (d-\bar{d}) \Big]^2 = 0 \ ,
\ee
riflette il fatto che le D5-brane sono accomodate, in questo caso, 
su un solo punto fisso, mentre le ulteriori D5-brane, ottenute da 
D9-brane magnetizzate, vedono tutti i punti fissi. 
Inoltre esso si annulla identicamente per gruppi di gauge unitari.

I $tadpole$  NS-NS {\it twisted}  hanno origine da $O_4 S_2 C_2$ e 
$O_4 C_2 S_2$ e danno luogo a nuovi accoppiamenti per i campi di NS-NS.
La loro struttura \`e piuttosto convoluta,
\beq
 & 15  &   \Big[   \frac{(n-\bar{n})}{4} \nonumber \\  
 & +   &   \frac{1}{4 \sqrt{(1+q^2H_1^2) 
(1+q^2H_2^2)}}[ (m-\bar{m}) (1+q^2 H_1 H_2) 
\pm   i q (H_2-H_1) (m+\bar{m}) ]^2   \Big]  \nonumber \\  
 & +   &   \Big[  \frac{(n-\bar{n})}{4}      \nonumber \\
 & +   &   \frac{1}{4 \sqrt{(1+q^2H_1^2) 
(1+q^2H_2^2)}}[ (m-\bar{m}) (1+q^2 H_1 H_2) 
\pm i q (H_2-H_1) (m+\bar{m})  \nonumber \\
 & -   &   (d-\bar{d}) ]^2 \Big] \ , 
\eeq
ma dopo l'identificazione delle molteplicit\`a coniugate, si annullano 
identicamente per $H_1=H_2$.

A questo punto possiamo descrivere alcuni modelli supersimmetrici 
corrispondenti al caso speciale $H_1=H_2$.
Per essi lo spettro chiuso proiettato \`e sempre lo stesso, e corrisponde a 
quello del modello $U(16)_9 \times U(16)_5$ descritto nel capitolo
precedente. I modi a massa nulla del settore aperto si possono invece
dedurre  da
\beq
\mc{A}_0 + \mc{M}_0 & = &  Q_o (0;0) [ \ n \bar{n} + m \bar{m} + d \bar{d} \ ]
\nonumber \\
                    & + &  Q_v (0;0) \Big[ \ \frac{1}{2} n (n+1) + \frac{1}{2} 
\bar{n} (\bar{n}+1) + \frac{1}{2} d (d+1) + \frac{1}{2} 
\bar{d} (\bar{d}+1) \nonumber   \\
                    & + & [ \frac{1}{2} m (m+1) + \frac{1}{2} 
\bar{m} (\bar{m}+1) ] ( \ k_1 k_2 + 1 \ ) \nonumber \\
                    & + & k_1 k_2 \frac{1}{2} (m + \bar{m}) (n + \bar{n})
+ (m + \bar{m}) (n - \bar{n}) \nonumber \\
                    & + & \frac{1}{2} (n \bar{d} + d \bar{n}) +
\frac{1}{2} (m \bar{d} + d \bar{m}) \ \Big] \ ,
\eeq
che mostra come il gruppo di gauge sia $ U(n)_9 \times U(m)_9 \times U(d)_5$.

Dalle condizioni di $tadpole$ si ottiene
\beq
n +  m       & = & 16  \ ,     \nonumber \\
k_1 k_2 m + d & = & 16  \ .
\label{tadpT4suZ24}
\eeq
Le minime  degenerazioni  dei livelli di Landau ammesse per questo
orbifold  sono $k_1 = k_2 = 2$, e quindi le possibili soluzioni dei 
$tadpoles$ di R-R (\ref{tadpT4suZ24}) sono \\
1) $ n=13 $ , $ m=3 $ , $ d=4 $\\
che corrisponde al gruppo di gauge $ U(13)_9 \times U(3)_9 \times U(4)_5$, di
rango 20, con ipermultipletti carichi nelle rappresentazioni 
$(78+\bar{78},1;1)$, in cinque copie della $(1,3+\bar{3};1)$, in una copia 
della $(1,1;6+\bar{6})$, in una copia della $(13,1;\bar{4})$, in quattro
copie della $(\bar{13},3;1)$, ed in una copia della $(1,\bar{3};4)$.\\
2) $ n=14 $ , $ m=2 $ , $ d=8 $\\
che corrisponde al gruppo di gauge $ U(14)_9 \times U(2)_9 \times U(8)_5$ di
rango 24 con ipermultipletti carichi nelle rappresentazioni 
$(91+\bar{91},1;1)$, in  cinque copie della $(1,1+\bar{1};1)$, in una copia 
della $(1,1;28+\bar{28})$, in una copia della $(14,1;\bar{8})$, in quattro
copie della $(\bar{14},2;1)$, ed in una copia della $(1,\bar{2};8)$.\\
3) $ n=12 $ , $ m=4 $ , $ d=0 $\\
che descrive un modello senza D5-brane e con gruppo di gauge 
$U(12) \times U(4)$, di rango 16, con ipermultipletti carichi 
nelle rappresentazioni $(66+\bar{66},1)$, in  cinque copie della 
$(1,6+\bar{6})$, ed in quattro copie della $(\bar{12},4)$.

Come si pu\`o notare, la particolarit\`a di questi modelli sta nel fatto che
alcuni multipletti appaiono in famiglie, mentre  il rango dei gruppi di gauge,
 oltre ad essere ridotto, non \`e  una potenza di due.

\section{Shift-Orbifolds}
L'estensione di questa tecnica a modelli in dimensione inferiore a sei con
supersimmetria minimale, 
necessita la compattificazione su $orbifolds$ pi\`u complessi che contengono 
D9 e D5 brane. In particolare, nel seguito considereremo 
cosiddetti \emph{\bf  shift-orbifolds} per il caso $ Z_2 \times Z_2$.
Mentre per compattificazioni su $orbifolds$ le coordinate di stringa vengono
identificate sotto inversioni interne (il caso $Z_2$ descritto in precedenza, 
ad esempio, corrisponde a nrotazioni combinate  di $ \pi$), gli 
\emph{shift-orbifolds} si ottengono combinandole con
 $shifts$ discreti dei vettori di base del reticolo di compattificazione,
o, pi\`u in dettaglio,  combinando   $shifts$ con simmetrie interne.
Il ruolo giocato da un'operazione di questo genere \`e di particolare 
interesse, poich\'e consente di implementare il meccanismo di Scherk-Schwarz 
per la rottura spontanea della supersimmetria in teoria di stringa.

In teoria dei campi il meccanismo di Scherk-Schwarz per la rottura spontanea 
della supersimmetria \`e essenzialmente una generalizzazione dell'usuale
riduzione dimensionale: si permette ai campi di dipendere in modo opportuno 
dalle coordinate interne, rispettando la consistenza  della compattificazione. 
Data la presenza delle sole  eccitazioni di Kaluza-Klein in
teoria dei campi, il corrispondente 
meccanismo di Scherk-Schwarz coinvolge  $shifts$ degli impulsi interni di
Kaluza-Klein.

In teoria di stringhe questo meccanismo pu\`o essere esteso deformando 
funzioni di partizione di stringhe chiuse in ragione dell'impulso e del 
$winding$ lungo  direzioni compatte, compatibilmente con l'invarianza
modulare. Dun\-que, a differenza della teoria dei campi, in questo caso si ha 
l'opzione di deformare i momenti o i $windings$, e quindi si hanno due 
meccanismi differenti di rottura della supersimmetria, che nel seguito
definiremo rispettivamente
 come \emph{rottura di Scherk-Schwarz} o \emph{rottura di M-teoria}
\cite{Antoniadis:1999ep}. 

A scopo propedeutico, analizzeremo il caso di uno \emph{shift-orbifold}
unidimensionale. Partendo dalla teoria di Tipo IIB compattificata su un cerchio
 di raggio $R$, costruiremo i discendenti aperti del modello con $shifts$
negli impulsi, noto come \emph{modello di Scherk-Schwarz}.
In questo caso la superstringa IIB deve essere proiettata con il generatore
di $Z_2$ $ (-)^F \d $, dove $F=F_L+F_R$ \`e il numero fermionico totale
spazio-temporale, mentre $\d$ \`e lo $shift$ $ X^9 \ra X^9 + \pi R$ lungo
la dimensione compatta di raggio $R$, che agisce sugli stati come $(-)^m$.
La funzione di partizione che ne deriva \`e
\beq
\mc{T}_{S-S} & = & \frac{1}{2} \sum_{m,n} \ 
\Big[  |V_8-S_8|^2 \L_{m,n} + |V_8+S_8|^2 (-)^m \L_{m,n}  \Big]  \nonumber \\
             & + & \frac{1}{2} \sum_{m,n} \ 
\Big[  |O_8-C_8|^2 \L_{m,n+1/2} + |O_8+C_8|^2 (-)^m \L_{m,n+1/2}  \Big].
\label{Ts-s}
\eeq
Introducendo le somme reticolari proiettate su momenti pari o dispari 
\beq 
\mc{E}_a & = & \sum_{m,n} \frac{1 + (-)^m}{2} \L_{m,n+a} \ , \nonumber \\
\mc{O}_a & = & \sum_{m,n} \frac{1 - (-)^m}{2} \L_{m,n+a} \ ,
\label{reticolarsums}
\eeq
l'ampiezza (\ref{Ts-s}) pu\`o essere riscritta in termini di una 
decomposizione ortogonale dello spettro chiuso del modello, ovvero
\beq
\mc{T}_{S-S} & = & \mc{E}_0 ( |V_8|^2 + |S_8|^2 ) + \mc{E}_{1/2} ( |O_8|^2 + 
|C_8|^2 )  \nonumber \\
             & - & \mc{O}_0 ( V_8 \bar{S}_8 +S_8 \bar{V}_8 ) - \mc{O}_{1/2}
( O_8 \bar{C}_8 +C_8 \bar{O}_8 ) \ .
\label{Tort}
\eeq
Si noti che per $R < \sqrt{\a'}$,  questo modello sviluppa un modo 
tachionico associato ad $|O_8|^2$, mentre per $ R \ra \infty$ esso 
si riduce alla stringa  IIB in dieci dimensioni.

La bottiglia di Klein riceve contributi da tutti i settori della (\ref{Tort})
invarianti sotto $\O$. Questi includono i soli  stati con 
$winding$ nullo ($n=0$), e quindi l'ampiezza corrispondente
\be
\mc{K}_{S-S} = \frac{1}{2} ( V_8 - S_8 ) P_m \ ,
\label{Ks-s}
\ee
non viene alterata dagli $shift$ negli impulsi. Nel passaggio al canale 
trasverso la (\ref{Ks-s}) diviene quindi
\be
\tld{\mc{K}}_{S-S} = \frac{2^{9/2}}{2} \frac{R}{\sqrt{\a'}} ( V_8 - S_8 ) 
W_{2m} \ .
\label{Ktlds-s}
\ee

In maniera analoga l'ampiezza di anello nel canale trasverso viene dedotta
da $\mc{T}_{S-S}$ restringendo la parte diagonale dello spettro al settore con
impulsi nulli ($m=0$). Quindi nel tubo possono fluire solamente gli stati 
associati a $V_8$ ed $S_8$ con $windings$ pari, e quelli associati ad $O_8$ e 
$C_8$ con $windings$ dispari. Di conseguenza possiamo introdurre quattro tipi
di cariche di Chan-Paton, parametrizzate dai quattro interi $n_1, n_2, n_3$
ed $n_4$, ottenendo cos\`\i 
\beq
\tld{\mc{A}}_{S-S} & = & \frac{2^{-11/2}}{2} \frac{R}{\sqrt{\a'}} \Big[ 
[(n_1+n_2+n_3+n_4)^2 V_8 - (n_1+n_2-n_3-n_4)^2 S_8] W_n  \nonumber \\
                   & + &  [(n_1-n_2+n_3-n_4)^2 O_8 - (n_1-n_2-n_3+n_4)^2 C_8] 
W_{n+1/2}  \Big].
\label{Atlds-s}
\eeq
Infine i caratteri comuni a $\tld{\mc{K}}_{S-S}$ ed $\tld{\mc{A}}_{S-S}$
determinano l'ampiezza di M\"obius nel canale trasverso
\be
\tld{\mc{M}}_{S-S} = - \frac{1}{\sqrt{2}} \frac{R}{\sqrt{\a'}} \Big[
(n_1+n_2+n_3+n_4)^2 \hat{V}_8 W_{2n} - (n_1+n_2-n_3-n_4)^2 \hat{S}_8 (-)^n
W_{2n} \Big] \ .
\label{Mtlds-s}
\ee

Le condizioni di {\it tadpole} si ottengono dalle (\ref{Ktlds-s}), 
(\ref{Atlds-s}) e (\ref{Mtlds-s}), ponendo a zero i coefficienti di 
riflessione per i modi a massa nulla, che hanno origine da $V_8$ ed $S_8$:
$$
\mbox{NS-NS} \ : \  \frac{2^{9/2}}{2} + \frac{2^{-11/2}}{2} 
(n_1+n_2+n_3+n_4)^2 -\frac{1}{\sqrt{2}} (n_1+n_2+n_3+n_4) = 0  \ ,
$$
$$
\mbox{R-R} \ : \  \frac{2^{9/2}}{2} + \frac{2^{-11/2}}{2} (n_1+n_2-n_3-n_4)^2 -
\frac{1}{\sqrt{2}} (n_1+n_2-n_3-n_4) = 0  \ ,
$$
ovvero:
$$
\mbox{NS-NS} \ : \ (n_1+n_2+n_3+n_4) = 32 \ ,
$$
$$
\mbox{R-R} \ : \  (n_1+n_2-n_3-n_4) = 32 \ .
$$
Da queste si evince che $n_1$ ed $n_2$ determinano il numero di D9-brane,
mentre $n_3$ ed $n_4$ determinano il numero di anti D9-brane.
Le condizioni di R-R fissano quindi il numero netto di brane del modello.
Imponendo anche i {\it tadpole}  NS-NS, l'introduzione delle anti-brane \`e 
vietata ($n_3=n_4=0$), e lo spettro risultante, privo di tachioni,  ha gruppo di
gauge $SO(n_1) \times SO(32-n_1)$.

Con procedimento analogo si pu\`o ottenere un altro tipo di $shift-orbifold$
unidimensionale, noto come \emph{modello di M-teoria}. Anche in questo caso il
punto di partenza \`e la teoria di tipo IIB compattificata su un cerchio di 
raggio $R$, ma l'azione dello $shift$ $\d$ lungo la direzione compatta \`e
la T-duale del caso precedente,  e dunque agisce sugli stati come $(-)^n$.
L'ampiezza di toro \`e quindi
\beq
\mc{T}_{M-th} & = & \tld{\mc{E}}_0 ( |V_8|^2 + |S_8|^2 ) + 
\tld{\mc{E}}_{1/2} ( |O_8|^2 + 
|C_8|^2 )  \nonumber \\
             & - & \tld{\mc{O}}_0 ( V_8 \bar{S}_8 +S_8 \bar{V}_8 ) - 
\tld{\mc{O}}_{1/2}
( O_8 \bar{C}_8 +C_8 \bar{O}_8 ) \ ,
\eeq
dove le somme reticolari $ \tld{\mc{E}}$ ed $\tld{\mc{O}}$ sono 
identiche alle (\ref{reticolarsums}) ma con impulsi e {\it windings} 
scambiati. Si osservi che in questo caso l'ampiezza di toro sviluppa
un'instabilit\`a tachionica per $ R > \sqrt{\a'}$, mentre per $R\ra 0$
la supersimmetria viene ripristinata.
In qusto caso la bottiglia di Klein nel canale diretto risente
della deformazione nei {\it windings}, e diventa
\beq
\mc{K}_{M-th} = \frac{1}{2} ( V_8 - S_8 ) P_m + 
\frac{1}{2} ( O_8 - C_8 ) P_{m+1/2} \ .
\eeq
La corrispondente ampiezza nel canale trasverso
\beq
\tld{\mc{K}}_{M-th} = \frac{2^{9/2}}{2} \frac{R}{\sqrt{\a'}} 
( \ V_8 W_{4n} - S_8 W_{4n+2} \ )   \ ,
\eeq
riceve, di conseguenza, contributi a massa nulla solo dagli stati
che derivano da $V_8$,  e quindi gli {\it shift-orientifold} nei
{\it windings} coinvolgono numeri uguali di O9 e O$\bar{9}$-piani,
la cui carica globale di R-R si annulla.

L'ampiezza di anello nel canale trasverso \`e
\beq
\tld{\mc{A}}_{M-th} & = & \frac{2^{-11/2}}{2} \frac{R}{\sqrt{\a'}} 
[  ( n_1 + n_2 + n_3 + n_4 )^2 V_8 -
 ( n_1 + n_2 - n_3 - n_4 )^2 S_8  ]  W_{4n} \nonumber \\
                    & + &  [  ( n_1 - n_2 + n_3 - n_4 )^2 V_8 -
 ( n_1 - n_2 - n_3 + n_4 )^2 S_8  ]  W_{4n+2} \ .
\eeq
Dai segni relativi delle molteplicit\`a di Chan-Paton
nei coefficienti di $S_8$, si vede che $n_1$ ed $n_2$ contano il 
numero di D9-brane, mentre $n_3$ ed $n_4$ quello di D$\bar{9}$-brane.
Con l'usuale procedimento, da $\tld{\mc{A}}_{M-th}$ e $\tld{\mc{K}}_{M-th}$
si pu\`o ricavare l'espressione per l'ampiezza di M\"obius 
nel canale trasverso
\beq
\tld{\mc{M}}_{M-th} & = &  - \frac{2}{2} \frac{R}{\sqrt{\a'}} \ 
[ \ ( n_1 + n_2 + n_3 + n_4 )^2  \ \hat{V}_8 \ W_{4n} \nonumber \\
                    & - & ( n_1 - n_2 - n_3 + n_4 )^2 \ \hat{S}_8  
\ W_{4n+2} \ ] \ ,
\eeq
e quindi le condizioni di {\it tadpole}
$$  n_1 + n_2 + n_3 + n_4 = 32 \ , \quad  n_1 + n_2 = n_3 + n_4 \ . $$
Ma nel limite $R \ra 0$, il modello sviluppa ulteriori 
{\it tadpole}, che derivano dai settori con somme shiftate $W_{4n+2}$,
che collassano a massa nulla:
$$ n_1 + n_3 = n_2 + n_4 \ , \quad   n_1 - n_2 - n_3 + n_4 = 32 \ . $$
Imponendo le precedenti condizioni, si giunge all'unica soluzione
$$ n_1 = 16 = n_4 \ , \quad n_2 = 0 = n_3  \ . $$
Lo spettro a massa nulla che ne risulta
\beq
\mc{A}_0 + \mc{M}_0 & = &  \Big[ \  \frac{n_1 (n_1 - 1)}{2} + 
\frac{n_4 (n_4 - 1)}{2} \ \Big] \ V_8 \nonumber \\
                    & - &  \Big[ \  \frac{n_1 (n_1 - 1)}{2} + 
\frac{n_4 (n_4 - 1)}{2} \ \Big] \ S_8 \ ,
\eeq
\`e quindi supersimmetrico, e contiene infatti
multipletti vettoriali per il gruppo di gauge $SO(16) \times SO(16)$.
Le eccitazioni massive per\`o non sono supersimmetriche, a causa
delle differenti proiezioni di M\"obius dei modi fermionici e bosonici,
e per la presenza dei settori $O_8$ e $C_8$:
\beq
\mc{A}_{M-th} & = & \frac{(n_1^2 + n_4^2)}{2} ( V_8 - S_8 ) P_m +
n_1 n_4 ( O_8 - C_8 ) P_{m+1/2} \ , \nonumber \\
\mc{M}_{M-th} & = & - \frac{1}{2} [ \ (n_1 + n_4 ) \hat{V}_8 P_m -
 ( n_1 + n_4 ) \hat{S}_8 (-)^m P_m \ ] \ .
\eeq
Questo modello fornisce inoltre un esempio del fenomeno noto come
\emph{brane supersymmetry}: le eccitazioni di massa pi\`u bassa di una brana
immersa in uno spazio-tempo non supersimmetrico possono essere 
supersimmetriche \cite{Antoniadis:1999ep, Antoniadis:1999xk}.

\section{Modelli in quattro dimensioni}
I risultati originali di questa tesi sono legati allo studio delle 
deformazioni 
magnetiche di modelli in quattro dimensioni, costruiti come discendenti di
 $orbifolds$ $Z_2 \times Z_2$.
Verranno quindi illustrati alcuni esempi di classi di modelli di Tipo I, sia
nel caso di proiezioni di $orbifold$ standard, sia nel caso in cui queste
ultime siano accompagnate da $shifts$ negli impulsi o nei $windings$
\cite{Antoniadis:2000ux}.

La teoria di partenza \`e la Tipo IIB compattificata su tre doppi tori,
$T_{45}, T_{67}, T_{89}$, proiettata dall'azione delle quattro operazioni di
$Z_2 \times Z_2$, che indicheremo come $o, g, f$ ed $h$.
Queste, a parte l'identit\`a $o$, agiscono come  rotazioni di $\pi$ su coppie
 tori interni, come segue
\be
g \ : \ (+,-,-) \ , \quad f \ : \ (-,+,-) \ , \quad h \ : \ (-,-,+) \ .
\ee

Per la descrizione delle ampiezze di questi modelli, \`e utile introdurre le
sedici quantit\`a $T_{ij}$ (con $i=o,g,f,h$):
\beq
T_{io}  =  \t_{io}+\t_{ig}+\t_{ih}+\t_{if} \ , \quad 
T_{ig}  =  \t_{io}+\t_{ig}-\t_{ih}-\t_{if} \ , \nonumber \\
T_{ih}  =  \t_{io}-\t_{ig}+\t_{ih}-\t_{if} \ ,  \quad
T_{if}  =  \t_{io}-\t_{ig}-\t_{ih}+\t_{if} \ ,
\eeq
dove i $\t_{ij}$ sono i sedici caratteri di $Z_2 \times Z_2$ definiti come:
\beq
\t_{oo} & = & V_2 O_2 O_2 O_2 + O_2 V_2 V_2 V_2 - S_2 S_2 S_2 S_2 - C_2 C_2 C_2
 C_2 \ , \nonumber \\
\t_{og} & = & O_2 V_2 O_2 O_2 + V_2 O_2 V_2 V_2 - C_2 C_2 S_2 S_2 - S_2 S_2 C_2
 C_2 \ , \nonumber \\
\t_{oh} & = & O_2 O_2 O_2 V_2 + V_2 V_2 V_2 O_2 - C_2 S_2 S_2 S_2 - S_2 C_2 C_2
 S_2 \ , \nonumber \\
\t_{of} & = & O_2 O_2 V_2 O_2 + V_2 V_2 O_2 V_2 - C_2 S_2 C_2 S_2 - S_2 C_2 S_2
 C_2 \ , \nonumber \\
\t_{go} & = & V_2 O_2 S_2 C_2 + O_2 V_2 C_2 S_2 - S_2 S_2 V_2 O_2 - C_2 C_2 O_2
 V_2 \ , \nonumber \\
\t_{gg} & = & O_2 V_2 S_2 C_2 + V_2 O_2 C_2 S_2 - S_2 S_2 O_2 V_2 - C_2 C_2 V_2
 O_2 \ , \nonumber \\
\t_{gh} & = & O_2 O_2 S_2 S_2 + V_2 V_2 C_2 C_2 - C_2 S_2 V_2 V_2 - S_2 C_2 O_2
 O_2 \ , \nonumber \\
\t_{gf} & = & O_2 O_2 C_2 C_2 + V_2 V_2 S_2 S_2 - S_2 C_2 V_2 V_2 - C_2 S_2 O_2
 O_2 \ , \nonumber \\
\t_{ho} & = & V_2 S_2 C_2 O_2 + O_2 C_2 S_2 V_2 - C_2 O_2 V_2 C_2 - S_2 V_2 O_2
 S_2 \ , \nonumber \\
\t_{hg} & = & O_2 C_2 C_2 O_2 + V_2 S_2 S_2 V_2 - C_2 O_2 O_2 S_2 - S_2 V_2 V_2
 C_2 \ , \nonumber \\
\t_{hh} & = & O_2 S_2 C_2 V_2 + V_2 C_2 S_2 O_2 - S_2 O_2 V_2 S_2 - C_2 V_2 O_2
 C_2 \ , \nonumber \\
\t_{hf} & = & O_2 S_2 S_2 O_2 + V_2 C_2 C_2 V_2 - C_2 V_2 V_2 S_2 - S_2 O_2 O_2
 C_2 \ , \nonumber \\
\t_{fo} & = & V_2 S_2 O_2 C_2 + O_2 C_2 V_2 S_2 - S_2 V_2 S_2 O_2 - C_2 O_2 C_2
 V_2 \ , \nonumber \\
\t_{fg} & = & O_2 C_2 O_2 C_2 + V_2 S_2 V_2 S_2 - C_2 O_2 S_2 O_2 - S_2 V_2 C_2
 V_2 \ , \nonumber \\
\t_{fh} & = & O_2 S_2 O_2 S_2 + V_2 C_2 V_2 C_2 - C_2 V_2 S_2 V_2 - S_2 O_2 C_2
 O_2 \ , \nonumber \\
\t_{ff} & = & O_2 S_2 V_2 C_2 + V_2 C_2 O_2 S_2 - C_2 V_2 C_2 O_2 - S_2 O_2 S_2
 V_2 \ .
\label{taus}
\eeq
$O_2, V_2, S_2$ e $C_2$ sono i quattro caratteri di livello 1 di $O(2)$,
mentre l'ordinamen\-to dei quattro fattori corrisponde alle otto dimensioni
trasverse dello spazio-tempo.

Esistono due classi di modelli supersimmetrici $Z_2 \times Z_2$ che 
differiscono  per la presenza di ``torsione discreta''. 
Nella funzione di partizione ci\`o si riflette in un segno, che  ha
effetti cruciali sia sul contenuto dei campi a massa nulla che sulla struttura
dei discendenti.

Le ampiezze di toro per tali modelli supersimmetrici $Z_2 \times Z_2$ sono
\beq
\mc{T} & = & \frac{1}{4} \Big[ \ |T_{oo}|^2 \L_1 \L_2 \L_3 + |T_{og}|^2 \L_1
\Big| \frac{4 \eta^2}{\vth_2^2} \Big|^2 +  |T_{of}|^2 \L_2
\Big| \frac{4 \eta^2}{\vth_2^2} \Big|^2 +  |T_{oh}|^2 \L_3
\Big| \frac{4 \eta^2}{\vth_2^2} \Big|^2   \nonumber \\
       & + & |T_{go}|^2 \L_1 \Big| \frac{4 \eta^2}{\vth_4^2} \Big|^2 + 
|T_{gg}|^2 \L_1 \Big| \frac{4 \eta^2}{\vth_3^2} \Big|^2 + 
|T_{ho}|^2 \L_3 \Big| \frac{4 \eta^2}{\vth_4^2} \Big|^2  \nonumber \\  
       & + & |T_{hh}|^2 \L_3 \Big| \frac{4 \eta^2}{\vth_3^2} \Big|^2 + 
|T_{fo}|^2 \L_2 \Big| \frac{4 \eta^2}{\vth_4^2} \Big|^2 +   
|T_{ff}|^2 \L_2 \Big| \frac{4 \eta^2}{\vth_3^2} \Big|^2   \nonumber \\
       & + & \e  ( |T_{gh}|^2 + |T_{gf}|^2 + |T_{fg}|^2 + |T_{fh}|^2 +
|T_{hg}|^2 + |T_{hf}|^2 ) \Big| \frac{8 \eta^3}{\vth_2 \vth_3 \vth_4} \Big|^2
 \Big]  ,
\label{TZ2perZ2}
\eeq
dove per semplicit\`a sono stati lasciati impliciti i contributi dei bosoni
trasversi. Inoltre $ \L_1, \L_2$ e $\L_3$ sono le somme reticolari associate ai
tre tori interni, e $ \e = \pm1$ discrimina tra i due modelli: quello 
\emph{con torsione discreta} ($ \e=-1$), e quello {\it senza} torsione 
discreta ($\e=1$).

Le ampiezze $\mc{K}, \mc{A}$ ed $\mc{M}$, necessarie per la determinazione 
dello spettro dei discendenti aperti, vengono costruite seguendo l'usuale 
procedura. Per quanto riguarda il settore chiuso proiettato, esistono diverse
scelte per $\mc{K}$, che definiscono la proiezione per gli
stati twistati. 
Nel seguito discuteremo per semplicit\`a i soli  modelli {\it senza} torsione 
discreta, con la pi\`u semplice funzione di Klein

Nel canale diretto, l'ampiezza di bottiglia di Klein \`e
\beq
\mc{K} & = & \frac{1}{8}\  \Big[ \ ( P_1 P_2 P_3 + P_1 W_2 W_3 + W_1 P_2 W_3 +
W_1 W_2 P_3 ) \ T_{oo}  \nonumber \\
       & + & 2 \times 16 \ [ \ (P_1 + W_1) \ T_{go} + (P_2 + W_2) \ T_{fo} 
\nonumber \\
       & + & (P_3 + W_3) \ T_{ho} \ ] \ \Big( \frac{\eta}{\vth_4} \Big)^2 
\ \Big] \  ,
\label{KZ2perZ2}
\eeq
dove $P_i$ e $W_i$ sono i sottoreticoli delle somme reticolari $\L_i$
ristretti rispettivamente ai soli impulsi e $windings$.
Dopo una trasformazione modulare $S$, l'ampiezza (\ref{KZ2perZ2}) diviene
\beq
\tld{\mc{K}} & = & \frac{2^5}{8} \ \Big[ \ ( v_1 v_2 v_3 W_1^e W_2^e W_3^e + 
\frac{v_1}{v_2 v_3} W_1^e P_2^e P_3^e + \frac{v_2}{v_1 v_3} P_1^e W_2^e P_3^e
\nonumber \\ 
              & + & \frac{v_3}{v_1 v_2} P_1^e P_2^e W_3^e ) T_{oo}  
              + 2 [ ( v_1 W_1^e + \frac{P_1^e}{v_1} ) T_{og} +
( v_2 W_2^e + \frac{P_2^e}{v_2} ) T_{of} \nonumber \\
              & + & ( v_3 W_3^e +  \frac{P_3^e}{v_3} ) T_{oh} ] 
\Big( \frac{2 \eta}{\vth_2}  \Big)^2  \ \Big] \ ,
\label{tldKZ2perZ2}  
\eeq 
ed all'origine delle somme reticolari i vari termini si arrangiano 
come al solito in quadrati perfetti:
\beq
\tld{\mc{K}_0} & = & \frac{2^5}{8} \ \Big[ \ \Big( \  \sqrt{v_1 v_2 v_3} + 
 \sqrt{\frac{v_1}{v_2 v_3}} +  \sqrt{\frac{v_2}{v_1 v_3}} +  
 \sqrt{\frac{v_3}{v_1 v_2}} \ \Big)^2 \t_{oo}   \nonumber \\
               & + & \Big( \  \sqrt{v_1 v_2 v_3} + 
 \sqrt{\frac{v_1}{v_2 v_3}} -  \sqrt{\frac{v_2}{v_1 v_3}} -  
 \sqrt{\frac{v_3}{v_1 v_2}} \ \Big)^2 \t_{og}   \nonumber \\
               & + & \Big( \  \sqrt{v_1 v_2 v_3} - 
 \sqrt{\frac{v_1}{v_2 v_3}} +  \sqrt{\frac{v_2}{v_1 v_3}} -  
 \sqrt{\frac{v_3}{v_1 v_2}} \ \Big)^2 \t_{of}   \nonumber \\
               & + & \Big( \  \sqrt{v_1 v_2 v_3} - 
 \sqrt{\frac{v_1}{v_2 v_3}} -  \sqrt{\frac{v_2}{v_1 v_3}} +  
 \sqrt{\frac{v_3}{v_1 v_2}} \ \Big)^2 \t_{oh}   \ \Big] \ .
\eeq
Per quanto riguarda i discendenti aperti, gli unici termini che contribuiscono
all'ampiezza di anello nel canale trasverso, sono quelli $untwisted$, in
quanto quelli $twisted$ coinvolgono, in (\ref{TZ2perZ2}) per $\e=1$,
 combinazioni diagonali   dei caratteri $\t_{ij}$ che  non sono
autoconiugati, e quindi gli stati $twisted$  non possono fluire nel tubo. 
Pertanto l'ampiezza di anello del modello con $\e=1$ non contiene termini di proiezione,
ed \`e data da
\beq
\mc{A} & = & \frac{1}{8} \ \Big[ \ (N^2 P_1 P_2 P_3 + D_1^2 P_1 W_2 W_3 + 
D_2^2 W_1 P_2 W_3 +  D_3^2 W_1 W_2 P_3 ) \ T_{oo}  \nonumber \\
       & + & 2 \ ( N D_1 P_1 T_{go} +  N D_2 P_2 T_{fo} 
        +   N D_3 P_3 T_{ho} ) 
\ \Big( \frac{\eta}{\vth_4} \Big)^2   \nonumber \\
       & + & 2 \ ( D_2 D_3 W_1 T_{go} +  D_1 D_3 W_2 T_{fo} +  D_1 D_2 P_3 
T_{ho} ) \ \Big( \frac{\eta}{\vth_4} \Big)^2  \  \Big] \ .
\label{AZ2perZ2}
\eeq
Si osservi come in questa espressione, a parte i termini standard $N^2, D_i^2$,
associati rispettivamente alle 9-brane ed alle 5-brane, ed i corrispondenti 
termini $ND_i$, siano presenti altri tre tipi di stringhe, la cui natura \`e
intimamente connessa alla struttura dell'$orbifold$ $Z_2 \times Z_2$.
Relativamente ai tre tori interni infatti, le tre operazioni di $twist$
di $Z_2 \times Z_2$ danno origine a stringhe aperte con estremi caratterizzati
da quattro tipi di condizioni al bordo: $ NNN, NDD, DND $ e $ DDN$, le cui 
possibili combinazioni conducono precisamente a tutti i tipi di stringhe 
presenti nel modello.

Operando una trasformazione $S$, la (\ref{AZ2perZ2}) diventa
\beq
\tld{\mc{A}} & = & \frac{2^{-5}}{8} \ \Big[ \ ( N^2 v_1 v_2 v_3 W_1 W_2 W_3 +
\frac{D_1^2 v_1}{v_3 v_2} W_1 P_2 P_3 + \frac{D_2^2 v_2}{v_1 v_3} 
P_1 W_2 P_3   \nonumber \\
             & + & \frac{D_3^2 v_3}{v_1 v_2} P_1 P_2 W_3 ) T_{oo} 
              + 2  ( N D_1 v_1 W_1 T_{og} + N D_2 v_2 W_2 T_{of}  \nonumber \\
             & + & N D_3 v_3 W_3 T_{oh} ) \Big( \frac{2 \eta}{\vth_2} \Big)^2
               + 2  ( D_12 D_3 \frac{1}{v_1} P_1 T_{og} + D_1 D_3 
\frac{1}{v_2}  P_2  T_{of}   \nonumber \\
             & + &  D_1 D_2 \frac{1}{v_3}  P_3 T_{oh} ) 
\Big( \frac{2 \eta}{\vth_2} \Big)^2 \ \Big] \ ,
\label{tldAZ2perZ2}
\eeq
la cui struttura ha un'interpretazione pi\`u chiara all'origine delle somme
reticolari, dove i vari termini si assemblano in quadrati perfetti.
Infatti
\beq
\tld{\mc{A}_0} & = & \frac{2^{-5}}{8} \ \Big[ \ \Big( \ N \sqrt{v_1 v_2 v_3} + 
D_1 \sqrt{\frac{v_1}{v_2 v_3}} + D_2 \sqrt{\frac{v_2}{v_1 v_3}} +  
D_3 \sqrt{\frac{v_3}{v_1 v_2}} \ \Big)^2 \t_{oo}   \nonumber \\
               & + & \Big( \ N \sqrt{v_1 v_2 v_3} + 
D_1 \sqrt{\frac{v_1}{v_2 v_3}} - D_2 \sqrt{\frac{v_2}{v_1 v_3}} -  
D_3 \sqrt{\frac{v_3}{v_1 v_2}} \ \Big)^2 \t_{og}   \nonumber \\
               & + & \Big( \ N \sqrt{v_1 v_2 v_3} - 
D_1 \sqrt{\frac{v_1}{v_2 v_3}} + D_2 \sqrt{\frac{v_2}{v_1 v_3}} -  
D_3 \sqrt{\frac{v_3}{v_1 v_2}} \ \Big)^2 \t_{of}   \nonumber \\
               & + & \Big( \ N \sqrt{v_1 v_2 v_3} - 
D_1 \sqrt{\frac{v_1}{v_2 v_3}} - D_2 \sqrt{\frac{v_2}{v_1 v_3}} +  
D_3 \sqrt{\frac{v_3}{v_1 v_2}} \ \Big)^2 \t_{oh}   \ \Big] \ ,
\eeq
presenta quattro diversi coefficienti di riflessione legati alle 9-brane ed ai
tre tipi di 5-brane.

A partire da $\tld{\mc{K}_0}$ e da $\tld{\mc{A}_0}$, seguendo la procedura 
standard, si calcola $\tld{\mc{M}_0}$ e da questa l'ampiezza di M\"obius 
completa nel canale trasverso, ovvero
\beq
\tld{\mc{M}} & = & - \frac{2}{8} \ \Big[ \ ( N v_1 v_2 v_3 \ W_1^e W_2^e W_3^e 
+ \frac{D_1 v_1}{v_2 v_3} \ W_1^e P_2^e P_3^e + \frac{D_2 v_2}{v_1 v_3} \ P_1^e
 W_2^e P_3^e \nonumber \\ 
              & + & \frac{D_3 \ v_3}{v_1 v_2} P_1^e P_2^e W_3^e ) \ 
\hat{T}_{oo} +  [ \ (N + D_1) v_1 W_1^e + (D_2 + D_3) \frac{P_1^e}{v_1}\ ]  \ 
\hat{T}_{og} \  \Big( \frac{2 \hat{\eta}}{\hat{\vth}_2} \Big)^2  \nonumber \\  
              & + & [ \ (N + D_2) v_2 W_2^e + (D_3 + D_1) \frac{P_2^e}{v_2} ] 
\ \hat{T}_{of}  \ \Big( \frac{2 \hat{\eta}}{\hat{\vth}_2} \Big)^2  \nonumber \\
              & + & [ \ (N + D_3) v_3 W_3^e + (D_2 + D_1) \frac{P_3^e}{v_3} ] 
\ \hat{T}_{oh}  \ \Big( \frac{2 \hat{\eta}}{\hat{\vth}_2} \Big)^2 \  \Big] \ ,
\eeq
dalla quale, tramite una trasformazione modulare $P$ si ottiene l'ampiezza
di M\"obius nel canale diretto
\beq
\mc{M} & = & - \frac{1}{8} \ \Big[ \ ( N P_1 P_2 P_3 + D_1 P_1 W_2 W_3 + 
D_2 W_1 P_2 W_3 + D_3 W_1 W_2 P_3 ) \ \hat{T}_{oo} \nonumber \\
       & - & [ \ (N + D_1) P_1 + (D_2 + D_3) W_1  \ ]  \ 
\hat{T}_{og} \  \Big( \frac{2 \hat{\eta}}{\hat{\vth}_2} \Big)^2  \nonumber \\
       & - & [ \ (N + D_2) P_2 + (D_3 + D_1) W_2 \ ]  \ 
\hat{T}_{of} \  \Big( \frac{2 \hat{\eta}}{\hat{\vth}_2} \Big)^2  \nonumber \\
       & - & [ \ (N + D_3) P_3 + (D_2 + D_1) W_3  \ ]  \ 
\hat{T}_{oh} \  \Big( \frac{2 \hat{\eta}}{\hat{\vth}_2} \Big)^2  \ \Big] \ .
\label{MZ2perZ2}
\eeq
Si noti che $\mc{M}$, a differenza di $\mc{A}$, contiene quattro proiezioni,
corrispondenti alle quattro operazioni di $Z_2 \times Z_2$.
Dalla (\ref{AZ2perZ2}) e dalla (\ref{MZ2perZ2}) \`e possibile leggere lo 
spettro a massa nulla. Le condizioni di $tadpole$ fissano
\be
N = D_1 = D_2 = D_3 = 32 \ ,
\ee
ma a causa dell'assenza dei termini di rottura, per garantire un'appropriata
interpretazione degli stati di stringa aperta, \`e necessario riscalare le 
molteplicit\`a di Chan-Paton per un fattore 2. 
Quindi, ponendo $ N = 2 n$ e $ D_i = 2 d_i$, si ottiene
\beq
\mc{A}_0 + \mc{M}_0 & = &  \Big[ \ \frac{n(n+1)}{2} +  \frac{d_1(d_1+1)}{2} + 
\frac{d_2(d_2+1)}{2} + \frac{d_3(d_3+1)}{2} \ \Big] \ \t_{oo} \nonumber \\
                    & + &   \Big[ \ \frac{n(n-1)}{2} +  \frac{d_1(d_1-1)}{2} + 
\frac{d_2(d_2-1)}{2} + \frac{d_3(d_3-1)}{2} \ \Big] \  \t_{og}   \nonumber \\
                    & + &  \Big[ \ \frac{n(n-1)}{2} +  \frac{d_1(d_1-1)}{2} + 
\frac{d_2(d_2-1)}{2} + \frac{d_3(d_3-1)}{2} \ \Big] \  \t_{of}   \nonumber \\
                    & + &  \Big[ \ \frac{n(n-1)}{2} +  \frac{d_1(d_1-1)}{2} + 
\frac{d_2(d_2-1)}{2} + \frac{d_3(d_3-1)}{2} \ \Big] \   \t_{oh}   \nonumber \\
                    & + &  [ n d_1 + d_2 d_3 ] \ ( \t_{go} + \t_{gg} + 
\t_{gh} + \t_{gf} ) \nonumber \\
                    & + &  [ n d_2 + d_1 d_3 ] \ ( \t_{fo} + \t_{fg} + 
\t_{fh} + \t_{ff} ) \nonumber \\
                    & + &  [ n d_3 + d_1 d_2 ] \ ( \t_{ho} + \t_{hg} + 
\t_{hh} + \t_{hf} ) \ ,
\label{openspectrumZ2Z2}
\eeq
dalla quale si deduce che il gruppo di gauge \`e $USp(16)^4$, con multipletti
vettoriali nell'aggiunta, tre multipletti chirali nella $(120,1,1,1), 
(1,120,1,1)$, $(1,1,120,1)$ e $(1,1,1,120)$, sei multipletti chirali nella
$(16,16,1,1)$ ed in cinque rappresentazioni bifondamentali che differiscono
l'una dall'altra per la permutazione dei fattori. 

Questi  \emph{orbifolds} supersimmetrici  $Z_2 \times Z_2$,
 possono essere deformati combinando le proiezioni standard di $orbifold$ 
con $shifts$ negli impulsi o nei $windings$ degli stati di 
reticolo. I modelli risultanti sono caratterizzati da discendenti 
aperti che presentano rottura
parziale della supersimmetria, e coinvolgono peculiari configurazioni di
D5-brane.  Le brane si arrangiano tipicamente in multipletti
di immagini, scambiati da alcune delle operazioni di \emph{orbifold}, 
causando  cos\`\i \ la riduzione del rango dei corrispondenti gruppi di gauge.
Come accennato nei paragrafi precedenti, quando le operazioni convenzionali
di \emph{orbifold} vengono combinate con  $shifts$, i modelli risultanti
forniscono la realizzazione, a livello di stringa, del meccanismo di
Scherk-Schwarz.
Vogliamo quindi analizzare gli aspetti salienti degli effetti combinati di
$shifts$ $(\d_L,\d_R)$ e delle operazioni di \emph{orbifold} $Z_2 \times Z_2$
sui discendenti aperti delle compattificazioni della Tipo IIB 
\cite{Antoniadis:2000ux}.

Sia $ X_i \ra \s(X_i)$ l'azione di \emph{orbifold} sulle coordinate complesse
dei tre tori interni. Esistono diversi modi di combinare, in maniera 
consistente con la struttura del gruppo $Z_2 \times Z_2$, le tre operazioni
$g, f$ ed $h$ con gli $shifts$.
In ogni caso, a meno di T-dualit\`a e relativa ridefinizione della proiezione
$\O$, le possibilit\`a pi\`u interessanti per questa classe di modelli sono
riassunte dalle matrici
\be
\s_1 (\d_1,\d_2,\d_3) = \left(
\ba{ccc}
\d_1  & -\d_2 & -1 \\
-1    &  \d_2 & -\d_3 \\
-\d_1 & -1    &\d_3
\ea
\right) \ ,
\
\s_2 (\d_1,\d_2,\d_3) = \left(
\ba{ccc}
\d_1  & -1    & -1\\
-1    & \d_2  & -\d_3\\
-\d_1 & -\d_2 & \d_3
\ea
\right) \ ,
\label{s1s2}
\ee
dove le tre righe si riferiscono alle nuove operazioni, che continueremo ad
indicare con $g, f$ ed $h$, e dove $- \d_i$ indica la combinazione di uno
$shift$ nella direzione $i$ con un'inversione di \emph{orbifold}, ovvero 
$ - \d_i$ : $X_i \ra -X_i + \d_i$.

Fissando l'attenzione sulla proiezione $\O$ convenzionale, \`e possibile 
descrivere dieci diversi modelli di \emph{shift-orbifolds}  $Z_2 \times Z_2$,
corrispondenti a varie scelte di $shifts$ negli impulsi e nei $windings$.
In ogni caso, la costruzione di ampiezze di vuoto per i discendenti
aperti ha come punto di partenza l'ampiezza di toro dell'orbifold 
supersimmetrico (\ref{TZ2perZ2}), dove gli $shifts$ inducono delle modifiche
sui corrispondenti contributi twistati.
In tutti questi modelli deformati, l'orbita modulare legata alla 
torsione discreta \`e  assente, perch\'e almeno una somma reticolare
associata ai termini $twisted$ si alza in massa.

Nel seguito, la classe di modelli a cui faremo riferimento in dettaglio
sar\`a quella ottenuta introducendo nella matrice $\s_2$ di eq.(\ref{s1s2}), 
$shifts$ negli impulsi $p_2$ e $p_3$ sul secondo e terzo toro interno. 
In questo caso la supersimmetria $N=2$ viene rotta a $N=1$, e la
 corrispondente ampiezza di toro \`e  
\beq
\mc{T} & = & \frac{1}{4} \ \Big[ \ |T_{oo}|^2 \L_1 \L_2 \L_3 + |T_{og}|^2 \L_1
\Big| \frac{4 \eta^2}{\vth_2^2} \Big|^2 +  |T_{of}|^2  (-)^{m_2} \L_2
\Big| \frac{4 \eta^2}{\vth_2^2} \Big|^2  \nonumber \\
       & + & |T_{oh}|^2  (-)^{m_3}  \L_3\Big| \frac{4 \eta^2}{\vth_2^2} \Big|^2
 + |T_{go}|^2 \L_1 \Big| \frac{4 \eta^2}{\vth_4^2} \Big|^2 + 
|T_{gg}|^2 \L_1 \Big| \frac{4 \eta^2}{\vth_3^2} \Big|^2  \nonumber \\
       & + & |T_{ho}|^2 \L_3^{n_3+1/2} \ \Big| \frac{4 \eta^2}{\vth_4^2} 
\Big|^2  + |T_{hh}|^2 (-)^{m_3} \L_3^{n_3+1/2} \ \Big| 
\frac{4 \eta^2}{\vth_3^2} \Big|^2  \nonumber \\  
       & + & |T_{fo}|^2 \L_2^{n_2+1/2} \ \Big| \frac{4 \eta^2}{\vth_4^2} 
\Big|^2 + |T_{ff}|^2 (-)^{m_2} \L_2^{n_2+1/2} \ \Big| \frac{4 \eta^2}{\vth_3^2}
 \Big|^2  \ \Big] \ .  
\label{Tp23}
\eeq    
Come al solito, i discendenti aperti sono determinati essenzialmente 
dall'am\-piezza di Klein nel canale diretto $\mc{K}$ e dall'ampiezza 
di anello nel canale trasverso$\tld{\mc{A}}$.

L'ampiezza di bottiglia di Klein viene deformata dagli $shifts$, che possono
rendere massivi i contributi del canale trasverso.
Come nel caso supersimmetrico,  $\mc{K}$
coinvolge  termini proporzionali a $P_1 P_2 P_3, P_1 W_2 W_3, W_1 P_2 W_3$
e $W_1 W_2 P_3$, che riflettono la struttura  $Z_2 \times Z_2$ di questi
modelli, e determinano il contenuto di $D9, D5_1, D5_2$ e $D5_3$ brane. 
Quando questi termini sono accompagnati da fattori di fase
indotti dagli $shifts$, i corrispondenti contributi nel canale trasverso
vengono alzati in massa e le condizioni di $tadpole$ eliminano le brane
corrispondenti.
Nel caso in esame quindi, in cui gli $shifts$ nei momenti agiscono sul 
secondo e terzo toro, l'ampiezza di Klein nel canale diretto
\beq
\mc{K} & = & \frac{1}{8} \ \Big[ \ ( P_1 P_2 P_3 + P_1 W_2 W_3 + (-)^{m_2}
W_1 P_2 W_3 + (-)^{m_3} W_1 W_2 P_3 ) \ T_{oo} \nonumber \\
       & + & 2 \times 16 \ T_{go} P_1 \Big( \frac{\eta}{\vth_4} \Big)^2 
\ \Big] \ ,
\label{Kp23}
\eeq
rivela la presenza di solo $D9$ e $D5_1$ brane.

Come al solito, il comportamento dell'ampiezza di Klein nel canale trasverso
\beq
\tld{\mc{K}} & = & \frac{2^5}{8} \ \Big[ \ ( v_1 v_2 v_3 W_1^e W_2^e W_3^e + 
\frac{v_1}{v_2 v_3} W_1^e P_2^e P_3^e + \frac{v_2}{v_1 v_3} P_1^e W_2^o P_3^e
\nonumber \\ 
              & + & \frac{v_3}{v_1 v_2} P_1^e P_2^e W_3^o ) T_{oo}  
              + 2  v_1 W_1^e  T_{og}  \Big( \frac{2 \eta}{\vth_2} \Big)^2  
\ \Big] \ ,
\label{tldKp23}  
\eeq 
all'origine delle somme reticolari fissa il contributo $twisted$ di $\mc{K}$,
che assicura che i vari settori indipendenti  dello spettro abbiano 
coefficienti di riflessione che si arrangino in quadrati perfetti. Infatti
\be
\tld{\mc{K}}_0 = \frac{2^5}{8} \Big[ \ \Big( \sqrt{v_1 v_2 v_3} + 
\sqrt{\frac{v_1}{v_2 v_3}} \Big)^2 (\t_{oo} + \t_{og}) +  
\Big( \sqrt{v_1 v_2 v_3} - \sqrt{\frac{v_1}{v_2 v_3}} \Big)^2 
(\t_{of} + \t_{oh}) \ \Big]  .
\label{Ktld0p23}
\ee

L'altro ingrediente fondamentale per la costruzione del modello \`e 
l'ampiez\-za di anello nel canale trasverso $\tld{\mc{A}}$. 
Nel caso in esame, lo $shift-orbifold$ impone dei vincoli, di origine
geometrica, sui modi \emph{untwisted} che possono fluire nel tubo.
In generale, i punti fissi di un'operazione formano
multipletti le cui componenti devono accomodare gli stessi insiemi di brane.
Di conseguenza le brane sono accompagnate da un certo numero di immmagini,
le cui configurazioni sono determinate appunto, dalle restrizioni sulla forma
di $\tld{\mc{A}}$. 
La presenza di multipletti di brane nella configurazione di vuoto si 
riflette nella presenza, in $\tld{\mc{A}}$, di corrispondenti operatori di
proiezione. Per il modello $p_2 p_3$, il proiettore che moltiplica gli stati
nel canale trasverso, e che nel canale diretto da' origine a doppietti di
brane, \`e
\be
\Pi = \frac{1 + (-)^{m_2+m_3}}{2} \ ,
\ee 
come pu\`o essere visto da una dettagliata analisi degli operatori di vertice
\cite{Antoniadis:2000ux}.
L'ampiezza di anello nel canale trasverso assume quindi la forma 
\beq
\tld{\mc{A}} & = &  \frac{2^{-5}}{8} \Big[ \ [ N^2 v_1 v_2 v_3 W_1 W_2 
W_3 + D_1^2 \frac{v_1}{v_2 v_3} W_1 ( P_2^e P_3^e + P_2^o P_3^o ) ] \  T_{oo}
\nonumber \\
             & + & 4 ( G^2 + 2 G_1^2 ) T_{go} W_1 v_1 \Big( \frac{2 \eta}{\vth_4} \Big)^2 + 4  F^2 T_{fo} W_2^{n+1/2} v_2 \Big( \frac{2 \eta}{\vth_4} \Big)^2
\nonumber \\
             & + & 4 H^2 T_{ho}  W_3^{n+1/2} v_3 \Big( \frac{2 \eta}{\vth_4} \Big)^2 + 2 N D_1 T_{og}  W_1 v_1 \Big( \frac{2 \eta}{\vth_2} \Big)^2 
\nonumber \\
             & + & 4 G G_1 T_{gg}  W_1 v_1 \Big( \frac{2 \eta}{\vth_3} \Big)^2 
\ \Big] \ ,
\eeq
e dopo una trasformazione $S$, da' luogo, nel canale diretto, a
\beq
\mc{A} & = &  \frac{1}{8} \Big[ \ [ N^2 P_1 P_2 P_3 + \frac{D_1^2}{2} P_1 ( W_2
 W_3 + W_2^{n+1/2} W_3^{n+1/2} ) ] \  T_{oo}  
\nonumber \\
       & + &  ( G^2 + 2 G_1^2 ) T_{og} P_1  \Big( \frac{2 \eta}{\vth_2} \Big)^2
 + F^2 T_{of} (-)^{m_2}  P_2 \Big( \frac{2 \eta}{\vth_2} \Big)^2
\nonumber \\
       & + &  H^2 T_{oh} (-)^{m_3}  P_3 \Big( \frac{2 \eta}{\vth_2} \Big)^2
2 N D_1 T_{go}  P_1  \Big( \frac{\eta}{\vth_4} \Big)^2 
\nonumber \\
        & + & 4 G G_1 T_{gg}  P_1  \Big( \frac{\eta}{\vth_3} \Big)^2 
\ \Big] \ .
\label{AZ2perZ2shift}
\eeq
Si noti che, mentre i termini che descrivono  stringhe 99 hanno la forma 
usuale, con i corrispondenti tre termini di 
proiezione (anche se alterati dagli $shifts$ negli impulsi), la configurazione
delle stringhe di tipo 55 \`e piuttosto particolare, ed ammette un'interessante
interpretazione geometrica: essa corrisponde ad un doppietto di brane associate
ad una coppia di tori fissi sotto l'operazione $g$, ma scambiati da 
$f$ ed $h$.

La presenza di multipletti di brane \`e una caratteristica generale di tali 
\emph{shift-orbifolds}. Infatti, se anche si provasse a disporre tutte le
brane su un unico punto fisso, queste verrebbero spostate dalle operazioni
del gruppo, dando origine cos\`\i \  ad insiemi di immagini.
Questo argomento suggerisce la struttura (geometrica) dei termini di rottura
nel canale trasverso: le $D5_1$ brane occupano una coppia di tori fissi, 
lasciando vuoti tutti gli altri. Quindi, come richiesto dalla consistenza 
dell'ampiezza di anello nel canale diretto, nel canale trasverso i termini
di rottura devono avere la struttura  seguente:
\be
\frac{2}{4} \ ( G \mp 4 G_1 )^2 + \frac{14}{4} G^2 \ ,
\ee
e questo determina le proiezioni in eq. (\ref{AZ2perZ2shift}).
Mediante la procedura standard, l'ampiezza di M\"obius nel canale trasverso
viene determinata da $\tld{\mc{K}}_0$ e da
\beq
\tld{\mc{A}}_0 & = & \frac{2^{-5}}{8} \Big[ \ \Big( N \   \sqrt{v_1 v_2 v_3} + D_1 \ \sqrt{\frac{v_1}{v_2 v_3}} \Big)^2 \ (\t_{oo} + \t_{og}) \nonumber \\
               & + &  \Big( N \  \sqrt{v_1 v_2 v_3} - D_1 \
\sqrt{\frac{v_1}{v_2 v_3}} \Big)^2 \  (\t_{of} + \t_{oh}) \ \Big]  ,
\label{Atld0p23}
\eeq
ed il risultato \`e
\beq
\tld{\mc{M}} & = & \frac{2}{8} \Big[ \ \hat{T}_{oo} ( N  v_1 v_2 v_3 W_1^e 
W_2^e W_3^e + D_1 \frac{v_1}{v_2 v_3} W_1^e P_2^e P_3^e )
              +  \hat{T}_{og} D_1  W_1^e v_1 
\Big( \frac{2 \hat{\eta}}{\hat{\vth}_2} \Big)^2  \nonumber \\
             & + & \hat{T}_{og} N  W_1^e v_1 
\Big( \frac{2 \hat{\eta}}{\hat{\vth}_2} \Big)^2 + N \hat{T}_{of} W_2^o v_2 
\Big( \frac{2 \hat{\eta}}{\hat{\vth}_2} \Big)^2 - N \hat{T}_{oh} W_3^o v_3 
\Big( \frac{2 \hat{\eta}}{\hat{\vth}_2} \Big)^2  \ \Big] \ .
\eeq
Infine, applicando una trasformazione $P$ all'espressione precedente, si
ottiene
\beq
\mc{M} & = & - \frac{1}{8} \Big[ \ \hat{T}_{oo} ( N  P_1 P_2 P_3 + D_1 P_1 W_2
 W_3 ) -  \hat{T}_{og} ( N + D_1 ) P_1 \Big( \frac{2 \hat{\eta}}{\hat{\vth}_2} 
\Big)^2  \nonumber \\
       & - & N \hat{T}_{of} (-)^{m_2} P_2  
\Big( \frac{2 \hat{\eta}}{\hat{\vth}_2} \Big)^2 + N \hat{T}_{oh} (-)^{m_3} P_3 
\Big( \frac{2 \hat{\eta}}{\hat{\vth}_2} \Big)^2  \ \Big] \ .
\eeq

Le condizioni di $tadpole$ \emph{untwisted} fissano
\be
N = D_1 = 32 \ ,
\ee
mentre quelle \emph{twisted} danno
\be
G = G_1 = 0 \ ,
\ee
possono essere identicamente soddisfatte parametrizzando le cariche come segue:
\beq
N   & = &  o + g + \bar{o} + \bar{g} \ , \quad  
G   \ = \  i \ (o + g + \bar{o} + \bar{g} ) \ ,  \nonumber \\
H   & = &  o - g + \bar{o} - \bar{g} \ , \quad  
F   \ = \  i \ (o - g - \bar{o} + \bar{g} ) \ ,  \nonumber \\
D_1 & = &  2 \ d + 2 \ \bar{d} \ , \qquad
G_1 \ = \  i \  ( d - \bar{d} ) \ ,
\eeq
ed a patto che $ o + g = 16 $ e $ d = 8 $.

I modi a massa nulla del settore aperto sono descritti da
\beq
\mc{A}_0 + \mc{M}_0 & = &  \t_{oo} ( o \bar{o} + g \bar{g} + d \bar{d} ) + 
\t_{og} ( o \bar{g} + g \bar{o} + d \bar{d} ) + \t_{oh}  ( o g + \bar{o} 
\bar{g} ) \nonumber \\ 
                    & + & ( \t_{oh} + \t_{of} ) \Big[ \frac{d (d-1)}{2} + 
\frac{\bar{d} (\bar{d}-1)}{2}  \Big]  \nonumber \\
                    & + &  \t_{of} \Big[  \frac{o (o-1)}{2} + \frac{\bar{o} 
(\bar{o}-1)}{2} +\frac{g (g-1)}{2} + \frac{\bar{g} (\bar{g}-1)}{2}  \Big] 
\nonumber \\
                    & + & ( \t_{gh} + \t_{gf} ) (o d + g d + \bar{o} \bar{d} +
 \bar{g} \bar{d} ) \ .
\label{spettrop23}
\eeq    
I gruppi di gauge compatibili con le condizioni di $tadpole$ sono quindi 
$ [ U(o) \times U(g) ]_9 \times U(8)_{5_1}$, mentre
lo spettro a massa nulla \`e non chirale, poich\'e  accanto ad  ogni 
rappresentazione compare anche la sua coniugata, ed include multipletti chirali
99 nelle rappresentazioni $ (\frac{o (o-1)}{2},1,1), 
(1,\frac{g (g-1)}{2},1)$, $(o, \bar{g}, 1),$ $(o, g, 1)$, una coppia di 
ipermultipletti 55 nella rappresentazione (1, 1, $\frac{d (d-1)}{2}$) ed
ipermultipletti 59 nelle rappresentazioni $(o, 1, d)$ e $ (1, g, d)$. 
Si noti che lo spettro a massa nulla delle D5-brane presenta il fenomeno di 
{\it brane supersymmetry}, ovvero \`e composto da multipletti di $N=2$ globale.

\section{Modelli magnetizzati in D = 4}
In questo paragrafo verr\`a descritta la costruzione di due nuovi modelli 
deformati magneticamente in quattro dimensioni, ottenuti seguendo la 
procedura utilizzata in sei dimensioni. Questa sar\`a dapprima 
applicata al mo\-dello $ Z_2 \times Z_2 $ convenzionale senza 
torsione discreta, in modo da chiarire, in schemi pi\`u complessi, il 
meccanismo di magnetizzazione. In secondo luogo verr\`a applicata al modello
$p_2 p_3$ con $shifts$ negli impulsi descritto nel paragrafo precedente.
Ricordando che i campi magnetici sui tori interni non alte\-rano
le ampiezze di vuoto di stringa chiusa $\mc{T}$ e $\mc{K}$, possiamo 
consi\-derare direttamente l'ampiezza di anello di eq. (\ref{AZ2perZ2}),
ponendo $D_1=0$. L'introduzione di due campi magnetici interni $H_2$ ed 
$H_3$ sul secondo e terzo toro, conduce quindi alla seguente 
ampiezza di anello:
\beq
\mc{A} & = & \frac{1}{2} \Big[ ( n^2 P_1 P_2 P_3 + 2 m \bar{m} P_1
\tld{P}_2 \tld{P}_3 + d_2^2 W_1 P_2 W_3 + d_3^2 W_1 W_2 P_3 ) T_{oo}(0;0;0)
\nonumber \\
       & + &  2  \ d_2 d_3 W_1 T_{go}(0;0;0) \ \Big( \frac{ \eta}{\vth_4(0)}
 \Big)^2 + 2 n d_2 P_2 T_{fo}(0;0;0) \ \Big( \frac{ \eta}{\vth_4(0)}
 \Big)^2 T_{go}(0;0;0) \ \Big( \frac{ \eta}{\vth_4(0)}
 \Big)^2 +
\nonumber \\
       & + & 2 n d_3 P_3 T_{ho}(0;0;0) \ \Big( \frac{ \eta}{\vth_4(0)}
 \Big)^2 
\nonumber \\
       & - & 2 n [ m T_{oo}(0;z_2 \t;z_3 \t) + \bar{m} 
T_{oo}(0;- z_2 \t;-z_3 \t) ]
P_1 \frac{k_2 \eta}{\vth_1(z_2 \t)} \frac{k_3 \eta}{\vth_1(z_3 \t)}
\nonumber \\
       & - & 2 i  d_2 [ m T_{fo}(0;z_2 \t;z_3 \t) + \bar{m} 
T_{fo}(0;-z_2 \t;-z_3 \t) ]
\frac{\eta}{\vth_4(0)}  \frac{k_2 \eta}{\vth_1(z_2 \t)} 
\frac{\eta}{\vth_4(z_3 \t)}
\nonumber \\
       & - & 2 i  d_3 [ m T_{ho}(0;z_2 \t;z_3 \t) + \bar{m} 
T_{ho}(0;-z_2 \t;-z_3 \t) ]
\frac{\eta}{\vth_4(0)}  \frac{\eta}{\vth_4(z_2 \t)}  
\frac{k_3\eta}{\vth_1(z_3 \t)}
       \nonumber \\
       & - & [ m^2 T_{oo}(0;2 z_2 \t;2 z_3 \t) + \bar{m}^2 
T_{oo}(0;- 2 z_2 \t;- 2 z_3 \t) ]
P_1 \frac{2 k_2 \eta}{\vth_1(2 z_2 \t)} \frac{2 k_3 \eta}{\vth_1(2 z_3 \t)} 
\Big] ,
\nonumber \\
\eeq
dove le molteplicit\`a di Chan-Paton $N, D_1$ e $D_2$ 
sono state riscalate per un fattore 2  (come nella (\ref{openspectrumZ2Z2})),
e dove $n \ra n + m + \bar{m}$.

Con procedura analoga si pu\`o scrivere la deformazione magnetica 
dell'am\-piezza di M\"obius di eq. (\ref{MZ2perZ2})
\beq 
\mc{M} & = & - \frac{1}{4} \Big[ ( n P_1 P_2 P_3 + d_2 W_1 P_2 W_3 + d_3 W_1
W_2 P_3 ) \hat{T}_{oo}(0;0;0) \nonumber \\
       & - & [ n P_1 + ( d_2 + d_3 ) W_1 ] \hat{T}_{og} 
\Big( \frac{2 \hat{\eta}}{\hat{\vth}_2(0)} \Big)^2 \nonumber \\
       & - & [ ( n + d_2 ) P_2 + d_2  W_2 ] \hat{T}_{of} 
\Big( \frac{2 \hat{\eta}}{\hat{\vth}_2(0)} \Big)^2 \nonumber \\
       & - & [ ( n + d_3 ) P_3 + d_3  W_3 ] \hat{T}_{oh} 
\Big( \frac{2 \hat{\eta}}{\hat{\vth}_2(0)} \Big)^2 
\nonumber \\
       & - & [ m \hat{T}_{oo}(0;2 z_2 \t;2 z_3 \t) + 
\bar{m} \hat{T}_{oo}(0;-2 z_2 \t;-2 z_3 \t) ] P_1 
\frac{2 k_2 \hat{\eta}}{\hat{\vth}_1(2 z_2 \t)} 
\frac{2 k_3 \hat{\eta}}{\hat{\vth}_1(2 z_3 \t)} 
\nonumber \\
       & - & [ m \hat{T}_{og}(0;2 z_2 \t;2 z_3 \t) + 
\bar{m} \hat{T}_{og}(0;-2 z_2 \t;-2 z_3 \t) ] P_1
\frac{2 \hat{\eta}}{\hat{\vth}_2(2 z_2 \t)}  
\frac{2 \hat{\eta}}{\hat{\vth}_2(2 z_3 \t)} 
\nonumber \\
       & - & i \   m \hat{T}_{of}(0;2z_2 \t;2z_3 \t)
\frac{2 \hat{\eta}}{\hat{\vth}_2(0)}  
\frac{2 k_2 \hat{\eta}}{\hat{\vth}_1(2z_2 \t)} 
\frac{2 \hat{\eta}}{\hat{\vth}_2(2z_3 \t)} 
\nonumber \\
       & - & i \ \bar{m} \hat{T}_{of}(0;-2z_2 \t;- z_3 \t)  
\frac{2 \hat{\eta}}{\hat{\vth}_2(0)} 
\frac{2 k_2 \hat{\eta}}{\hat{\vth}_1(2z_2 \t)} 
\frac{2 \hat{\eta}}{\hat{\vth}_2(2z_3 \t)} 
\nonumber \\
       & - & i \  m \hat{T}_{oh}(0;2z_2 \t;2z_3 \t) 
\frac{2 \hat{\eta}}{\hat{\vth}_2(0)}
\frac{2 \hat{\eta}}{\hat{\vth}_2(2z_2 \t)}  
\frac{2 k_3 \hat{\eta}}{\hat{\vth}_1(2z_3 \t)} 
\nonumber \\
      & - & i \ \bar{m} \hat{T}_{oh}(0;-2z_2 \t;-2z_3 \t) 
\frac{2 \hat{\eta}}{\hat{\vth}_2(0)}
\frac{2 \hat{\eta}}{\hat{\vth}_2(2z_2 \t)}  
\frac{2 k_3 \hat{\eta}}{\hat{\vth}_1(2z_3 \t)}  \ \Big]  \  .
\eeq
Si noti che $\mc{A}$  non contiene termini di proiezione,
mentre $\mc{M}$ contiene quattro contributi, corrispondenti alle proiezioni
$o$, $g$, $f$ ed $h$. Questi danno luogo ad un numero netto di fermioni 
chirali nello spettro a quattro dimensioni, come si pu\`o vedere dai
contributi a massa nulla
\beq
\mc{A}_0 + \mc{M}_0 & = & \t_{oo} \  \Big( \frac{n (n+1)}{2} + 
\frac{d_2 (d_2+1)}{2} + \frac{d_3 (d_3+1)}{2} + m \bar{m} \Big)
\nonumber \\
                   & + &  \t_{og} \  
\Big[ \ 3 \  \Big( \frac{n (n-1)}{2} + 
\frac{d_2 (d_2-1)}{2} + \frac{d_3 (d_3-1)}{2} + m \bar{m} \Big)
\nonumber \\
                   & + & n ( d_2 + d_3 ) + d_2 d_3  + | k_2 k_3 | n ( m + \bar{m} )
 \nonumber \\
                   & + & 2 \  | k_2 k_3 | \Big( \frac{m (m-1)}{2} + 
\frac{\bar{m} (\bar{m}-1)}{2} \Big) \ \Big]
\nonumber \\
                   & + & \t_{gf} \  \Big[ \  \frac{m (m-1)}{2}  
( | k_2 k_3 | + 1 - | k_2 | + | k_3 | ) \nonumber \\
                   & + & \frac{\bar{m} (\bar{m}-1)}{2} 
( | k_2 k_3 | + 1 + | k_2 | - | k_3 | ) \nonumber \\
                   & + &  \frac{m (m+1)}{2}  ( | k_2 k_3 | - 1 + | k_2 | - | k_3 | ) 
\nonumber \\
                   & + & \frac{\bar{m} (\bar{m}+1)}{2} 
( | k_2 k_3 | - 1 - | k_2 | + | k_3 | ) 
\nonumber \\
                  & + & \bar{m} d_2 | k_2 | + m d_3 | k_3 | \ \Big] \nonumber \\
                  & + & \t_{gh} \ \Big[ \  \frac{m (m-1)}{2}  
( | k_2 k_3 | + 1 - | k_2 | + | k_3 | ) \nonumber \\
                  & + & \frac{\bar{m} (\bar{m}-1)}{2} 
( | k_2 k_3 | + 1 + | k_2 | - | k_3 | ) \nonumber \\
                  & + &  \frac{m (m+1)}{2}  
( | k_2 k_3 | - 1 - | k_2 | + | k_3 | ) \nonumber \\
                  & + & \frac{\bar{m} (\bar{m}+1)}{2} 
( | k_2 k_3 | - 1 + | k_2 | - | k_3 | ) \nonumber \\
                  & + & m d_2 | k_2 | + \bar{m} d_3 | k_3 | \  \Big] \ .
\eeq

In realt\`a le condizioni di $tadpole$ restringono la carica totale e,
per ristabilire la chiralit\'a con gruppi non abeliani occorre introdurre
delle coppie brana-antibrana.

Passiamo ora alla trattazione del modello $p_2 p_3$. In questo caso poniamo 
nell'ampiezza di anello di eq. (\ref{AZ2perZ2shift}) $D_1=0=G_1$, e scindiamo 
le cariche $N, G, H$ ed $F$ delle D9-brane nella maniera seguente:
\beq
& N & \ra   (o + m) + (\bar{o} + \bar{m}) + (g + \bar{g}) 
= N + (m + \bar{m}) \ , \nonumber \\
& G & \ra i (o + m) - i (\bar{o} + \bar{m}) + i (g - \bar{g}) 
= i G + i (m - \bar{m}) \ , \nonumber \\
& H & \ra (o + m) + (\bar{o}+\bar{m}) - (g + \bar{g}) 
= H + (m+\bar{m}) \ , \nonumber \\
& F &\ra i (o + m) - i (\bar{o}+\bar{m}) - i (g - \bar{g}) 
= i F +  i (m - \bar{m}) \ .
\eeq
Questo equivale all'introduzione di due campi magnetici interni, $H_2$ ed 
$H_3$, sul secondo e terzo toro, allineati lungo lo stesso
 sottogruppo $U(1)$ di $U(o)$.

L'ampiezza di anello e di striscia di M\"obius che ne risultano nel canale 
diretto sono rispettivamente
\beq
\mc{A} & = & \frac{1}{8} \Big[ \  N^2 T_{oo}(0;0;0) P_1 P_2 P_3 + 2 m \bar{m}
T_{oo}(0;0;0) P_1 \tld{P}_2  \tld{P}_3 \nonumber \\
       & - & 2 N [ m T_{oo}(0;z_2\t;z_3\t) + \bar{m} T_{oo}(0;-z_2\t;-z_3\t) ]
P_1 \frac{k_2 \eta}{\vth_1(z_2\t)} \frac{k_3 \eta}{\vth_1(z_3\t)} 
\nonumber \\
       & - &  [ m^2  T_{oo}(0;2z_2\t;2z_3\t) + \bar{m}^2  T_{oo}(0;-2z_2\t;-2z_3\t) ]
P_1 \frac{2 k_2 \eta}{\vth_1(2 z_2\t)} \frac{2 k_3 \eta}{\vth_1(2 z_3\t)} 
\nonumber \\
       & + & (- G^2 + 2 m \bar{m}) P_1 T_{og}(0;0;0) 
\Big(\frac{2 \eta}{\vth_2(0)}\Big)^2 \nonumber \\
       & - & 2 G [ m T_{og}(0;z_2\t;z_3\t) - \bar{m} T_{og}(0;-z_2\t;-z_3\t) ]
P_1 \frac{2 \eta}{\vth_2(z_2\t)} \frac{2 \eta}{\vth_2(z_3\t)} 
\nonumber \\
       & - &  [ m^2  T_{og}(0;2z_2\t;2z_3\t) + \bar{m}^2  T_{og}(0;-2z_2\t;-2z_3\t) ]
P_1 \frac{2 \eta}{\vth_2(2 z_2\t)} \frac{2 \eta}{\vth_2(2 z_3\t)} 
\nonumber \\
       & + & (- F^2 (-)^{m_2} P_2 + 2 m \bar{m} (-)^{m_2} \tld{P}_2) T_{of}(0;0;0) 
\Big(\frac{2 \eta}{\vth_2(0)}\Big)^2 \nonumber \\
       & + & (H^2 (-)^{m_3} P_3 + 2 m \bar{m} (-)^{m_3} \tld{P}_3) T_{oh}(0;0;0) 
\Big(\frac{2 \eta}{\vth_2(0)}\Big)^2 \ \Big]   \ ,
\eeq
\beq
\mc{M} & = &  - \frac{1}{8} \Big[ \  N \hat{T}_{oo}(0;0;0)  P_1 P_2 P_3 \nonumber \\
       & - & [ m \hat{T}_{oo}(0;2z_2\t;2z_3\t) + 
\bar{m} \hat{T}_{oo}(0;-2z_2\t;-2z_3\t) ] P_1 
\frac{2 k_2 \hat{\eta}}{\hat{\vth}_1(2 z_2\t)} 
\frac{2 k_3 \hat{\eta}}{\hat{\vth}_1(2 z_3\t)}  \nonumber \\
      & - &  [ m \hat{T}_{og}(0;2z_2\t;2z_3\t) + 
\bar{m} \hat{T}_{og}(0;-2z_2\t;-2z_3\t) ] P_1 
\frac{2 \hat{\eta}}{\hat{\vth}_2(2 z_2\t)} \frac{2 \hat{\eta}}{\hat{\vth}_2(2 z_3\t)} 
\nonumber \\
     & - & N [ \hat{T}_{of}(0;0;0) (-)^{m_2} P_2 - \hat{T}_{oh}(0;0;0) (-)^{m_3} P_3 ]
\Big(\frac{2 \hat{\eta}}{\hat{\vth}_2(0)}\Big)^2  \Big]  \ .
\eeq 
Si osservi che sia in $\mc{A}$ che in $\mc{M}$ i termini di rottura legati 
alle 
proiezioni $F$ ed $H$ sono solo quelli scarichi ($\propto F^2$, $H^2$), e 
quelli globalmente neutri ($\propto 2 m \bar{m}$). I termini con la stessa 
struttura ma con carica $\pm1$ e $\pm2$, non contribuiscono in quanto
contengono termini di rottura, che sarebbero incompatibili con la struttura di
$\tld{\mc{A}}$.
L'assenza di tali proiezioni, oltre a rendere lo spettro non chirale,
stabilendo un parallelo con il modello non magnetizzato,
mette in luce un aspetto fisico particolarmente interessante: i doppietti di
D5-brane 
sono sostituiti da orbite di Landau.
Ogni orbita, descritta dagli estremi carichi delle stringhe, \`e sempre
accompagnata da una sua immagine, e la loro sovrapposizione coerente
da' origine alla generica orbita. Questo dimezza il conteggio degli stati,
consistentemente con l'assenza delle proiezioni, ma da' luogo ad uno spettro
con molteplicit\`a intere poich\'e, come vedremo, i $k_i$  risultano essere 
pari.

Dall'analisi dello spettro aperto a massa nulla, che pu\`o essere letto dalle
seguenti espressioni:
\beq
\mc{A}_0 & = &  \frac{1}{8} \Big[ N^2 T_{oo}(0;0;0) - G^2 T_{og}(0;0;0) - F^2 
T_{of}(0;0;0) + H^2  T_{oh}(0;0;0) \nonumber \\
         & + & 2 m \bar{m} (  T_{oo}(0;0;0) + T_{og}(0;0;0) + T_{of}(0;0;0) +
T_{oh}(0;0;0)  ) \nonumber  \\
         & - & 2 N [  m T_{oo}(0;z_2\t;z_3\t) + \bar{m} 
T_{oo}(0;-z_2\t;-z_3\t) ]
 \frac{k_2 \eta}{\vth_1(z_2\t)} \frac{k_3 \eta}{\vth_1(z_3\t)} 
\nonumber \\
       & - &  [ m^2  T_{oo}(0;2z_2\t;2z_3\t) + \bar{m}^2  T_{oo}(0;-2z_2\t;-2z_3\t) ]
\frac{2 k_2 \eta}{\vth_1(2 z_2\t)} \frac{2 k_3 \eta}{\vth_1(2 z_3\t)} 
\nonumber \\
       & - & 2 G [ m T_{og}(0;z_2\t;z_3\t) - \bar{m} T_{og}(0;-z_2\t;-z_3\t) ]
P_1 \frac{2 \eta}{\vth_2(z_2\t)} \frac{2 \eta}{\vth_2(z_3\t)} 
\nonumber \\
       & - &  [ m^2  T_{og}(0;2z_2\t;2z_3\t) + \bar{m}^2  T_{og}(0;-2z_2\t;-2z_3\t) ]
P_1 \frac{2 \eta}{\vth_2(2 z_2\t)} \frac{2 \eta}{\vth_2(2 z_3\t)} \Big] \ ,
\nonumber    \\      
\eeq
\beq
\mc{M}_0 & = & - \frac{1}{8} \Big[ \  N [ \hat{T}_{oo}(0;0;0) - 
\hat{T}_{og}(0;0;0) -
\hat{T}_{of}(0;0;0) + \hat{T}_{oh}(0;0;0)  ]   \nonumber \\
       & - & [ m \hat{T}_{oo}(0;2z_2\t;2z_3\t) + 
\bar{m} \hat{T}_{oo}(0;-2z_2\t;-2z_3\t) ] 
\frac{2 k_2 \hat{\eta}}{\hat{\vth}_1(2 z_2\t)} 
\frac{2 k_3 \hat{\eta}}{\hat{\vth}_1(2 z_3\t)}  \nonumber \\
      & - &  [ m \hat{T}_{og}(0;2z_2\t;2z_3\t) + 
\bar{m} \hat{T}_{og}(0;-2z_2\t;-2z_3\t) ] 
\frac{2 \hat{\eta}}{\hat{\vth}_2(2 z_2\t)} 
\frac{2 \hat{\eta}}{\hat{\vth}_2(2 z_3\t)} 
 \Big]  \ ,
\nonumber \\
\eeq
si deduce che per questo modello il punto istantonico, nel quale viene 
mantenuta la supersimmetria, \`e $H_2=-H_3$ ($z_2=-z_3$). 
Lo spettro aperto dei modi non massivi \`e quindi descritto da 
\beq
\mc{A}_0 + \mc{M}_0  & = & \t_{oo} (o \bar{o} + g \bar{g} + m \bar{m}) 
                        +  \t_{og} \Big[ \
(o g + \bar{o} \bar{g} + o \bar{g} + \bar{o} g) \nonumber \\
                     & + & \frac{o(o+1)}{2} + \frac{\bar{o}(\bar{o}+1)}{2} 
+ \frac{g(g+1)}{2} + \frac{\bar{g}(\bar{g}+1)}{2} \nonumber \\
                      & + & \frac{|k_2||k_3|}{4} (o + g + \bar{o} 
+ \bar{g})(m + \bar{m}) + (o + g - \bar{o} - \bar{g})(m - \bar{m}) \nonumber \\
                       & + &  ( |k_2| |k_3| + 1 ) [\frac{m(m+1)}{2} + 
\frac{\bar{m}(\bar{m}+1)}{2}] \   \Big] \ .
\eeq 
Da questa espressione si pu\`o desumere che il gruppo di gauge \`e 
$U(o) \times U(g) \times U(m)$, e che $k_2$ e $k_3$ devono soddisfare la
condizione di consistenza $|k_2| |k_3| \geq 4$ affinch\`e la relazione tra spin
e statistica non venga violata. Inoltre $k_2$ e $k_3$ devono essere entrambi 
pari. La spiegazione di questa condizione \`e piuttosto sottile.
Un parametro di quantizzazione legato al volume del quadri-toro $T^4$, 
raddoppierebbe sull'$orbifold$, il quale ha la met\`a del volume, ma questo sarebbe
 ristretto a quattro-forme. Nel caso in esame invece, sono presenti solo due-forme, che 
vedono in maniera indipendente i due tori nei loro flussi. 
 Il fatto che $k_2$ e $k_3$ debbano
essere entrambi pari, \`e consistente con il fatto che la magnetizzazione di questo 
modello  mima i doppietti di $D5_1$-brane presenti nel modello 
originale, in termini di doppietti di orbite di Landau.
Come nel caso a sei dimensioni, i $tadpole$ di R-R del settore $twisted$, che
derivano per esempio da $S_2 C_2 O_2 O_2$
\be
[ \ G + (m + \bar{m}) \ ]^2 \ ,
\ee
sono consistenti con il fatto che
i doppietti di $D5_1$-brane vengono sostituiti dalle D9-brane magnetizzate
(o ``ruotate''), le quali  invadono l'intero primo toro interno.

Ricordiamo che le condizioni di {\it tadpole} si deducono dalle
ampiezze di vuoto nel canale trasverso 
$\tld{K}_0$, $\tld{A}_0$ ed $\tld{M}_0$,
che per il modello in considerazione hanno la forma
\beq
\tld{A}_0 & = & \frac{2^{-5}}{8} \ \Big[ \ [ N^2 + 2 m \bar{m} (1+q^2 H_2^2)
(1+q^2 H_3^2) ] v_1 v_2 v_3 T_{oo}(0;0;0) \nonumber \\
            & + & 4 \ [ 2 N m T_{oo}(0;z_2;z_3) + 2 N \bar{m} T_{oo}(0;-z_2;-z_3)
 ] v_1 \frac{k_2 \eta}{\vth_1(z_2)} \frac{k_3 \eta}{\vth_1(z_3)} \nonumber \\
            & + & 4 \ [  m^2 T_{oo}(0;2z_2;2z_3) + \bar{m}^2 
T_{oo}(0;-2z_2;-2z_3) ] v_1 \frac{2 k_2 \eta}{\vth_1(2 z_2)} 
\frac{2 k_3 \eta}{\vth_1(2 z_3)} \nonumber \\
            & + & 16 \Big[ \ (-G^2 + 2 m \bar{m}) T_{go}(0;0;0) v_1 
\Big( \frac{\eta}{\vth_4(0)} \Big)^2  \\
            & + &  [ - 2 G m T_{go}(0;z_2;z_3) + 2 G \bar{m} 
T_{go}(0;-z_2;-z_3) ] v_1 \frac{\eta}{\vth_4(z_2)} \frac{\eta}{\vth_4(z_3)}
 \nonumber \\
            & - &  [  m^2 T_{go}(0;2z_2;2z_3) + \bar{m}^2 
T_{go}(0;-2z_2;-2z_3) ] v_1 \frac{\eta}{\vth_4(2 z_2)} 
\frac{\eta}{\vth_4(2 z_3)} \ \Big] \ \Big] \ , \nonumber 
\eeq
\beq
\tld{M}_0 & = & - \frac{2}{8} \ \Big[ \ N \hat{T}_{oo}(0;0;0) v_1 v_2 v_3
+ N \hat{T}_{og}(0;0;0) v_1 \Big( \frac{2 \hat{\eta}}{\hat{\vth}_2(0)} \Big)^2
\nonumber \\
          & + &  [ \ m \hat{T}_{oo}(0;z_2;z_3) +  \bar{m} 
\hat{T}_{oo}(0;-z_2;-z_3) \ ] v_1 \frac{2 k_2 \hat{\eta}}{\hat{\vth}_1(z_2)} 
\frac{2 k_3 \hat{\eta}}{\hat{\vth}_1(z_3)} \nonumber \\
          & + &  [ \ m \hat{T}_{og}(0;z_2;z_3) +  \bar{m} 
\hat{T}_{og}(0;-z_2;-z_3) \ ] v_1 \frac{2 \hat{\eta}}{\hat{\vth}_2(z_2)} 
\frac{2 \hat{\eta}}{\hat{\vth}_2(z_3)}  \ \Big] \ . \nonumber \\
\eeq
Come nel caso a sei dimensioni, il $tadpole$  NS-NS $untwisted$, legato a
$V_2 O_2 O_2 O_2$ ($tadpole$ del dilatone)
\beq
\Big[ \ [ N & + & (m + \bar{m}) \sqrt{(1 + q^2 H_2^2) (1 + q^2 H_3^2)} \ ] 
\sqrt{v_1 v_2 v_3}  \nonumber \\ 
            & - & 32 (\sqrt{v_1 v_2 v_3} + \sqrt{\frac{v_1}{v_2 v_3}}) \ \Big]^2 \ ,
\label{diltadshift}
\eeq
pu\`o essere messo in relazione all'azione di Born-Infeld. Inoltre, ponendo
$H_2=-H_3$ e sfruttando la relazione di quantizzazione di Dirac, la 
(\ref{diltadshift}) diventa identica all'espressione per il tadpole di R-R
$untwisted$ che deriva da $S_2 S_2 S_2 S_2$
\beq
\Big[ \ [ N & + & (m + \bar{m}) + i q (H_2 + H_3)(m + \bar{m}) -32 ] 
\sqrt{v_1 v_2 v_3}  \nonumber \\ 
           & - & [ k_2 k_3 (m + \bar{m}) + 32 ] \ \Big]^2 = 0 \ ,
\label{tadimpor}
\eeq
e quindi si annulla identicamente. Le condizioni di {\it tadpole} risultanti sono
\beq
(o + g) + m & = & 16 \ , \nonumber \\
k_2 k_3 m   & = & 16 \ ,
\label{RRtad}
\eeq
dove \`e stata utilizzata la condizione di $tadpole$ $twisted$ $G=0$ e 
l'identificazione delle molteplicit\`a coniugate ($m, \bar{m}$).

Fissando l'attenzione sul caso $k_2=k_3=2$, che corrisponde alla scelta 
minimale per la degenerazione dei livelli di Landau di questo $orbifold$ 
$Z_2 \times Z_2$, le condizioni di $tadpole$ (\ref{RRtad}) fissano
$m=4$. Pertanto lo spettro aperto \`e descritto dal gruppo di gauge
$U(o) \times U(g) \times U(4)$, con multipletti chirali in due copie 
delle rappresentazioni $(o,1,4), (\bar{o},1,\bar{4})$ e $(1,g,4+\bar{4})$, 
ed in una copia delle rappresentazioni $(o,g+\bar{g},1), 
(\frac{o(o+1)}{2},1,1), (1,\frac{g(g+1)}{2},1)$ e delle loro coniugate.

Lo spettro risulta quindi non chirale, anche se caratterizzato da un 
pi\`u ampio numero di multipletti di materia rispetto al caso non 
deformato. Si perde inoltre il fenomeno di \emph{brane supersymmetry},
in quanto i modi  a massa nulla hanno supersimmetria $N=1$ e non
pi\`u $N=2$ come descritto nel paragrafo precedente.

\end{document}